\documentstyle[preprint,aps]{revtex}
\font\Bbbfont=msbm10
\def\Bbb#1{\hbox{\Bbbfont#1}}
\font\frakfont=eufm10
\def\frak#1{\hbox{\frakfont#1}}

\begin{document}
\title{THE GEOMETRY OF DYNAMICAL TRIANGULATIONS}

\author{J. Ambj\o rn\cite{nbi}}
\address{The Niels Bohr Institute, Blegdamsvej 17,
DK-2100 Copenhagen \O , Denmark}
\author{M. Carfora\cite{dar}}
\address{S.I.S.S.A.-I.S.A.S.,
Via Beirut 2-4, 34013 Trieste, Italy, and\\
Istituto Nazionale di Fisica Nucleare, Sezione di
         Pavia, via A. Bassi 6, I-27100 Pavia, Italy}
\author{A. Marzuoli\cite{mac}}
\address{Dipartimento di Fisica Nucleare e Teorica, Universit\`a degli
\nobreak Studi di Pavia, via A. Bassi 6, I-27100 Pavia, Italy, and\\
Istituto Nazionale di Fisica Nucleare, Sezione di
         Pavia, via A. Bassi 6, I-27100 Pavia, Italy}
\maketitle
\begin{abstract}
The express purpose
of these Lecture Notes is to go through some aspects of the simplicial
quantum
gravity model known as  the Dynamical Triangulations
approach. Emphasis has been on lying the foundations of the theory and
on illustrating its subtle and often unexplored connections with many
distinct mathematical fields ranging from global riemannian geometry,
moduli theory, number theory, and topology. Our exposition will
concentrate on these points so that graduate students may find in
these notes a
useful exposition of some of the rigorous results one can establish in
this field and hopefully a source of inspiration for new exciting
problems. We
also  illustrate the deep and beautiful interplay between the
analytical aspects of dynamical triangulations and the results of
MonteCarlo simulations. The techniques described here are rather novel
and  allow us to address successfully many high points of great
current interest in the subject of simplicial quantum gravity while
requiring very little in the way of fancy field theoretical
arguments. As a consequence, these notes contains mostly original and
yet
unpublished material of great potential interest both to the expert
practitioner  and to graduate students entering in the field.
Among the topic addressed here in considerable details there are:
{\it (i)} An analytical discussion of the geometry of dynamical
triangulations in
dimension $n=3$ and $n=4$; {\it (ii)} A constructive characterization
of entropy estimates for
dynamical triangulations in dimension $n=3$, $n=4$,  and
a comparision of the analytical results we obtain
with tha data coming from Monte Carlo simulations for the $3$-sphere
${\Bbb S}^3$ and the $4$-sphere ${\Bbb S}^4$; {\it (iii)}
A proof that in the four-dimensional
case  the
analytical data  and the numerical data provide the same critical line
$k_4(k_2)$ characterizing the infinite-volume limit of simplicial
quantum gravity; {\it (iv)} An analytical characterization of the
critical point $k_2^{crit}$ which is in an extremely good agreement
with the location of the critical point obtained by MonteCarlo
simulations; {\it (v)} A proof that corresponding to such a critical
point, $4$-dimensional (simply connected) manifolds undergo a higher
order phase transition possibly concealed by a bistable behavior:
this  settles down in the affirmative a long standing debate as to the
nature of the phase transition in $4$-dimensional simplicial quantum
gravity; {\it (vi)} A geometrical  analysis of the branched polymer
phase of $4D$-gravity; {\it (vii)} We also show that in the three-
dimensional case the confrontation between the analytical and
numerical data is very satisfactory; {\it (viii)} A proof that in
dimension $n=3$ there are pronounced hysteresis effects which are
possibly responsible for the observed first-order  nature of the phase
transition at criticality.
\end{abstract}
\pacs{0460N}
\narrowtext
\vfill\eject
\tableofcontents
\vfill\eject
\section{Introduction}

Recent years witnessed a massive introduction of methods of
statistical field theory in attempts to quantize gravity along the
lines of a conventional field theory\cite{Houches},\cite{Dav}. Such an
emphasis is the direct outspring of the possibility of discretizing
gravity in a way consistent with the underlying reparametrization
invariance of general relativity\cite{Williams},\cite{Dav}. The basic
idea is to use a variant of the standard Regge calculus\cite{Regge},
first suggested by Weingarten\cite{Weingarten}, in an attempt to make
quantum gravity well defined as a lattice statistical field theory.
This discretization is known as the theory of Dynamical
Triangulations (DT) or simplicial quantum gravity. It consists of a
replacement of the (Euclidean) functional integral over all
equivalence classes of metrics with a summation over all abstract
triangulations of the given manifold. The fixed edge-length of the
links of the triangulations plays here the role of {\it lattice
spacing} of the more familiar lattice gauge theories. The basic idea
is to search, in the parameter space of these discretized quantum
gravity models, for critical points where the {\it lattice spacing}
can be taken to zero and contact can be made to continuum physics.
Credit to this picture is lend by noticing that
 within this framework, Euclidean two-dimensional quantum gravity can
be formulated as the scaling limit of a dynamically triangulated
collection of surfaces thought of as an ordinary statistical system.
Contact with the continuum, as described by
Liouville theory\cite{Distler}, can successfully  be made, and the
robust alliance
of the theory of critical phenomena and Monte Carlo simulations even
sheds light on many aspects of the theory not liable of an analytical
approach.
\vskip 0.5 cm
If we extend the theory to higher dimensions we have to face the fact
that we have no continuum theory of Euclidean Quantum Gravity with
which we can compare, nonetheless many aspects of the theory can be
investigated by computer-assisted simulations with a good degree of
precision. These numerical methods allow us to get a rather complete
picture of the phase diagram of the discretized theory and they serve
as an important inspiration for analytical studies of higher
dimensional models of quantum gravity. These lecture notes attempts to
partially fill the gap between such numerical studies and a fully-
fledged analytical approach to higher dimensional dynamical
triangulations. Our emphasis here is on the applications of elementary
ideas of Piecewise-Linear (PL) geometry, global riemannian geometry,
and number theory in order to understand  dynamically triangulated
models of quantum gravity in dimension $n=3,4$. The occasion for such
an analytical approach comes by exploiting the techniques developed
for the recent proof\cite{Carfora} of the conjecture concerning the
existence of an exponential bound
to the number of distinct triangulations on a PL-manifold $M^n$,
($n\geq 3$), of given volume and topology. This exponential bound is
germane to the definition of dynamical triangulations as a lattice
statistical field theory model. Analytical proofs are well known in
two dimensions\cite{Pari},
and computer simulations supported very strongly the existence of such
bound also in higher dimensions\cite{Houches}. The actual proof for
higher dimensions draws from techniques of controlled topology and
geometry initiated by Cheeger, Gromov, Grove, and
Petersen\cite{Gromov},\cite{Grov}, and it suggests that such
mathematical methods may bear relevance to the whole program of
dynamical triangulations. A check with  Monte Carlo simulations (in
dimension $n=4$) shows an excellent agreement between the numerical
data and the analytical results
obtained by pursuing this geometrical approach to higher dimensional
dynamically triangulated models. It is not yet possible to reach the
level of sophistication of the two-dimensional case, but as we show in
these notes, the analytical methods we develop in dimensions $n=3$
and $n=4$, provide a good start for an analytical understanding of
dynamical triangulations in such dimensions. In particular, we
characterize with greater precision the entropy
estimates\cite{Carfora}, and  provide the corresponding
generating functions. Among other properties common to both the $3$-
dimensional and $4$-dimensional case, we prove that in $4$-
dimensions dynamically triangulated models it makes sense to sum over
all simply connected manifolds. We also discuss the existence of a
polymer phase in these models\cite{Jurk}.  We analytically prove that
the system admits a critical point $k_2=k_2^{crit}$ corrisponding to
which a dynamically triangulated $4$-manifold undergoes a higher order
phase transition possibly concealed by a bystable behavior. This
is  an important result suggesting that dynamical
triangulations provide a reliable candidate to a consistent quantum
gravity theory, and shows that the importance of
the interplay between analytical approaches and Monte Carlo
simulations continues unabated.
\vskip 0.5 cm
The arguments presented here should be amenable to considerable
improvements, and hopefully these notes can be used as a working tool.
For this reason, we have presented a detailed summary of the more
important and useful results in section {\bf III}. In section {\bf II}
 we review
the definition of the model. In section {\bf IV} we recall few
elementary
aspects of the PL-geometry of dynamical triangulations.
In section {\bf V} we
establish the connection with
Gromov's spaces of bounded geometry. This connection provides
mathematical foundation to the
heuristic arguments which motivate the approximation of distinct
riemannian
structures with distinct dynamical triangulations.
In section {\bf VI} we introduce the {\it Moduli space} associated
with
dynamical triangulations. These moduli spaces parametrize the set of
inequivalent
{\it deformations} of dynamical triangulations needed to approximate
riemannian structures with large symmetries. In sections {\bf VII}
and {\bf VIII}
we provide all entropy estimates needed for the higher dimensional DT
models. Here we exploit elementary number theory in order to provide
such estimates and construct the associated generating functions.
We discuss also  the $2$-dimensional case which fully agrees
with the known results.  The $4$-dimensional
case, and  the
$3$-dimensional case are discussed in section {\bf IX}. The agreement
between the analytical results we obtain and the Monte Carlo
simulations
is impressive in dimension $n=4$. We discuss in detail  an analytical
proof of
the existence of the phase transition
whose presence is strongly suggested by the computer simulations.
As recalled we prove that definitively it is not a first order
transition as hinted by some recent simulations.\par
Finally, we show that also in dimension $n=3$ the analysis of
the data coming from the two approaches is quite satisfactory.
In particular, we prove that the correct asymptotics for the canonical
partition function has a bistable behavior yielding for the hysteresis
effects which are believed to signal a first order nature of the
transition point.\par

\section{The model: Simplicial Quantum Gravity}

Let $M$ be an n-dimensional, ($n\geq 2$), manifold of given topology,
and with a finite number of fixed $(n-1)$-dimensional boundaries
$\Sigma_k$, $k=1,2,\ldots$.
Let $Riem(M)$ and $Diff(M)$ respectively denote the space of
riemannian  metrics $g$ on $M$, and the group of
diffeomorphisms on $M$. In the continuum formulation of quantum
gravity the task is to perform a path integral over equivalence
classes of metrics:
\vskip 0.5 cm
\begin{eqnarray}
Z(\Lambda, G,\Sigma_k,h)=\sum_{Top(M)}\int_{Riem(M)/Diff(M)}{\cal
D}[g(M)]e^{-S_{g}[\Lambda,G,\Sigma]}
\label{uno}
\end{eqnarray}
\vskip 0.5 cm
\noindent weighted with the Einstein-Hilbert action associated with
the riemannian manifold $(M,g)$, {\it viz.},
\vskip 0.5 cm
\begin{eqnarray}
S_g[\Lambda,G,\Sigma]=
\Lambda\int_M d^n\xi\sqrt{g}-\frac{1}{\sqrt{16\pi{G}}}\int_{M}
d^n\xi\sqrt{g}\, R +\, boundary\, terms
\end{eqnarray}
\vskip 0.5 cm
\noindent
The boundary terms depend on the metric $h$ and on the extrinsic
curvature induced on the boundaries $\Sigma_k$, and they are such that
for the
action obtained by gluing any two manifolds
$(M_1,g_1)$ and $(M_2,g_2)$ along a common boundary
$(\Sigma, h)$, we get
$S_{g_1+g_2}[\Lambda,G]=S_{g_1}[\Lambda,G,\Sigma]+
S_{g_2}[\Lambda,G,\Sigma]$. In this way, the partition function
(\ref{uno}) associated with a manifold $M$ with two boundaries
$(\Sigma_1,h_1)$ and $(\Sigma_2,h_2)$,
satisfies the basic composition law
\begin{eqnarray}
Z(\Sigma_1,h_1;\Sigma_2,h_2)=\sum_{\Sigma_i,h_i}
Z(\Sigma_1,h_1;\Sigma_i,h_i)Z(\Sigma_i,h_i;\Sigma_2,h_2)
\end{eqnarray}
which describes how $M$ can interpolate between its fixed boundaries
$(\Sigma_1,h_1)$ and $(\Sigma_2,h_2)$ by summing over all possible
intermediate states $(\Sigma_i,h_i)$.
\vskip 0.5 cm
Even without pretense of rigor, it is clear that the formal {\it a
priori}
measure ${\cal D}[g(M)]$
characterizing such sort of path integration should satisfy some basic
properties. In particular:\par
\noindent {\it (i)} ${\cal D}[g(M)]$ should be defined on the space of
riemannian
structures $Riem(M)/Diff(M)$ so to avoid counting as distinct any two
riemannian metrics $g_1(M)$ and $g_2(M)$
which differ one from the other simply by the action of a
diffeomorphism
of $\phi\colon{M}\to{M}$, {\it viz.}, such that
$g_2(M)=\phi^*g_1(M)$;\par
\noindent {\it (ii)} The measure ${\cal D}[g(M)]$ should play a
kinematical role and not a dynamical one. More explicitly,
for a given $r\in{\Bbb R}^+$, let
$[Riem(M)/Diff(M)]_{x_1,x_2;r}$ denote the set of all riemannian
structures
on a manifold $M$ with two marked points $x_1$ and $x_2$ with
preassigned distance $d_{g(M)}(x_1,x_2)=r$. When restricted to this
set
of
riemannian structures, ${\cal D}[g(M)]$
should factorize: fluctuations in the geometry of $M$
which are localized in widely separated regions,  should be
statistically
independent. This implies that  ${\cal D}[g(M)]$ should not describe
the {\it a priori} existence of long range correlations on the set of
riemannian
manifolds considered. Such correlations should be generated by the
spectrum of fluctuations of (\ref{uno})
by means of a statistical
suppression-enhancement mechanism similar to the {\it energy} versus
{\it entropy} argument familiar in statistical mechanics. In a field
theoretic formalism, this property is translated  in the familiar
requirement of ultra-locality, namely in the absence of derivatives in
the formal $L^2$ norm on the space of metrics, (the {\it De
Witt
supermetric}). \par
\noindent {\it (iii)} Finally,  ${\cal D}[g(M)]$ should be so
constructed
as to allow for the introduction of a $Diff(M)$-invariant short
distance cut-off
representing the shortest wavelength allowed in discussing
fluctuations of the
geometry of $M$. This cut-off should be removable under appropriate
circumstances,
in particular when, in a critical phase, long range correlations are
generated.
\vskip 0.5 cm
\noindent {\bf Summing over topologies}.
Long before addressing
the physical and mathematical characterizations of such requirements
on
${\cal D}[g(M)]$, both in the continuum
field-theoretic formalism
and in its possible discretized versions, (in this connection see the
recent work of Menotti and Peirano\cite{Houches}),
one must also
discuss the proper meaning to attribute to the formal sum
over
topologies appearing in (\ref{uno}):
are we summing over homotopy types,
homeomorphism types, or  over smooth
types of manifolds?

The homotopy type of a manifold is a topological
notion which (with the exception of dimension two) is too weak to
appear of some immediate utility.
The homeomorphism type of a manifold $M$ is the notion which is
more directly related to a natural summation over topologies, however
it
is difficult to handle in dimension larger than two, (see
below). Finally, the smooth type yields for a summation over the
possible distinct differentiable structures that a manifold $M$ can
carry. This latter summation appears more relevant to a field-
theoretic
formalism since in an expression such as (\ref{uno}) the
diffeomorphism
group of the underlying manifold $Diff(M)$ plays a basic role, but
as we shall see momentarily, a summation over the smooth types arises
more naturally in simplicial gravity. Thus,
this latter interpretation for $\sum_{Top(M)}$ is perhaps the most
appealing since it suggests an unexpected bearing of simplicial
quantum
gravity on the field-theoretic formalism, and in this sense the two
approaches are not alternative to each other. Simplicial quantum
gravity
may be necessary in addressing the important issue of summation over
distinct topologies, whereas the field-theoretic formalism is
necessary
in order to understand the nature of the continuum limit (if any
exists)
of the discretized models.

In dimension two there is equivalence between smooth structures and
the homeomorphism types, and the summation over topologies can be
given
the unambigous meaning of a summation over the Euler characteristic
$\chi(M)$ of the manifolds, since this invariant characterizes
surfaces.
However, even in such a simple case it is very difficult to obtain
clear-cut results for what concerns a reliable procedure for summing
over topologies. The use of matrix models of
$2D$-gravity and the associated
{\it double scaling} limit have shed some light on this issue, but we
are
still far from a clear understanding\cite{Houches}.\par

In dimension three topological manifolds are uniquely
smoothable\cite{Moise}, and the question of summing over topologies is
again
synonimous of summation over smooth structures. Thus, as long as we
confine our attention to the topological category or to the smooth
category, summation over topologies is  reduced to the yet unsolved
problem of enumerating the homeomorphism types of three-manifolds.
This enumeration cannot be realized as long as the Poincar\'e
conjecture is not proved. For instance, if there were a fake three-
sphere then  one could prove\cite{Ferry} that there cannot be finitely
many
homeomorphism types of three-manifolds even under
bounds on curvatures of the manifold, (typically, under curvature,
volume and diameter bounds one gets topological finiteness theorems
for homeomorphism and diffeomorphism types\cite{Grov} for
dimension $n\not= 3,4$).
\vskip 0.5 cm
In dimension four the situation is even more complex since smoothing
theory is not yet completely known. In open contrast to the theory of
$3$-manifolds, there is no equivalence between
topological manifolds and smooth structures, and Donaldson-Freedman's
theory shows that there are topological manifolds which are not
smoothable as well as manifolds admitting uncountably many
inequivalent smooth structures\cite{Freed}. The situation is further
worsened
by the fact that the topological classification  of all
(compact, orientable) $4$-manifolds is logically impossible, (again,
the fundamental group should be blamed for this). As stressed by
Fr\"{o}hlich\cite{Frohlich}, this circumstance has even be used to
foster the
credence
that only simply-connected, spinable $4$-manifolds should contribute
to the gravitational path integral. But clearly this point of view, if
not better
substained, cannot be advocated. Suppression of a class of manifolds
from path integration can be justified only on a dynamical ground, as
a form of statistical suppression driven by the spectrum of
fluctuations
of the theory.
\vskip 0.5 cm
This situation does not improve even if we confine our attention to a
manifold $M$ of fixed topology and with a given smooth structure, for,
in that case we have the additional problems associated with: {\it
(i)} The unboundedness of the Euclidean action; {\it (ii)} The
mathematical difficulties in defining a proper path integration over
the stratified manifold of riemannian structures; {\it (iii)} The
Einstein-Hilbert action is not renormalizable.
\vskip 0.5 cm
\noindent {\bf Simplicial quantum gravity}. The hope behind simplicial
quantum gravity is that some of the above
problems, concerning both the characterization of the measure
${\cal D}[g(M)]$ or the issues related to $\Sigma_{Top(M)}$,
can be properly addressed, in a non-perturbative setting, by
approximating the path integration over inequivalent riemannian
structures with a summation over combinatorially equivalent piecewise
linear manifolds.
The first attempt of using PL geometry in relativity dates back to the
pioneering work of Regge\cite{Regge}. His proposal was to approximate
Riemannian
(Lorentzian) structures by PL-manifolds in such a way as to obtain a
coordinate-free formulation of general relativity. The basic
observation in this approach
is that parallel transport and the (integrated) scalar
curvature have natural counterparts on PL manifolds once one gives
consistently the lengths of the links of the triangulation defining
the PL structure. The link length is the dynamical variable in
Regge calculus, and classically the PL version of the Einstein
field equations is obtained by fixing a suitable triangulation
and by varying the length of the links so as to find the
extremum of the Regge action. If the original triangulation is
sufficiently fine, this procedure consistently provides a good
approximation to the smooth spacetime manifold which is the
corresponding smooth solution of the Einstein equations. This
approach can also be successfully extended
so as to provide a quantum analogue to Regge
calculus\cite{Williams}, in which we replace the formal path
integration over the space of Riemannian structures with an
integration over the link variables.
\vskip 0.5 cm
Dynamical triangulations are a variant of Regge calculus in the sense
that in this formulation the
summation
over the length of the links is replaced by a direct summation over
abstract triangulations where the length of the links is fixed to a
given value $a$. In this way the elementary simplices of the
triangulation provide a {\it Diff-invariant} cut-off and each
triangulation is a representative of a whole equivalence class of
metrics. Regge calculus still mantain its validity and it provides
both the metric assignement for the PL manifold, obtained by gluing
the
simplices, and the corresponding action. Since all simplices now are
identical, the action will only depend on the numbers, respectively
$N_n$ and $N_{n-2}$, of $n$- and $(n-2)$-dimensional simplices of the
$n$-dimensional PL manifold. In this way we get that the Einstein-
Hilbert action for
$n$-dimensional (Euclidean) gravity
formally goes into the combinatorial action
\begin{eqnarray}
S[k_{n-2},k_n]=k_nN_n(T)-k_{n-2}N_{n-2}(T)+\frac{1}{2}k_{n-2}N_{n-
2}(\partial{T})
\label{four}
\end{eqnarray}
where $\frac{1}{2}N_{n-2}(\partial{T})$ is the boundary term,
and $k_n$, and $k_{n-2}$ are (bare) coupling constants related
to the cosmological constant $\Lambda$ and to the gravitational
coupling $G$, respectively. In particular, we can view $1/k_{n-2}$ as
a bare gravitational coupling constant.\par
The partition function associated with such discretized action is
\begin{eqnarray}
Z[k_{n-2},k_n]=\sum_{T\in {\cal T}}
e^{-k_nN_n+k_{n-2}N_{n-2}-\frac{1}{2}k_{n-2}N_{n-2}(\partial{T})}
\label{five}
\end{eqnarray}
where the summation is over {\it distinct} triangulations $T$ in a
suitable class of triangulations ${\cal T}$. Roughly speaking, we
consider any two triangulations (say with the some number of vertices)
distinct if there is no map between the vertices which is compatible
with the assignements of links, triangles, etc., while ${\cal T}$
restricts the class of triangulations to those
satisfying suitable topological constraints. The proper choice of the
class ${\cal T}$ is strictly connected to the difficult question of
how to sum over topologies in quantum gravity mentioned above.

The interpretation of
(\ref{five}) as providing also a sum over topologies is based on the
observation that a PL manifold is uniquely smoothable in low dimension
(in particular for $n=4$, by Cerf theorem, see {\it
e.g.},\cite{Freed}), (it is also worth stressing that every smooth
manifold admits a natural $C^1$-smooth PL-structure). In particular,
in dimension $n=4$ there is a bijective correspondence between
(isotopy) classes of smooth structures and PL-structures. Thus
(\ref{five}), if no restrictions are imposed
on ${\cal T}$,
subsumes in a rather natural
way a summation
over topologies if we decide that $\sum_{Top}$ actually
means summation over smooth structures. Even if not ambiguous in its
definition, (\ref{five}) blends summation over metric structures and
summation over smooth structures only in a formal way. Already in
dimension two, the sum (\ref{five}) is divergent since if the
topology is not fixed, the number of distinct triangulations grows
factorially with the volume of the manifold ({\it i.e.}, with the
number of simplices). However, if one fixes the topology, the number
of distinct triangulations of a $2$-dimensional PL manifold is
exponentially bounded\cite{Pari} according to
\begin{eqnarray}
({\Lambda})^{N_2}
{N_2}^{(\frac{{\chi}(\Sigma)}{2})({\gamma}_{str}-
2)-1}
\label{eight}
\end{eqnarray}
where  $\Lambda$ is a  suitable constant and ${\gamma}_{str}$,
the {\it string exponent}, is a  topological suscettivity
generated by the quantum fluctuations of the metric.
In this way, $\sum_{T\in\cal{T}}e^{-S(T)}$, where $\cal{T}$ is a given
PL surface, is well defined. Topology is then allowed to fluctuate
({\it i.e.}, one attempts to  extend the summation to all PL
surfaces), by a delicate limiting procedure ({\it Double Scaling
Limit}). In higher dimension is not yet known how to perform the
summation (\ref{five}), but  numerical simulation as well as the
results of the two-dimensional theory, suggests that (\ref{five})
makes sense if one fixes the topology.
This issue is clearly related to a rigorous characterization of the
discretized counterpart of the formal a priori measure
${\cal D}[g(M)]$. As a matter of fact, a {\it
necessary condition} for attributing a meaning to (\ref{five}), {\it
for fixed topology}, is to require that the number of distinct
triangulations of a given PL manifolds is exponentially bounded as a
function of the number of (top-dimensional) simplices\cite{Houches}.
Recently\cite{Carfora}, the existence of the exponential bound in all
dimensions and for all (fixed) topologies has been proved. The
existence of the exponential bound implies that for a fixed $k_{n-2}$
there is a critical lower value $k^{crit}_{n}(k_{n-2})$ of $k_n$ such
that the partition function  is well defined for
$k_n>k_n^{crit}(k_{n-2})$ and divergent  for $k_n<k_n^{crit}(k_{n-
2})$. To be more specific this implies that we can introduce the
(canonical) partition function for fixed (lattice) volume: {\it The
effective
Entropy}
\vskip 0.5 cm
\begin{eqnarray}
W(N,k_{n-2})_{eff}\doteq\sum_{T\in{\cal T}(N_n)}e^{k_{n-2}N_{n-2}(T)}
\label{doppiow}
\end{eqnarray}
\vskip 0.5 cm
\noindent where ${\cal T}(N_n)$ denotes the class of distinct
triangulations of
fixed volume ($N_n$), fixed topology and boundary conditions. This
effective entropy will characterize the
{\it infinite (lattice) volume} limit of the theory, defined by the
approach to the critical line $k_n\to k_n^{crit}(k_{n-2})$ in the
$(k_{n-2},k_n)$ coupling constant plane. The existence of the infinite
volume limit is a necessary {\it but not sufficient} condition for the
existence of a physically significant {\it continuum limit} of the
theory, and we only expect interesting critical behavior, {\it i.e.},
the onset of long range correlations, at certain
critical values of $k_{n-2}$. In the rest of these notes we will be
mainly interested in the analytical characterization of the
canonical partition function (the effective
entropy) $W(N,k_{n-2})_{eff}$, and in comparing the obtained results
with the existing numerical data coming from Monte Carlo simulations.
\vskip 0.5 cm
It is not difficult to see the origin of the difficulties in dealing
with the set of dynamically triangulated manifolds (of given volume
and topology) considered as a statistical system described by
(\ref{doppiow}). Roughly speaking we have a collection of identical
simplices whose only interaction is basically associated with
unpenetrability and gluings according to certain rules. The energetic
term is very simple and controls the average curvature. Thus
the free energy of the system is basically characterized by
the entropic factor enumerating the number of distinct dynamical
triangulations of given topology, volume, and average curvature.
Such an estimate is deeply non-local, ({\it e.g.}, see (\ref{eight})
for $n=2$),
and as such its characterization is conceptually different from the
standard approaches used to enumerate distinct configurations, say of
spins, on a rigid lattice. This situation is not new in statistical
mechanics since it is reminiscent of what happens
when dealing with the {\it hard sphere gas}: a set of identical
spherical particles whose only interaction is associated with
unpenetrability. Here too, the free energy is of entropic origin, and
the characterization of such an entropy is a highly non local problem
since it is related to the characterization of the densest packings in
spheres: the insertion of a sphere may change the packing up to a
distance proportional to the inverse of the average separation between
the surfaces of the spheres.
For dynamical triangulations, as we shall see, the source of non-
locality is subtler  since there is a delicate feedback with the
topology, ({\it e.g.}, see (\ref{eight}), where the critical exponent
$\gamma_s$ depends from the Euler characteristic of the surface
obtained by gluing the simplices). But these topological difficulties
would be present also in the hard sphere gas had we considered the
spheres evolving in a topological non-trivial ambient space. Thus, it
is perhaps not so surprising to note that the first proof of the
existence of an exponential bound for the entropy of dynamical
triangulations was carried out\cite{Carfora}  by exploiting the
geometry of sphere packings in riemannian manifolds. This interplay
between the geometry of packings, topology, and the metric properties
of riemannian manifolds, (or more general metric spaces), can be
traced back to M. Gromov's ideas. These ideas have sprung a
renaissance in recent developments in Riemannian geometry, and we will
exploit them here by showing that they are a good source of
inspiration also in simplicial quantum gravity and in the fascinating
field of the statistical mechanics of extended objects.

\section{Summary of Results}

For the convenience of the reader we present here a summary of the
constructive results that we prove in these notes and  that can be
useful in applications to higher dimensional dynamical triangulations.
If not otherwise stated, we refer to dynamically triangulated
manifolds, $M$, in dimension $n\geq 2$. The dynamical triangulations
in question have $N_{n-2}(T)\doteq\lambda+1$ bones $\sigma^{n-2}$,
$N_n(T){\doteq}N$
top-dimensional simplices $\sigma^n$, a given average number,
$b(n,n-2)\doteq\frac{1}{2}n(n+1)(N/\lambda+1)$, of $n$-simplices
incident on a bone, and a minimum number $\hat{q}$ of $n$-dimensional
simplices incident on a bone, (typically $\hat{q}=3$).
Sometimes we need to mark $n$ of these bones and a top-dimensional
simplex $\sigma^n_0$ incident on one of them. We refer collectively to
triangulations with such
markings as {\bf rooted} triangulations.
The topology of the triangulations explicitly enters our results
through the {\it representation variety} $\frac{Hom(\pi_1(M),G)}{G}$:
the set of conjugacy classes of representations of the fundamental
group of the manifold, $\pi_1(M)$ into a (compact) Lie group $G$. This
representation variety parametrizes the set of inequivalent
deformations of dynamical triangulations needed to approximate
riemannian structures endowed with $G$-structures, (for instance, this
parameter space tells us how to deform a $3$-dimensional dynamical
triangulation so as to approximate unambiguosly inequivalent flat $3$-
tori). Finally, in what follows we also introduce a normalizing factor
$0<c_n\leq1$, of the form
\begin{eqnarray}
c_n\doteq C_0 N_n^{\tau(n)}\cdot e^{\alpha_n N_n}
\end{eqnarray}
where the constants $0<C_0\leq 1$, $\tau(n)\geq0$, $\alpha_n\leq0$,
depend only on the dimension $n$. These normalizing constants are
necessary since our enumeration procedure (in establishing entropy
estimates) overcounts the number of distinct dynamical triangulations
admitted by a manifold of given topology. In dimension $n=2$, $c_n$
reduces to a constant, whereas in dimension $n=4$, $\tau(n=4)=5/2$ and
$\alpha_{n=4}=-\frac{1}{2}\ln(\cos^{-1}(1/4))$ and similarly in
dimension $n=3$. With
these notational remarks along the way, we have:
\vskip 1 cm
\noindent{\bf The generating function}.
{\it Let $0\leq t\leq 1$ be a generic indeterminate, and let
$p_{\lambda}(h)$ denote the number of partitions of the generic
integer $h$ into (at most) $\lambda+n-1$ parts, each $\leq (b(n,n-2)-
\hat{q})(\lambda+1)$.
In a given holonomy represention
$\Theta\colon\pi_1(M;\sigma^n_0)\to G$, and for a given value of the
parameter $b=b(n,n-2)$,
the generating function for the number of distinct rooted dynamical
triangulations with $N_{n-2}(T_a^{(i)})=\lambda+1$
bones and given number ($\propto{h}$) of $n$-dimensional simplices
incident on the $n$ marked bones is given by
\begin{eqnarray}
{\cal G}[W(\Theta,\lambda,b;t)]=c_n\cdot\sum_{h\geq
0}p_{\lambda}(h)t^h=c_n\cdot
\left[ \begin{array}{c}
(b(n,n-2)-\hat{q})(\lambda+1)+(\lambda+1)-n\\ (\lambda+1)-n
\end{array}  \right]
\end{eqnarray}
where $0<c_n\leq 1$ is the above-mentioned normalizing factor, and
\begin{eqnarray}
\left[ \begin{array}{c}
n\\ m
\end{array}  \right]=\frac{(1-t^n)(1-t^{n-1})\ldots(1-t^{m+1})}{(1-
t^{n-m})(1-t^{n-m-1})\ldots(1-t)}\doteq\frac{(t)_n}{(t)_{n-m}(t)_m}
\end{eqnarray}
is the Gauss polynomial in the variable $t$.}
\vskip 1 cm

\noindent {\bf The Rooted Entropy}.
{\it In a given  represention
$\Theta\colon\pi_1(M;\sigma^n_0)\to G$, and for a given value of the
parameter $b=b(n,n-2)$,
the number $W(\Theta,\lambda,b)$  of distinct rooted dynamical
triangulations
with $N_{n-2}(T_a^{(i)})=\lambda+1$
bones, is given by}
\begin{eqnarray}
W(\Theta,\lambda,b)= c_n\cdot
\left( \begin{array}{c}
(b(n,n-2)-\hat{q})(\lambda+1)+(\lambda+1)-n\\ (\lambda+1)-n
\end{array}  \right)
\label{bounds}
\end{eqnarray}
\vskip 1 cm
\noindent{\bf Asymptotics for the Entropy}.
{\it The number of
distinct dynamical triangulations, with $\lambda+1$ bones,
and with an average number, $b\equiv b(n,n-2)$, of
$n$-simplices incident on a
bone, on an $n$-dimensional, ($n\leq 4$), PL-
manifold $M$ of given fundamental group $\pi_1(M)$, can be
asymptotically estimated
according to
\vskip 0.5 cm
\begin{eqnarray}
W(\lambda,b)\simeq\nonumber
\end{eqnarray}
\vskip 0.5 cm
\begin{eqnarray}
W_{\pi}\cdot
\frac{c_n}{\sqrt{2\pi}}
\sqrt{\frac{(b-\hat{q}+1)^{1-2n}}{(b-\hat{q})^3}}
{\left [ \frac{(b-\hat{q}+1)^{b-\hat{q}+1}}{(b-\hat{q})^{b-\hat{q}}}
\right ] }^{\lambda+1}
(\frac{b(n,n-2)}{n(n+1)} \lambda)^{D/2}
{\lambda}^{-\frac{2n+3}{2}}
\end{eqnarray}
\vskip 0.5 cm
\noindent where
$W_{\pi}$ is a topology dependent parameter, and $D{\doteq}
dim[Hom(\pi_1(M),G)]$.}

\vskip 1 cm

\noindent{\bf The Canonical Partition Function}.
{\it
Let us consider the set of all simply-connected
$n$-dimensional, ($n=3,4$), dynamically triangulated manifolds.
Let us set
\begin{eqnarray}
& & A(k_{n-2})\doteq\nonumber\\
& & { \left [\frac{27}{2}e^{k_{n-2}}+1+
\sqrt{(\frac{27}{2}e^{k_{n-2}}+1)^2-1} \right ]}^{1/3}+
{ \left [\frac{27}{2}e^{k_{n-2}}+1-
\sqrt{(\frac{27}{2}e^{k_{n-2}}+1)^2-1} \right ]}^{1/3}-1,
\end{eqnarray}
and let $k_{n-2}^{crit}$ denote the unique solution
of the equation
\begin{eqnarray}
\eta^*(k_{n-2})\doteq\frac{1}{3}(1-\frac{1}{A(k_{n-2})})=\eta_{max}
\end{eqnarray}
where $\eta_{max}=1/4$, (for $n=4$) and $\eta_{max}=2/9$, (for $n=3$).
Let $0<\epsilon<1$ small enough, then
for all values of  the inverse gravitational coupling $k_{n-2}$ such
that
\begin{eqnarray}
k^{crit}_{n-2}-\epsilon< k_{n-2}<+\infty,
\end{eqnarray}
the large $N$-behavior of the canonical partition function
(effective entropy) is given  by the uniform asymptotics
\vskip 0.5 cm
\begin{eqnarray}
W(N, k_{n-2})_{eff}\doteq\sum_{T\in{\cal
T}(N_n)}e^{k_{n-2}N_{n-2}(T)}=\nonumber
\end{eqnarray}
\begin{eqnarray}
\frac{c_n}{\sqrt{2\pi}}{\hat{N}}^{-n-1/2}
e^{(\hat{N}+1)f(\eta_{max})}
\left[\frac{\xi_0}{\sqrt{\hat{N}}}
w_0(\psi_{max}(k_{n-2})
\sqrt{\hat{N}})+\frac{\xi_1}{\hat{N}}
w_{-1}(\psi_{max}(k_{n-2})\sqrt{\hat{N}})
\right]
\end{eqnarray}
\vskip 0.5 cm
\noindent where $w_r(z)\doteq\Gamma(1-r)e^{z^2/4}D_{r-1}(z)$,
($r<1$), $D_{r-
1}(z)$ and $\Gamma(1-r)$ respectively denote the parabolic cylinder
functions and the Gamma function, and where the constants $\xi_0$
and
$\xi_1$ are given by}
\vskip 0.5 cm
\begin{eqnarray}
\xi_0\doteq -
\frac{\eta_{max}^{-1/2}(1-
2\eta_{max})^{-n}}{\sqrt{(1-3\eta_{max})(1-
2\eta_{max})}}\frac{\psi_{max}(k_{n-2})}{f_{\eta}(\eta_{max})}
\end{eqnarray}
\vskip 0.5 cm
\begin{eqnarray}
\xi_1\doteq \frac{\xi_0}{\psi_{max}(k_{n-2})}
-\frac{\eta*^{-1/2}(1-
2\eta^*)^{-n}}{\sqrt{(1-3\eta^*)(1-
2\eta^*)}} \left[
\frac{1}{\psi_{max}(k_{n-2})\sqrt{-
f_{\eta\eta}(\eta^*)}} \right]
\end{eqnarray}
\vskip 0.5 cm
\noindent Where, for notational convenience we have set
\begin{eqnarray}
f(\eta)\doteq
-\eta\log\eta+(1-2\eta)\log(1-2\eta)-
(1-3\eta)\log(1-3\eta)+ k_{n-2}\eta
\end{eqnarray}
with $f_{\eta}\doteq df/d\eta$, and $f_{\eta\eta}{\doteq}
d^2f/d\eta^2$,and
\begin{eqnarray}
\psi_{max}(k_{n-2})\doteq sgn(\eta^*-\eta_{max})
\sqrt{2[f(\eta^*)-f(\eta_{max})]}.
\end{eqnarray}
\vskip 1 cm
\noindent (Note that for the $3$-dimensional case the above expression
for $W(N,k_{n-2})$ is slightly more general if one uses the variables
$(N_0,N_3)$-see below for details).
\vskip 1 cm
\noindent {\bf The infinite volume limit}.
{\it The critical
value of the coupling $k^{crit}_n$ corresponding to the {\it
infinite-volume} limit for simply-connected $n$-dimensional
dynamically triangulated manifolds is given by
\vskip 0.5 cm
\begin{eqnarray}
k_n^{crit}(k_{n-2})=\frac{1}{2}n(n+1)\left[
\ln \frac{A(k_2)+2}{3}+\lim_{N_n\to\infty}\frac{\ln{c_n}}{N_n} \right]
\end{eqnarray}
\vskip 0.5 cm
\noindent which, for $0<\epsilon<1$ small enough, holds for all $k_{n-
2}^{crit}-\epsilon<k_{n-2}<k_{n-
2}^{crit}$.
Whereas, for $k^{crit}_{n-2}<k_{n-2}<+\infty$ we get}
\vskip 0.5 cm
\begin{eqnarray}
k_n^{crit}(k_{n-2})=\nonumber
\end{eqnarray}
\begin{eqnarray}
\frac{1}{2}n(n+1)
\left\{
\log\left[\frac{(1-
2\eta_{max})^{(1-2\eta_{max})}}{\eta_{max}^{\eta_{max}}(1-
3\eta_{max})^{(1-3\eta_{max})}} \right] +k_{n-2}\eta_{max}
+\lim_{N_n\to\infty}\frac{\ln{c_n}}{N_n}
\right\}
\end{eqnarray}
\vskip 0.5 cm
\vskip 1 cm
\noindent {\bf The four-dimensional case}.
{\it
The critical
value of the coupling ${\tilde K}^{crit}_4$ corresponding to the {\it
infinite-volume} limit for a simply-connected $4$-manifold, (e.g., the
$4$-sphere ${\Bbb S}^4$),
is given by
\vskip 0.5 cm
\begin{eqnarray}
{\tilde k}_4^{crit}(k_2)=10
\left\{
\log\left[\frac{(1-
2\eta_{max})^{(1-2\eta_{max})}}{\eta_{max}^{\eta_{max}}(1-
3\eta_{max})^{(1-3\eta_{max})}} \right] +k_2\eta_{max}
-\log\left( \frac{e}{2.066} \right)
\right\}
\end{eqnarray}
\vskip 0.5 cm
\noindent (with $k^{crit}_2<k_2<+\infty$), and
$\eta_{max}=1/4$, and
\vskip 0.5 cm
\begin{eqnarray}
k_2^{crit}\simeq 1.387
\end{eqnarray}
\vskip 0.5 cm
\noindent (note that this is a theoretical upper bound; for instance
for $\eta_{max}=1/4.1$ one would get $k_2^{crit}\simeq 1.2$), whereas
one obtains
\vskip 0.5 cm
\begin{eqnarray}
{\tilde k}^{crit}_4=10 \ln\frac{A(k_2)+2}{3e/2.066}
\end{eqnarray}
\vskip 0.5 cm
\noindent when $k_2^{crit}-\epsilon<k_2< k_2^{crit}$.
Corresponding to the critical value $k_2\to k_2^{crit}$ the system
undergoes a higher order phase transition.}
\vskip 1 cm

\noindent {\bf The three-dimensional case}.
{\it Let us consider the set of all simply-connected
$3$-dimensional  dynamically triangulated manifolds described
in terms of the variables $N_3$, $N_0$.
Let $k_{0}^{inf}$, and $k_0^{crit}$ respectively denote the unique
solutions
of the equations
\vskip 0.5 cm
\begin{eqnarray}
\frac{1}{3}(1-\frac{1}{A(k_{0})})+\frac{1}{6}=\eta_{min}=\frac{1}{6}
\end{eqnarray}
\begin{eqnarray}
\frac{1}{3}(1-\frac{1}{A(k_{0})})+\frac{1}{6}=\eta_{max}=\frac{2}{9}
\end{eqnarray}
\vskip 0.5 cm

\noindent Let $0<\epsilon<1$ small enough, then
for all values of  the inverse gravitational coupling $k_{0}$ such
that
\begin{eqnarray}
k^{inf}_{0}+\epsilon< k_{0}<k^{crit}_{0}-\epsilon,
\end{eqnarray}
the large $N$-behavior of the canonical partition function $W(N,
k_{0})_{eff}$ for $3$-dimensional simplicial quantum gravity on a
simply connected manifold
is given by
the uniform asymptotics
\vskip 0.5 cm
\begin{eqnarray}
\frac{c_n}{\sqrt{2\pi}}{\hat{N}}^{-7/2}
e^{(\hat{N}+1)f(\eta_{min})}
\left[\frac{\alpha_0}{\sqrt{\hat{N}}}
w_0(\psi_{min}(k_{0})
\sqrt{\hat{N}})+\frac{\alpha_1}{\hat{N}}
w_{-1}(\psi_{min}(k_{0})\sqrt{\hat{N}})
\right]+\nonumber
\end{eqnarray}
\begin{eqnarray}
+\frac{c_n}{\sqrt{2\pi}}{\hat{N}}^{-7/2}
e^{(\hat{N}+1)f(\eta_{max})}
\left[\frac{\xi_0}{\sqrt{\hat{N}}}
w_0(\psi_{max}(k_{0})
\sqrt{\hat{N}})+\frac{\xi_1}{\hat{N}}
w_{-1}(\psi_{max}(k_{0})\sqrt{\hat{N}})
\right]
\label{lunga}
\end{eqnarray}
\vskip 0.5 cm
\noindent  where the constants $\alpha_0$,
$\alpha_1$, and $\xi_0$, $\xi_1$ are given by
\vskip 0.5 cm
\begin{eqnarray}
\alpha_0\doteq -
\frac{\eta_{min}^{-1/2}(1-2\eta_{min})^{-n}}{\sqrt{(1-3\eta_{min})(1-
2\eta_{min})}}\frac{\psi_{min}(k_{0})}{f_{\eta}(\eta_{min})}
\end{eqnarray}
\vskip 0.5 cm
\begin{eqnarray}
\alpha_1\doteq \frac{\alpha_0}{\psi_{min}(k_{0})}
-\frac{\eta*^{-1/2}(1-2\eta^*)^{-n}}{\sqrt{(1-3\eta^*)(1-
2\eta^*)}} \left[
\frac{1}{\psi_{min}(k_{0})\sqrt{-
f_{\eta\eta}(\eta^*)}} \right]
\end{eqnarray}
\vskip 0.5 cm
\begin{eqnarray}
\xi_0\doteq -
\frac{\eta_{max}^{-1/2}(1-2\eta_{max})^{-n}}{\sqrt{(1-3\eta_{max})(1-
2\eta_{max})}}\frac{\psi_{max}(k_{0})}{f_{\eta}(\eta_{max})}
\end{eqnarray}
\vskip 0.5 cm
\begin{eqnarray}
\xi_1\doteq \frac{\xi_0}{\psi_{max}(k_{0})}
-\frac{\eta*^{-1/2}(1-2\eta^*)^{-n}}{\sqrt{(1-3\eta^*)(1-
2\eta^*)}} \left[
\frac{1}{\psi_{max}(k_{0})\sqrt{-
f_{\eta\eta}(\eta^*)}} \right]
\end{eqnarray}
\vskip 0.5 cm
\noindent The critical
value of the coupling ${k}^{crit}_3$ corresponding to the {\it
infinite-volume} limit for the $3$-sphere ${\Bbb S}^3$
is given by
\vskip 0.5 cm
\begin{eqnarray}
k_3^{crit}(k_0)=6\ln \frac{A(0.16k_0)+2}{3[\cos^{-1}(1/3)]^{1/5}}-
0.16k_0
\end{eqnarray}
\vskip 0.5 cm
\noindent and corresponding to the critical point
\vskip 0.5 cm
\begin{eqnarray}
k_0\to k_0^{crit}\simeq 3.845
\end{eqnarray}
\vskip 0.5 cm
\noindent the system described by (\ref{lunga}) exhibits a bistable
behavior leading to hysteresis effects.}
\vfill\eject

\section{Triangulations}

In this section we recall few basic notions of Piecewise-Linear (PL)
topology in order to provide a self-contained set up for the some of
the mathematical
problems we wish to discuss in simplicial quantum gravity. Together
with standard
material, (for which we refer freely to \cite{Rourke}), we also
collect few definitions and results which are not readily accessible
in textbooks.
\subsection{Preliminaries: simplicial manifolds and pseudo-manifolds}

We first list some well known, but necessary, preliminaries.\par
By an $n$-simplex $\sigma^n\equiv(x_0,\ldots,x_n)$ with vertices
$x_0,\ldots,x_n$ we mean the following subspace of $R^d$,
(with $d>n$),
\begin{eqnarray}
\sigma^n\doteq\sum_{i=0}^n\lambda_ix_i
\label{nine}
\end{eqnarray}
where $x_0,\ldots,x_n$ are $n+1$ points in general position in $R^d$,
and $\sum_{i=0}^n\lambda_i=1$ with $\lambda_i\geq{0}$.

A face of an $n$-simplex $\sigma^n$ is any simplex whose vertices are
a subset of those of $\sigma^n$, and
a simplicial complex $K$ is a finite collection of simplices in $R^d$
such that if $\sigma_1^n\in K$ then so are all of its faces, and if
$\sigma_1^n$, $\sigma_2^m\in K$ then $\sigma_1^n\cap\sigma_2^m$ is
either a face of $\sigma_1^n$ or is empty. The $h$-skeleton of $K$ is
the subcomplex $K_h\subset K$ consisting of all simplices of $K$ of
dimension $\leq{h}$.
\vskip 0.5 cm
Let $K$ be a (finite) simplicial complex. Consider the set theoretic
union
$|K|\subset R^d$ of all simplices from $K$
\begin{eqnarray}
|K|\doteq \cup_{\sigma\in K}\sigma
\label{ten}
\end{eqnarray}
Introduce on the set $|K|$ a topology that is the strongest  of all
topologies in which the embedding of each simplex into $|K|$ is
continuos, (the set $A\subset |K|$ is closed iff $A\cap\sigma^k$ is
closed in $\sigma^k$ for any $\sigma^k\in K$).
The space $|K|$ is the underlying {\it polyhedron}, geometric carrier
of the simplicial complex $K$, it provides the topological space
underlying the simplicial complex. The topology of $|K|$
can be more conveniently described in terms
of the {\it star of a simplex} $\sigma$,
$star(\sigma)$, the union of all simplices of which $\sigma$ is a
face.
The open subset of the
underlying polyhedron $|K|$ provided by the interior of the carrier of
the star of $\sigma$ is the {\it open star} of $\sigma$.
Notice that the open star is a subset of the
polyhedron $|K|$, while the star is a sub-collection of simplices in
the simplicial complex $K$. It is immediate to verify that the open
stars can be used to define the topology of $|K|$. The polyhedron
$|K|$ is said to be triangulated by the simplicial complex $K$. More
generally, a triangulation of a topological space $M$ is a simplicial
complex $K$ together with a homeomorphism $|K|\to M$.
\vskip 0.5 cm
\noindent {\bf Simplicial maps}. A simplicial map $f\colon K\to L$
between two
simplicial complexes $K$
and $L$ is a continuous map $f\colon |K|\to|L|$ between the
corresponding underlying polyhedrons which takes $n$-simplices to $n$-
simplices for all $n$, (piecewise straight-line segments are mapped to
piecewise straight-line segments). The map $f$ is a simplicial
isomorphism if
$f^{-1}\colon L\to K$ is also a simplicial map. Such maps preserve the
natural combinatorial structure of ${\Bbb R}^n$.
Note that a simplicial map is determined by its values on vertices. In
other words, if $f\colon K^0\to L^0$ carries the vertices of each
simplex of $K$ into some simplex of $L$, then $f$ is the restriction
of a unique simplicial map.\par
\vskip 0.5 cm
\noindent {\bf Combinatorial properties}. Sometimes, one refers to a
simplicial
complex $K$ as a simplicial
division of $|K|$. A
subdivision $K'$ of $K$ is a simplicial complex such
that $|K'|=|K|$ and each $n$-simplex of $K'$  is contained in an $n$-
simplex of $K$, for every $n$. A property of simplicial complex $K$
which is invariant under subdivision is a combinatorial property or
{\it Piecewise-Linear} (PL) property of
$K$, and
a Piecewise-Linear  homeomorphism $f\colon K\to L$ between two
simplicial complexes is a map which is a simplicial isomorphism for
some subdivisions $K'$ and $L'$ of $K$ and $L$.
\vskip 0.5 cm
\noindent {\bf Piecewise-Linear manifolds}. A PL manifold of dimension
$n$ is a
polyhedron $M=|K|$ each point of
which
has a neighborhood, in $M$, PL homeomorphic to an open set in $R^n$.
PL
manifolds are realized by simplicial manifolds under the equivalence
relation generated by PL homeomorphism. Any piecewise linear manifold
can be triangulated, however
notice that any particular triangulation of a piecewise linear
manifold is not well defined up to piecewise linear homeomorphism,
(because such homeomorphisms involve subdivisions of the given
triangulations).
A triangulated space can be characterized as a PL-manifold according
to the\cite{Rourke}
\newtheorem{Gaiat}{Theorem}
\begin{Gaiat}
A simplicial complex $K$ is a simplicial manifold of dimension
$n$ if for all $r$-simplices $\sigma^r\in K$, the link of $\sigma^r$,
$link(\sigma^r)$ has the topology of the boundary of the standard $(n-
r)$-simplex, {\it viz.} if $link(\sigma^r)\simeq{\Bbb S}^{n-r-1}$.
\end{Gaiat}
Recall that the link of a simplex $\sigma$ in a simplicial complex $K$
is the union of all faces $\sigma_f$ of all simplices in
$star(\sigma)$ satisfying $\sigma_f\cap\sigma=\emptyset$, also recall
that the {\it Cone} on the link $link(\sigma^r)$,
$C(link(\sigma^r))$, is the product $link(\sigma^r)\times[0,1]$ with
$link(\sigma^r)\times\{1\}$ identified to a point.
The above theorem follows\cite{Rourke}  by noticing that a point
in the interior of an $r$-simplex $\sigma^r$ has a neighborhood
homeomorphic to $B^r\times C(link(\sigma^r))$, where
$B^r$ denotes the ball in ${\Bbb R}^n$. Since
$link(\sigma^r)\simeq{\Bbb S}^{n-r-1}$, and $C({\Bbb S}^{n-r-
1})\simeq
B^{n-r}$, we get that $|K|$ is covered by neighborhoods homeomorphic
to $B^r\times B^{n-r}\simeq B^n$ and thus it is a manifold. Note that
the theorem holds whenever the links of vertices are $(n-1)$-spheres.
As long as the dimension $n\leq 4$, the converse of this theorem is
also true. But this is not the case in larger dimensions, and there
are examples of triangulated manifolds where the link of a simplex is
not a sphere. In general, necessary and sufficient conditions for
having a manifolds out of a simplicial complex require that the link
of each cell has the homology of a sphere, and that the link of every
vertex is simply-connected, (see {\it e.g.}, Thurston's
notes\cite{Rourke}). Since we are interested in dimension $2\leq n\leq
4$, we  can simply disregard such subtler
characterizations. However (see the next comments on gluings), this
issue is actually bypassed by the particular way
triangulations are generated in simplicial quantum gravity.
\vskip 0.5 cm

\noindent {\bf Gluings}.
In dynamical triangulations theory one is not given from the outset a
triangulation of a manifold, rather one generates an $n$-dimensional
PL-manifold by gluing a finite set of $n$-simplices $\{\sigma^n\}$. A
rather detailed analysis of such gluing procedures is given in
Thurston's notes\cite{Rourke}, (clearly for different purposes), and
here we simply recall the most relevant facts.
\vskip 0.5 cm
Given a finite set of simplices and the associated collection of
faces, a {\it gluing} is a choice of pairs of faces together with
simplicial identifications maps between faces  such that each face
appears in exactly one of the pairs. The identification space, $K$,
resulting from the quotient of the union of the simplices by the
equivalence relation generated by the identification maps, is
homeomorphic to the polyhedron of a simplicial complex. The gluing
maps are linear, and consequently the simplicial complex $K$ obtained
by gluing face-by-face $n$-simplices has the structure of a PL-
manifold in the complement of the $(n-2)$-skeleton. Since the link of
an $(n-2)$-simplex is a circle, it is not difficult to prove that the
PL-structure actually extend to the complement of the $(n-3)$-
skeleton, and that the identification space of a gluing among finite
$n$-simplices is a PL-manifold if and only if  the link of each cell
is PL-homeomorphic to the standard PL-sphere.
\vskip 0.5 cm
 \noindent {\bf Pseudo-manifolds}. It is important to stress that in
dimension
$n>2$, not every
simplicial
complex $M=|T|$ obtained by gluing simplices along faces is a
simplicial
manifold, and in general one speaks of {\it pseudo-manifolds}.
\newtheorem{Gaiad}{Definition}
\begin{Gaiad}
$M=|T|$ is an $n$-dimensional pseudomanifold if:
(i) every simplex of $T$ is either an $n$-simplex or a face of an $n$-
simplex; (ii) each $(n-1)$-simplex is a face of at most two $n$-
simplices; (iii) for any two simplices $\sigma^n$, $\tau^n$ of $T$,
there exists a finite sequence of $n$-simplices
$\sigma^n=\sigma_0^n,\sigma_1^n,\ldots,\sigma_j^n=\tau^n$ such that
$\sigma_i^n$ and $\sigma_{i+1}^n$ have an $(n-1)$-face in common,
({\it i.e.}, there is a simplicial path connecting $\sigma^n$ and
$\tau^n$).
\end{Gaiad}
Recall that a regular point $p$ of a polyhedron $|T|$ is a point
having a neighborhood in $|T|$ homeomorphic to an $n$-dimensional
simplex, otherwise $p$ is called a {\it singular point}. Absence of
singular points in a pseudo-manifolds characterizes triangulated
manifolds. Moreover, an $n$-dimensional polyhedron $|T|$ is a pseudo-
manifold if and only if the set of regular point in $|T|$ is dense and
connected and the set of all singular points is of dimension less than
$n-1$. Thus, in the applications to gravitational physics, the
occurrence of pseudo-manifolds is associated with the presence of
local irregularities that should not have a sensible effect in the
continuum limit of the theory, since the set of regular points has the
density properties mentioned above.
\vskip 0.5 cm
\noindent {\bf Dehn-Sommerville relations}. In
order to construct a simplicial manifold $T$ by gluing simplices
$\sigma^n$, through their $n-1$-dimensional faces, the
following  constraints must be satisfied
\begin{eqnarray}
\sum_{i=0}^n(-1)^iN_i(T)=\chi(T)
\label{eleven}
\end{eqnarray}
\begin{eqnarray}
\sum_{i=2k-1}^n(-1)^i\frac{(i+1)!}{(i-2k+2)!(2k-1)!}N_i(T)=0
\label{twelve}
\end{eqnarray}
\vskip 0.5 cm
\noindent if $n$ is even, and  $1\leq k\leq n/2$. Whereas if
$n$ is odd
\vskip 0.5 cm
\begin{eqnarray}
\sum_{i=2k}^n(-1)^i\frac{(i+1)!}{(i-2k+1)!2k!}N_i(T)=0
\label{thirteen}
\end{eqnarray}
\vskip 0.5 cm
\noindent with $1\leq k\leq (n-1)/2$.
These relations are known as the Dehn-Sommerville equations. The
first, (\ref{eleven}), is just the Euler-Poincar\'e equation for the
triangulation
$T$ of which $N_i(T)$ denotes the number of $i$-dimensional
simplices, the {\it f-vector} of the triangulation $T$. The conditions
(\ref{twelve}) or (\ref{thirteen}), are a
consequence of the fact that in a simplicial manifold, constructed by
gluings,
the link of
every $(2k-1)$-simplex (if $n$ is odd) or $2k$-simplex (if $n$ is
even), is an odd-dimensional sphere, and hence it has Euler number
zero.
\vskip 0.5 cm
Note also that in order to
generate an $n$-dimensional polyhedron there is a {\bf minimum number}
$\hat{q}(n)$ of simplices $\sigma^n$ that must join together at a
bone, (for instance, in dimension $2$, we have
$\hat{q}(2)=3$, otherwise we have no polyhedral surface).
In general, since we will usually consider only the class of regular
simplicial manifolds we assume that
the number, $q(\sigma^i)$, of $n$-dimensional simplices $\sigma^n$
which share the subsimplex $\sigma^i$ is such that
\begin{eqnarray}
q(\sigma^i)\geq n-i+1
\label{solidbound}
\end{eqnarray}
with $i\leq n-2$.
\vskip 0.5 cm
There is a  simple but important relation which must be
satisfied by the numbers, $q(\sigma^{n-2})$, of $n$-dimensional
simplices sharing the $(n-2)$-dimensional subsimplices, {\bf the
bones} of $T$, $\sigma^{n-2}$, {\it viz.},
\begin{eqnarray}
\sum_{\sigma^{n-2}}q(\sigma^{n-2})=\frac{n(n+1)}{2}N_{n}
\label{sub}
\end{eqnarray}
which follows by noticing that the number of $(n-2)$-dimensional sub-
simplices in an $n$-
dimensional simplex is $\frac{1}{2}n(n+1)$. We shall
exploit this relation quite intensively, and
for later convenience, let us introduce
the average number of simplices $\sigma^n$, in $T$, incident
on the $(n-2)$-dimensional subsimplices, $\sigma^{n-2}$,  namely
\vskip 0.5 cm
\begin{eqnarray}
b(n,n-2)\doteq\frac{1}{N_{n-2}(T)} \sum_{\sigma^{n-2}}q(\sigma^{n-2})=
\frac{1}{2}n(n+1)\left(
\frac{N_n(T)}{N_{n-2}(T)} \right)
\label{avnumber}
\end{eqnarray}
\vskip 0.5 cm
\noindent It is easily verified that the relations (\ref{eleven}),
(\ref{twelve}), (\ref{thirteen})
leave $\frac{1}{2}n-1$, ($n$
even) or $\frac{1}{2}(n-1)$ unknown quantities  among the $n$ ratios
$N_1/N_0,\ldots,N_n/N_0$, \cite{Dav}. Thus, in dimension
$n=2,3,4$, the datum of $b(n,n-2)$, (trivial for $n=2$), and
of the number of bones $N_{n-2}$, fixes, through the Dehn-Sommerville
relations
all the $f$-vectors $N_i(T)$ of the dynamical triangulation
considered.
\vskip 0.5 cm
Together with the Dehn-
Sommerville relations, equation (\ref{sub}) implies the following
bounds for the average incidence number $b(n,n-2)$,
when $N_{n-2}>>1$:\par
\noindent {\it (i)} For $n=2$:
\begin{eqnarray}
b(2,0)=6;
\end{eqnarray}
\noindent {\it (ii)} for $n=3$:
\begin{eqnarray}
3\leq b(3,1)\leq6;
\label{bitre}
\end{eqnarray}
\noindent {\it (iii)} for $n=4$:
\begin{eqnarray}
 3\leq b(4,2)\leq 5.
\label{biquattro}
\end{eqnarray}

Notice however that the lower bounds in (\ref{bitre}) and
(\ref{biquattro}) are not optimal. This fact was noticed during
computer simulations (see {\it e.g.} \cite{Bakker}). The actual
characterization of the optimal lower bounds for $b(n,n-2)$ in
dimension $n=3$ and $n=4$ it is rather important since it
bears relevance to the location of the critical point in simplicial
quantum gravity. It is not trivial to characterize such bounds since
they are a consequence of
the following result from the combinatorial topology of $3$-
dimensional $PL$-manifolds proved by D. Walkup \cite{Walkup}:

\begin{Gaiat}
For any combinatorial $3$-manifold the inequality
\begin{eqnarray}
N_1\geq 4N_0-10
\label{quattro}
\end{eqnarray}
holds with equality if and only if it is a stacked sphere
\end{Gaiat}
(a stacked sphere is by definition a triangulation of a sphere which
can be constructed from the boundary of a simplex by successive adding
of pyramids over some facets).

By exploiting this result we can prove the following
\newtheorem{Gaial}{Lemma}
\begin{Gaial}
For any triangulation $T\to M^n$ of a closed $n$-dimensional $PL$-
manifold $M$, the actual lower bound for the average incidence number
$b(n,n-2)$ is given for $n=3$ by
\begin{eqnarray}
b(n,n-2)|_{n=3}=\frac{9}{2}
\end{eqnarray}
and for $n=4$ by
\begin{eqnarray}
b(n,n-2)|_{n=4}=4
\end{eqnarray}
\label{walkup}
\end{Gaial}
\vskip 0.5 cm
\noindent {\bf Proof}. Let us consider first the three-dimensional
case. The Dehn-Sommerville relations for  $3$-manifolds are
\begin{eqnarray}
N_0-N_1+N_2-N_3=0
\end{eqnarray}
\begin{eqnarray}
N_2=2N_3
\end{eqnarray}
from which we immediately get $N_3=N_1-N_0$. On the other hand,
Walkup's theorem implies $N_0\leq\frac{1}{4}N_1+\frac{5}{2}$. Thus,
\begin{eqnarray}
N_3\geq\frac{3}{4}N_1-\frac{5}{2}
\end{eqnarray}
which yields the stated result for $n=3$.
\vskip 0.5 cm
In dimension $n=4$, the Dehn-Sommerville relations read
\begin{eqnarray}
N_0-N_1+N_2-N_3+N_4=\chi
\label{dsuno}
\end{eqnarray}
\begin{eqnarray}
2N_1-3N_2+4N_3-5N_4=0
\label{dsdue}
\end{eqnarray}
\begin{eqnarray}
5N_4=2N_3
\label{dstre}
\end{eqnarray}
Let us apply Walkup's theorem to to the link
of every vertex $v$ in our $4$-dimensional triangulation. In this way
we get
\begin{eqnarray}
N_1[link(v)]\geq 4_0[link(v)]-10
\label{dscinque}
\end{eqnarray}
and by summing over all vertices one obtains
\begin{eqnarray}
3N_2\geq 8N_1-10N_0
\label{dssei}
\end{eqnarray}
which is the basic inequality that we are going to exploit.
From it, (multiplying  by $3$), we obtain
\begin{eqnarray}
9N_2\geq 24N_1-30N_0
\label{dssette}
\end{eqnarray}
There is another independent relation of this sort which can be
obtained if we multiply (\ref{dsuno}) by $30$, and exploit
(\ref{dstre}), {\it viz.},
\begin{eqnarray}
30N_0-30N_1+30N_2-18N_3=30\chi
\label{dsotto}
\end{eqnarray}
From (\ref{dsdue}) one gets $2N_1-3N_2+2N_3=0$ which, (upon
multiplication by
$9$), yields $18N_3=27N_2-18N_1$. Inserting this latter expression in
(\ref{dsotto}) we eventually get
\begin{eqnarray}
3N_2=30\chi+12N_1-30N_0
\label{dsnine}
\end{eqnarray}
which together with (\ref{dssette}) provides a direct proof of
the  bound for $b(n,n-2)|_{n=4}$. Explicitly, let us subtract
(\ref{dsnine}) to
(\ref{dssette}). We get
\begin{eqnarray}
6N_2\geq 12N_1-30\chi,
\end{eqnarray}
 namely
\begin{eqnarray}
N_2\geq 2N_1-5\chi
\label{dsdieci}
\end{eqnarray}
Since $2N_1-3N_2+2N_3=0$, we get $2N_3-2N_2=N_2-2N_1$, thus, according
to (\ref{dsdieci}), we have
\begin{eqnarray}
2N_3-2N_2\geq -5\chi
\label{dsundici}
\end{eqnarray}
And from (\ref{dstre}) we obtain
\begin{eqnarray}
5N_4\geq 2N_2-5\chi
\label{dsdodici}
\end{eqnarray}
which, for $N_2>>\chi$, yields for $b(n,n-2)|_{n=4}\geq 4$. $\Box$
\vskip 0.5 cm
It is also interesting to note that according to relation (\ref{sub})
the incidence numbers
$\{q(\sigma^{n-2})\}$ of a simplicial manifold cannot be provided by a
monotonically increasing sequence  of integers, for otherwise
$\Sigma_{\sigma^{n-2}}q(\sigma^{n-2})$ would be $O(N^2_n)$ in plain
contrast with (\ref{sub}). The actual distribution of the incidence
numbers $q(\sigma^{n-2})$ of a triangulation has typically a peak
around a rather small value of $q(\sigma^{n-2})$ ($\geq 3$) and only a
few of the bones $\sigma^{n-2}$ can support a large number  ({\it
i.e.}, $O(N_n)$) of incident $n$-simplices. In particular, from
(\ref{sub}) its is immediate to prove that if we
let $q_{max}\doteq \sup_{T(N,B)}\sup_{\sigma^{n-2}}\{q(\sigma^{n-
2})\}$
denote the maximum value of $q(\sigma^{n-2})$, as $\sigma^{n-2}$
varies
in the set of triangulations with $N$ simplices $\sigma^n$ and given
value of $b(n,n-2)$, then, for $N_{n-2}>>1$,
\vskip 0.5 cm
\begin{eqnarray}
q_{max}=\frac{n(n+1)}{2F(n)}\left(\frac{b(n,n-2)-\tilde{q}}{b(n,n-2)}
\right)N_n
\label{qmax}
\end{eqnarray}
\vskip 0.5 cm
\noindent where $F(n)$ is the number of bones over which
$q_{max}$ can be attained, and $\tilde{q}$ is the average incidence
over the remaining bones, {\it i.e.},
\begin{eqnarray}
\tilde{q}\doteq \frac{1}{N_{n-2}}\sum_{F(n)+1}^{N_{n-2}}q(\sigma^{n-
2}_{\alpha})
\end{eqnarray}
\vskip 0.5cm
An elementary example of such a configuration is easily constructed in
dimension $n=2$ by considering two distinct vertices $v_1$ and $v_2$
on each one of which $N/2$ triangles are incident. The link of both
vertices is an $N/2$-gon, and by gluing the stars of $v_1$ and $v_2$
through such links we get a triangulation of the sphere ${\Bbb S}^2$
with $N$ triangles, with $q(v_1)=q(v_2)=\frac{N}{2}$.
\vskip 0.5 cm

Actually, the
existence of this sort of triangulation  is a rather general
fact of some relevance to us. Indeed it provides, for large
$N_n$, a triangulation minimizing the average $\tilde{q}$,
($\tilde{q}=\hat{q}+1=4$ in the above example), with few vertices
carrying a large incidence. In general, the existence of such
triangulations follows from the construction of
a triangulation $T^{n+1}\to B^{n+1}$ of the $(n+1)$-dimensional ball
$B^{n+1}$ inducing on ${\Bbb S}^n\simeq\partial{B}^{n+1}$ a
triangulation $T^n\to{\Bbb S}^n$ of the $n$-dimensional sphere ${\Bbb
S}^n$ with $N_n$ simplices and $N_{n-2}=\frac{1}{3!}(n-1)(n+1)N_n+n$
bones. $\frac{N_n}{n}$ of the $n$-dimensional simplices are incident
on $n$ distinguished bones $\{{\sigma^*}^{n-2}_i\}_{i=1,\ldots,n}$
spanning an embedded ${\Bbb S}^{n-2}$ in $T^n$. Thus, for $n=2$,
$\{{\sigma^*}^{n-2}_i\}_{i=1,\ldots,n}$ consists of two vertices ({\it
viz.}, ${\Bbb S}^0$ which is indeed topologically identified with a
double point). For $n=3$, $\{{\sigma^*}^{n-2}_i\}_{i=1,\ldots,n}$
consists of $3$ edges bounding a triangle ($\simeq{\Bbb S}^1$),
whereas for $n=4$ the supporting bones $\{{\sigma^*}^{n-
2}_i\}_{i=1,\ldots,n}$ yields for $4$ triangles bounding a tetrahedron
({\it viz.}, an ${\Bbb S}^2$).\par
 In order to construct such triangulation of ${\Bbb S}^n$ consider the
$(n+1)$-dimensional ball $B^{n+1}$ triangulated by a sequence of
$N_{n+1}$, $(n+1)$-dimensional simplices $\{\sigma^{n+1}\}$ all
sharing a common $(n-1)$-dimensional simplex $\sigma^{n-
1}_*\doteq\sigma_*$, {\it i.e.},
\begin{eqnarray}
\sigma_1^{n+1}(\sigma_*), \sigma_2^{n+1}(\sigma_*),\ldots,
\sigma_{N+1}^{n+1}(\sigma_*)=\sigma_1^{n+1}(\sigma_*).
\end{eqnarray}
and such that
\begin{eqnarray}
F[\sigma_j^{n+1}(\sigma_*),\sigma_{j+1}^{n+1}(\sigma_*)]\cap
F[\sigma_{j+1}^{n+1}(\sigma_*),\sigma_{j+2}^{n+1}(\sigma_*)]=
\sigma_*
\end{eqnarray}
where $F[\sigma_j^{n+1}(\sigma_*),\sigma_{j+1}^{n+1}(\sigma_*)]\doteq
\sigma_j^{n+1}(\sigma_*)\cap\sigma_{j+1}^{n+1}(\sigma_*)$ is the $n$-
dimensional face through which $\sigma_j^{n+1}(\sigma_*)$ and
$\sigma_{j+1}^{n+1}(\sigma_*)$ are glued together. The boundary of
such $T^{n+1}\to B^{n+1}$ is a triangulation of the sphere
$T^n\to{\Bbb S}^n$ with $N_n=nN_{n+1}$ top-dimensional simplices
$\sigma^n$, and $N_{n-2}=\frac{1}{3!}(n-1)(n+1)N_n+n$ bones. The
intersection $\sigma_*\cap{T^n}$ defines
an embedded ${\Bbb S}^{n-2}$ in $T^n$ generated by
$n$ bones, $\{{\sigma^*}^{n-2}_i\}_{i=1,\ldots,n}$, on each of which
$\frac{N_n}{n}$ simplices $\sigma^n$ are incident.
\vskip 0.5 cm

These
results can be naturally extended to $PL$-manifolds of arbitrary
topology by gluing, the above triangulation of the $n$-sphere into a
{\it small} triangulation of the given PL-manifod $M$, (say a
triangulation of $M$ with a minimal number of vertices).
Explicitly, let $T_{small}(M)$ be a triangulation with few
simplices, $N_n(T_{small})$ , of an $n$-dimensional Pl-manifold  $M$
of given topology, (for a characterization of such type of
triangulations and for examples see the paper by W. K\"{u}hnel in
\cite{Frohlich}). Consider the large $N_n$ triangulation $T({\Bbb
S}^n)$ of the $n$-sphere ${\Bbb S}^n$ defined above, let
$\sigma_1^n(M)$ be a simplex in $M$ and $\sigma_2^n({\Bbb S}^n)$ a
simplex in ${\Bbb S}^n$. Set
\begin{eqnarray}
T(M)\doteq\left( T_{small}(M)-\hat{\sigma}_1(M) \right)\cup_f
\left(T({\Bbb S}^n)-\hat{\sigma}_2^n({\Bbb S}^n) \right)
\end{eqnarray}
where $\hat{\sigma}(\cdot)$ denotes the interior of the given
$\sigma$'s, and $f$ is an homeomorphism from the boundary of
$\sigma_1^n(M)$ to the boundary of $\sigma_2^n({\Bbb S}^n)$. If
$N_n({\Bbb S}^n)>>N_n(T_{small})$, such $T(M)$ provides a large $N_n$
triangulation of $M$ whose distribution of incidence numbers
$\{q(\sigma^{n-2})\}$ has the properties of the collection of
incidence numbers of $T({\Bbb S}^n)$.
\vskip 0.5 cm
Anticipating a little bit, when this sort of triangulations
entropically dominate in the canonical partition function
(\ref{doppiow}) we should expect that the resulting statistical system
as described by (\ref{doppiow}) exhibits a very large {\it effective
dimensionality} as $N_n\to\infty$. A useful clue to this is to note
that an effective dimensionality on a triangulated manifold can be
defined by counting the number of vertices visited in a walk of $k$
steps, starting from a typical vertex. If the triangulation is
sufficiently regular then, for $k$ large, the number of visited
vertices grows as $k^n$, $n$ being the topological dimension of the
underlying PL-manifold.
On the other hand, triangulations such as the ones described above are
such that all vertices can be visited in a few steps: as $N_n$
increases the system effectively behaves as if we  were dealing with a
regular {\it infinite-dimensional} lattice.
\vskip 0.5 cm

Such remarks are elementary, nonetheless they play an important
role in what follows since the incidence numbers $\{q(\sigma^{n-2}\}$
are directly related with  curvature when discussing dynamical
triangulations. Thus there is an intimate relations between the
distributions of the incidence numbers $\{q(\sigma^{n-2}\}$,
metric properties, (curvature), and effective dimension of the
resulting statistical system.

\vskip 0.5 cm
\subsection{Distinct triangulations of the same PL manifold}
By the very
definition of PL manifolds, it
follows that  there exist different triangulations, $T^{(i)}$, of
the some PL manifold M. For later convenience, it is better to
formalize this remark, and recall the following standard
characterization by W. Tutte\cite{Pari}:

\begin{Gaiad}
Two  triangulations, $T^{(1)}$ and $T^{(2)}$ of
the some underlying PL
manifold $M$ are identified if there is a one-to-one mapping of
vertices, edges,
faces, and higher dimensional simplices of $T^{(1)}$  onto vertices,
edges,
faces, and higher dimensional simplices of $T^{(2)}$ which preserves
incidence relations. If no such mapping exists the corresponding
triangulations
are said to be {\it distinct}.
\end{Gaiad}
\vskip 0.5 cm
Notice that sometimes, ({\it e.g.}, W. Thurston\cite{Rourke},p.105),
such triangulations are said to be {\it combinatorially equivalent}.
However, we shall avoid this terminology since, in simplicial quantum
gravity, combinatorial equivalence is used as synonimous of PL-
equivalence.

\vskip 0.5 cm
Not surprisingly, in connection to the simplicial quantum gravity
program natural questions to ask are among the classics of PL-
Topology: {\it (i)} When can a
topological manifold be triangulated? {\it (ii)} Do any two
triangulations of
a manifold admit isomorphic subdivisions? The answer to the first
question is, for general manifolds, negative\cite{Rourke}: Kirby,
Siebenmann and Wall showed that there exist topological manifolds
which do not admit combinatorial structures, and that
there are non-standard structures for PL-manifolds of dimension at
least $5$. What Kirby and Siebenmann proved\cite{Rourke} is that for
$n\geq 5$ there is a single obstruction $k(M)\in H^4(M,{\Bbb Z}_2)$,
(where $H^i(M,{\Bbb Z}_2)$ denotes the $i$-th cohomology group of $M$
with ${\Bbb Z}_2$ coefficients field), to
the existence of a piecewise linear structure on a topological
manifold $M$, and that the vanishing of this obstruction ($k(M)=0$)
allows for the existence of a  worth of $|H^3(M,{\Bbb Z}_2)|$ distinct
PL structures. Moreover, if $n\leq 7$,
every PL manifold admits a compatible differentiable structure which
is unique, (up to diffeomorphisms), if $n\leq 6$.
The {\it Hauptvermutung } is the subject of  question {\it (ii)}, and
it is false in general. For instance, as hinted before, the $5$-sphere
${\Bbb S}^5$
admits PL-inequivalent triangulations. In the  dimensions of interest
to simplicial quantum gravity, the situation
is more encouraging. Indeed, for dimension $n\leq 3$ both questions
{\it (i)}, and {\it (ii)},
have an affermative answer. However, it is not yet known if every $n$-
dimensional topological manifold is homeomorphic to a polyhedron for
$n\geq 4$. Actually, (see the paper by T.
Januszkiewicz\cite{Ballmann}), there is a closed topological four-
manifold $M$ with the following properties: {\it (i)}
$M$ is not homotopy equivalent to a $PL$ manifold; {\it (ii)}
$M$ is not homemorphic to a simplicial complex. The construction of
this example is rather involved and exotic (the basic ingredient being
the $E_8$ homology manifold), but shows once more, if necessary, that
the interplay between the topology of $4$-dimensional manifolds and
quantum gravity cannot be easily tackled. For this reason when
discussing the $4$-dimensional case,
we shall not consider explicitly topological $4$-manifolds, and work
directly in the PL-category. When discussing deformations of dynamical
triangulations we shall need to consider smooth structures, but this
is simply a technical issue since as already recalled the category of
smooth $4$-manifolds is equivalent to the category of PL
$4$-manifolds.
\vskip 0.5 cm

Another natural question to ask is
under what conditions a given
set of integers $N_i$ is the $f$-vector of a triangulated polyhedron.
This problem is of some  relevance to simplicial quantum gravity,
since it
can be related to the counting of distinct triangulations, ({\it the
Entropy problem}).
Indeed, a possible way of stating the entropy problem is to enumerate
all distinct triangulations, on a  manifold of given topology, for a
given choice of possible $f$-vectors.
However,
for general triangulated $n$-manifolds the characterization of all
possible $f$-vectors is beyond reach. Partial answers can be obtained
only under rather severe  constraints. Typically,
one has to assume that the polyhedron in question is realized as a
{\it convex} polyhedron
in Euclidean $d$-dimensional space, ${\Bbb R}^d$. In this case, the
possible $f$-vectors are characterized by Stanley's
theorem\cite{Stanley}, according to which
the Dehn-Sommerville relations are necessary but not sufficient (they
suffice in dimension $2$ if we add the obvious condition $N_0\geq 4$
expressing the fact that to bound a solid takes more than three
vertices). In general, for a sequence of integers
$N_0,N_1,\ldots,N_{d-1}$, set $N_{-1}=1$ and $m=[\frac{d}{2}]$. For
$0\leq p\leq d$, define $h_p$ by

\begin{eqnarray}
h_p\doteq\sum_{i=p}^d(-1)^{i-p} \left( \begin{array}{c}
i\\ p
\end{array}  \right)N_{d-i-1}
\end{eqnarray}
 and for, $1\leq p\leq m$, set
\begin{eqnarray}
g_p{\doteq} h_p-h_{p-1}
\end{eqnarray}
then we have\cite{Stanley}
\begin{Gaiat}
A sequence of integers $N_0,N_1,\ldots,N_{d-1}$ occurs as the number
of vertices, edges, $\ldots$, $(d-1)$- faces of a $d$-dimensional
convex simplicial polytope if and only if the following conditions
hold: $h_p=h_{d-p}$, for $0\leq p\leq m$, (these are just a rewriting
of Dehn-Sommerville
relations); $g_p\geq 0$ for $1\leq p \leq m$; and, if one writes
\begin{eqnarray}
g_p=\left( \begin{array}{c}
n_p\\ p
\end{array}  \right)+ \left( \begin{array}{c}
n_{p-1}\\ p-1
\end{array}  \right)+\ldots+
\left( \begin{array}{c}
n_r\\ r
\end{array}  \right)
\end{eqnarray}
with $n_p>n_{p-1}>\ldots>n_r\geq r\geq 1$, (there is a unique way of
decomposing an integer in this way), then
\begin{eqnarray}
g_{p+1}\leq\left( \begin{array}{c}
n_p+1\\ p+1
\end{array}  \right)+ \left( \begin{array}{c}
n_{p-1}+1\\ p
\end{array}  \right)+\ldots+
\left( \begin{array}{c}
n_r+1\\ r+1
\end{array}  \right)
\end{eqnarray}
for $1\leq p\leq m-1$.
\end{Gaiat}
\vskip 0.5 cm
This result was conjectured  by McMullen\cite{McMullen}, sufficiency
and later necessity was proven respectively by Billera and
Lee\cite{Billera},  and Stanley and
McMullen\cite{Stanley},\cite{Smullen}.
Finally, it is  worth noticing  that for the class of all spheres the
determination of all $f$-vectors has
been conjectured by Stanley\cite{Sphere}. Owing to the strong
limitations imposed by the convexity constraint, it is not very
convenient to formulate the entropy problem in terms of the
characterization of the possible $f$-vectors of simplicial manifolds,
and in enumerating distinct triangulations we shall follow a different
strategy. However, it must be stressed that
progress in Stanley's theory, (and its unexpected connections
with algebraic geometry) may shed light on this question.
\vskip 0.5 cm

\subsection{Dynamical Triangulations}

Let $T\to M=|T|$ be a simplicial manifold, (henceforth, if there is no
danger of confusion, we shall use the shorthand notation $M=|T|$ for
denoting such a triangulated manifold).
A triangulation $T$
can be used to produce a metric on the underlying PL manifold $M=|T|$
by declaring that all simplices in the triangulation are isometric to
the standard simplex of the given dimension. This procedure provides a
collection of compatible metrics on $M$, which can be extended to a
piecewise-flat metric on $M$ in an
obvious way, here we follow the elegant construction provided by
Fr\"{o}hlich\cite{Frohlich}.
Let us choose an
orthonormal frame ${\{\vec{e}_a(\sigma^n)\}}_{a=1}^n$
in the interior of each simplex $\sigma^n\in T$, (this frame can be
conveniently located at the barycentre of $\sigma^n$), and
introduce another frame by choosing $n$ directed edges
$\{\vec{E}_{\mu}(\sigma^n)\}_{\mu=1}^n$ in the boundary of $\sigma^n$.
We can write
$\vec{E}_{\mu}(\sigma^n)=$
$\sum_{a=1}^nt^a_{\mu}(\sigma^n){\vec{e}}_a(\sigma^n)$,
where the $n\times{n}$ matrix, the {\it $n$-bein},
$(t^a_{\mu}(\sigma^n))$ is, for each $\sigma^n$, a regular matrix,
$\sigma^n$ being a non-degenerate simplex.\par
\vskip 0.5 cm
\noindent {\bf Fixed edge-length metric}. In terms of the chosen edges
$\vec{E}_{\mu}(\sigma^n)$, we can write
the remaining edges in the
boundary of $\sigma^n$ as
$\vec{E}_{\mu\nu}(\sigma^n)=\vec{E}_{\mu}(\sigma^n)-
\vec{E}_{\nu}(\sigma^n)$, with $1\leq\mu<\nu\leq{n}$. The metric
inside each simplex $\sigma^n$ is assumed to be the usual Euclidean
metric, $\delta$. It can be expressed in terms of the matrix
$\delta(\sigma^n)=g_{\mu\nu}(\sigma^n)=\langle\vec{E}_{\mu},
\vec{E}_{\nu}\rangle_{\delta}$.
We  assume that each edge
$\vec{E}_{\mu}(\sigma^a)$ or $\vec{E}_{\mu\nu}(\sigma^n)$ has a {\it
fixed length}
$a=|\vec{E}_{\mu}(\sigma^n)|=|\vec{E}_{\mu\nu}(\sigma^n)|$, this
characterizes fixed edge-length
triangulations as compared to the variable edge-length triangulations
typical of Regge calculus.  In this way
we get a metric in each simplex $\sigma^n\in M$ which is extended to
$M$ by specifying that the metric is continuous when one crosses the
faces of the $\sigma^n$'s, and that each  face of a simplex is a
linear flat subspace of the simplex in question.
Thus the metric on a fixed edge-length PL manifold is entirely
specified by the fixed distance elements $a$ between nearest neighbor
vertices of the corresponding simplicial manifold, {\it viz.},
\begin{Gaiad}
For any $a\in {\Bbb{R}}^+$,
a Dynamical Triangulation $T_a$ of a
polyhedron
$M=|T|$,  with distance cut-off $a$, and with $N_n(T_a)$ simplices
$\sigma^n$, is a division of
a simplicial
manifold $M=|T|$ generated by gluing, along their adjacent faces,
$N_n(T_a)$ equilateral simplices
$\{\sigma^n\}$ with edge-length
$|\vec{E}_{\mu}(\sigma^n)|=|\vec{E}_{\mu\nu}(\sigma^n)|=a$,
$\forall\sigma^n\in T$.
\end{Gaiad}

\vskip 0.5 cm
Note that in order to avoid overcounting dynamical triangulations
differing only
by a rescaling of the edge-lengths,
one always refers to a given value of the cut-off, (typically $a=1$),
when characterizing distinct triangulations.
\vskip 0.5 cm
\subsection{Dynamical triangulations as length-spaces}
The metric $g_{\mu\nu}$ generated by extending over $M=|T_a|$ the
flat metrics $g_{\mu\nu}(\sigma^n)$ of the simplices $\{\sigma^n\}$,
is not the object of primary interest for unravelling the metric
geometry of a dynamically triangulated manifold. For one thing, this
metric is manifestly singular around the $(n-2)$-dimensional
simplices, (which have conical neighborhoods), for the other there is
nothing sacred in $g_{\mu\nu}$ and one can capture more explicitly the
geometry of $M$
by looking at all curves between given points $p_1$ and $p_2$ in $M$
and by setting the distance $d(p_1,p_2)$ equal  to the infimum of the
length of these curves. Incidentally one can reconstruct $g$ at any
(regular) point $p$ from the distance function. Setting
$f(\hat{p})\doteq [d(p,\hat{p})]^2$, it is easily verified that  $g$
is provided, to leading order, by the second differential
$D^2f(\hat{p})|_{\hat{p}=p}$.\par
In order to discuss some aspects of the interior geometry of a
dynamical triangulation thought of as a metric space, we introduce few
preliminary notions hinting to a characterization of {\it geodesics}
on a dynamical triangulation without using calculus\cite{Ballmann}.
\vskip 0.5 cm
A {\it curve} in a dynamically triangulated manifold $M$ is a
continuous map $c\colon I\to M=|T_a|$, where $I$ is an interval. The
length $L(c)$ of the curve $c\colon [a,b]\to M$ is then defined
according to
\begin{eqnarray}
L(c)\doteq \sup\sum_{h=1}^jd(c(t_{h-1}),c(t_h))
\end{eqnarray}
where the supremum is taken over all subdivisions
$a=t_0<t_1<\ldots<t_j=b$ of $[a,b]$, and $d(,)$ denotes the distance
function associated with the given dynamical triangulation $T_a\to
M=|T_a|$. We assume that $c$ is rectifiable, ({\it i.e.},
$L(c)<\infty$), and introduce the
{\it arc length} parametrization as the non-decreasing continuous
surjective map
\begin{eqnarray}
s\colon [a,b]\to[0,L(c)]
\end{eqnarray}
defined by $s(t)\doteq L(c|[a,t])$. In terms of the arc-length, the
curve $c$ is  {\it travelled} at unit speed, {\it viz.},
\begin{eqnarray}
\tilde{c}\colon [0,L(c)]\to M
\end{eqnarray}
with $\tilde{c}(s(t))\doteq c(t)$, and
\begin{eqnarray}
L(\tilde{c}|[s_1,s_2])=|s_1-s_2|
\end{eqnarray}
Similarly we can characterize a curve $c\colon{I}\to{M}$ travelled at
speed $v\geq0$ if the curve is such that
\begin{eqnarray}
L(c|[t_1,t_2])=v|t_1-t_2|
\end{eqnarray}
for all $t_1,t_2\in I$. If a curve $c\colon{I}\to{M}$ has constant
speed $v\geq0$ and if for any $t\in I$ the curve $c$ has a
neighborhood $U$ in $I$ such that
\begin{eqnarray}
d(c(t_1),c(t_2))=v|t_1-t_2|
\label{geodesic}
\end{eqnarray}
for all $t_1,t_2\in U$, then such a $c$ is called a {\it geodesic} in
the dynamically triangulated manifold $M=|T_a|$.
Such a geodesic is said to be {\it minimizing} if (\ref{geodesic})
holds for all $t_1,t_2\in I$.
\vskip 0.5 cm
The {\it interior metric} $d_{int}$ associated to a dynamical
triangulation $T_a$ is defined according to\cite{Ballmann}
\begin{eqnarray}
d_{int}(x,y)\doteq\inf\{L(c)\}
\end{eqnarray}
where $\inf$ is taken over all curves from $x$ to $y$. It is easily
verified that for a dynamical triangulation such distance function
coincides with the original distance function $d$,
$d_{int}=d$, and in such a sense a dynamical triangulation is a {\it
Length space}\cite{Gromov}. It is also a {\it geodesic
space}\cite{Ballmann} since for any pair of points $x,y$ in $M$ there
is a minimizing geodesic from $x$ to $y$. \par
\vskip 0.5 cm
At first sight these facts may seem trivialities, (at least for a
dynamically triangulated manifold), however they have important
consequences when discussing rigorously the way a dynamically
triangulated manifold approximate smooth riemannian manifolds.
A first application will be encountered when discussing curvature in
the PL-setting.
\vskip 0.5 cm

\noindent {\bf PL connections and the Incidence Matrix}. In general,
associated with a PL
metric
$g_{\mu\nu}(\sigma^n)=\langle\vec{E}_{\mu},
\vec{E}_{\nu}\rangle_{\delta}$, $\sigma^n\in T$,
 there is also a unique connection (the
Levi-Civita connection) which can be characterized by the set of
matrices, $\Gamma(\sigma^n_{i},\sigma^n_{i+1})$, describing the change
of bases
$\{\vec{E}_{\mu}(\sigma_i^n)\}_{\mu=1}^n\to
\{\vec{E}_{\mu}(\sigma_{i+1}^n)\}_{\mu=1}^n$, (regarded as different
bases of ${\Bbb R}^n$), in passing between
two  adjacent simplices in
$M$, sharing a common face $F(\sigma^n_i,\sigma^n_{i+1})$.
In Regge calculus\cite{Frohlich}, the Levi-Civita connection so
defined is uniquely determined by the length of the edges and by the
incidence matrix of $T$, the $N_n(T)\times N_n(T)$ matrix
$I(\sigma^n_i,\sigma^n_j)$, whose entries are $1$ if $\sigma^n_i$ and
$\sigma^n_j$ are glued along a common face, $0$ otherwise. It follows
that when the metric $g$ is generated by the fixed edge length
prescription, the corresponding Levi-Civita connection is uniquely
determined by $I(\sigma^n_i,\sigma^n_j)$.
\begin{Gaiad}
The family $\alpha_g\doteq
\{ \Gamma(\sigma_i^n,\sigma_{i+1}^n)\}_{i=1,\ldots}$ with $\sigma_i^n$
and $\sigma_{i+1}^n$ adjacent in $T_a$, is the Levi-Civita
connection associated with the dynamical triangulation $T_a$ of the PL
manifold $M=|T|$.
\end{Gaiad}
Notice that if in place of the change of bases
$\{\vec{E}_{\mu}(\sigma_i^n)\}_{\mu=1}^n\to
\{\vec{E}_{\mu}(\sigma_{i+1}^n)\}_{\mu=1}^n$ we consider the
corresponding change of orthonormal bases
${\{{\vec{e}}_{a}(\sigma_i^n)\}}_{a=1}^n\to
{\{{\vec{e}}_{a}(\sigma_{i+1}^n)\}}_{a=1}^n$, in passing from one
simplex
to another, then the connection $\alpha_g$ can be equivalently
expressed in terms of  orthogonal matrices
$O(\sigma_i^n,\sigma_{i+1}^n)$ given by
\begin{eqnarray}
\Gamma(\sigma_i^n,\sigma_{i+1}^n)=t^{-
1}(\sigma^n_i)O(\sigma_i^n,\sigma_{i+1}^n)t(\sigma^n_{i+1})
\end{eqnarray}

\vskip 0.5 cm
\noindent {\bf Simplicial paths and parallel
transporters}. A trasparent definition of curvature in a simplicial
manifold can be provided if together with the standard notion of curve
on $M$, we introduce the combinatorial notion of
simplicial path. A {\it simplicial path} in a
simplicial manifold $M$ is defined as a
sequence of
simplices $\{\sigma^n_j\}_{j=1}^l$, $l=2,3,\ldots$ with $\sigma^n_j\in
M$ and such that $\sigma^n_j$ and $\sigma^n_{j+1}\not=\sigma^n_j$
share a common face. If $\sigma^n_1=\sigma^n_l$ then we speak of a
{\it simplicial loop} in the  manifold $M$.
With any  simplicial loop
$\omega$, we can associate the parallel transporter
\begin{eqnarray}
R_{\omega}=\prod_{\sigma_j,\sigma_{j+1}\in\omega}
\Gamma(\sigma^n_j,\sigma^n_{j+1})
\label{sixteen}
\end{eqnarray}
where the product is  path-ordered.
Let $B$ denote the generic bone in $T_a$, and
let $\omega(B)$
that unique loop winding around the bone $B$. If $\sigma_j^n(B)$
denotes the generic simplex containing $B$, then\cite{Frohlich}
\begin{eqnarray}
\omega(B)=\{\sigma_1^n(B),\sigma_2^n(B),\ldots,\sigma_m^n(B)=
\sigma_1^n(B)\}
\end{eqnarray}
with $\sigma_1^n(B)\cap\ldots\cap\sigma_m^n(B)=B$.
The corresponding rotation matrix
\begin{eqnarray}
R_{\omega(B)}=\prod_i R(\sigma_i(B),\sigma_{i+1}(B))
\end{eqnarray}
 is such that
all vectors in the $(n-2)$-dimensional hyperplane spanned by the bone
$B$ are eigenvectors with eigenvalue 1, and in the $2$-dimensional
plane orthogonal to the bone $B$,  $R_{\omega(B)}$ reduces to a
rotation, where the rotation angle $\phi(B)$ is
given by
\begin{eqnarray}
\phi(B)=\sum_{j=1}^{q(B)}\delta_j(B)
\label{deficit}
\end{eqnarray}
$q(B)$ being  the number of simplices $\sigma^n$ incident on the
bone $B$, and $\delta_j(B)$ is the angle between the unique two faces
of $\sigma^n_j$ containing $B$.
\vskip 0.5 cm

\noindent {\bf Curvature}. In Regge calculus, the quantity
$r(B)\equiv 2\pi-\phi(B)$ is the deficit angle at the bone $B$.
Since the angle between two faces sharing a common bone $B$,
in an equilateral simplex $\sigma^n$, is given by
$\cos^{-1}1/n$, we can write
\begin{eqnarray}
\phi(B)=q(B)\cos^{-1}\frac{1}{n}
\label{angle}
\end{eqnarray}
where $q(B)$ denotes the number of simplices $\sigma^n$ incident
on the bone $B$. In defining curvature for dynamically triangulated
manifolds one has to change  the perspective on Regge calculus
somewhat (as advocated by Hamber and Williams\cite{Ruth}) and view the
geometry of the triangulation as representing an approximation to some
smooth geometry and the local curvature as some average curvature for
a small volume, (in the next section we formalize somewhat this point
of view by adopting the Gromov-Hausdorff topology on the space of
dynamically triangulated manifolds). Thus, to each $(n-2)$-dimensional
bone $B$ we assign
an $n$-dimensional volume density given by
\begin{eqnarray}
vol_n(B)\doteq
\frac{2}{n(n+1)}\sum_{\sigma^n_j\in\omega(B)}vol(\sigma^n_j)
\end{eqnarray}
where $\{\sigma^n_j\}$ is the set of $n$-dimensional simplices in the
unique simplicial loop $\omega(B)$ winding around the given bone $B$.
The volume $vol_n(B)$ can be considered as the natural share of the
volumes of the $n$-dimensional simplices to which the bone belongs. It
follows that every bone $B$ in a dynamical
triangulation $T_a$ carries a
curvature given by
\begin{eqnarray}
K(B)=2\left [ 2\pi-q(B)\cos^{-1}\frac{1}{n} \right ]\frac{vol_{n-
2}(B)}{vol_n(B)}
\label{curv}
\end{eqnarray}
where $vol_{n-2}(B)$ is the usual $n-2$-dimensional (Euclidean)
measure of the bone $B$. Recall that the Euclidean volume of a $j$-
dimensional equilateral simplex of edge-length $a$ is
\begin{eqnarray}
vol(\sigma^j)=\frac{a^j\sqrt{(j+1)}}{j!\sqrt{2^j}}
\end{eqnarray}
thus, we get
\begin{eqnarray}
K(B)=\frac{n^2(n^2-1)}{a^2}\sqrt{\frac{n-1}{n+1}}
\left [\frac{2\pi-q(B)\cos^{-1}\frac{1}{n}}{q(B)} \right ]
\end{eqnarray}
\vskip 0.5 cm
\noindent For instance, in dimension four we have
$vol_{n-2}(B)=\frac{1}{4}\sqrt{3}a^2$,
$vol_n(\sigma^n)=\frac{1}{96}\sqrt{5}a^4$, and (\ref{curv}) reduces to
\begin{eqnarray}
K(B)=\frac{240}{a^2}\sqrt{\frac{3}{5}}\left[ \frac{2\pi-q(B)\cos^{-
1}\frac{1}{n}}{q(B)} \right ]
\end{eqnarray}
\vskip 0.5 cm

\noindent {\bf Curvature assignements}. Up to the choice of the cut-
off edge-
length, $a$, and of a
numerical factor depending only from the dimension $n$,
curvature in a dynamical triangulation $T$ with $N_{n-2}(T)$ bones
$B(\alpha)$, $\alpha=1,2,\ldots,N_{n-2}(T)$, is
directly provided by the sequence of integers $\{q(\alpha)\}$, where
$q(\alpha)$ denotes the number of simplices $\sigma^n$ incident on the
bone $B(\alpha)$. Henceforth, when speaking of {\it curvature
assignements} to the bones $\{B(\alpha)\}$ of a dynamical
triangulation $T_a$, we shall explicitly refer to the sequence of
integers $\{q(\alpha)\}$.
\vskip 0.5 cm
\noindent {\bf The Einstein-Hilbert action for dynamical
triangulations}. The
Regge version of the Einstein-Hilbert action (with cosmological term)
for a PL-manifold $M$, (without boundary), is given
by\cite{Regge},\cite{Houches},\cite{Frohlich}
\vskip 0.5 cm
\begin{eqnarray}
S_{Regge}(\Lambda,G)\doteq \Lambda\sum_{\sigma^n}vol(\sigma^n)-
\frac{1}{16\pi G}\sum_B r(B)vol(B)
\label{reggeaction}
\end{eqnarray}
\vskip 0.5 cm
When the PL-manifold $M$ is dynamically triangulated $T_a\to{M}$,
$S_{Regge}(\Lambda,G)$ simplifies considerably reducing to
\vskip 0.5 cm
\begin{eqnarray}
S[k_{n-2},k_n]\doteq k_nN_n(T_a)-k_{n-2}N_{n-2}(T_a)
\end{eqnarray}
\vskip 0.5 cm
\noindent where we have set according to standard usage
\vskip 0.5 cm
\begin{eqnarray}
k_n\doteq \Lambda vol(\sigma^n)+\frac{1}{2}n(n+1)\frac{\cos^{-
1}\frac{1}{n}}{16\pi G}vol(\sigma^{n-2})
\end{eqnarray}
\vskip 0.5 cm
\begin{eqnarray}
k_{n-2}\doteq \frac{vol(\sigma^{n-2})}{8G}
\end{eqnarray}
Thus, for dynamical triangulations the gravitational action reduces to
a combinatorial object with two (running) couplings
$k_{n-2}$ and $k_n$. The former proportional to the inverse
gravitational coupling, while the latter is a linear combination of
$1/16\pi{G}$ with the cosmological constant $\Lambda$.
Accordingly, the formal path integration (\ref{uno}) is replaced (see
(\ref{five})) on a dynamically triangulated Pl manifold $M$, (of fixed
topology), by the (grand-canonical) partition function
\begin{eqnarray}
Z[k_{n-2},k_n]=\sum_{T\in {\cal T}(M)}
e^{-k_nN_n+k_{n-2}N_{n-2}}
\label{grandpartition}
\end{eqnarray}
\vskip 0.5 cm
The action $S[k_{n-2},k_n]$ and the associated partition function are
so deceiptively simple that one may wonder under which rug we are
dumping  all the notorious problems affecting the Einstein-Hilbert
action when coming to a path-integral quantization. Obviously most of
these problems are now shifted in characterizing the summation over
distinct triangulations $\sum_{T\in {\cal T}(M)}$ which replaces the
formal path integration over distinct riemannian structures, namely in
characterizing the dynamical triangulation counterpart of the formal
measure $D[g(M)]$. In general, these sort of deals hardly  are true
bargains. However, as we shall see, this trade off in difficulties
will pay off
since a rather complex problem (Path Integration over
$Riem(M)/Diff(M)$), the mathematical formulation of which is rather
ambiguous, has been transformed into a well-defined counting problem:
{\it enumerate all distinct dynamical triangulations $T\in {\cal
T}(M)$
admitted by a PL manifold of given topology}. However, before
addressing this problem, we have to formalize in more precise terms in
which sense dynamical triangulations do approximate a riemannian
structure on a smooth manifold.

\vfill\eject

\section{Approximating riemannian manifolds with
dynamically triangulations of bounded geometry}

The fact that curvature for dynamical triangulations is a
combinatorial object has a number of deep consequences, mainly of
topological nature, having an important bearing on simplicial quantum
gravity. These properties follow from the fact
that a dynamically triangulated manifold $M=|T_a|$, with $N_n(T_a)$
$n$-simplices and with a given average number $b(n,n-2)$ of $n$-
simplices incident on the bones $\sigma^{n-2}$, (see
(\ref{avnumber})), is a piecewise-flat singular metric space of {\it
bounded
geometry}, a concept which goes back to A.D.
Alexandrov\cite{Alexandrov}. There is a definite way in which
such singular metric spaces approximate (or are approximated by
suitably smooth) manifolds, and this will provide the rationale  for
approximating
riemannian manifolds with dynamical triangulations.
\vskip 0.5 cm
\subsection{Dynamical triangulations as singular metric spaces}

To start with an intuitive description, let us consider two geodesic
rays starting from the some point $p$ in a Riemannian manifold $V$.
Thus, suppose that  $l_1\colon[0,1]\to V$  and $l_2\colon[0,1]\to V$
are (minimal)
geodesics with $l_1(0)=l_2(0)=p$. Fix $t,s\in(0,1)$ and construct the
plane Euclidean triangle with (linear-parametrized) lines
$L_1\colon[0,1]\to {\Bbb R}^2$, $L_2\colon[0,1]\to {\Bbb R}^2$ such
that $L_1(0)=L_2(0)=(0,0)$, $||L_1(s)||=d(l_1(0),l_1(s))$,
$||L_2(t)||=d(l_1(0),l_2(t))$, and
\begin{eqnarray}
||L_1(s)-L_2(t)||=d(l_1(s),l_2(t)).
\end{eqnarray}
If the sectional curvatures of $V$ are everywhere non-negative, then
the geodesics $l_1$ and $l_2$ tend to come together compared with the
corresponding lines $L_1$, $L_2$; {\it i.e.},
\begin{eqnarray}
||L_1(1)-L_2(1)||\geq d(l_1(1),l_2(1))
\end{eqnarray}
whereas non-positive sectional curvatures force the goedesics to
diverge faster than in the Euclidean situation:
$||L_1(1)-L_2(1)||\leq d(l_1(1),l_2(1))$.
\vskip 0.5 cm
Once one has a notion of geodesics, the above characterization of
positive or negative curvature makes sense for any length-spaces, and
in that case we speak of a space of positive (or negative)
{\it Alexandrov curvature}. Actually, similar characterizations can be
carried out in order to define spaces of Alexandrov curvature bounded
above of below by a constant $k\in{\Bbb R}$, by comparing the
geodesics $l_1$, $l_2$ with the corresponding geodesics on the
appropriate simply connected surface of constant Gauss curvature $k$.
These comparision techniques are a standard tool in global riemannian
geometry, (see {\it e.g.}, Chavel's book\cite{Ballmann}).
\vskip 0.5 cm
\noindent {\bf Geodesic triangles}. For $k\in{\Bbb R}$, let $\Sigma_k$
denote the {\it model surface} of constant Gauss curvature $k$.
Motivated by the above remarks, we characterize dynamical
triangulations as singular metric spaces with bounded Alexandrov
curvature, by comparing suitable triangles in $M=|T_a|$ with triangles
in model surfaces $\Sigma_k$. Since in a dynamical triangulation
curvature is localized around bones $\sigma^{n-2}$, we have to
consider geodesic triangles which encircles bones. This can be done
according to the following procedure. \par
\vskip 0.5 cm
Let $\alpha\doteq\sigma^{n-2}$ be the generic bone in a dynamical
triangulation $M$. Consider its associated link, $link(\sigma^{n-
2})\simeq{\Bbb S}^1$, and three distinct vertices, labelled $1$, $2$,
$3$ in $link(\sigma^{n-2})$. If $q(\alpha)$ simplices $\sigma^n$ are
incident on the bone $\alpha$, then $link(\sigma^{n-2})$ has a length
\begin{eqnarray}
||link(\sigma^{n-2})||=q(\alpha)\cdot{a}
\end{eqnarray}
$a$ being the edge-length of the given dynamical triangulation. The
vertices $1$,$2$,$3$ are chosen (up to a rotation) in such a way that
their pairwise distance, $d(i,k)$, $i\not=k=1,2,3,$, measured along
the link $link(\sigma^{n-2})$, satisfies
\begin{eqnarray}
d(i,k)\geq[\frac{q(\alpha)}{3}]
\end{eqnarray}
where the bracket function $[x]$ stands for the greatest integer not
exceeding $x$. Consider now (in the closure of the open star of the
bone $\alpha$) the three geodesic segments pairwise joining the
vertices so chosen, {\it viz.}, $c_1\doteq{l(2,3)}$,
$c_2\doteq{l(1,3)}$, and  $c_3\doteq{l(1,2)}$, then
\begin{Gaiad}
The geodesic triangle $\Delta(\alpha)\doteq(c_1,c_2,c_3)$ consisting
of the three geodesic segments $c_1$, $c_2$, $c_3$ in $M$, (called the
sides of the triangle), is the geodesic triangle
encircling the bone $\alpha\in M$. A geodesic triangle
$\bar{\Delta}\doteq(\bar{c_1},\bar{c_2},\bar{c_3})$ in $\Sigma_k$ is
called an Alexandrov (or comparision) triangle for $\Delta(\alpha)$ if
$length(c_i)=length(\bar{c_i})$, $i=1,2,3$.
\end{Gaiad}
It is not difficult to verify that  such a comparision
triangle exists, (and it is unique up to a congruence), and we have
the following
\begin{Gaial}
A dynamically triangulated manifold $M=|T_a|$, with $N_n(T_a)$
$n$-simplices and with a given average number $b(n,n-2)$ of $n$-
simplices incident on the bones $\sigma^{n-2}$, is a piecewise-flat
singular metric
space of bounded Alexandrov curvature $k$, with $k_{min}\leq k\leq
k_{max}$, where $k_{min}$, $k_{max}$ are characterized by
\begin{eqnarray}
\frac{1}{2\pi}a\cdot\hat{q}\cos^{-
1}(1/n)=\frac{\sin(a\sqrt{k_{max}})}{\sqrt{k_{max}}}
\end{eqnarray}
and
\begin{eqnarray}
\frac{1}{2\pi}a\cdot{q_{max}}\cos^{-1}(1/n)=\frac{\sinh(a\sqrt{-
k_{min}})}{\sqrt{-k_{min}}}
\end{eqnarray}
where, (see (\ref{qmax}))
\begin{eqnarray}
q_{max}=\frac{n(n+1)}{2F(n)}\left(\frac{b(n,n-2)-\tilde{q}}{b(n,n-2)}
\right)N_n.
\end{eqnarray}
\label{alexandrov}
\end{Gaial}
\vskip 0.5 cm
{\bf Proof}. We start by showing that comparision triangle exists for
any bone of the given dynamical triangulation. In order to prove this,
let us first remark  that the geodesic triangle $\Delta(\alpha)$ has
sides satisfying the triangle inequality
\begin{eqnarray}
length(c_i)+length(c_{i+1})\geq length(c_{i+2})
\end{eqnarray}
indices taken modulo $3$. However, this is not sufficient to insure
the existence of a comparision triangle in $\Sigma_k$ since, if
$k>0$, the perimeter of $\Delta(\alpha)$ must
satisfy the bound
\begin{eqnarray}
length(c_1)+length(c_2)+ length(c_3)<\frac{2\pi}{\sqrt{k}}
\label{perimeter}
\end{eqnarray}
(this bound is trivially extended to the case $k\leq0$ by setting
$\frac{2\pi}{\sqrt{k}}\doteq\infty$ for $k\leq0$).
To prove that (\ref{perimeter}) holds, let $x$ be any point in the
relative interior of the given bone $\alpha$, and consider the set of
all  tangent vectors to the geodesic rays (lines) emanating from $x$,
of length $a$, pointing into $\sigma^n$, (the generic simplex incident
on $\alpha$), and orthogonal to $\alpha$. This set of vectors forms a
spherical (${\Bbb S}^1$)-simplex. Since this set is, up to an
isometry, independent of $x\in\alpha$, we shall denote this set as
${\Bbb S}^1link(\alpha,\sigma^n)$: the {\it spherical link} of
$\alpha$ in $\sigma^n$\cite{Ballmann}. Geometrically, the spherical
link
${\Bbb S}^1link(\alpha,\sigma^n)$ can be thought of as the arc of
circumference {\it
circumscribed} to $link(\alpha,\sigma^n)$, the link of the bone
$\alpha$ in $\sigma^n$.\par
\vskip 0.5 cm
If the Regge curvature around the given bone is positive, then
\begin{eqnarray}
q(\alpha)<\frac{2\pi}{\cos^{-1}(1/n)}
\end{eqnarray}
Since $\cos^{-1}(1/n)\geq\frac{\pi}{3}$ for $n\geq 2$, we get that
$q(\alpha)<6$. The union of the spherical links
${\Bbb S}^1link(\alpha,\sigma^n)$ over all $q(\alpha)$ simplices
$\sigma^n$ incident on $\alpha$ is then a circumference of length
$a\cdot{q}(\alpha)\cos^{-1}(1/n)<2\pi{a}$. Consider now the sphere,
${\Bbb S}^2(r)$, of radius $r$, such that a great circle of radius
$a$ on ${\Bbb S}^2(r)$ has a circumference of length
$a\cdot{q}(\alpha)\cos^{-1}(1/n)$. An elementary computation
characterizes the radius $r$ of such ${\Bbb S}^2(r)$ in terms of
$q(\alpha)$ as
\begin{eqnarray}
\frac{1}{2\pi}a\cdot{q}(\alpha)\cos^{-1}(1/n)=r\sin(\frac{a}{r})
\end{eqnarray}
or, by introducing the curvature $k\doteq\frac{1}{r^2}$ of
${\Bbb S}^2(r)$
\begin{eqnarray}
\frac{1}{2\pi}a\cdot{q}(\alpha)\cos^{-
1}(1/n)=\frac{\sin(a\sqrt{k})}{\sqrt{k}}
\label{jacobi}
\end{eqnarray}
\vskip 0.5 cm
\noindent Since the geodesic triangle $\Delta(\alpha)$ is inscribed in
the circumference of length $a\cdot{q}(\alpha)\cos^{-1}(1/n)$, we get
that (\ref{perimeter}) holds with $k$ given in terms of $q(\alpha)$ by
(\ref{jacobi}). By developing (\ref{jacobi}) to leading order in $a$,
we get, not unexpectedly, that $k$ is given by
\begin{eqnarray}
k\simeq \frac{6}{a^2}\left( 2\pi -q(\alpha)\cos^{-1}(1/n)  \right)
\end{eqnarray}
namely, up to a numerical factor, by
the Regge curvature associated with the bone $\alpha$. This
elementary analysis can be extended to the case of negative Regge
curvature obtaining
\begin{eqnarray}
\frac{1}{2\pi}a\cdot{q}(\alpha)\cos^{-1}(1/n)=\frac{\sinh(a\sqrt{-
k})}{\sqrt{-k}}
\label{jacobi2}
\end{eqnarray}
thus proving that a dynamically triangulated manifold
$M=|T_a|$, with $N_n(T_a)$
$n$-simplices and with a given average number $b(n,n-2)$ of $n$-
simplices incident on the bones $\sigma^{n-2}$, admits comparision
triangles on  model surfaces, $\Sigma_k$, of constant curvature
$k$,  where $k$ is provided by (\ref{jacobi}) and
(\ref{jacobi2}). Since on such set of dynamical triangulations the
possible incidence numbers $\{q(\alpha)\}$ are bounded between
\begin{eqnarray}
\hat{q}\leq q(\alpha)\leq q_{max},
\end{eqnarray}
(see (\ref{qmax})),
the above argument shows that a dynamical triangulation
$M=|T_a|$, as a singular metric space, is an Alexandrov space with
bounded curvature $k_{min}\leq k\leq k_{max}$ where $k_{min}$ and
$k_{max}$
are respectively provided by
\begin{eqnarray}
\frac{1}{2\pi}a\cdot\hat{q}\cos^{-
1}(1/n)=\frac{\sin(a\sqrt{k_{max}})}{\sqrt{k_{max}}}
\end{eqnarray}
and
\begin{eqnarray}
\frac{1}{2\pi}a\cdot{q_{max}}\cos^{-1}(1/n)=\frac{\sinh(a\sqrt{-
k_{min}})}{\sqrt{-k_{min}}},
\end{eqnarray}
as stated. $\Box$

\vskip 0.5 cm

The notion of bounded geometry associated with the characterization of
dynamical triangulations as Alexandrov spaces
becomes especially interesting if one considers the set of all
dynamical triangulations satisfying the hypotheses of lemma
\ref{alexandrov}. Guided by the intuition of what happens in similar
circumstances for ordinary riemannian
manifolds\cite{Gromov},\cite{Grov},
we expect that the metric space structures
of such  dynamical triangulations is controlled by such curvature
bounds. To
formulate this kind of controlled behavior, it is useful to introduce
a topology in the set of dynamical triangulations
which arises directly from the classical idea of Hausdorff distance
between compact sets in a metric space\cite{Gromov}.
The basic rationale is that
given a length cut-off $\epsilon$, two metric spaces are to be
considered near in this topology, (one is the $\epsilon$-Gromov-
Hausdorff approximation of the other), if their metric properties are
similar at length scales $L\geq{\epsilon}$, namely
if there is a way of fitting them in a
larger matric space so that they are close to each other. To formalize
this intuition, we introduce the notion of  {\it rough
isometry}.\par
\vskip 0.5 cm

\noindent {\bf $\epsilon$-rough isometries}. Consider two compact
(finite
dimensional) metric spaces
$M_1$   and $M_2$, let $d_{M_1}(\cdot,\cdot)$ and
$d_{M_2}(\cdot,\cdot)$
respectively denote the corresponding distance functions, and let
$\phi\colon M_1\to M_2$ be a map between $M_1$ and $M_2$,
(this map is not required to be continuous ). If
$\phi$ is such that: {\it (i)}, the $\epsilon$-neighborhood of
${\phi}(M_1)$ in $M_2$ is equal to $M_2$, and {\it (ii)}, for
each $x$, $y$ in $M_1$ we have
\begin{eqnarray}
|d_{M_1}(x,y)-d_{M_2}({\phi}(x),{\phi}(y))|<\epsilon
\end{eqnarray}
then $\phi$ is said to be an $\epsilon$-{\it Hausdorff approximation}.
The  {\it Gromov-Hausdorff distance}
between the two
metric spaces $M_1$ and  $M_2$,  $d_G(M_1,M_2)$, is then
defined
according
to\cite{Gromov}
\begin{Gaiad}
  $d_G(M_1,M_2)$ is the lower bound of the positive numbers $\epsilon$
such that there exist $\epsilon$-Hausdorff approximations from $M_1$
to $M_2$ and from $M_2$ to $M_1$.
\label{miauno}
\end{Gaiad}
\vskip 0.5 cm
Alternatively\cite{Petersen}, $d_G(M_1,M_2)\leq\epsilon$ if there is a
metric, $d(\cdot,\cdot)$, on the disjoint union
$M_1\coprod M_2$ extendings the metrics on $M_1$ and $M_2$ and such
that $M_1\subset\{x\in M_1\coprod M_2\colon d(x,M_2)\leq\epsilon \}$,
$M_2\subset\{x\in M_1\coprod M_2\colon d(x,M_1)\leq\epsilon \}$.
Namely, two compact metric spaces $M_1$ and $M_2$ are Gromov-Hausdorff
nearby if we can find metrics on the disjoint union of $M_1$ and $M_2$
such that $M_1$ and $M_2$ look like the same when they are close to
each other.
\vskip 0.5 cm

The notion of $\epsilon$-Gromov-Hausdorff approximation is the weakest
large-scale equivalence relation between metric spaces of use in
geometry, and
is manifestly adapted to the needs of
simplicial quantum gravity, (think of a manifold and of a simplicial
approximation to it).\par
Notice that $d_G$ is not, properly
speaking,  a distance since it
does not satisfy the triangle inequality, but it rather gives rise to
a metrizable uniform structure in which the set of isometry classes
of all compact metric spaces,
is Hausdorff and complete. This enlarged space does
naturally contain riemannian manifolds and metric
polyedra. As stressed in \cite{Petersen}, the importance of this
notion  lies not so much in the fact that we have a
distance function, but in that we have a way of measuring when metric
spaces look alike.\par
\vskip 0.5 cm
\subsection {Approximating Riemannian manifolds through dynamical
triangulations}
In this section we prove that,
dynamical triangulations $T_a$, with a
given number $N_n(T_a)$ of $n$-simplices and a given ratio $b(n,n-2)$,
are in the Gromov-Hausdorff closure of the $\infty$-dimensional
compact  metric space characterized  by the following
\begin{Gaiad}
For $r$ a real number, $D$ a positive real number, and
$n$ a natural number, let us define the associated space of Bounded
Geometries,
${\cal R}(n,r,D,V)$, as the
Gromov-Hausdorff closure of the space
of isometry
classes of closed connected
 n-dimensional riemannian manifolds $(M,g)$ with
Ricci curvature bounded below by $r$, {\it viz.},
\begin{eqnarray}
\inf_{x\in M}\{\inf \{ Ric(u,u) \colon u \in T_xM,
|u_x|=1\}\}\geq r  \nonumber
\end{eqnarray}
\noindent and diameter bounded above by $D$, {\it viz.},
\begin{eqnarray}
diam(M)\equiv\sup_{(p,q)\in M\times M}d_M(p,q)\leq D\nonumber
\end{eqnarray}
\label{miadue}
\end{Gaiad}
\vskip 0.5 cm
It is rather immediate to recast this characterization in our current
setting so as to prove the following {\it approximation} property
(actually a statement of the Lebesgue covering property) which has
far reaching consequences for simplicial gravity:

\begin{Gaial}
Let ${\cal DT}_n(a,b,N)$ denote
the set of $n$-dimensional, ($n=2,3,4$),dynamically triangulated
manifolds $M=|T_a|$, with
$N_n(T_a)\equiv{N}$ top-dimensional simplices  and given average
incidence $b(n,n-2)$, and let
$k_{min}$ be the corresponding lower bound on the Alexandrov
curvature, (see lemma \ref{alexandrov}). Correspondingly, let
${\cal R}(n,r,D,V)$ denote the space of bouded geometries with
\begin{eqnarray}
r\geq k_{min}
\end{eqnarray}
\begin{eqnarray}
V= N\cdot{vol(\sigma^n)}
\end{eqnarray}
\begin{eqnarray}
D\leq c_1 [N\cdot{vol(\sigma^n)}]^{1/n}
\end{eqnarray}
where  $vol(\sigma^n)$  denotes
the
(Euclidean) volume of the equilateral
simplex $\sigma^n$ of edge-length $a$, and $c_1$,  is a constant
(independent from the
triangulation
considered). Then for any $n=2,3,4$, any
finite $b(n,n-2)\in
{\Bbb Q}^+$, and $N\in {\Bbb N}$, there is an $\epsilon=\epsilon(a)>0$
such that for any riemannian manifold $V\in {\cal R}(n,r,D,V)$ we have
\begin{eqnarray}
d_G(V,M=|T_a|)<\epsilon(a)
\end{eqnarray}
for some dynamical triangulation $M=|T_a|\in
{\cal DT}_n(a,b,N)$.
\label{density}
\end{Gaial}
In other words, the set of dynamical
triangulations
${\cal DT}_n(a,b,N)$ uniformly approximates riemannian manifolds of
bounded geometry.

Note that we restrict the dimension $n$ to the values
$n=2$, $3$, $4$, because in such a case, the Dehn-Sommerville
relations allow us to control the $f$-vector of the triangulation in
terms of the number of top-dimensional simplices $N_n(T_a)=N$ and of
the
average incidence number $b(n,n-2)$. By giving the appropriate
combinatorial data, the extension to $n\geq 5$ does not present
particular difficulties.
In order to prove lemma \ref{density}, we need to recall some standard
results related to the compactness properties of
${\cal R}(n,r,D,V)$.
\vskip 0.5 cm
A sequence of compact metric spaces
$\{M_i\}$ converges to a compact metric space $M$ provided
that $d_G(M_i,M)\to 0$ as $i\to\infty$. Thus
we have to prove that there is an $\epsilon=\epsilon(a)>0$ such that
the Gromov-
Hausdorff distance between the generic riemannian manifold $M\in{\cal
R}(n,r,D,V)$ and a suitably chosen dynamical triangulation $T_a$, is
lesser than or equal to $\epsilon$. Note that the $\epsilon$ involved
is a function of
the cut-off $a$ entering
the definition of $T_a$. The proof
of the above lemma
is a rather immediate consequence of the following well-known
properties\cite{Grov}:

\begin{Gaiat}
Let $M\in {\cal R}(n,r,D,V)$ be the generic riemannian manifold of
bounded
geometry, and let $d_M(\cdot,p)$ denote the corresponding distance
function from a chosen point $p\in M$.  Then, for any
given  $\epsilon >0$, it is always
possible to find an ordered set of points $\{p_1,\ldots,p_m\}$ in $M$,
 so that: {\it (i)} the open metric balls, (the {\it geodesic
balls}),
$B_{M}(p_{i},\epsilon) = \{x \in M \vert d(x, p_{i})<
\epsilon\}$, $i=1,\ldots,m$, cover $M$; {\it(ii)} the open balls
$B_{M}(p_{i},{\epsilon\over 2})$,
$i=1,\ldots,m$, are
disjoint, {\it i.e.}, $\{p_1,\ldots,p_m\}$
is a {\it minimal} $\epsilon$-net in $M$.\par
\end{Gaiat}

We recall that the {\it filling function} of a minimal $\epsilon$-net
is the number $m$ of points $\{p_1,\ldots,p_m\}$, while
the first order {\it intersection pattern} of a minimal $\epsilon$-net
in $M$ is the set of pairs $\{(i,j)|i,j=1,\ldots,m;
B(p_i,\epsilon)\cap B(p_j,\epsilon)\not=\emptyset  \}$.

It is important
to remark that  on ${\cal R}(n,r,D,V)$ neither the filling function
nor the
intersection pattern can  be arbitrary. The filling function is
always
bounded
 above for each  given $\epsilon$, and  the best  filling,
with geodesic  balls  of  radius  $\epsilon$, of
a riemannian manifold of diameter $diam(M)$, and Ricci curvature
$Ric(M)\geq (n-1)H$, is controlled by the
corresponding filling of the geodesic ball of radius
$diam(M)$ on the  space form  of  constant  curvature given by $H$,
the bound being of the form\cite{Grov} $N^{(0)}_{\epsilon}\leq
N(n,H(diam(M))^2,(diam(M))/\epsilon)$.\par
\vskip 0.5 cm
The multiplicity of the first
intersection pattern is  similarly  controlled  through  the
geometry
 of the  manifold to  the effect  that its  average degree,
({\it
i.e.}, the  average number of mutually intersecting balls), is bounded
above by  a constant  as the
radius of the balls defining  the covering  tend to  zero, ({\it
i.e.},
as $\epsilon \to 0$ ).  Such constant  is independent  from
$\epsilon$
and can be estimated\cite{Grov} in terms of the parameters $n$, and
$H(diam(M))^2$.\par
\vskip 0.5 cm
\begin{Gaiad}
Two metric spaces ({\it e.g.}, two riemannian manifolds) $M_1$ and
$M_2$ in ${\cal R}(n,r,D,V)$ are called $\epsilon$-equivalent
(or $\epsilon$-rough isometric) if and only if they can be equipped
with minimal $\epsilon$-nets $\{p_1,\ldots,p_m\}_1$ and
$\{p_1,\ldots,p_s\}_2$, respectively, such that $s=m$ and with the
same intersection pattern.
\end{Gaiad}
\vskip 0.5 cm

We can introduce a natural minimal geodesic ball covering associated
with
a dynamical  triangulation $T_a$ according to the following

\begin{Gaial}
Let $(M=|T_a|,T_a)$ denote a dynamically triangulated manifold  with
fixed edge
legth $a$. With each vertex $p_i\doteq \sigma^0_i$ belonging to the
triangulation,  we
associate
the largest  open  metric ball contained in the open star of
$p_i$. Then the metric ball covering of $M=|T_a|$ generated by such
balls
$\{B_i\}$ is a minimal geodesic
ball covering. It defines
the geodesic ball covering associated with
the dynamically triangulated manifold $(M,T_a)$.
\end{Gaial}
\vskip 0.5 cm

It is immediate to see that the set of balls considered defines
indeed a minimal geodesic ball covering. The open balls obtained from
$\{B_i\}$ by halfing their radius
are disjoint  being contained in the open stars
of $\{p_i\}$ in the baricentric subdivision of the triangulation.
The balls
with doubled
radius cover $(M,T)$, since they are the largest open balls
contained in the stars of the vertices $\{p_i\}$  of $T_a$.\par
\vskip 0.5 cm
We are now in a position for proving lemma \ref{density}.
\vskip 0.5 cm
{\bf Proof}. Given a riemannian manifold  $M\in {\cal R}(n,r,D,V)$,
endowed
with with a minimal $\epsilon$-net $\{q_i\}_{i=1,\ldots,m}$
we can associated with it the  dynamical triangulations $T_a$,
with $a=a(\epsilon)$, having the same combinatorial intersection
pattern $\{p_i=\sigma^0_i\}_{i=1,\ldots,m}$ and such that
$|d(q_i,q_j)-d(p_i,p_j)|<\epsilon$ for all $i,j=1,\ldots,m$. On the
disjoint union $M\coprod T_a$ we can define the metric
$d(x,y)=\min\{d(q,q_i)+\epsilon+d(p_i,p)\colon i=1,\ldots,m \}$
for $q\in M$, $p\in T_a$. More explicitly, this metric is constructed
by declaring $d(p_i,q_j)=\epsilon$, and by setting, for $p\in M$ and
$q\in T_a$, the distance $d(p,q)$ equal to the infimum of the sums
$d(p,x_1)+d(x_1,x_2)+\ldots+d(x_l,q)$ where each consecutive pair
$(x_i,x_{i+1})$, (or $(p,x_1)$, or $(x_l,q)$) either has both elements
in one of $M$ or $T_a$ or has $x_i=p_m$, $x_{i+1}=q_m$. The metric so
defined is such that
$M\subset\{x\in M\coprod T_a\colon d(x,T_a)\leq\epsilon \}$,
$T_a\subset\{x\in M\coprod T_a\colon d(x,M)\leq\epsilon \}$, and thus
$d_G(M,T_a)\leq 2\epsilon$, which implies the stated approximation
result.
$\Box$
\vskip 0.5 cm

 Since a given intersection pattern may be common to distinct
dynamical triangulations, there can be many distinct
dynamical triangulations that are $2\epsilon$-rough isometric
to a given riemannian
manifold. This is not surprising, the equivalence between riemannian
manifolds associated with minimal $\epsilon$-nets is very crude, and
there are distinct riemannian manifolds sharing the same $\epsilon$-
net, ({\it i.e.}, which are $2\epsilon$-rough isometric). The point is
that all such manifolds as well as all such
dynamical triangulations have a Gromov-Hausdorff distance between each
other which is smaller than or equal to $2\epsilon$, and as
$\epsilon\to 0$, {\it viz.}, as the dynamical triangulations become
finer and finer such dynamical triangulations converge to a well
defined metric space in ${\cal R}(n,r,D,V)$.
\vskip 0.5 cm
It is worth noticing that it is possible to obtain a converse of this
result,
indeed\cite{Grov}, any polyhedral compact subset of a Euclidean space
is the Gromov-Hausdorff limit of a sequence of closed Riemannian
manifolds.
\vskip 0.5 cm
The density result just obtained provides the mathematical rationale
for using dynamical triangulations in
simplicial quantum gravity. The original idea was related to the
intuition that fixed-edge length triangulations, $T_a$, (of fixed
topology $M=|T_a|$) provide a {\it grid} of reference manifolds laid
upon the $\infty$-dimensional space of riemannian structures allowed
on such $M$. The compactness results associated with the use of
Gromov-Hausdorff convergence support this picture, and the uniform
approximation
result of lemma \ref{density} makes precise
this intuition by proving that dynamically triangulated manifolds do
provide a reference grid as long as we look at riemannian manifolds
and dynamical triangulations of bounded geometry. At first sight
it may
appear surprising that no strict topology control is required on such
triangulations, but this is actually a bonus coming automatically
along with the requirement of bounded geometry, ({\it i.e.}, fixed
$N_n$ (Volume) and fixed average incidence $b(n,n-2)$ (curvature
bounds)). As we shall see in the next section these natural bounds
severely control topology, by avoiding such patologies as infinite
genus surfaces (in dimension $n=2$), or wild homotopy types, etc.
One may question the naturality of such bounds in the sense that
dynamical triangulations is an approach to quantizing gravity in
the sense of a statistical field theory, and in seeking the associated
critical phenomena we eventually need to go to the infinite-volume
limit. In this limit the bounds on $N_n$ and $b(n,n-2)$ are removed
and one consequently loses control on everything. The point we wish to
make is exactly that the above results tell us {\it how} to carry out
such sort of limits. One must rigorously establish estimates for
dynamical triangulations, the associated partition functions and $m$-
points correlation functions, in a given space of bounded geometry
${\cal D}T_n(a,b,N)\subset {\cal R}(n,r,D,V)$, and then consider a
nested sequence ${\cal D}T_n(a_1,b_1,N_1)\subset
{\cal D}T_n(a_2,b_2,N_2)\subset\ldots\subset{\cal
D}T_n(a_i,b_i,N_i)\ldots$ in which the convergence properties of the
partition function and the associated infinite-volume limit can be
safely discussed. As we shall see, the actual characterization of the
infinite volume limit does not call for such a sophisticated
machinery. However the discussion of a possible continuum limit may do
require the introduction of a suitable projective limit procedure for
${\cal D}T_n(a,b,N)$.

\subsection{Topological finiteness theorems for Dynamical
Triangulations of bounded geometry}

Since dynamical triangulations, (for given $N$ and $b(n,n-2)$),
are in a Gromov space of bounded geometries, ${\cal R}(n,r,D,V)$, we
can control to a considerable extent their topology in terms of the
parameters $n$, $r$, $D$, and $V$, (namely in terms of the number of
top-dimensional simplices $N$ and $b(n,n-2)$, in the range of
dimension
considered).
We start with a
result expressing finiteness of homotopy types of dynamically
triangulated  manifolds of
bounded geometry\cite{Grov}.

\begin{Gaiat}
For any dimension $2\leq n \leq 4$,
the number of different homotopy-types of dynamically triangulated
manifolds
realized in ${\cal DT}_n(a,b,N)\subset {\cal R}(n,r,D,V)$ is
finite and is a function of $n$,
$V^{-1}D^n$, and $rD^2$.
\label{homotopy}
\end{Gaiat}
(Two polyhedra $M_1=|T_a(1)|$ and $M_2=|T_a(2)|$ are said to have the
same
 homotopy type if there exists a continuous
 map $\phi$ of $M_1$ into $M_2$ and $f$ of $M_2$ into $M_1$,
such that both $f \cdot \phi$ and $\phi \cdot f$ are homotopic
to the respective identity mappings, $I_{M_1}$ and $I_{M_2}$.
 Obviously, two homeomorphic polyhedra are of
the same homotopy type, but the converse is not true).\par
\vskip 0.5 cm
This homotopy finiteness property of dynamically triangulated
manifolds is an obvious adaptation of a more general result proved in
a truly remarkable series of papers by P. Petersen, K. Grove, and J.
Y. Wu \cite{Grov}. Actually, for dynamical triangulations, the above
theorem can be sharpened so as to provide (rather weak) conditions
under which two distinct dynamically triangulated manifolds  share a
common homotopy type, and we can estimate the number of distinct
homotopy types. We have the following
\begin{Gaiat}
For any dimension $2\leq n \leq 4$, and for a given value $a$ of the
cut-off edge-length,
the number of different homotopy-types of dynamically triangulated
manifolds
realized in ${\cal DT}_n(a,b,N)\subset {\cal R}(n,r,D,V)$ is
bounded above by
\begin{eqnarray}
\sharp{Homotopy}[{\cal DT}_n(a,b,N)]\leq [C(n)^{-
1}\cdot{N}vol(\sigma^n)\cdot{a}^{-n}]^4
\end{eqnarray}
where $C(n)$ is some universal constant depending only on the
dimension $n$, and $vol(\sigma^n)$ is the euclidean volume of the
simplex $\sigma^n$ (of edge-length $a$).
\label{petersen1}
\end{Gaiat}
\vskip 0.5 cm
{\bf Proof}.  Recall\cite{Ferry},\cite{Grov} that a
continuous function
${\psi}\colon [0,\alpha)\to {\bf R}^+$, $\alpha >0$, with
${\psi}(0)=0$, and ${\psi}(\epsilon)\geq \epsilon$, for all
$\epsilon\in [0,\alpha)$, is a local geometric contractibility
function for a (finite dimensional) compact metric space $M$ if, for
each $x\in M$ and
$\epsilon\in (0,\alpha)$, the open ball $B(x,\epsilon)$, of radius
$\epsilon$ centered at $x\in M$, is
contractible in the larger ball $B(x,{\psi}(\epsilon))$,
(which says that in $M$ small metric balls are contractible relative
to  bigger
balls). On a dynamical triangulation $T_a\in{\cal DT}_n(a,b,N)$,
consider the smallest  open ball containing the star of the generic
vertex. This is a ball or radius $a$ which is contractible within
itself. Thus, for any such $T_a$ we may consider the local geometric
contractibility function given by
\begin{eqnarray}
\psi(\epsilon)=\epsilon
\end{eqnarray}
on $[0,a]$. We can now exploit in our setting a result of
P.Petersen\cite{Grov} to the effect that any two dynamical
triangulations $T_a^1$ and $T_a^2$ in ${\cal DT}_n(a,b,N)$ with
Gromov-Hausdorff distance $d_G(T_a^1,T_a^2)<\epsilon^*$ are homotopy
equivalent if $\epsilon^*\simeq a/(32n^2)$, (this bound is not
optimal, and similarly to what happens for riemannian manifolds with
criticality radius bounded below, one should get a sharper bound of
the form $a/[25(n+1)]$\cite{Grov}). Since $|T_a|$ can be covered by
the open metric balls of radius $a$ centered on the vertices of $T_a$,
it also admits a refinenement
of this covering generated by open metric balls of radius
$\epsilon^*/2$. If $N(\epsilon^*/2)$ denotes the number of such balls,
then an argument due to Yamaguchi\cite{Grov} can be used to show that
${\cal DT}_n(a,b,N)$ contains less than $[N(\epsilon^*/2)]^4$ distinct
homotopy types. In order to estimate $N(\epsilon^*/2)$, notice that
the volume of a metric ball, $B(\epsilon^*/2)$, of radius
$\epsilon^*/2$ in $|T_a|$  is bounded below by
\begin{eqnarray}
vol[B(\epsilon^*/2)]\geq \hat{q}(\sigma^0)\cdot
vol(\sigma^n)_{\epsilon^*/2}=(n+1)\cdot{c(n)}(\epsilon^*/2)^n
\end{eqnarray}
where $\hat{q}(\sigma^0)=(n+1)$ is the minimum number of $n$-
dimensional simplices sharing a vertex, and
$vol(\sigma^n)_{\epsilon^*/2}=c(n)(\epsilon^*/2)^n$ is the euclidean
volume of a standard euclidean simplex $\sigma^n$ with edge-length
$\epsilon^*/2$, (this characterizes the constant $c(n)$ as
$c(n)\doteq\sqrt{n+1}/(n!\sqrt{2^n})$). Since the
volume of $|T_a|$ is given by $N\cdot{vol(\sigma^n)}$, we get
\begin{eqnarray}
N(\epsilon^*/2)\leq
\frac{N\cdot{vol(\sigma^n)}}{(n+1)\cdot{c(n)}(\epsilon^*/2)^n}
\label{coverbound}
\end{eqnarray}
Since $\epsilon^*\simeq(a/32n^2)$, by
introducing (\ref{coverbound}) in the Yamaguchi estimate we get the
stated result with $C(n)\doteq\frac{c(n)(n+1)}{(64n^2)^n}$. $\Box$
\vskip 0.5 cm
A similar result, obtained by adapting to dynamical triangulations
results obtained by Yamaguchi, Grove, Petersen and Wu\cite{Grov},
allows also to control the Betti numbers, (with generic field
coeffiecients $F$), of any
$M=|T_a|\in{\cal DT}_n(a,b,N)$. Explicitly we have
\begin{Gaiat}
Let $b_i(M,F)$ denote the $i$-th Betti number with field coefficients
$F$, then
\begin{eqnarray}
\sum_{i=1}^nb_i(M,F)\leq(n+1)^{(n+1)}
\left[ C(n)^{-1}\cdot{N}vol(\sigma^n)\cdot{a}^{-n}  \right]^n
\end{eqnarray}
\label{petersen2}
\end{Gaiat}
\vskip 0.5 cm
{\bf Proof}. The proof is an obvious rewriting of a result of
P.Petersen, (see the corollary at p. 393 of P.Petersen's
paper\cite{Grov}).
\vskip 0.5 cm
The above results clearly show that when going to the infinite-volume
limit, $\lim_{N\to\infty,a\to0}<N>vol(\sigma^n)\simeq const.$, (where
$<N>$ denotes the statistical average of $N$ with respect to a
canonical ensemble of dynamical triangulations), more and more
distinct topological types of manifolds come into play. However the
resulting {\it topological complexity} is not too wild since the
number of distinct homotopy types (and/or the Betti numbers) grows
polynomially with the inverse cut-off $a^{-1}$.
\vskip 0.5 cm

By considering separately the dimension $n=2$, $n=3$, $n=4$,
we can obtain
a more specific topological finiteness
theorem. This result follows   again from the local geometric
contractibility which by construction characterizes dynamically
triangulated manifolds. Since sufficiently small distance balls on a
dynamically
triangulated manifold of bounded geometry are always contractible, we
get, by specializing a theorem of Grove, Petersen,
Wu\cite {Grov},
\begin{Gaiat}
The set of dynamically triangulated manifolds in
${\cal DT}_n(a,b,N)\subset {\cal R}(n,r,D,V)$ contains\par
\noindent {\it (i)} finitely many simple homotopy types (all $n$),\par
\noindent {\it (ii)} finitely many homeomorphism types if $n=4$,\par
\noindent {\it (iii)} finitely many diffeomorphism types if $n=2$ (or
$n\geq 5$, in case we remove the constraint $2\leq n\leq 4$).
\end{Gaiat}
 Note that quantitative estimates associated with these finiteness
results have the same structure of the ones associated with theorems
\ref{petersen1} and \ref{petersen2}, and can be easily worked out
along the same lines. Note also that
finiteness of the homeomorphysm types cannot be proved in dimension
$n=3$
as long as the Poincar\'e conjecture is not proved. If there were
a fake three-sphere then one could prove\cite{Ferry} that a statement
such
as {\it
(ii)} above is false for $n=3$.
Finally, the statement on finiteness of {\it simple homotopy} types,
(which actually holds in
any dimension), is particularly important for the applications in
quantum gravity we discuss in the sequel. Roughly speaking the notion
of simple
homotopy is a refinement of the notion of homotopy equivalence,
relating (in our case) triangulated manifolds  (more in general CW
complexes) which can be obtained one from the other by a sequence of
expansion or collapses of simplices, (for details see
\cite{Cohen}, \cite{Rourke}). Roughly speaking, it
may
be thought of as an intermediate step between homotopy equivalence and
homeomorphism.
\vskip 0.5 cm
\noindent {\bf Topology and enumeration of triangulations}. The fact
that for a
dynamical triangulations with a given number, $N$, of
$n$-dimensional simplices $\sigma^n$, and a given average bone
incidence $b(n,n-2)$ we
have a good a priori control of topology is the basic reason that
allows us to
enumerate
distinct dynamical triangulations. An important step in this
enumeration is to understand how we can reconstruct the triangulation
by the knowledge of easy accessible geometrical data. At this stage it
is worth recalling that  the Levi-Civita connection for a dynamical
triangulation is uniquely determined by the incidence matrix
$I(\sigma^n_i,\sigma^n_j)$ of the triangulation itself, thus in order
to enumerate distinct dynamical triangulations {\it we can
equivalently  characterize the set of distinct Levi-Civita
connections} $\alpha_g=
\{ \Gamma(\sigma_i^n,\sigma_{i+1}^n)\}_{i=1,\ldots}$. The problem we
have to face is thus reduced to a rather standard question, familiar
in gauge theories, where one needs to reconstruct the holonomy of a
gauge-connection from data coming from  the Wilson loop functionals.
Here, we need to reconstruct all discretized Levi-Civita connection
from Wilson-loop datas which  reduce to the possible set of curvature
assignements to the bones, and  by taking due care of the non-trivial
topological information coming from the moduli of
locally homogeneous manifolds, ({\it e.g.}, constant curvature
metrics), which need to be  described in terms of deformations of
dynamical triangulations.
In order to address this problem we exploit the holonomy
representation
associated with the Levi-Civita connection $\alpha_g$.

\vfill\eject

\section{Moduli spaces for dynamically triangulated
manifolds}

Let $(T_a, M=|T_a|)$ be a dynamically triangulated manifold.
If we
denote by $\Omega_{\sigma}(M)$ the family of all simplicial loops in
$M$, starting at a given simplex $\sigma^n_0$, it easily follows that
$\Omega_{\sigma}(M)$, modulo $\sigma^n_0$-based homotopic
equivalence, is isomorphic to the fundamental group of $M$,
$\pi_1(\sigma^n_0,M)$,  based at the given simplex $\sigma^n_0$.
Moreover, by factoring out the effect of loops homotopic to the
trivial $\sigma^n_0$-based loop, the mapping
\begin{eqnarray}
R\colon \omega\in\Omega_{\sigma}(M)\mapsto R_{\omega}
\label{seventeen}
\end{eqnarray}
yields a representation of the fundamental group
$\pi_1(\sigma^n_0,M)$. The following definition specilizes to the
PL setting some of the well known properties of the holonomy of
riemannian manifolds.

\begin{Gaiad}
The holonomy group of the (fixed edge length) PL manifold $(M,T_a)$
at the simplex $\sigma^n_0$, $Hol(\sigma^n_0)$ is the subgroup of the
orthogonal group $O(n)$ generated by the set of all parallel
transporters $R_{w}$ along loops of the simplicial manifold $T$ based
at
$\sigma^n_0$. The subgroup obtained by the loops which are homotopic
to
the identity, is the restricted holonomy group of $(M,T)$ at
$\sigma^n_0$, $Hol^0(\sigma^n_0)$.
\end{Gaiad}
If we change the base simplex $\sigma^n_0$ to the simplex
$\bar{\sigma^n_0}$ and fix some path $\rho$ from $\sigma^n_0$ to
$\bar{\sigma^n_0}$, then
$Hol(\bar{\sigma^n_0})=A_{\rho}Hol(\sigma^n_0)A^{-
1}_{\rho}$, (and similarly for the restricted holonomy group). As a
consequence the holonomy groups at the various simplices of $T_a$ are
all isomorphic. And we can speak of the holonomy and restricted
holonomy group of the (fixed edge length) PL manifold $M=|T_a|$:
$Hol(M)$, and
$Hol^0(M)$, respectively. Moreover, if $\omega_0$ and $\omega_1$ are
simplicial loops based at $\sigma_0^n$ which represent the same
element, $[\omega]$, of $\pi_1(M,\sigma_0^n)$, then $R_{\omega_0}$ and
$R_{\omega_1}$ belong to the same arc component of $Hol(M)$. From this
observation and from the definition of restricted holonomy, it
naturally follows that there exists a canonical homomorphism
\begin{eqnarray}
\pi_1(M)\to\frac{Hol(M)}{Hol^0(M)}
\label{eighteen}
\end{eqnarray}
which defines the {\it Holonomy Representation} of $\pi_1(M)$.
\vskip 0.5 cm
Holonomy representations of this sort are of relevance to dynamical
triangulations. This connection is very subtle and it is related to
the well-known fact that generally we cannot
triangulate locally homogeneous spaces
with equilateral flat simplices. For instance, in order to model a
space of constant curvature with a
dynamical triangulation we have to {\it slightly} deform the
triangulation (following a suggestion anticipated many years ago by
R\"{o}mer and
Z\"{a}hringer\cite{Frohlich}). This deformation procedure is
apparently trivial in the sense that a small alteration of the chosen
edge-length allows one to fit the triangulation on a constant
curvature background. However subtle problems arise if there are non-
trivial deformations ({\it i.e.}, moduli) of the underlying constant
curvature metric, ({\it e.g}, distinct flat tori). Formally, these
non-trivial deformations are described by the
cohomology of the manifold with coefficients in the sheaf of Killing
vector fields of the manifold, namely with local coefficients in the
(adjoint representation of the) Lie algebra $\frak g$ of the  group
$G$ of isometries considered. This cohomology is isomorphic to the
cohomology of the fundamental group $\pi_1(M)$ with value in the
holonomy representation of $\pi_1(M)$ in $\frak g$, and this is one of
the
basic mechanisms that calls into play topology in
dynamical triangulations. Such issues can be very effectively dealt
within the framework of deformation theory in algebraic geometry, (as
hinted in the above formal sheaf-theoretic remarks), however, we think
appropriate here a more pedagogical approach, also because this issue
has not really been considered previously in dynamical triangulation
theory.

\subsection{R\"{o}mer-Z\"{a}hringer
deformations of dynamically
triangulated manifolds}

As a warm up after the above introductory remarks, let us consider in
particular the {\it flat} Levi-Civita connections.
Since the connection is flat, the parallel transport along a closed
simplicial loop $\omega$ based at $\sigma^n_0$ depends only on the
homotopy class of $\omega$. In this case the parallel transport,
through its associated holonomy, gives rise to a
representation of the fundamental group
\begin{eqnarray}
\theta\colon\pi_1(M,\sigma^n_0)\to Hol(M)\subset O(n)
\label{flat}
\end{eqnarray}
the image of which is the holonomy group of the flat Levi-Civita
connection considered. Such $\theta$ is a homomorphism of
$\pi_1(M,\sigma_0^n)$ onto $Hol(M)$, and since $\pi_1(M)$ is
countable, $Hol(M)$ is totally disconnected. In this case one speaks
of $\theta$ as the {\it holonomy homomorphism}. \par
It is a well-known fact that connected compact flat $n$-dimensional
riemannian manifolds are covered by a flat $n$-torus
$p\colon {\Bbb T}^n\to M$, and that  the associated group of deck
transformations is isomorphic to $Hol(M)$, {\it viz.},
$M={\Bbb T}^n/Hol(M)$. Quite surprisingly, any finite group $G$ can be
realized as the holonomy group of some flat compact connected
riemannian manifold, (this is the Auslander-Kuranishi theorem, see
{\it e.g.}\cite{Wolf}). Accordingly, flat riemannian manifolds are
completely determined by homomorphisms of their fundamental group
$\pi_1(M)$ onto the (finite) group $G$, up to conjugation.  The kind
of
difficulties encountered in dynamically triangulating flat riemannian
manifolds are particularly clear when discussing the geometry of their
covering space.
\vskip 0.5 cm
\noindent {\bf Flat tori}. It is well known that at least in dimension
$n\geq
3$, it is not
possible to build up a flat space ({\it e.g.}, a flat three-torus) by
gluing flat equilateral simplices $\{\sigma^n\}$, this simply follows
from the observation that the angle $\cos^{-1}\frac{1}{n}$, $n\geq 3$,
is not an integer fraction of $2\pi$. Thus, for $n\geq 3$, dynamical
triangulations cannot directly describe flat tori.
Superficially, this does not seem to be a serious problem, since we
can deform a few simplices in such a way to match with the $2\pi$-
flatness constraint, (see next paragraph for the details). However,
this is hardly a sound prescription since, as follows from their
definition, there are distinct flat tori. More formally, let us fix a
basis
$({\vec{e}}_1,\ldots,{\vec{e}}_n)$ of ${\Bbb
R}^n$. Then with each $n$-ple of integers $(a^1,\ldots,a^n)\in{\Bbb
Z}^n$ we can associate a transformation, $A$, of ${\Bbb R}^n$ given by
$A(\vec{x})=\vec{x}+\sum_{j=1}^na^j{\vec{e}}_j$. This results in a
transformation group isomorphic to ${\Bbb Z}^n$ acting properly
discontinuously on ${\Bbb R}^n$. Upon quotienting ${\Bbb R}^n$ by this
action we get a manifold diffeomorphic to the $n$-dimensional torus
${\Bbb T}^n$. Since ${\Bbb R}^n$ is thought of as endowed with its
standard flat metric, the resulting ${\Bbb T}^n$ is likewise flat.
However, by changing base to ${\Bbb R}^n$ we change the group of
transformation associated with $A$, and the resulting flat torus is
not necessarily isometric to the original one. Thus, the naive
deformation procedure suggested above is ambiguous as long as it does
not specify which particular flat torus one is modelling.
 In particular,
in dimension $n=2$,
let $\vec{a}_1$ and $\vec{a}_2$ be the vectors generating the
lattice, and assume that
$\vec{a}_1$ is the first basis vector of ${\Bbb{R}}^2$,
$\vec{a}_1=\vec{e}_1$, and that $\vec{a}_2=(x,y)$ lies in the first
quadrant of  ${\Bbb{R}}^2$. Then the resulting flat tori are
parametrized by the
points in the plane region
\begin{eqnarray}
\{(x,y)|x^2+y^2\geq 1, 0\leq x\leq\frac{1}{2}, y>0  \}
\end{eqnarray}
to the effect that flat tori generated by lattices corresponding to
two
distinct
points of this region are distinct.
Thus, even in dimension $n=2$, where the flatness $2\pi$-constraint
can be met,
it follows that the lattice
generated by equilateral triangles, corresponding
to a dynamically triangulated torus, can only discretize (up to
dilation) the flat torus coming from exagonal lattices,
($x=\frac{1}{2}$, $y=\frac{\sqrt{3}}{2}$):  we cannot dynamically
triangulate with equilateral triangles all remaining flat tori, ({\it
e.g.}, all the rectangular tori $\vec{a}_2=(0,y)$), (obviously, this
obstruction persists only if we want the triangulation and the lattice
to be consistent, otherwise we can dynamically triangulate any $2$-
torus; in higher dimension we completely loose this freedom).
\vskip 0.5 cm
The difficulties in associating dynamical triangulations with
manifolds of large symmetries is not restricted to the case of  flat
tori just described. Similar problems are encountered in associating
dynamically triangulations to any riemannian manifold
endowed with a {\it geometric structure}, namely manifolds locally
modelled on homogeneous spaces,
for instance manifolds
 of constant
curvature, basically because such manifolds do not generally admit
regular tilings by euclidean equilateral simplices.
\vskip 0.5 cm
\noindent {\bf Deforming a triangulation}. Even if there is no regular
way to
generate locally homogeneous PL
manifolds by gluing equilateral simplices, ({\it e.g.}, flat tori), we
can
always assume that in
trying to model a locally homogeneous riemannian manifold $M$ with a
dynamical
triangulation $T_a$, this can be done in such a way that the Gromov-
Hausdorff distance between $M$ and $T_a$ is minimal. As a matter of
fact, we can, at least in line of principle,  construct an
approximating sequence
$\{T_{\mu}\}_{\mu=1,2,\ldots}$  of triangulated models of $M$,
interpolating between $M$ and $T_a$, by using triangulations of $M$
with simplices which are nearly equilateral, (the essence of this
remark was very clearly suggested by R\"{o}mer and
Z\"{a}hringer\cite{Frohlich}). With each triangulated manifold
$T_{\mu}$ of this sequence we can associate the edge-length
fluctuation functional

\begin{eqnarray}
s(T_{\mu})=\sum_{\sigma^1}\frac{[|\vec{E}_{ij}(\sigma^1)|-
a(T_{\mu})]^2}{N_1(T_{\mu})}
\end{eqnarray}
where
$a(T_{\mu})\equiv\frac{\sum_{\sigma^1}|\vec{E}_{ij}(\sigma^1)|}{N_1}$
is the average edge-length in $T_{\mu}$. The functional $s(T_{\mu})$
has reasonable continuity properties in the Gromov-Hausdorff topology,
and since
the Gromov space of bounded geometries is compact\cite{Gromov}, we can
always choose the
sequence $\{T_{\mu}\}_{\mu=1,2,\ldots}$ in such a way that it
minimizes $s(T_{\mu})$.
\vskip 0.5 cm
However, if there are non-trivial (infinitesimal) deformations of the
riemannian structure $M$, preserving the local homogeneous geometry,
({\it e.g.}, inequivalent flat tori, or  moduli of Riemann surfaces),
the
R\"{o}mer-Z\"{a}hringer procedure may associate distinct
moduli to the same approximating dynamical triangulation. Thus,
the question remains of how to keep track of which locally homogeneous
manifold is Gromov-Hausdorff approximated by the dynamical
triangulation considered. This can be done by exploiting the Gromov-
Hausdorff continuity of the homotopy type of metric spaces of bounded
geometry and by generalizing the previous remark relating the holonomy
of flat
connections to the  representations of the fundamental
group,
(see eqn. (\ref{flat})).
\vskip 0.5 cm

\subsection{Dynamical triangulations and locally homogeneous
geometries}

In order to formalize the above remarks, we need some preparatory
material on riemannian
metrics on homogeneous spaces and the way a dynamical triangulation
can approximate such particular riemannian structures. From the
mathematical point of view,
most of this material is standard and as a guide, the reader may found
profitable to consult the papers by W.Thurston\cite{Rourke} and by
W.M.Goldman\cite{Goldman} where the classical  geometry of
deformations of locally homogeneous riemannian manifolds is
beautifully discussed. As far as applications to dynamical
triangulation are concerned, the point of view developed here and in
the following paragraphs is quite new and full of deep implications
for unravelling the subtle connection between topology and simplicial
quantum gravity.

\vskip 0.5 cm
\noindent {\bf Locally homogeneous geometries}. Let $G$ be a Lie
group,
$H\subset G$ a closed (compact) subgroup, we
denote by
$X=G/H\doteq \{gH\colon g\in G\}$ the set of cosets. To each $g\in G$
we can associate the left translation $\tau_g\colon G/H\to G/H$ given
by $\tau_g(g'H)=(gg')H$ which makes $G$ act as a transitive Lie
transformation group on the homogeneous space $X$. Such a space $X$ is
riemannian homogeneous if $G/H$ is endowed with a riemannian metric
relative to which $\tau_g$ is an isometry for every $g\in G$.
A {\it locally
homogeneous} riemannian  manifold $M$ is a riemannian manifold locally
modelled on the geometry of a homogeneous space
$(X,G)$. Constant curvature metrics are a typical example. They are
locally
isometric  either to euclidean
space, the sphere, or hyperbolic space, depending on wether the
curvature is zero, positive, or negative respectively.
More generally, locally homogeneous riemannian manifolds include all
geometric structures used to uniformize surfaces or to discuss
Thurston's geometrization program for  $3$-dimensional
manifolds\cite{Scott}.
\vskip 0.5 cm
To be more precise, let $X$ be a real analytic manifold and let $G$ be
a finite dimensional Lie group acting analytically and faithfully
({\it i.e.} $gx=x$ for all $x\in X$ implies $g=id$) on $X$. We say
that a manifold $M$ has a $(G,X)$ structure if we have an open
covering $\{U_i\}$ of $M$, and coordinate charts
$\phi_i\colon U_i\to X$ with transition functions
\begin{eqnarray}
\gamma_{ij}\doteq\phi_i\cdot\phi_j^{-1}\colon
\phi_j(U_i\cap U_j)\to\phi_i(U_i\cap U_j)
\end{eqnarray}
where each $\gamma_{ij}$ agrees locally with an element of the group
$G$.
\vskip 0.5 cm
Let $M$ be a $(X,G)$ manifold and $\hat{M}$ its universal covering. If
we fix a base point $x_0$ in $M$,
$\hat{M}$ can be identified with the quotient of the set of pairs
$(x,\beta)$, where $x\in{M}$ and $\beta$ is a path from $x_0$ to $x$,
with respect to the equivalence relation
\begin{eqnarray}
(x,\beta)\simeq (x^*, \beta^*)
\end{eqnarray}
if and only if $x=x^*$ and $\beta^*\beta^{-1}\simeq 1$ in
$\pi_1(M,x_0)$. The projection $p\colon\hat{M}\to M$ being defined by
$p[(x,\beta)]=x$. We wish to recall the characterization of the {\it
Developing map} $D\colon\hat{M}\to X$ associated with a locally
homogeneous manifold $M$. The basic idea is that given a $(X,G)$
structure on a manifold M it is natural to try to make the charts
$\{(U_i,\phi_i)\}$ match up exactly and compatibly with the topology
of $M$ so as to tesselate $M$ (or portion of $M$) for suitably chosen
$\{(U_i,\phi_i)\}$, (in our case the $\{(U_i,\phi_i)\}$
will be associated with equilateral simplices or suitable deformations
of them).\par
The developing map is defined through {\it analytic continuation}
along a path $\alpha$. Fix a basepoint $x_0\in M$ and a chart
$(U_0,\phi_0)$
the domain of which contains the basepoint $x_0$. Given a path
$\alpha\colon [0,1]\to M$ starting at the basepoint subdivide it at
points $x_0=\alpha(t_0)$, $x_1=\alpha(t_1),\ldots,x_n=\alpha(t_n)$,
($t_0=0$ and $t_n=1$), in such a way that each subpath is contained in
the domain of a single coordinate chart $(U_i,\phi_i)$. Starting with
the given $\phi_0$ we adjust $\phi_1$ at $x_1$ in such a way that it
agrees with the value of $\phi_0$ at $x_q$. Since
$\gamma_{ij}\doteq\phi_i\cdot\phi_j^{-1}$ is a locally constant map
into the group $G$, the adjustement of $\phi_1$ is easily obtained by
replacing $\phi_1(x_1)$ with $\psi_1\doteq\gamma_{01}\phi_1(x_1)$. By
proceeding in this way, the generic $k$-th adjusted chart is obtained
by replacing $\phi_k(x_k)$ with
\begin{eqnarray}
\psi_k\doteq\gamma_{01}(x_1)\gamma_{12}(x_2)\ldots\gamma_{k-
1,k}(x_k)\phi_k(x_k).
\end{eqnarray}
The adjusted charts $\psi_1,\psi_2,\ldots, \psi_n$ form the analytic
continuation
\begin{eqnarray}
\phi_0^{[\alpha]}\doteq \psi_n
\end{eqnarray}
of the chosen $\phi_0$ along the path $\alpha$. It can be verified
that this analytic continuation only depends on the homotopy class
$[\alpha]$ of the path, in particular it does not depend from the open
sets $\{U_i\}$, but {\it only from the choice of the first open set}
$(U_0,\phi_0)$ containing the basepoint $x_0$. We get in this way a
well-defined application
$D\colon\hat{M}\to X$. Notice that different choices of
$(U_0,\phi_0)$ characterize applications $D$ differing by a $G$-
translation in $X$. Formally we have\cite{Rourke}
\begin{Gaiad}
For a given choice of the initial chart $(U_0,\phi_0)$ and of the
basepoint $x_0\in U_0$, the developing map of a locally homogeneous
manifold $M$ is the map $D\colon\hat{M}\to X$ defined by
$D\doteq\phi_0^{[\alpha]}\cdot{p}$ in a neighborhood of
$(x,\alpha)\in\hat{M}$.
\end{Gaiad}
Note that in the definition of developing map it is not strictly
necessary that the action of $G$ on $X$ is transitive.
\vskip 0.5 cm
\noindent {\bf Holonomy homomorphism}. We can now recall the
characterization of the {\it Holonomy} of a
locally homogeneous manifold $M$, (the holonomy representation
(\ref{flat}) is a particular case of
such a construction). Let $\sigma$ be an element of the fundamental
group of $M$, $\pi_1(M,x_0)$. To such a loop there is associated a
covering transformation $T_{\sigma}\colon\hat{M}\to\hat{M}$ in the
universal cover defined by $T_{\sigma}[(x,\tau)]\doteq
[(x,\sigma\tau)]$. Analytic continuation along the loop $\sigma$
provides $\phi_0^{[\sigma]}=\psi_n$ which is well-defined at the
basepoint $x_0$, (since $x_n=x_0$), and depends only on the homotopy
class of the loop $\sigma$. We call {\it Holonomy} of $M$ the map
$H\colon\pi_1(M)\to{G}$ defined by $\sigma\mapsto{h(\sigma)}$ where
$h_{\sigma}$ is that element  of $G$ such that
\begin{eqnarray}
\phi_0^{[\sigma]}=h_{\sigma}\phi_0
\end{eqnarray}
Since $D\cdot T_{\sigma}=h_{\sigma}\cdot D$, the holonomy map $H$ is a
group homomorphism from $\pi_1(M)$ into the structure group of the
geometry  $G$, {\it the holonomy
homomorphism}. The holonomy is determined up to conjugacy in $G$
according to the fact that the developing map is determined only up to
left composition with elements of $G$.
\vskip 0.5 cm
Note that if the developing map is a covering map, $M$ is a complete
$(G,X)$ manifold and its holonomy group characterizes $M$, in the
sense that $M$ can be reconstructed as the quotient space $X/Hol(M)$,
(see Thurston's notes\cite{Rourke} for more details).
\vskip 0.5 cm
More generally\cite{Goldman}, the developing map $D\colon\hat{M}\to X$
can be used to pull back the $(X,G)$ structure from $X$ to $\hat{M}$
which thus inherits a natural structure of locally homogeneous
riemannian manifold. Moreover, the holonomy homomorphism
$H\colon\pi_1(M)\to{G}$ determines the action of $\pi_1(M)$ on the
universal cover $\hat{M}$ by $(X,G)$ automorphism. This implies that
the pair $(D,H)$ associated with the developing map $D$ and a holonomy
representation $H$ in the structure group of the geometry, the so-
called {\it Development pair}, is a useful
globalization of a locally homogeneous structure $(X,G)$ defined on a
manifold $M$ by local coordinate charts $\{U_i,\phi_i\}$: the pair
$(D,H)$ completely determines the locally homogeneous geometry on $M$.
\vskip 0.5 cm
\noindent {\bf Moduli of locally homogeneous geometries}.
For our purposes we need to consider not just a particular locally
homogeneous structure $(M,g^*)$ on a compact manifold $M$, but rather
a {\it space} of such structures. To this end\cite{Goldman}, mark a
basepoint $x_0\in{M}$ and let $\pi_1(M,x_0)$ and $\hat{M}\to{M}$ be
the corresponding fundamental group and universal covering space. We
consider the set of homomorphisms of $\pi_1(M,x_0)$ into the structure
group of the geometry $G$, $Hom(\pi_1,G)$ endowed with the pointwise
convergence topology, {\it i.e.}, a sequence $\theta_n\in
Hom(\pi_1(M),G)$ is said to converge to a homomorphism $\theta$ if
$\lim_{n\to\infty}\theta_n(\omega)=\theta(\omega)$ for each
$\omega\in\pi_1(M,x_0)$. Let us consider, as before, the coordinate
map $\phi_0\colon{U_0\subset M}\to{X}$, in particular its value at
$x_0\in U_0$. This defines the {\it germ} at $x_0$ of the locally
homogeneous geometry of $M$.
\vskip 0.5 cm
The space
${\cal I}_{(X,G)}(M)$ of {\it isotopy classes} of based locally
homogeneous structures on $(M,x_0)$ having the same germ
$\phi_0\colon{U_0\subset M}\to{X}$ at $x_0$
can be defined\cite{Goldman} as the set
\begin{eqnarray}
{\cal I}_{(X,G)}(M)=\{(D,\theta)|\theta\in{Hom(\pi_1(M),G)}\}
\end{eqnarray}
and where $D\colon\hat{M}\to{X}$ is a $\theta$-equivariant nonsingular
smooth map, {\it viz.}, a developing map.
If we topologize ${\cal I}_{(X,G)}(M)$ using the $C^{\infty}$ topology
on the developing map $D$, then the map
\begin{eqnarray}
\hat{hol}\colon {\cal I}_{(X,G)}(M)\to Hom(\pi_1(M),G)
\end{eqnarray}
which associates  to the developing pair $(D,\theta)$ the
corresponding holonomy representation $\theta$ is continuous.
Notice that if $Diff_0(M,x_0)$ is the identity component in the group
of all diffeomorphisms $M\to M$ fixing $x_0$, then\cite{Goldman}
$Diff_0(M,x_0)$ acts properly and freely on
${\cal I}_{(X,G)}(M)$ and the map $\hat{hol}$ is invariant
under this action. It is also easily verified that $\hat{hol}$ is $G$-
equivariant also respect to the natural $G$-actions of $G$ on
the developing map and on the representation space $Hom(\pi_1(M),G)$.
In the quoted Goldman's paper\cite{Goldman}, the intersted reader can
find the statement and the proof of an equivariance slice theorem
which allows to parametrize the space
${\cal I}_{(X,G)}(M)$ of based geometric structures on $M$ in terms of
a representation $\theta\in{Hom(\pi_1(M),M)}$ and of a diffeomorphism
$h\in{Diff_0(M,x_0))}$, (see the deformation theorem at p. 178
in\cite{Goldman}). We state this result in a somehow less technical
fashion
\vskip 0.5 cm
\begin{Gaiat}
The map $\hat{hol}\colon {\cal I}_{(X,G)}(M)\to Hom(\pi_1(M),G)$ is an
open map, and for any given locally homogeneous structure $u\in {\cal
I}_{(X,G)}(M)$ there is a corresponding neighborhood $W\subset {\cal
I}_{(X,G)}(M)$ such that for any two (based) locally inhomogeneous
structures $u_1, u_2\in W$ such that
$\hat{hol}(u_1)=\hat{hol}(u_2)=\theta\in{Hom(\pi_1(M),G)}$
there is a (based) diffeomorphism $h\in Diff_0(M,x_0)$ and $\gamma\in
G$ with $(h,\gamma)\cdot u_1=u_2$.
\label{Goldman}
\end{Gaiat}
In other words, in a sufficiently small neighborhood of  a given
locally homogeneous structure, any two locally homogeneous structures
with the same holonomy representations are equivalent, and we can use
the holonomy homomorphism as a sort of coordinate map for the space of
inequivalent locally homogeneous structures that we can obtain from
the given one by local deformation.
\vskip 0.5 cm
It also follows from the above theorem that the map
\begin{eqnarray}
hol\colon\hat{I}_{(X,G)}(M)\doteq\frac{{\cal
I}_{(X,G)}(M)}{Diff_0(M,x_0)}\to Hom(\pi_1(M),G)
\end{eqnarray}
is a local homeomorphism. If we define\cite{Goldman} the {\it
Deformation Space} of locally homogeneous structures on $M$ as the
quotient
\begin{eqnarray}
Def_{(X,G)}(M)=\frac{\hat{I}_{(X,G)}(M)}{G}
\end{eqnarray}

then one would expect that the map $hol$ extends to a local
homeomorphism from $Def_{(X,G)}(M)$ and the orbit space
$Hom(\pi_1(M),G)/G$ (endowed with the quotient topology). In general
this is the case only if we have more information on the
structure of the orbit space under the $G$-action. Typically $hol$ is
well-behaved if we restrict our attention to the subspace in
$Hom(\pi_1(M),G)/G$ generated by the conjugacy classes of {\it stable}
representations. In any case we shall work mainly with
$hol\colon\hat{I}_{(X,G)}(M)\to Hom(\pi_1(M),G)$, and simply assume,
at least at this stage, that the representations considered are such
that the local properties of the representation variety
$Hom(\pi_1(M),G)/G$ describe, through the holonomy map $hol$, the
local properties of the deformation space
$Def_{(X,G)}(M)$ of locally homogeneous structures on $M$.
This is an important point for it will imply that the representations
in
$\frac{Hom(\pi_1(M),G)}{G}$ provide the missing piece of information
needed for describing the inequivalent deformations of dynamical
triangulations  approximating a given locally homogeneous riemannian
manifold.
\vskip 0.5 cm
By exploiting the preparatory material so introduced, we now describe
in details the procedure for parametrizing the deformation space of
dynamical triangulations.

\vskip 0.5 cm
\subsection{Moduli of dynamical triangulations}

Consider on a dynamical triangulation $(M,T_a)$ the set of metric
balls of radius $a$, $\{U_i(M)\}$, centered at the baricenters, $p_i$,
of the $N_{n-2}(M)$ bones $\{\sigma^{n-2}\}$ of $M$. Each such
$U_i(M)$ is homeomorphic to $B^{n-2}\times{C(link(\sigma^{n-2}))}$,
where $C(link(\sigma^{n-2}))$ is the cone on the link of $\sigma^{n-
2}$, and $B^k$ denotes the topological $k$-dimensional ball. Since
$C(link(\sigma^{n-2}))\simeq{B^2}$, we get that
$U_i(M)\simeq{B^2\times B^{n-2}}\simeq{B^n}$, {\it viz.} each metric
ball  $U_i(M)$ is indeed a topological ball. Since their radius is
$a$, the collection of such $U_i(M)$, $i=1,\ldots,N_{n-2}(M)$,
generates an open covering of $M$.
\vskip 0.5 cm
Assume that the dynamically triangulated manifold
$(M,T_a)$, endowed with the  covering $\{U_i(M)\}$,
is Gromov-Hausdorff near to a locally
homogeneous $(X,G)$ riemannian manifold (of bounded geometry)
$(\Xi,g)$.
This assumption can be translated into the requirement
that the riemannian manifold $(\Xi,g)$ admits a geodesic ball covering
$\{q_i\}_{i=1,\ldots,m}\in\Xi$  generated by metric balls  $U_i(\Xi)$,
of
radius $\epsilon=a$,  such that
\begin{eqnarray}
|d_M(p_i,p_j)-d_{\Xi}(q_i,q_j)|<a.
\label{near}
\end{eqnarray}
since, by a standard argument already exploited various times, this
bound
implies that  the Gromov-Hausdorff distance between $M$ and $\Xi$ is
such that $d_G(M,\Xi)\leq 2a$. By itself, the dynamically triangulated
manifold $M$ cannot be endowed, for any fixed value of the cut-off
$a$, with a locally homogeneous metric geometry, (with the exception

of some particular cases in dimension $n=2$). As already stressed this
may be
a source of serious ambiguities. In particular, let us assume that
the given locally homogeneous riemannian manifold $(\Xi,g)$ admits
non-trivial infinitesimal deformations (moduli) $Def_{(X,G)}(\Xi)$,
then, up to the cut-off length scale $a$, the dynamically triangulated
manifold $(M,T_a)$ is roughly-isometric to any of the inequivalent
locally homogeneous manifolds associated with the moduli of $(\Xi,g)$,
{\it i.e.}, according to (\ref{near}),
\begin{eqnarray}
d_G(M,\Xi^*)\leq 2a
\end{eqnarray}
for any $\Xi^*\in{Def_{(X,G)}(M)}$.
Thus, $(M,T_a)$ can be isometric to any locally
homogeneous structure in $Def_{(X,G)}(M)$. In order to avoid this
ambiguity, we need to {\it declare} which locally homogeneous geometry
$(M,T_a)$ is actually approximating.
\vskip 0.5 cm
To enforce such control, let us consider  $(M,T_a)$ endowed with the
covering $\{U_i(M)\}$. Choose a basepoint $p_0$ among the baricenters
$\{p_i\}$ of the bones of $M$, and let $U_0(M)$ denote the
corresponding metric ball in the cover $\{U_i(M)\}$. Finally, mark an
$n$-dimensional simplex $\sigma^n_0$ among those incident on the bone
$\sigma^{n-2}_0$ associated with the given choice of $p_0$. Notice
that with these markings we can associate to $U_0(M)$ a well-
defined element, $R_0(\sigma^{n-2}_0)$, of the orthogonal group
$O(n)$: $R_0(\sigma^{n-2}_0)$ is the rotation around the bone
$\sigma^{n-2}_0$ of the vectors defining the marked simplex
$\sigma^n_0$.
Together with these elements, let us also choose a representation
$\theta\colon\pi_1(M,p_0)\to G\supseteq{O(n)}$ of $\pi_1(M,p_0)$ into
the structure
group $G\supseteq{O(n)}$ of the locally homogeneous geometry
approximated by
$(M,T_a)$.

\vskip 0.5 cm
Since $(M,T_a)$ is Gromov-Hausdorff near to a locally homogeneous
manifold $(\Xi,g)$, there is by hypothesis an open set $U_0(\Xi)$ on
$\Xi$, corresponding to $U_0(M)$, centered around $q_0\in\Xi$, and a
corresponding map $\phi_0\colon U_0(\Xi)\to (X,G)$ associated with the
element, $R_0(\sigma_0^{n-2})$, of the orthogonal group describing the
rotation around the
bone $\sigma_0^{n-2}$. If we think of  $R_0(\sigma_0^{n-2})$ as an
element of $G\supseteq{O(n)}$ associated with the chosen marked
simplex $\sigma_0^n$, we can generate $\phi_0$ by analytical
continuation of  $\phi(q_0)\doteq R_0(\sigma^{n-2}_0)$ in $U_0(\Xi)$
along the unique simplicial loop winding around $\sigma_0^{n-2}$.
Thus, associated
with the markings of $(U_0(M),\sigma^n_0)$ on $M$ there is a local
chart $(U_0(\Xi),\phi_0)$ for the locally homogeneous geometry of
$\Xi$. Since Gromov-Hausdorff nearby metric spaces share a common
homotopy type, it also follows that
$\pi_1(M,p_0)\simeq\pi_1(\Xi,q_0)$, (provided that $a$ is small enough
so that $M$ and $\Xi$ are sufficiently near in the Gromov-Hausdorff
distance), and the representation of $\pi_1(M,p_0)$ induces
a corresponding representation that we still denote by $\theta$,
$\theta\colon\pi_1(\Xi,q_0)\to G$.
\vskip 0.5 cm
The relevance of such interplay between the markings of
$(U_0(M),\sigma^n_0)$ on $M$ and the corresponding local chart
$(U_0(\Xi),\phi_0)$ on $\Xi$, is twofold and follows from the
properties of locally homegeneous geometries recalled in the previous
paragraph. First,
the local chart $(U_0(\Xi),\phi_0)$  can be used as a basepoint for
the
development pair $(D,\theta\colon\pi_1(\Xi,q_0)\to G)$
defining the locally homogeneous structure $(\Xi,g)$ considered.
Moreover, according to the Goldman-Thurston deformation theorem
\ref{Goldman}, if the geometry $(\Xi,g)$ admits non-trivial
deformations $Def_{(X,G)}(\Xi)$, such deformations are locally
parametrized by the representation variety $Hom(\pi_1(\Xi),G)/G$.
Thus, the data needed in order to specify unambiguosly which locally
homogeneous geometry we are approximating with $(M,T_a)$ is basically
contained in {\it the association of a
holonomy homomorphism} $\theta\colon\pi_1(M,p_0)\to G$ to the
dynamical
triangulation $(M,T_a)$.

\vskip 0.5 cm
Summing up, we have the following
\vskip0.5 cm
\begin{Gaiat}
Let $(D,\hat{\theta}\colon\pi_1(\Xi,q_0)\to G)$ be the developing pair
of a locally homogeneous geometry $(\Xi,g)$ with
a non-trivial deformation space $Def_{(X,G)}(\Xi)$, and let
$(M,T_a)\in{\cal D}T_n(a,b,N)$ denote a dynamical triangulation
approximating $(\Xi,g)$. Then
the distinct locally homogeneous riemannian manifolds in
$Def_{(X,G)}(\Xi)$ are Gromov-Hausdorff approximated
by the pairs $[(M,T_a)_{root},\theta]$ where:
{\it (i)} $(M,T_a)_{root}$ is the {\bf rooting} of the dynamical
triangulation $(M,T_a)$ generated by  the marking of a bone
$\sigma_0^{n-2}$ of $(M,T_a)$, of  a simplicial loop $\omega_0$
winding around $\sigma_0^{n-2}$, and of a base simplex
$\sigma^n_0\in\omega_0$, while {\it (ii)} $\theta$ is a holonomy
homomorphism \begin{eqnarray}
\theta\colon\pi_1(M,p_0)\to G\supseteq{O(n)}
\end{eqnarray}
varying in a suitable neighborhood of $\hat{\theta}$ in
the representation variety $\frac{Hom(\pi_1(M),G}{G}$.
\end{Gaiat}
\vskip 0.5 cm
{\bf Proof}. The markings in {\it (i)} are simply the markings
characterizing the data
$(U_0(M),\sigma^n_0)$ on $M$ inducing  a local chart
$(U_0(\Xi),\phi_0)$ for the locally homogeneous geometry of $\Xi$,
through which $\Xi$ can be {\it unrolled} by means of the developing
map $D$. These data are necessary for providing a well-defined
holonomy representation. Henceforth we shall refer to dynamical
triangulations with
such marked simplices as {\bf rooted triangulations} for short.\par
As we have recalled\cite{Goldman}, the map
\begin{eqnarray}
\hat{hol}\colon{Def_{(X,G)}(\Xi)}\to Hom(\pi_1(\Xi),G)
\end{eqnarray}
which associates to the developing pair $(D,\theta)$, ({\it i.e.}, to
the locally homogeneous manifold $(\Xi,g)$), the holonomy
representation $\theta$ is a continuous and open map.
Two locally homogeneous manifolds $\Xi_1$ and $\Xi_2$ in
$Def_{(X,G)}(\Xi)$ are near in the $C^{\infty}$ topology (on the
respective developing maps $D_1$ and $D_2$) and the corresponding
geodesic ball coverings $\{U_i(\Xi_1)\}$ and $\{U_i(\Xi_2)\}$ are such
that
$d_G(\Xi_1,\Xi_2)\leq \epsilon$ where $\epsilon$  depends on the
$C^{\infty}$-norm on the respective developing pairs $||(D,\theta_1)-
(D,\theta_2)||_{\infty}$. It follows that $(M,T_a)$ is Gromov-
Hausdorff near any locally homogeneous manifolds in
$Def_{(X,G)}(\Xi)$.

By choosing, if necessary, a sufficiently small neighborhood of
$(\Xi,g)$ in $Def_{(X,G)}(\Xi)$, we can assume that $(M,T_a)$
shares a common homotopy type with all $(\Xi^*,g)$ in
$Def_{(X,G)}(\Xi)$. Thus
an immediate
application of the Goldman-Thurston deformation theorem \ref{Goldman}
implies that we can establish a one-to-one correspondence between
each pair  $[(M,T_a)_{root},\theta]$ and a corresponding locally
homogeneous manifold in $Def_{(X,G)}(\Xi)$, simply by declaring that a
given  $[(M,T_a)_{root},\theta]$ is associated with the locally
homogeneous manifold in $Def_{(X,G)}(\Xi)$ defined by the holonomy
homomorphism $\theta$, and this correspondence is well-defined
for $\theta$ varying in a suitable neighborhood of
$\hat{\theta}$ in $\frac{Hom(\pi_1(M),G}{G}$. $\Box$
\vskip 0.5 cm
Thus, with a slight abuse of language, it seems appropriate to
consider the representation variety $\frac{Hom(\pi_1(M),G}{G}$
as the {\bf Moduli Space} for a dynamically triangulated manifold
$M\in{\cal D}T_n(a,b,N)$.
\vskip 0.5 cm
Since we are interested in closed manifolds $M$, the study of this
moduli space is basically equivalent to the study of discrete
cocompact subgroups $\Gamma$ of the maximal isometry group $G$ that
can be ammitted by $M$, (recall that a discrete subgroup
$\Gamma\subset G$ is called cocompact if the quotient space $G/\Gamma$
is compact). The underlying geometric rationale is that the Lie group
in question is the isometry group of the simply connected homogeneous
riemannian manifold $X$ which can be approximated by the (universal
cover of the) dynamical triangulation $M$. The discrete subgroup
$\Gamma$ appears (if it has no torsion) as the deck transformation
group of the covering $X\to\Xi$ where $\Xi=X/\Gamma$ is the smooth
riemannian manifold approximated by $M$, thus $\Gamma\simeq \pi_1(M)$,
(if $\Gamma$ has torsion we will be dealing with orbifolds). It is
intersting to remark that even if not every semisimple Lie group $G$
may be realized as the isometry group of some symmetric space $X$,
every such $G$ is locally isomorphic, ({\it viz.} the isomorphism is
at the level of the universal coverings of the groups in question), to
the isometry group of some symmetric space $X$. The existence of a
non-trivial moduli space for dynamical triangulations (of closed
manifolds) is thus reduced to the study of the {\it rigidity of
lattices} in such groups $G$\cite{Gromovpansu}.
\vskip 0.5 cm
For the convenience of the reader, we have collected in an Appendix
some basic material on the geometry of the representation variety
$\frac{Hom(\pi_1(M),G}{G}$. Here we
stress that since $\pi_1(M)$ is
a finitely generated group, (say with $m$ generators),
$Hom({\pi}_1(M),G)$ is an analytic subvariety\cite{Goldman} of $G^m$,
(as a variety
is defined by the $m$-tuples $(\Theta(1),\ldots,\Theta(m))\in G^m$
such that $\rho_i(\Theta(1),\ldots,\Theta(m))=1$ for each relation
$\rho_i$ associated with the given presentation of the group
$\pi_1(M)$). Note that on applying a theorem due to X.
Rong\cite{Rong}, one can actually provide a upper bound to the minimal
number of generators for $\pi_1(M)$ for $M\in{\cal D}T_n(a,b,N)$, (the
bound depends on the parameters $a$,$b$, and $N$).
\vskip 0.5 cm
\subsection{Gauge-Fixing of the  moduli of a
Dynamical Triangulation}

Far most of our efforts will be bent toward an enumeration of the
distinct dynamical triangulations a PL manifold can carry. Since
distinct dynamical triangulations (of bounded geometry) are meant to
approximate distinct riemannian structures, we are actually addressing
the characterization of a measure
on the space of riemannian structures. As we have seen in the previous
sections, the correspondence between dynamical
triangulations and riemannian structure ceases to be one-to-one nearby
locally homogeneous
manifolds, no
matter how fine is the triangulation. Distinct moduli of locally
homogeneous riemannian manifolds
may indeed correspond, through a R\"{o}mer-Z\"{a}hringer deformation,
to a same dynamical triangulation. In order to carry over an
enumeration of distinct dynamical triangulations while preserving a
one-to-one correspondence between riemannian structures and dynamical
triangulations, we have have to impose a suitable {\it gauge fixing}
so as to remove this possible source of ambiguity.
The situation is akin to the one familiar in gauge theories where, in
order to introduce  a suitable path integration on the configuration
space of the theory, we have to remove gauge freedom, (by using a
slice theorem for the action of the gauge group), to obtain a local
(formal) measure on the orbit space labelling the gauge equivalent
configurations.  Eventually, as in any gauge fixing
procedure, in order to obtain the correct count of the number of
distinct dynamical triangulations, we have to divide the number of
distinct holonomies by the {\it local volume} of the moduli space
parametrizing the conjugacy classes of representation
$\Theta\colon\pi_1(M;\sigma^n_0)\to G$ in a neighborhood of the chosen
representation. From a combinatorial point of
view, the necessity of this factorization comes from the fact that the
number of distinct holonomies provides the number of distinct
dynamical triangulation {\it in a given holonomy representation of the
fundamental group}, namely it is a labelled or {\it rooted}
enumeration. The unrooted enumeration can be obtained from the rooted
one
by dividing by all possible labellings of the inequivalent
representations.
\vskip 0.5 cm
\noindent {\bf A heuristic remark}. In connection with this counting
problem for
dynamical
triangulation,
it is worth remarking a subtle point. Had we addressed the counting of
just distinct (in the sense of Tutte) dynamical triangulations on a PL
manifold of given volume and topological type, the above gauge fixing
problem would have not explicitly appeared. The fact is that in the
definition of equivalence according to Tutte, no mention is made of
the metric properties of the triangulation considered, and thus in the
process of counting, we enumerate also triangulations potentially
fitting on locally homogeneous spaces. The requirement of
triangulating with equilateral simplices, is a restriction of no
consequence as long as we consider generic riemannian manifold, but
quite effective when coming to triangulating riemannian manifolds of
large symmetry.
We enumerate distinct triangulations basically by considering distinct
curvature assignements, {\it viz.}, namely using as labels for
distinct triangulations the distinct Levi-Civita connections generated
by distinc curvature assignements.
This strategy can be considered as the discretized counterpart of the
techniques underlying  uniformization theory for $2$-dimensional
surfaces or the geometrization program for $3$-dimensional manifolds.
The simplest example is provided by $2$-dimensional surfaces. In such
a case, a given riemannian metric can be conformally deformed to a
metric of constant curvature. The conformal factor is related (via the
coformal Laplacian) to the gaussian curvature of the surface, and in
this sense the datum of the local curvature seems sufficient to
reconstruct the original surface. This fails since there are non-
equivalent infinitesimal deformations of the base constant curvature
metrics (moduli), and in order to reconstruct the metric, we need
curvature assignements and the datum of which constant curvature
metric, ({\it e.g.}, which flat metric, if the surface in question is
a $2$-torus), is used. Actually, there is a further subtle problem
related to the fact that the action of the semi-direct product
of the group of diffeomorphisms and of the conformal group  makes
difficult an actual reconstruction of the metric from the datum of the
local  curvature and of the moduli. But at this heuristic level, the
parallel between the geometric setup of uniformization theory and the
reconstruction of dynamical triangulations from the
curvature assignements is appropriate.

\vskip 0.5 cm
{\bf A measure on  the moduli space}. According to the above remarks,
in
order to complete our gauge-fixing
procedure, we have to introduce a measure on the moduli space ${\cal
M}(T_a)$ of a given dynamical triangulation.
This procedure is related to the characterization of measures
on the representation variety
$\frac{Hom({\pi}_1(M),G)}{G}$. In
dimension two, the study of such measures  has
been considerably developed by exploiting its connection with the
semiclassical limit of Yang-Mills theory on Riemann
surfaces\cite{Witten}. For dynamical triangulations in dimension
greater than two, the connection with measure theory on
$\frac{Hom({\pi}_1(M),G)}{G}$ has been stressed
in\cite{Carfora} as the origin of the possible critical behavior in
higher dimensional simplicial quantum gravity.
\vskip 0.5 cm
Let us consider the
representation, $\theta$, of $\pi_1(M)$ on the Lie algebra ${\frak g}$
generated by
composing the given  representation
${\Theta}\colon{\pi}_1(M)\to{G}$ with the adjoint action of G on
${\frak g}$, {\it viz.},
${\theta}\colon{\pi}_1(M)\to_{\Theta}{G}\to_{Ad} End({\frak g})$,
(henceforth we
will always  refer to this representation ).
 The  tangent space to
$\frac{Hom({\pi}_1(M),G)}{G}$ corresponding to the conjugacy
class of representations
$[\theta]$ is provided by $H^1(\pi_1(M),{\frak g})$, the first
cohomology group of
$\pi_1(M)$ with values in the Lie algebra ${\frak g}$. It is a known
fact that this cohomology group is isomorphic to
the first cohomology group of the dynamically triangulated manifold
$M=|T_a|$, with
coefficient in the adjoint  bundle $ad(\theta)$ defined by the
representation $\theta$, {\it i.e.}, $H^1(M,ad(\theta))$,
(for more details see the Appendix). Notice that
$H^1(M,ad(\theta))$ is only a formal tangent space to
$\frac{Hom({\pi}_1(M),G)}{G}$, since the representation variety
has singularities in correspondence to
reducible representations and/or linearization unstable
representations.
\vskip 0.5 cm

The choice of a well-defined holonomy representation amounts to the
choice of a map (basically a section)
$S\colon\frac{Hom({\pi}_1(M),G)}{G}\to Hom({\pi}_1(M),G)$
which associates to
a conjugacy class, $[\theta]$, of representations, the holonomy
representation $\theta\colon\pi_1(M,p_0)\to \frak{g}$, based at
the chosen basepoint.
\vskip 0.5 cm

Let $\delta G$ denote the generic (infinitesimal)
deformations, at $\theta$, of the chosen slice $S$. Note that
$\delta G\in Z^1(\pi_1(M);\frak{g})$, the space of $1$-cocycle on
$\pi_1(M)$ with coefficients in the Lie algebra $\frak{g}$.
This is
the (formal) tangent space to $Hom(\pi_1(M),G)$ at the chosen
representation
$\theta$. A component of the deformation $\delta G$ may be trivial
in the sense
that it may lie in the tangent space to the $Ad$-orbit containing
$\theta$. This part of the deformation maps, by $Ad$-conjugation,
$S(\theta)$ to a deformed representation $\tilde{\theta}$ in a
nearby slice $S({\tilde{\theta}})$ with
$[{\tilde{\theta}}]=[\theta]$. The non-trivial part of the deformation
lies in the tangent space to the chosen slice $S$, {\it viz.},
$TS[\frac{Hom({\pi}_1(M),G)}{G}]\simeq H^1(\pi_1(M),\frak{g})$. This
latter part of the deformation maps
$\theta$ in a neighboring representation which cannot be obtained by
$Ad$-conjugation from $\theta$. As recalled, deformations in
$TS[\frac{Hom({\pi}_1(M),G)}{G}]$ can be interpreted as inequivalent
R\"{o}mer-Z\"{a}hringer deformations of the given dynamical
triangulation.
\vskip 0.5 cm
A natural\cite{Witten},\cite{Forman} volume element on the space of
such
deformations is  the product of a volume form on the Lie algebra
$\frak{g}$ and of a
volume on $TS[\frac{Hom({\pi}_1(M),G)}{G}]$. In order to define
the former,
consider an Haar measure on the group $G$, with total volume $Vol(G)$,
and let $y_1,y_2,\ldots,y_t$ be Euclidean coordinates on the Lie
algebra $\frak{g}$. We
choose the $y_h$ in such a way that the measure $\prod_h^tdy_h$ on
$\frak{g}$ is the chosen Haar measure at the identity element of $G$.
Analogously, we can introduce a
measure, $d\mu$, on the tangent space to the
chosen representative of the (smooth component of the)
variety $S[\frac{Hom({\pi}_1(M),G)}{G}]\subset
Hom({\pi}_1(M),G)$. Recall that  $\frac{Hom({\pi}_1(M),G)}{G}$ is an
analytic subvariety of $G^m$, $m$ being the number of generators of
$\pi_1(M)$. The fixed Ad-invariant inner product on $\frak{g}$ induces
a Riemannian structure on $G^m$, and hence on
$TS[\frac{Hom({\pi}_1(M),G)}{G}]$, (at least on its smooth
component,
to which we are restricting our considerations). The corresponding
inner product between tangent vectors in $H^1(M,ad(\theta))$, will be
denoted by $<<\cdot,\cdot>>$, and $d\mu$ can be thought of as the
associated volume measure. We shall denote by
\begin{eqnarray}
{\cal V}{\doteq} vol[\frac{Hom({\pi}_1(M),G)}{G}]
\end{eqnarray}
the volume of the moduli space
$\frac{Hom({\pi}_1(M),G)}{G}$ in this measure.
We cannot use $d\mu$ as such since we want to localize, as the
representation variety becomes smaller and smaller, the choice of the
representations $\theta$ around the trivial representation. This is
dictated by the requirement that as
${\cal V}\to 0$, ({\it e.g.}, for simply connected manifolds),
the measures should converge to the  atom concentrated on the
representation $[\omega, p_0]\mapsto Id_G$, where
$[\omega, p_0]$ denotes the trivial loops based at the chosen
basepoint $p_0$.

To this end,  we normalize the product measure
$d\mu\times\frac{\prod_h^t dy_h}{Vol(G)}$ on
$\frak{g}\times TS[\frac{Hom({\pi}_1(M),G)}{G}]$,
with the heat kernel on the analytic manifold
$Hom({\pi}_1(M),G)\subset G^m$, and set
\begin{eqnarray}
d\aleph\doteq  K\left (N_n vol(\sigma^n){\cal V},\theta,
Id\right )
d\mu\times\frac{\prod_h^t dy_h}{Vol(G)}
\end{eqnarray}
where
$\theta$ is the given representation
in $Hom({\pi}_1(M),G)$, $Id$ is the trivial representation,
$vol(\sigma^n)$ is the volume of the elementary simplex $\sigma^n$,
and
\begin{eqnarray}
K(t,x,y)\colon {\Bbb R}^+\times Hom({\pi}_1(M),G)\times
Hom({\pi}_1(M),G)\to {\Bbb R}
\end{eqnarray}
is the fundamental solution of the heat equation on
$Hom({\pi}_1(M),G)$, (since $Hom({\pi}_1(M),G)$ is an analytic
subvariety of $G\times G\times\ldots\times G$, $K(t,x,y)$ can be
constructed with the heat kernel on $G$). Notice that $N_n
vol(\sigma^n){\cal V}$ can be interpreted as the volume of the
dynamically triangulated manifold $M$ as {\bf expressed in terms of
the
volume of the representation variety} $Hom({\pi}_1(M),G)/G$.

\vskip 0.5 cm
Thus, the natural normalizing
volume element we consider in the procedure of gauge fixing
the
R\"{o}mer-Z\"{a}hringer deformations of a dynamical triangulation
is
\vskip 0.5 cm
\begin{eqnarray}
Vol[{\cal  M}(T_a)]\doteq \int_{\frak{g}\times
S[\frac{Hom({\pi}_1(M),G)}{G}]}
K\left (N_n vol(\sigma^n){\cal V},\theta, Id\right )
d\mu\times\frac{\prod_h^t dy_h}{Vol(G)}
\label{loc}
\end{eqnarray}
\vskip 0.5 cm

\noindent {\bf Moduli asymptotics}. If we exploit the standard
properties of the
heat kernel, we can get an explicit asymptotic expression
for $Vol[{\cal  M}(T_a)]$ in the limit
of small volume for the representation variety.
\begin{Gaiat}
For a given value of $N_n$ and for
${\cal V} vol(\sigma^n)$ near zero,
the leading contribution to $Vol[{\cal  M}(T_a)]$ is
given by
\vskip 0.5 cm
\begin{eqnarray}
Vol[{\cal  M}(T_a)]\to_{{\cal V} vol(\sigma^n)<<1}\to
\nonumber
\end{eqnarray}
\begin{eqnarray}
 A({\cal V})\cdot (2\pi N_n{\cal V} vol(\sigma^n))^{-D/2}
(1+O(N_n{\cal V} vol(\sigma^n)))
\label{variety}
\end{eqnarray}
where $D={\doteq} dim [Hom({\pi}_1(M),G)]$, and $A({\cal V})$ is a
function of ${\cal V}$
\label{modfactor}
\end{Gaiat}
 (The explicit expression of $A({\cal V})$ is
provided in the proof).\par
\noindent {\bf Proof}. We shall consider $Hom({\pi}_1(M),G)\subset
G^m$ as a
compact manifold with a Riemannian structure induced by the Cartan-
Killing metric on $\frak{g}$.
According to standard estimates\cite{Davies} on the heat kernel on a
compact Riemannian manifold, there is a neighborhood $U$ of the
diagonal in $Hom({\pi}_1(M),G)\times Hom({\pi}_1(M),G)$ such that for
any pair of representations $(\theta_1,\theta_2)\in U$ and
$Vol[Hom({\pi}_1(M),G)/G]$ near $0$, we  can write
\begin{eqnarray}
K\left (N_n{\cal V} vol(\sigma^n),
\theta_1,\theta_2\right)=\nonumber
\end{eqnarray}
\begin{eqnarray}
(2\pi N_n{\cal V} vol(\sigma^n))^{-D/2}
\exp
\left(-\frac{|\theta_1-\theta_2|^2}{2N_n{\cal V} vol(\sigma^n)}
\right)(1+O(N_n{\cal V} vol(\sigma^n)))
\label{heat}
\end{eqnarray}
where
$|\theta_1-\theta_2|$ denotes the riemannian distance, (in the induced
Cartan-
Killing norm), on $Hom({\pi}_1(M),G)\subset G^m$.

\vskip 0.5 cm
Set $t{\doteq} N_n{\cal V} vol(\sigma^n)$, and let us
consider\cite{Millson},\cite{Goldman}
the representations $\theta$, for $t$ near $0$, as  one-parameter
family of
representations $\theta^t$
\begin{eqnarray}
\theta^t=\exp[tu(a)+O(t^2)]\theta(a)
\end{eqnarray}
where $a\in {\pi}_1(M)$, and where $u\colon{\pi}_1(M)\to{\frak g}$.
Given $a$ and $b$ in ${\pi}_1(M)$, if we differentiate the
homomorphism condition $\theta^t(ab)=\theta^t(a)\theta^t(b)$, we
get that the
$u$ are  one-cocycles of ${\pi}_1(M)$ with coefficients
in the ${\pi}_1(M)$-module ${\frak g}_{\theta}$,
\begin{eqnarray}
u(ab)=u(a)+ [Ad({\theta}(a))]u(b)
\end{eqnarray}
namely, $u\in Z^1(\pi_1(M),{\frak g})$, the tangent space to
$Hom{({\pi}_1(M),G)}$.  Among the vectors $u\in Z^1(\pi_1(M),{\frak
g})$, those tangent to the $Ad$-orbit are of the form
$u(a)=[Ad({\theta}(a))]h-h$, for some $h\in\frak{g}$, these are $1$-
coboundaries on $\pi_1(M)$ with coefficients in the representation,
{\it i.e.}, elements of $B^1(\pi_1(M),{\frak g})$. We can decompose
$u$ in a component  $P_Hu\in H^1(\pi_1(M),{\frak
g})=Z^1(\pi_1(M),{\frak g})/B^1(\pi_1(M),{\frak g})$ and in a
component $P_{\frak{g}}u\in\frak{g}$, $P_H$ and $P_{\frak{g}}$
denoting the respective projection operators.
Thus
\begin{eqnarray}
\theta^t-Id=t (P_Hu+P_{\frak{g}}u)+O(t^2)
\end{eqnarray}
Introducing this in (\ref{heat}) we get
\begin{eqnarray}
K\left (N_n{\cal V} vol(\sigma^n),
\theta,Id\right)=\nonumber
\end{eqnarray}
\begin{eqnarray}
(2\pi N_n{\cal V} vol(\sigma^n))^{-D_H/2-D_{\frak{g}/2}}
\exp
\left(-\frac{t(|P_Hu|^2+|P_{\frak{g}}u|^2)}{2N_n{\cal V}
vol(\sigma^n)}
\right)(1+O(t))
\label{components}
\end{eqnarray}
where $D_H$ is the dimension of the smooth component of
$\frac{Hom({\pi}_1(M),G)}{G}$ and $D_{\frak{g}}=dim G$.
Integrating (\ref{components}) over
$\frak{g}\times S[\frac{Hom({\pi}_1(M),G)}{G}]$ and evaluating the
gaussian integral over $\frak{g}$ we get to leading
order:
\begin{eqnarray}
\int_{\frak{g}\times S}(2\pi N_n{\cal V} vol(\sigma^n))^{-D_H/2-
D_{\frak{g}/2}}
\exp
\left(-\frac{t(|P_Hu|^2+|P_{\frak{g}}u|^2)}{2N_n{\cal V}
vol(\sigma^n)}
\right)=\nonumber
\end{eqnarray}
\begin{eqnarray}
(2\pi N_n{\cal V} vol(\sigma^n))^{-D_H/2-D_{\frak{g}}/2}\int_{S}
[det P_{\frak{g}}\cdot det P_{\frak{g}}^*]^{-1/2}\exp
\left(-\frac{|P_Hu|^2}{2N_n{\cal V} vol(\sigma^n)}
\right)
\label{int}
\end{eqnarray}
which, upon setting
\begin{eqnarray}
A({\cal V})\doteq
\int_{S}
[det P_{\frak{g}}\cdot det P_{\frak{g}}^*]^{-1/2}\exp
\left(-\frac{|P_Hu|^2}{2N_n{\cal V} vol(\sigma^n)}
\right)
\end{eqnarray}
yields the stated result. $\Box$
\vskip 0.5 cm
Note that the computation of the integral $A({\cal V})$, extended over
the moduli space $\frac{Hom({\pi}_1(M),G)}{G}$, can be related to the
theory of Reidemeister torsion\cite{Witten},\cite{Forman}.
\vskip 0.5 cm

\noindent {\bf Moduli asymptotics for DT-surfaces}. Both the
presence of the
volume of the representation variety
and of its dimension, are an important aspect of (\ref{variety}).
In dimension $n=2$, we can be more explicit, and
if we gather information on the dimension of
$\frac{Hom({\pi}_1(M),G)}{G}$\cite{Goldman},\cite{Bergery}, then in
the case of $2$-dimensional surfaces theorem \ref{modfactor}
specializes to the following

\begin{Gaial}
For $n=2$ and for
${\cal V} vol(\sigma^n)$ near zero, the leading contribution to
$Vol[{\cal M}(T_a)]$ is given as a function of surface topology by:
\vskip 0.5 cm
(i) The sphere ${\Bbb S}^2$:
\begin{eqnarray}
Vol[{\cal M}(T_a)]\to_{{\cal V} vol(\sigma^n)<<1}\to 1
\end{eqnarray}
\vskip 0.5 cm
(ii) The torus ${\Bbb T}^2$:
\begin{eqnarray}
Vol[{\cal M}(T_a)]\to_{{\cal V} vol(\sigma^n)<<1}\to
 A({\cal V})\cdot  (2\pi N_n{\cal V} vol(\sigma^n))^{-5/2}
\end{eqnarray}
\vskip 0.5 cm
(iii) Orientable surfaces $\Sigma_h$ of genus $h>1$, and for
representations of $\pi_1(\Sigma_h)\to G$ whose center has a constant
dimension $dim Z$:
\begin{eqnarray}
Vol[{\cal M}(T_a)]\to_{{\cal V} vol(\sigma^n)<<1}\to
 A({\cal V})\cdot  (2\pi N_n{\cal V} vol(\sigma^n))^{-[(2h-1)dim
G+dim Z]/2}
\end{eqnarray}
\label{boh}
\end{Gaial}
 \vskip 0.5 cm
(If we have a non-trivial centralizer for the representation, then we
have to slightly alter the argument in theorem \ref{modfactor}).
\vskip 0.5 cm
\noindent {\bf Proof}. For a sphere $dim[H^1(\pi_1(M),{\frak g})]=0$,
and
trivially ${\cal V}=0$. As ${\cal V}\to 0$, the measure $d\aleph$
weakly converges to $\delta(Id)$, where $\delta(Id)$ is
the Dirac distribution concentrated on the trivial representation.
This immediately yields the stated result.

The torus ${\Bbb T}^2$, can be actually considered as a limiting case
of (iii). From \cite{Goldman} it follows that if
$z(\theta)$ denotes the centralizer
of
${\theta}({\pi}_1(\Sigma_h))$ in $G$, then the dimension of
$Hom({\pi}_1(\Sigma_h),G)$ is given by
 \begin{eqnarray}
(2h-1)dim(G)+dim(z(\theta))
\label{dimension}
\end{eqnarray}
By taking $G=SL(2,{\Bbb R})$, and setting $h=1$  we immediately get
the stated result for ${\Bbb T}^2$. The general result for
$h>1$ is a trivial consequence of the dimensionality formula
(\ref{dimension}). $\Box$.
\vskip 0.5 cm
We wish to stress that these results (in particular the asymptotics of
the volume of moduli space) refer only to the smooth component of the
representation space $Hom({\pi}_1(M),G)/G$. Singularities are
present in correspondence with reducible representations and around
representations which are linearization unstable \cite{Goldman}.
Also,  $Hom({\pi}_1(M),G)/G$ may have many distinct connected
components. For example, in the relevant case
of $G=PSL(2,{\Bbb R})$, $Hom({\pi}_1(M),PSL(2,{\Bbb R}))$ has
$2^{2h+1}+2h-3$ connected components,\cite{Goldman}. In $2$d-Yang-
Mills, it is a known fact that, at least in the semiclassical limit,
the
singular component of the representation variety does not contribute
to the volume of the moduli space\cite{Forman}, \cite{Witten},
\cite{King}. Since there are many points of contact between our
approach in evaluating the asymptotics of the volume of the moduli
space of R\"{o}mer-Z\"{a}hringer deformations of a dynamical
triangulation, and $2$d-Yang-Mills, we may reasonably assume that the
singular part of the representation variety is not playing any
relevant
role in our case, at least in dimension $n=2$. For dimension $n>2$
this is an open issue under current investigation.
As we shall see in due course, the asymptotic
of the volume of the moduli space of R\"{o}mer-Z\"{a}hringer
deformations plays a basic role in simplicial quantum gravity being
related
to the entropy exponent of the
theory, thus these matters are quite relevant in addressing on
analytical grounds the existence of the continuum limit of the theory.
\vskip 0.5 cm
\noindent {\bf A compact formula for surfaces}. We conclude this
section by
providing a compact expression for the the asymptotics of the volume
of
the moduli space of R\"{o}mer-Z\"{a}hringer deformations of a
dynamical triangulated surface. In the two-dimensional case,
according to lemma \ref{boh} the small volume behavior of
$Vol[{\cal M}(T_a)]$ is
\begin{eqnarray}
 \tilde{A}_h({\cal V})\cdot N_n^{\xi(\Sigma_h)}
\end{eqnarray}
where the constant $\tilde{A}_h({\cal V}){\doteq}
A({\cal V})(2\pi {\cal V} vol(\sigma^n))^{-[(2h-1)dim G+dim Z]/2}$,
(for genus $h\geq 1$, whereas $\tilde{A}_h({\cal V}){\doteq} A({\cal
V})$ for the sphere), and $\xi(\Sigma_h)$ is an exponent depending
from the genus, and the group $G$. This exponent can be given a more
compact expression if we introduce
an {\it effective dimension}, $D_{eff}$, of $Hom({\pi}_1(M),G)$
according to
\vskip 0.5 cm
\begin{eqnarray}
D_{eff}{\doteq} (2h-1)\alpha dim(G)+\alpha dim(z(\theta)) +\beta
\label{effdimension}
\end{eqnarray}
\vskip 0.5 cm
\noindent where the parameters $\alpha$ and $\beta$ are determined by
fitting (\ref{effdimension}) with the dimensions of the singular case
provided by the sphere ${\Bbb S}^2$ and the limiting case of the torus
${\Bbb T}^2$. Explicitly, according to lemma \ref{boh}, we have
\begin{eqnarray}
\xi(\Sigma_h)=-(2h-1)\frac{\alpha}{2} dim(G)-\frac{\alpha}{2}
dim(z(\theta)) -\frac{\beta}{2}+1
\end{eqnarray}
with
\begin{eqnarray}
\frac{\alpha}{2} dim(G)-
\frac{\alpha}{2} dim(z(\theta))-\frac{\beta}{2}+1&= & \xi({\Bbb S}^2)=
1\nonumber\\
-\frac{\alpha}{2} dim(G)-
\frac{\alpha}{2} dim(z(\theta))-\frac{\beta}{2}+1&= & \xi({\Bbb
T}^2)=  1-\frac{5}{2}
\end{eqnarray}

In this way we get
\begin{eqnarray}
\alpha &= & \frac{5}{2dim G}\nonumber\\
\beta &= & \frac{5}{2}(1-\frac{dim(z(\theta))}{dim G})
\end{eqnarray}
and
\vskip 0.5 cm
\begin{eqnarray}
\xi(\Sigma_h)=\frac{5}{4}\chi(\Sigma_h)-\frac{5}{2}
\label{expeffettivo}
\end{eqnarray}
where $\chi(\Sigma_h){\doteq} 2-2h$ is the Euler-Poincare
characteristic of the surface $\Sigma_h$.
These observations establish the following
\begin{Gaial}
The small {\it volume} limit
 asymptotics of the volume of the moduli space of R\"{o}mer-
Z\"{a}hringer deformations of a dynamical triangulated surface is
given by
\begin{eqnarray}
Vol[{\cal M}(T_a)]\to_{{\cal V} vol(\sigma^n)<<1}\to
 \tilde{A}_h({\cal V})\cdot N_n^{\frac{5}{4}\chi(\Sigma_h)-
\frac{5}{2}}
\label{compact}
\end{eqnarray}
\label{twocompact}
\end{Gaial}
\vfill\eject

\section{Curvature assignements for dynamical triangulations}

According to the analysis of the previous sections, the topological
information associated with a dynamical triangulation is encoded in
the moduli space of its R\"{o}mer-Z\"{a}hringer deformations. Here,
our purpose is to characterize the metric properties of the
triangulations in terms of the (restricted) holonomy of its natural
Levi-Civita connections, and reduce the enumeration of distinct
dynamical triangulations to the enumeration of distinct curvature
assignements.
\vskip 0.5 cm
\noindent {\bf Simplicial Lassos}. We start by introducing the
following
\begin{Gaiad}
A simplicial loop $\nu$, based at a simplex $\sigma^n_0$, is
a (small) {\it lasso with nose} at the bone $B$, if it
can be
decomposed into three simplicial curves
$\nu(B)=\tau^{-1}\cdot\omega(B)\cdot\tau$,
where $\tau$ is a simplicial
path from $\sigma^n_0$ to one of the simplices $\sigma^n(B)$
containing
the bone $B$, $\tau^{-1}$ is the same curve going backward, and
$\omega(B)$ is the unique loop winding around the bone $B$.
\end{Gaiad}

Since the open stars of the bones
provide an open covering of $M=|T_a|$,  any simplicial loop in $M$, if
it
is homotopic to zero, is a product of {\it lassos} whose nose is
always contained in the  open star of some bone $B$, (this is the
{\it Lasso lemma}\cite{Besse}). Thus, in order to evaluate the
holonomy
around a generic contractible simplicial loop based at a marked
simplex $\sigma_0^n$ we can proceed as follows.\par
\vskip 0.5 cm
Let us fix a holonomy representation of the fundamental group
of $M$, $\theta\colon\pi_1(M)\to G\to
End(\frak{g})$. Recall that the choice of such $\theta$
calls for the marking of a bone $\sigma^{n-2}_0$, (whose barycenter
provides the basepoint $p_0$), of a
simplex $\sigma^n_0\supset\sigma^{n-2}_0$, and of the holonomy around
the marked bone.
As the notation suggests, we naturally identify the marked simplex
$\sigma^n_0$ with the simplex on which the simplicial lassos are
based, also, for notational simplicity, we
let $\lambda+1= N_{n-2}(T)$ denote the number of  bones in the
dynamical triangulation $T_a$.
Let us consider the set of simplicial lassos, $\{\nu(\alpha),
\sigma_0^n\}$
based at the marked $\sigma^n_0$ and with their noses
corresponding to the $\lambda+1$ bones
$B_0, B(1),\ldots,B(\lambda)$. The effect of parallel transporting,
according to the Levi-
Civita connection associated with $T_a$, along any of these
lassos is a rotation in a two-
dimensional plane, (orthogonal to the bone considered).
If $\omega$ is the generic (contractible) simplicial
loop  based at $\sigma_0^n$, then the holonomy along the generic loop
$\omega$ based at
$\sigma_0$, can be
written as
\begin{eqnarray}
R_{\omega}(\sigma_0)=
\prod_{\alpha=0}^{\lambda}
\{ A(\alpha)^{-1}R(m(\alpha)\phi(\alpha))A(\alpha) \}
\label{momentum}
\end{eqnarray}
where
$m(\alpha)\in {\Bbb Z}$ are the integers providing the winding numbers
of $\omega$ around the various bones,  $R[m(\alpha)\phi(\alpha)]$
denote the rotations in the planes orthogonal to the bones
$B(\alpha)$, and finally, $A(0)$, $A(\alpha)$ denote the orthogonal
matrices describing the parallel transport along the ropes,
$\tau_{\alpha}$, of the lassos,
($A(0)$ is associated with the trivial path $\sigma^n_0$).\par
\vskip 0.5 cm
On rather general grounds, it is
known\cite{Anandan} that through holonomy we can reconstruct the
underlying connection, (up to conjugation). In our case, as already
stressed, this reconstruction procedure is particularly relevant since
distinct connections are in correspondence with distinct
triangulations. It follows that the conjugacy class of the
(restricted) holonomy $R_{\omega}(\sigma_0)$ is $G$-invariant and
fully describes the underlying dynamical triangulation in the given
representation $\Theta\colon\pi_1(M,p_0)\to G$. The natural
class functions of the holonomy (and hence of the triangulation) are
linear combination of the group characters, namely of the traces of
$R_{\omega}(\sigma_0)$, (these latters being thought of as elements of
the fundamental representation of the restricted holonomy group).
These
traces depend only from the given set of incidence numbers
$\{q(\alpha)\}$ of the triangulation, and a question of relevance is
to what extent such traces, and hence the curvature assignements,
characterize the triangulation itself.
In order to answer such a question we exploit the connection between
curvature assignements an the theory of partitions of integers.
\vskip 0.5 cm
\subsection{Partitions of integers and curvature assignements}

According to (\ref{sub}) the curvature
assignements
$\{q(\alpha)\}$ of a dynamical triangulation $T_a$ cannot be
arbitrary, since they are constrained by the total number
$N_n(T^{(i)})$ of $n$-simplices in $T_a$. We have
\begin{eqnarray}
q(o)+\sum_{\alpha=1}^{\lambda}q(\alpha)=\frac{1}{2}n(n+1)N_n(T^{(i)})
\end{eqnarray}
where the summation extends over the bones of $T^{(i)}_a$, (notice
that
we
have included in the summation the marked bone, by considering
also
the incidence number $q(0)$).
Another natural constraint comes about from the fact that in order to
generate an $n$-dimensional polyhedron there is a minimum number
$\hat{q}(n)$ of simplices $\sigma^n$ that must join together at a
bone, (according to the introductory remarks, in dimension $n=2$,
$n=3$, $n=4$ we have
$\hat{q}(n)=3$, otherwise we have no polyhedral manifold. Rather than
use this explicit value, we keep in using the general notation
$\hat{q}(n)$).
 If we
define
\begin{eqnarray}
Q(0)&{\doteq} &  q(0)-\hat{q}(n)\nonumber\\
Q(\alpha)&{\doteq} &  q(\alpha)-\hat{q}(n)
\end{eqnarray}
then we can put these two constraints together by writing
\vskip 0.5 cm
\begin{eqnarray}
Q(0)+\sum_{\alpha=1}^{\lambda}Q(\alpha)=
\left[\frac{1}{2}n(n+1)\left(
\frac{N_n(T^{(i)})}{N_{n-2}(T^{(i)})} \right) -\hat{q}(n)
\right](\lambda+1)
\end{eqnarray}
\vskip 0.5 cm
\noindent (recall that $N_{n-2}(T^{(i)})=\lambda+1$, also recall that
$b(n,n-2)\equiv \frac{1}{2}n(n+1)(
N_n(T^{(i)})/N_{n-2}(T^{(i)}))$
is the average number of simplices $\sigma^n$, in $T^{(i)}$, incident
on a bone).
\vskip 0.5 cm
Thus, we have
\vskip 0.5 cm
\begin{eqnarray}
Q(0)+\sum_{\alpha=1}^{\lambda}Q(\alpha)=
\left[b(n,n-2) -\hat{q}(n)
\right](\lambda+1)
\label{constraint}
\end{eqnarray}
\vskip 0.5 cm
Recall that a {\it partition}
of a positive integer $n$ is a representation of $n$ as the sum
$n=1\cdot\eta_1+2\cdot\eta_2+\ldots+n\cdot\eta_n$, where
$\eta_1,\eta_2,\ldots,\eta_n$ are non-negative integers, and
$\eta_1+\eta_2+\ldots+\eta_n$ determines the number of summands of the
partition.\par
This remark suggests that for each given value of $Q(0)$, $b(n,n-2)$,
and $\lambda$, the set of curvature assignements
$q(\alpha)=Q(\alpha)+\hat{q}$ satisfying the constraint
(\ref{constraint}) may be related to (some of) the possible partitions
of the integer $\left[b(n,n-2) -\hat{q}(n)
\right](\lambda+1)$. Among the attractive features of such an
interpretation is the possibility of exploiting the large body of
knowledge we have on partition theory, a well-established chapter of
combinatorics and analytic number theory. In order to exploit these
techniques, we need to modify a little (\ref{constraint}) so that it
resembles more directly a partition.\par
\vskip 0.5 cm
First of all note that it is natural to assume that the marked $q(o)$,
(or equivalently $Q(0)$), is the largest curvature assignement in the
dynamical triangulation considered $T_a$, {\it i.e.},
$q(0)\geq q(\alpha)$, ( thus $Q(0)$ is a sort of {\it root blob} of
curvature). As we have seen in the introductory chapter,
(see (\ref{qmax})), the largest curvature assignements
can well be of the order of $\lambda$, and  the largest curvature
assignements which simultaneously minimize the remaining curvature
assignements on the other bones are realized on particular
triangulations, (see the discussion following (\ref{qmax})).  In such
triangulations, most of the top-dimensional simplices are
incident on $n$ distinguished bones, (bounding an embedded ${\Bbb
S}^{n-2}$). Notice that, when generated
with equilateral simplices, these  particular triangulations have a
discrete simmetry associated with the interchange of the $n$
distinguished bones carrying most of the curvature. Since we have to
consider (\ref{constraint}) as
$q(0)$ varies from $\hat{q}$ to $q_{max}$, we have to factor out these
residual symmetries, and we do so by marking not just one bone, (the
one providing the basepoint $p_0$), but rather  $n$ bones, $\sigma^{n-
2}(0),\sigma^{n-2}(1),\ldots,\sigma^{n-2}(n-1)$, which support, in the
extreme case described above, the largest curvatures (with
$q(0)=q(1)=\ldots=q(n-1)$) while minimising the incidence on the
remaining bones $\{\sigma^{n-2}(\alpha)\}_{\alpha=n,\ldots,\lambda}$.
The original marked bone $\sigma^{n-2}_0$ is one of those $n$
distinguished bones, and can be selected by the
required marking of an $n$-simplex $\sigma^n_0$ associated with the
gauge fixing procedure for the moduli space of dynamical
triangulations. Thus, we start by replacing (\ref{constraint}) with
\vskip 0.5 cm
\begin{eqnarray}
nQ(0)+\sum_{\alpha=n}^{\lambda}Q(\alpha)=
\left[b(n,n-2) -\hat{q}(n)
\right](\lambda+1)
\label{constraint2}
\end{eqnarray}
\vskip 0.5 cm
Since $nQ(0)\geq Q(\alpha)$, we can refer the curvature assignements
$Q(\alpha)$ to the chosen value for the curvature assignement $nQ(0)$
around the marked bones. To this end, let us add and subtract to
(\ref{constraint2}) the expression $nQ(0)(\lambda+1)$, which allows us
to eventually rewrite (\ref{constraint}) as
\vskip 0.5 cm
\begin{eqnarray}
\sum_{\alpha=n}^{\lambda}\tilde{Q}(\alpha)=
\left[nQ(0)-b(n,n-2)+\hat{q} \right](\lambda+1)-n(n-1)Q(0)
\label{partizione}
\end{eqnarray}
where
\begin{eqnarray}
\tilde{Q}(\alpha){\doteq} nQ(0)-Q(\alpha)
\label{nuovoq}
\end{eqnarray}
\vskip 0.5 cm
This form (\ref{partizione}) of (\ref{constraint}) can be interpreted
as providing a partition of
the integer $[nQ(0)-b+\hat{q}](\lambda+1)-n(n-1)Q(0)$ in at most
$\lambda+1-n$
integers $\tilde{Q}(\alpha)$, (some of the $\tilde{Q}(\alpha)$ can be
zero), each bounded above by
$(b-\hat{q})(\lambda+1)$.
\vskip 0.5 cm
Notice that is not {\it a priori} true that
{\it each} partition of the integer
$[nQ(0)-b+\hat{q}](\lambda+1)-n(n-1)Q(0)$ can be generated by
curvature
assignements through the $\tilde{Q}(\alpha)$. Thus, we shall
denote by
\begin{eqnarray}
p_{\lambda}^{curv}\left( [nQ(0)-b+\hat{q}](\lambda+1)-n(n-1)Q(0)
\right)
\end{eqnarray}
the number of partitions of $[nQ(0)-b(n,n-2)+\hat{q}](\lambda+1)-n(n-
1)Q(0)$ into at most $\lambda+1-n$ parts, (each
$\leq (b(n,n-2)-\hat{q})(\lambda+1)$), arising from actual curvature
assignements.
\vskip 0.5 cm

\noindent {\bf A notational remark}. Notice that in comparing with
standard formulae for partition theory\cite{Andrews}, a more standard
notation would have been
\begin{eqnarray}
p_{\lambda+1-n}^{curv}\left( [nQ(0)-b+\hat{q}](\lambda+1)-n(n-1)Q(0)
\right)
\end{eqnarray}
however, in order not to burden the notation further, we
shall adopt the above simplified version; in some cases, if there is
no danger of confusion, we shall simply write
$p_{\lambda}^{curv}$ for the above expression and for other similar
expressions involving partitions.
\vskip 0.5 cm

\subsection{Curvature assignements of distinct dynamical
triangulations}

 For small $\lambda$'s the actual distribution of
partitions corresponding to curvature assignements is rather
accidental, however for $\lambda>>1$ we have that
a large fraction of partitions
corresponds to actual curvature assignements.  We start by proving
that {\it generically} distinct curvature assignements correspond to
distinct dynamical triangulations. We can prove this by the elementary
probabilistic methods which are typical in discrete mathematics
whenever attention is on asymptotic
properties. We also bound the size of the {\it non-generic} set of
distict dynamical triangulations associated to the same set of
curvature assignements. The tails associated both to partitions of
integers which are not curvature assignements and to distinct
dynamical triangulations having the same curvature assignements
provide corrections to the naive counting
through unrestricted partitions of integers. Such corrections are
difficult to control in details, however they have the structure of a
rescaling that will allow us to implement a very straightforward
counting strategy which will prove very effective.
\vskip 0.5 cm
Let us start by noticing that it is obviously false that every
possible set of distinct curvature
assignement corresponds to triangulations having distinct
incidence matrices.
One may easily construct particular examples violating such statement.
Consider
for instance two dynamically triangulated
(exagonal) flat tori. By flipping links, we may generate two curvature
bumps on each torus, (on each torus, the bumps may correspond to
distinct curvature assignements) . By inserting in each torus a copy
of these curvature bumps in such a way that their distance is
different in the two tori, we get distinct triangulations with the
same curvature assignements. As another more general example,
(again in dimension $2$),
consider a two-dimensional triangulation with a large number of
vertices, let $abc$, $acd$, and $efg$, $egh$ four triangles pairwise
sharing a common edge ($ac$ for the former pair, and $eg$ for the
latter), but otherwise largely separated,
(so that their incidence numbers on the respective vertices are
uncorrelated). Let us assume that the corresponding incidence number
are given by: $q(a)=\alpha$, $q(b)=\gamma$, $q(c)=\beta$,
$q(d)=\delta$, for the first pair of triangles, while for the second
we set: $q(e)=\gamma+1$, $q(f)=\alpha-1$, $q(g)=\delta+1$,
$q(h)=\beta-1$. It is immediate to check that a flip move will
interchange the curvature assignements of the first pair of adjacent
triangles with those of the second pair, thus changing the incidence
relations in the triangulation. Nonetheless, the sequence of curvature
assignements remains unchanged, and again we have two distinct
triangulations with the same sequence of curvature assignements.
\vskip 0.5 cm
Clearly, these counterexamples work either because we carefully chose
a particularly symmetric triangulation or because we
adjusted the curvature assignements to the particular action of the
flip move. However, when moving from a dynamical triangulation to
another, with a set of ergodic moves (Pachner (\cite{Migdal})), the
curvature
assignements are not fixed, and in this sense  the above
counterexamples are
not generic, and if we denote by $\{T_a^{(i)}\}_{curv}$ the set of
distinct triangulations sharing a common set of curvature
assignements, we can prove that as $\lambda>>1$,
$\{T_a^{(i)}\}_{curv}$ is in a suitable sense a small subset in
$\{T_a^{(i)}\}$. We can formalize this remark according to the
\vskip 0.5 cm
\begin{Gaial}
For a PL-manifold $M$, let
$\Theta\colon\pi_1(M;p_0)\to G$, be a
given holonomy representation, and let $\{T_a^{(i)}\}_{\lambda}$
denote the set of distinct  dynamical triangulations of
$M=|T_a^{(i)}|$ with $N_{n-2}(T_a^{(i)})=\lambda+1$
bones, and let $\{T_a^{(i)}\}_{\lambda\in {\Bbb Z}}$  be the
associated
inductive limit space generated as $\lambda$ grows. Let ${\cal
P}_T\doteq (\{T_a^{(i)}\}_{\lambda\in {\Bbb Z}},{\cal B},m_T)$ be the
probability space endowed with the normalized counting
measure $m_T$ on every non-empty open set ${\cal B}\subset
\{T_a^{(i)}\}_{\lambda\in {\Bbb Z}}$.
For any $T_a\in\{T_a^{(i)}\}_{\lambda}$, let  $\{q(\alpha)\}$ describe
the
sequence of integers providing
the
incidence of the simplices $\{\sigma_i^n\}_{i=1}^{N_n(T_a)}$ over
the $\lambda+1$ bones $\{\sigma_k^{n-2}\}_{k=1}^{N_{n-2}(T_a)}$.
Then, as $\lambda\to\infty$, the curvature assignements
$\{q(\alpha)\}$ characterize with probability one
the dynamical triangulation $T_a[q(\alpha)]$.
\label{triangoli}
\end{Gaial}
\vskip 0.5 cm
{\bf Proof}.
We start by creating a probability space ${\cal P}_T$ whose elements
are the distinct triangulations in $\{T_a^{(i)}\}_{\lambda\in {\Bbb
Z}}$, and
whose Borel probability measure, $m_T$, is the normalized counting
measure on every non-empty open set ${\cal B}\subset
\{T_a^{(i)}\}_{\lambda\in {\Bbb Z}}$. Note that we have to consider
triangulation with variable $\lambda$, since the known set of ergodic
moves \cite{Migdal} mapping a triangulation $T_a^{(i)}(1)_{\lambda}$
into a distinct
triangulation $T_a^{(i)}(2)_{\lambda}$, (with the same $\lambda$), go
through intermediate steps which do not preserve volume (hence the
number of bones $\lambda$ at a given $b(n,n-2)$).
If, for a given value of $\lambda$,
$\{T_a^{(i)}\}_{curv}$ is the subset of distinct triangulations with a
same set of curvature assignements, we can write ($\lambda$ fixed)
\begin{eqnarray}
m_T[\{T_a^{(i)}\}_{curv}]=
\frac{Card \{T_a^{(i)}\}_{curv}}{Card \{T_a^{(i)}\}}
\end{eqnarray}
Recall \cite{Migdal} that on $({\cal P}_T,{\cal B},m_T)$, the Pachner
moves
characterize a continuous and ergodic dynamical system with respect to
$m_T$. If
\begin{eqnarray}
\lim_{\lambda\to\infty}\frac{Card \{T_a^{(i)}\}_{curv}}{Card
\{T_a^{(i)}\}}>0
\end{eqnarray}
then by Birkhoff ergodic theorem\cite{Walters} we would get
\begin{eqnarray}
m_T[\{T_a^{(i)}\}_{curv}]>0
\end{eqnarray}
and the Pachner moves would fix $m_T$-almost everywhere the set
$\{T_a^{(i)}\}_{curv}$ thus contradicting their ergodicity. Hence
we must have
\begin{eqnarray}
\lim_{\lambda\to\infty}\frac{Card \{T_a^{(i)}\}_{curv}}{Card
\{T_a^{(i)}\}}=0
\end{eqnarray}
which reflects the fact that Pachner moves can fix the curvature
assignements only under special circumstances. $\Box$
\vskip 0.5 cm
Even if this formal result may at first sight appear very reassuring
we still have to face the fact that distinct dynamical triangulations
associated to a same set of curvature assignements may give rise to
significant subleading corrections to the counting of distinct
triangulations of a given $PL$-manifold.
Thus we need a way for controlling how large the set
$\{T_a^{(i)}\}_{curv}$ can be. This control is provided by the
following
result
\vskip 0.5 cm
\begin{Gaial}
There are constants, $\mu(n)>0$, and $\tau(n)\geq 0$, depending only
on the dimension $n$, such that
\begin{eqnarray}
Card\{T_a^{(i)}\}_{curv}\leq \mu(n)\cdot\lambda^{\tau(n)}
\end{eqnarray}
\label{equicurv}
\end{Gaial}
\vskip 0.5 cm
\noindent {\bf Proof}. For a given set of curvature assignements
over the $\lambda+1$ bones of a dynamically triangulated $PL$-manifold
$M$, let $\{T_a^{(i)}\}_{curv}$ denote the set of distinct dynamical
triangulations (with $\lambda+1$ bones), having the same curvature
assignements. Consider the corresponding dual complexes, {\it viz.},
for a given $T_a\in\{T_a^{(i)}\}_{curv}$, consider its barycentric
subdivision, then the closed stars (in the subdivision) of the
vertices of $T_a$ form the collection of $n$-cells of the complex dual
to the given $T_a$. Distinct triangulations in $\{T_a^{(i)}\}_{curv}$
correspond to distinct dual complexes. The use of the dual complex
rather than of the triangulation itself is related to the fact
that the incidence matrix of the dual complex is directly related to
the curvature assignements of the original triangulation. This is most
easily seen in dimension $n=2$, whereas in dimension $n>2$ the
curvature assignements of a dynamical triangulation $T_a$, over the
respective bones, can be read from the incidence matrix
$A\doteq\{a_{ij}\}_{0\leq i,j\leq\lambda}$ associated with the $2$-
skeleton of the dual complex, ({\it e.g.}, in dimension $n=3$ think of
a froth
pattern in space). Explicitly,  one has that the line sums of the $2$-
skeleton incidence matrix $A$ are such that
\begin{eqnarray}
\sum_{j=0}^{\lambda}a_{kj}=q(k)
\label{linesum}
\end{eqnarray}
where $q(k)$ is the number of simplices in $\{T_a^{(i)}\}_{curv}$
which are incident on the bone $\sigma^{n-2}(k)$. A similar relation
holds for the column sums. In order to estimate
$Card\{T_a^{(i)}\}_{curv}$ for a given set of curvature assignements
$\{q(k)\}_{0\leq k\leq\lambda}$ we have to enumerate the set of such
distinct incidence matrices. The generic matrix $A$ of such a kind, is
a symmetric $(0,1)$ matrix with vanishing diagonal entries satisfying,
for each line, at least the relation (\ref{linesum}), (other
constraints, associated with the Dehn-Sommerville relations, have to
be satisfied). In order to enumerate such matrices we can proceed as
follows. We have
\begin{eqnarray}
\left( \begin{array}{c}
\lambda-1\\ q(1)
\end{array}  \right)
\end{eqnarray}
choices for the first row $\{a_{1j}\}$ of $A$, and in general, since
$A$ is a symmetric traceless matrix, once chosen
the $k-1$-th row, we have
\begin{eqnarray}
\left( \begin{array}{c}
\lambda-k\\ q(k)-(\sum_{j=1}^{k-1}a_{kj})
\end{array}  \right)
\end{eqnarray}
choices for the $k$-th row.
\vskip 0.5 cm
$Card\{T_a^{(i)}\}_{curv}$
is bounded above by the number of distinct $(0,1)$ matrices satisfying
the constraint (\ref{linesum}), thus we get
\vskip 0.5 cm
\begin{eqnarray}
Card\{T_a^{(i)}\}_{curv}\leq\prod_{k=0}^{\lambda}
\left( \begin{array}{c}
\lambda-k\\ q(k)-(\sum_{j=1}^{k-1}a_{kj})
\end{array}  \right)
\end{eqnarray}
\vskip 0.5 cm
\noindent By taking logarithms and on applying Jensen's inequality,
($\ln\sum_k
h_k\geq\sum_k\ln h_k$), we get
\vskip 0.5 cm
\begin{eqnarray}
\ln[Card\{T_a^{(i)}\}_{curv}]\leq\ln\sum_{k=0}^{\lambda}
\left( \begin{array}{c}
\lambda-k\\ q(k)-(\sum_{j=1}^{k-1}a_{kj})
\end{array}  \right)
\label{loga}
\end{eqnarray}
\vskip 0.5 cm
\noindent We set
\begin{eqnarray}
\tau(n)\doteq \max_{k}[q(k)-\sum_{j=1}^{k-1}a_{kj}]
\end{eqnarray}
and  bound the binomial coefficient in (\ref{loga})
according to
\vskip 0.5 cm
\begin{eqnarray}
\left( \begin{array}{c}
\lambda-k\\ q(k)-(\sum_{j=1}^{k-1}a_{kj})
\end{array}  \right) \leq
\frac{(\lambda-k)!}{(\lambda-k-\tau(n))!}
\end{eqnarray}
\vskip 0.5 cm
On applying Stirling's formula
$\lambda!=(\lambda/e)^{\lambda}\sqrt{2\pi\lambda}e^{\alpha_{\lambda}}$
, where $1/(12\lambda+1)<\alpha_{\lambda}<1/(12\lambda)$, we easily
get from (\ref{loga}) that for
$\lambda>>1$
\begin{eqnarray}
Card\{T_a^{(i)}\}_{curv}\leq \mu(n)\cdot\lambda^{\tau(n)},
\end{eqnarray}
as stated. $\Box$
\vskip 0.5 cm
The above bound on $Card\{T_a^{(i)}\}_{curv}$ is quite rough but
sufficient for our purposes, since it implies that
$Card\{T_a^{(i)}\}_{curv}$ can, at worst, give rise to
a subleading polynomial correction to the  counting associated
with the curvature assignements.

\vskip 0.5 cm
This remark implies that the knowledge of the asymptotics of the
number of partitions
$p^{curv}_{\lambda}$
of the integer $[nQ(0)-b+\hat{q}](\lambda+1)-n(n-1)Q(0)$ which are
curvature assignements provides a way of counting distinct
triangulations of a given $PL$-manifold, at least when $\lambda$ is
large.
\vskip 0.5 cm
Trivially $p_{\lambda}^{curv}<p_{\lambda}$, and a first estimate of
the
asymptotic behaviour of $p_{\lambda}^{curv}$ can be obtained by the
asymptotics of $p_{\lambda}$:
\begin{Gaial}
Let ${\cal D}T_n(a,b,\lambda)$ denote the set of distinct dynamical
triangulations, of an $n$-dimensional PL manifold $M$,
with $N_{n-2}(T_a)=\lambda+1$ bones, and average incidence $b(n,n-
2)=b$. Then, for large $\lambda$ the number of distinct partitions of
the integer
 $[nQ(0)-b+\hat{q}](\lambda+1)-n(n-1)Q(0)$
(into $\lambda+1-n$ parts) coming from curvature assignements
of a dynamical triangulation $T_a\in {\cal D}T_n(a,b,\lambda)$
is exponentially bounded above according to
\begin{eqnarray}
p_{\lambda}^{curv}\leq e^{[2+\ln(b-\hat{q})]\lambda}{\lambda}^{-2}
\label{expcurv}
\end{eqnarray}
\label{curvexp}
\end{Gaial}
{\bf Proof}.
The asymptotic behavior of
$p_{\lambda}([nQ(0)-b+\hat{q}](\lambda+1)-n(n-1)Q(0))$
follows
by noticing that (see {\it e.g}\cite{Hall}, pp. 32)
$p_{\lambda}([nQ(0)-b+\hat{q}](\lambda+1)-n(n-1)Q(0))$ depends only on
the value
of $[nQ(0)-b+\hat{q}](\lambda+1)-n(n-1)Q(0)$ modulo $(\lambda+1-n)!$,
and that it is a
polynomial of degree $\lambda-n$ whose leading term is
\begin{eqnarray}
\frac{\{[nQ(0)-b+\hat{q}](\lambda+1)-n(n-1)Q(0)\}^{\lambda-n
}}{(\lambda+1-n)!(\lambda-n)!}
\label{partialasy}
\end{eqnarray}
On applying Stirling's formula, we get for
$\lambda>>n$
\begin{eqnarray}
p_{\lambda}([nQ(0)-b+\hat{q}](\lambda+1)-n(n-1)Q(0))\simeq
\frac{e^{2\lambda}}{2\pi}
\frac{\{[nQ(0)-
b+\hat{q}]\lambda\}^{\lambda}}{\lambda^{2\lambda}[nQ(0)-b(n,n-
2)+\hat{q}]\lambda}
\end{eqnarray}
\vskip 0.5 cm
\noindent As
$Q(0)$ varies, $p_{\lambda}([nQ(0)-b+\hat{q}](\lambda+1)-n(n-1)Q(0))$
is bounded above by the right side of the above expression evaluated
for  $\max[nQ(0)]=(b-
\hat{q})\lambda$, which yields
$p_{\lambda}\simeq
e^{[2+\ln(b-\hat{q})]\lambda}{\lambda}^{-2}$. Since
$p^{curv}_{\lambda}<p_{\lambda}$, the result follows. $\Box$
\vskip 0.5 cm
The bound (\ref{expcurv}) may be  consistent as well with a
subexponential growth for $p_{\lambda}^{curv}$, a
subexponential growth that, if present, may make the actual control of
$p_{\lambda}^{curv}$ quite unassailable. However, more can be said on
the actual asymptotics of the curvature assignements in ${\cal
D}T_n(a,b,\lambda)$.  Since some of the ${\tilde{Q}}(\alpha)$ may be
zero,
the number of distinct partitions of the integer
 $[nQ(0)-b+\hat{q}](\lambda+1)-n(n-1)Q(0)$
into $\lambda+1-n$ parts coming from curvature assignements
is approximated with greater and greater accuracy, for $\lambda$
large, by the unrestricted partition of
 $[nQ(0)-b+\hat{q}](\lambda+1)-n(n-1)Q(0)$ into curvature
assignements, ({\it viz}, we no longer require to have a partition
into $\lambda+1-n$ parts). This latter partition function is in
general much  more difficult to estimate than the original one,
however its asymptotics can be qualitatively controlled through known
techniques of analytic number theory, and together with the elementary
bound (\ref{expcurv}) we can readily establish an exact exponential
asymptotics for $p_{\lambda}^{curv}$.
\vskip 0.5 cm
For a given value of $b$,  and $\lambda>>1$, let ${\cal C}(b,\lambda,
T_a)$, (${\cal C}$ for short), be the set of positive integers
occurring as curvature assignements of some rooted triangulation $T_a$
of bounded  geometry in
${\cal D}T_n(a,b,\lambda)$. Set
$k\doteq [nQ(0)-b+\hat{q}](\lambda+1)-n(n-1)Q(0)$, and for
$t$ denoting a generic indeterminate, let
\begin{eqnarray}
f_{curv}(t)\doteq \sum_{k>0}p^{curv}(k)t^k
\end{eqnarray}
be the generating function for the set of unrestricted partitions,
$p^{curv}(k)$,
of the integer $k$ into curvature assignements of some $T_a\in
{\cal D}T_n(a,b,\lambda)$. Then, by exploiting the formula for the sum
of a finite geometric series, one  can easily show that for $|t|<1$
\begin{eqnarray}
f_{curv}(t)=\prod_{k\in{\cal C}}(1-t^k)^{-1}
\end{eqnarray}
where $k\in{\cal C}$ is a shorthand notation for denoting a partition
of $k=(k_1,k_2,\ldots)$
whose parts belong to ${\cal C}$, (recall that for large $\lambda$
distinct partitions generically correspond to distinct
triangulations), and where
\begin{eqnarray}
\prod_{k\in{\cal C}}(1-t^k)^{-1}\doteq
(1-t^{k_1})^{-1}\times(1-t^{k_2})^{-1}
\times(1-t^{k_3})^{-1}\times\ldots
\end{eqnarray}
If we denote the partition $k\in{\cal C}$ in the form
\begin{eqnarray}
k=(1^{a_1}2^{a_2}3^{a_3}\ldots)
\end{eqnarray}
where exactly $a_i$of the $\{k_j\}$ are equal to $i$, we can
eventually rewrite $f_{curv}(t)$ as the unconstrained product
\begin{eqnarray}
f_{curv}(t)=\prod_{j=0}^{\infty}(1-t^j)^{-a_j}.
\end{eqnarray}
\vskip 0.5 cm
At this stage, we can exploit a general result due to G.
Meinardus\cite{Andrews} providing the structural asymptotics of
partition functions which are associated with infinite product
generating functions. Such a technique is neatly described in Andrews'
monography, and we shall follow closely his exposition, (see
\cite{Andrews}, Chap.VI).
\vskip 0.5 cm
Let us consider the Dirichlet series associated with the set of
multiplicities $\{a_j\}$ associated with a curvature assignement
partition $k\in{\cal C}$, {\it viz.},
\begin{eqnarray}
D(s)=\sum_{j=1}^{\infty}\frac{a_j}{j^s}
\label{Dirichlet}
\end{eqnarray}
with $s\doteq u+iv$, and let $\alpha\in{\Bbb R}^+$ be its abscissa of
convergence, ({\it i.e.}, (\ref{Dirichlet}) is assumed to converge for
$s>\alpha$). Let us further assume that $D(s)$ admits an analytic
continuation in the strip $u\geq -C_0$ with $0<C_0<1$, and that in
this region $D(s)$ is analytic with the exception of a pole of order
one located at $s=\alpha$ with residue
$Res[D(s)]_{s=\alpha}=Res(\alpha)$. Under a number of further
technical assumptions which are not relevant for our purposes, we get
the following result which is a straightforward application of
Meinardus' theorem:
\begin{Gaial}
As $k\to\infty$, the leading asymptotics of $p^{curv}(k)$ is given by
\vskip 0.5 cm
\begin{eqnarray}
p^{curv}(k)\simeq C k^{\eta}\exp\left\{
k^{\alpha/(\alpha+1)}(1+\frac{1}{\alpha})[Res(\alpha)
\Gamma(\alpha+1)\zeta(\alpha+1)]^{1/(\alpha+1)}
\right\}(1+\ldots)
\label{Meinardus}
\end{eqnarray}
\vskip 0.5 cm
\noindent  where $C$ is a suitable constant, (see {\it e.g.}, Theorem
6.2 in \cite{Andrews}), $(1+\ldots)$ stands for higher order terms,
$\Gamma(\alpha+1)$, $\zeta(\alpha+1)$ respectively denote the Euler
gamma function and the Riemann zeta function, and finally
\begin{eqnarray}
\eta\doteq \frac{D(0)-1-\frac{1}{2}\alpha}{1+\alpha}
\end{eqnarray}
\label{Meinasymptotics}
\end{Gaial}
\vskip 0.5 cm
In our case, $k$ stands for  $[nQ(0)-b+\hat{q}](\lambda+1)-n(n-1)Q(0)$
and for $nQ(0)=(b-\hat{q})\lambda$, we get from (\ref{Meinardus})
\begin{eqnarray}
p^{curv}\simeq C[(b-\hat{q})\lambda^2]^{\eta}
\exp\left\{
[(b-\hat{q})\lambda^2]^{\alpha/(\alpha+1)}(1+\frac{1}{\alpha})
[Res(\alpha)
\Gamma(\alpha+1)\zeta(\alpha+1)]^{1/(\alpha+1)}
\right\}(1+\ldots)
\end{eqnarray}
This asymptotics is consistent with the bound (\ref{expcurv}) only if
the abscissa of convergence of the series $D(s)$ is
$\alpha=1$, and if its residue $Res(\alpha)$ and the value $D(0)$ are
bounded according to
\begin{eqnarray}
Res(\alpha)\leq\sqrt{\frac{6}{b-\hat{q}}}\frac{2+\ln(b-\hat{q})}{2\pi}
\end{eqnarray}
\begin{eqnarray}
D(0)\leq -\frac{1}{2}
\end{eqnarray}
\vskip 0.5 cm
\subsection{The counting principle}

The bound (\ref{expcurv}) and the asymptotics of lemma
\ref{Meinasymptotics} establish that $p_{\lambda}^{curv}$ has an
exponential growth with $\lambda$, and that the coefficients $a_n$
characterizing the Dirichlet series $D(s)$, (and the partitions in
curvature assignements), have to satisfy some growth condition
(related to the above bounds). Clearly, more geometric information
would be necessary for getting a complete control of
$p_{\lambda}^{curv}$, however what really matters for our purposes is
the observation stemming from the above analysis that the ratio
\begin{eqnarray}
\frac{p_{\lambda}^{curv}}{p_{\lambda}}\doteq c_n
\end{eqnarray}
where $0\leq c_n\leq 1$, either is a constant or an exponentially
decreasing function of $\lambda$, and  $b(n,n-2)$ {\it viz.},
$c_n=\exp[(\alpha_1-\alpha_2b)\lambda+\alpha_0]$ for some suitable
constants $\alpha_0$, $\alpha_1$, $\alpha_2$. This functional
dependence of $c_n$ on $\lambda$ can be conveniently recast in the
form
\begin{eqnarray}
\frac{p_{\lambda}^{curv}}{p_{\lambda}}\doteq
c_n=C_0e^{\alpha_nN_n(T_a)}
\label{kscale}
\end{eqnarray}
where the constants $0<C_0\leq 1$, and $\alpha_{n}\leq 0$ depend only
on the dimension $n$.
\vskip 0.5 cm
Since $c_n$ has exactly the structure of the exponential of the {\it
volume part} of the
dynamical triangulation action, and since $p^{curv}_{\lambda}$
generically
enumerates distinct dynamical triangulations, we can as well as use
for such a purpose the much more manageable counting function
$p_{\lambda}$ provided that in the final results we allow for a {\it
renormalization} of the coupling
$k_n$, according to
\vskip 0.5 cm
\begin{eqnarray}
k_n \to k_n+\delta_n
\label{rinorm}
\end{eqnarray}
\vskip 0.5 cm
\noindent with $\delta_n\leq 0$, a  constant depending
only on the dimension $n$.
\vskip 0.5 cm
In other words, by using $p_{\lambda}$ as an enumerator,
we are using a discrete measure on a space (the space of distinct
partitions of the integer $[nQ(0)-b+\hat{q}](\lambda+1)-n(n-1)Q(0)$)
which contains the relevant space of curvature assignements, but which
is definitely larger. The discrete measure associated with
$p_{\lambda}$ induces a corresponding measure,
$p_{\lambda}^{curv}$ on the space of distinct dynamical
triangulations, and the two measure scales according to
(\ref{kscale}), (which appears as a the discrete Radon-Nykodim
derivative of the measure $p^{curv}_{\lambda}$ with respect
$p_{\lambda}$). The nature of this scaling allows us to use directly
the {\it overcounting} measure $p_{\lambda}$ since in seeking the
infinite volume limit and a possible scaling limit for the partition
function (\ref{grandpartition}) we have to vary both couplings $k_n$,
$k_{n-2}$. Roughly speaking, in using directly $p_{\lambda}$ in place
of $p_{\lambda}^{curv}$ it is more or less like including tadpoles and
self-energies in the triangulation counting. The shift (\ref{kscale})
and (\ref{rinorm})
allows us to remove, in the final results, the effect of such
inclusions.
We formalize the results of the lemmas \ref{equicurv},
\ref{curvexp}, \ref{Meinasymptotics} and of the associated remarks, in
the
following lemma providing
the rationale of our {\it counting strategy},
\vskip 0.5 cm
\begin{Gaial}
 There exists constants, $0<C_0\leq 1$,  $\alpha_n\leq 0$, and
$\tau(n)\geq 0$
 depending only on the dimension $n$, such that
the number of distinct curvature assignements of rooted dynamical
triangulations $T_a\in
{\cal D}T_n(a,b,\lambda)$ is enumerated, up to the  scaling
\vskip 0.5 cm
\begin{eqnarray}
\frac{Card\{T_a^{(i)}\}}{p_{\lambda}}\doteq
c_n=C_0N_{n}^{\tau(n)}e^{\alpha_nN_n(T_a)},
\label{corrections}
\end{eqnarray}
\vskip 0.5 cm
\noindent by the set of distinct partitions $p_{\lambda}$ of the
integer
$[nQ(0)-b+\hat{q}](\lambda+1)-n(n-1)Q(0)$ into at most $\lambda+1-n$
parts, (each $\leq (b-\hat{q})(\lambda+1)$).
\label{scalemma}
\end{Gaial}

\subsection{A geometrical remark}

According to the above results, curvature assignements {\it
generically}
characterize dynamical triangulations.
In the two-dimensional case this can be related to the fact that
curvature assignements provide the rotation matrices defining the
holonomy, and thus the triangulation. Let us verify that geometrically
also for
$n=3$, and $n=4$ curvature assignements provide most of the
information characterizing the
holonomy for dynamical triangulations. We work out in details the
three-dimensional
case by using an $SU(2)$ connection. The case for dimension $n=4$ can
be easily discussed along the same lines by
exploiting the isomorphism $SO(4)\approx \frac{SU(2)\times
SU(2)}{{\Bbb{Z}}_2}$. This analysis is due to J. Lewandowsky, to whom
we wish to express special gratitude.
\vskip 0.5 cm
Consider the marked simplex $\sigma^n_0$, (in this case a
tetrahedron), and fix an orthonormal basis $\{\vec{e}_{x,y,z}\}$ in
$\sigma^n_0$. Among the
six bones (edges) in the boundary of $\sigma^n_0$, we can choose
$\frac{1}{2}n(n-1)=3$ bones,
$\tau_i$, $i=0,1,2$,
 one of which is the marked one, $\tau_0$, and all sharing a common
vertex, (recall the need of marking $n$ bones in interpreting
(\ref{constraint}) in terms of partition of integers!). These bones
are defined by
the vectors $\vec{E}(\tau_i)\in {\Bbb R}^3$, (with components referred
to the fixed orthonormal basis $\{\vec{e}_{x,y,z}\}$). Set
\begin{eqnarray}
\vec{\tau}_i= \sigma_x\frac{E_x(\tau_i)}{|\vec{E}(\tau_i)|}+
\sigma_y\frac{E_y(\tau_i)}{|\vec{E}(\tau_i)|}+
\sigma_z\frac{E_z(\tau_i)}{|\vec{E}(\tau_i)|}
\end{eqnarray}
 with $\sigma_x$, $\sigma_y$, $\sigma_z$ denoting the Pauli matrices.
The holonomy matrix describing the $2$-dimensional plane
rotations generated by winding around the
bones $\tau_i$ can be written as the $SU(2)$ matrix
\begin{eqnarray}
U(\tau_i)=I\cos(\phi(i)/2)+\vec{\tau}_i\sin(\phi(i)/2)
\end{eqnarray}
where $I$ is the identity operator, and $\phi(i)=q(i)\cos^{-
1}\frac{1}{3}$ denotes the rotation angle around the bone $\tau_i$. In
general,  we
can reconstruct the holonomy matrix $U(\tau_i)$ for every
distinguished bone $\tau_i$ in the marked base simplex $\sigma_0^n$,
by giving the traces of $U(\tau_i)$, and the traces of
$U(\tau_i)U(\tau_j)$, $i\not= j$. The traces $tr[U(\tau_i)]$ are
proportional to the incidence numbers $q(i)$, $i=0,1,2$, while
$tr[U(\tau_i)U(\tau_j)]$ are proportional to the scalar product in
$SU(2)$
between the spinors $\vec{\tau}_i$ and $\vec{\tau}_j$,
respectively associated to the bones $\tau_i$, $\tau_j$. These latter
data are actually already known, since the base simplex is
equilateral. Thus the $SU(2)$-holonomies around the base simplex
$\sigma^n_0$ are trivially determined once we give the corresponding
curvature assignements.
The reconstruction of
the $SU(2)$- holonomies around the remaining $\lambda+1-
\frac{1}{2}n(n+1)=\lambda-5$ bones
$\tau_{\alpha}=\{\sigma^{n-2}_{\alpha}, 0<\alpha<\lambda,\sigma^{n-
2}_{\alpha}\notin\sigma^n_0 \}$, is similar.
Formally, for such a reconstruction, we  would need the assignement of
the
traces $\chi(\alpha)=tr[U(\tau_{\alpha})]$, and of
$\chi(\alpha,j)=tr[U(\tau_{\alpha})U(\tau_j)]$, with $j=0,1,2$.
Indeed, with the characters $\chi(\alpha)=tr[U(\tau_{\alpha})]=
2\cos\frac{\phi(\alpha)}{2}$
we can reconstruct the rotation angles $\phi(\alpha)$, while with
the characters $\chi(\alpha,j)=tr[U(\tau_{\alpha})U(\tau_j)]$,
{\it viz.},
\begin{eqnarray}
\chi(\alpha,j)= 2\left( \cos\frac{\phi(\alpha)}{2}
\cos\frac{\phi(j)}{2} -<\vec{\tau}_{\alpha},\vec{\tau}_j>
\sin\frac{\phi(\alpha)}{2}\sin\frac{\phi(j)}{2} \right)
\label{system}
\end{eqnarray}
we can reconstruct the $SU(2)$ inner products
\begin{eqnarray}
<\vec{\tau}_{\alpha},\vec{\tau}_j>\doteq -\frac{1}{2}Tr
[\vec{\tau}_{\alpha}\cdot\vec{\tau}_j]
\end{eqnarray}
and hence obtain the spinors $\vec{\tau}_{\alpha}$ through which
the $SU(2)$-holonomies
$U(\tau_{\alpha})=I\cos(\phi(\alpha)/2)+\vec{\tau}_{\alpha}
\sin(\phi(\alpha)/2)$ are defined.\par
The characters $\chi(a)=tr[U(\tau_{a})]$ with
$a=0,1,\ldots,\lambda$ are given (up to a constant factor) by the
incidence numbers $\{q(a)\}_{a=0}^{\lambda}$, and thus are known.
Since the simplices in the triangulations are all
equilateral, the bones $\vec{\tau}_j$ are not arbitrarily
distributed with respect to the reference bones
$\vec{\tau}_{\alpha}$, but are in correspondance with the $12$ points
on the unit sphere ${\Bbb S}^2$ marked by the vertices of the
inscribed Icosahedron, (arrange the icosahedron in space so that the
uppermost vertex coincides with $(0,0,1)$). Hence, the $SU(2)$ inner
products
$<\vec{\tau}_{\alpha},\vec{\tau}_j>$  are known, and
the characters $\chi(\alpha,j)=tr[U(\tau_{\alpha})U(\tau_j)]$ are
given in terms of the $\{q(a)\}_{a=0}^{\lambda}$. It follows that
the curvature assignements  $\{q(a)\}_{a=0}^{\lambda}$
provide {\it grosso modo} the $SU(2)$ holonomy matrices, (up to
conjugation), what is missing at this detailed level, (which is
independent from any asymptotics), is the combinatorial data of how to
put together the local holonomies so as to reconstruct
the underlying dynamical triangulation $T_a$.
\vfill\eject
\section{Entropy estimates}

According to the results of the previous section, dynamical
triangulations are {\it generically} characterized by the
corresponding curvature assignements $\{q(\alpha)\}$, and the simple
counting of the possible curvature assignements provides a very
accurate estimate of the number of distinct dynamical triangulations a
$PL$ manifolds (with given fundamental group) can support. The
previous results have already evidentiated the close connection
between the theory of partitions of integers and the curvature
assignements, our strategy is to further exploit this connection in
order to estimate the number of distinct dynamical triangulations.

\vskip 0.5 cm
\subsection{The generating functions for the enumeration of dynamical
triangulation}

Since the partitions of $[n Q(0)-b(n,n-2)+\hat{q}](\lambda+1)-n(n-
1)Q(0)$ into
at most $\lambda+1-n$ parts, (each
$\leq (b(n,n-2)-\hat{q})(\lambda+1)$), can be thought of as
enumerating, in the large $\lambda$ limit, distinct
curvature assignements, theorem \ref{triangoli} implies that they
also enumerate distinct rooted dynamical triangulations.
This observation establishes the following
\begin{Gaiat}
Let $\Theta\colon\pi_1(M;p_0)\to G$, be a
given holonomy representation of an $n$-dimensional PL-manifold $M$,
$n\geq 2$. The number $W(\Theta,\lambda,b,Q(0))$, of distinct rooted
dynamical triangulations on $M$,  with $\lambda+1$ bones and with a
given average number, $b{\doteq} b(n,n-2)$, of $n$-simplices incident
on a bone, is given, in the large $\lambda$ limit, by
\begin{eqnarray}
W(\Theta,\lambda,b,Q(0))=c_n\cdot p_{\lambda}([nQ(0)-
b+\hat{q}](\lambda+1)-n(n-1)Q(0))
\end{eqnarray}
where $c_n$ is the rescaling factor of lemma \ref{scalemma}.
\end{Gaiat}

\noindent {\bf Remark}.
Note that selecting the value $Q(0)$ of the
curvature at the $n$ marked bones is equivalent to considering
dynamical
triangulations with a
boundary, (the links of the marked bone), of variable volume,
(the length being proportional to the chosen $Q(0)$). Thus we have the
\begin{Gaial}
The function $W(\Theta,\lambda,b,Q(0))=c_n\cdot p_{\lambda}([ nQ(0)-
b+\hat{q}](\lambda+1)-n(n-1)Q(0))$ provides the number of distinct
dynamical
triangulations with boundary $\partial T_a$
consisting of the disjoint union of $n=$ spheres of dimension $n-1$,
${\Bbb S}^{n-1}$, each one of
volume $Vol[\partial T_a]=q(0)vol[\sigma^{n-1}]$.
\end{Gaial}
{\bf Proof}. This trivially follows from the above theorem by noticing
that by removing from $T_a$ the open star of the marked bones, (on
each of
which $q(o)$ simplices $\{\sigma^n\}$ are incident), we get dynamical
triangulations with a boundary $\partial T_a\simeq\coprod {\Bbb S}^{n-
1}$ of
volume $Vol[\partial T_a]=nq(0)vol[\sigma^{n-1}]$. $\Box$

\vskip 0.5 cm
\subsection{Gauss polynomials and dynamical triangulations}

 By exploiting
the properties of the partitions $p_{\lambda}([nQ(0)-
b+\hat{q}](\lambda+1)-n(n-1)Q(0))$ we can actually characterize the
generating
function for the number of distinct rooted dynamical triangulations.
\begin{Gaiat}
Let $0\leq t\leq 1$ be a generic indeterminate, and let
$p_{\lambda}(h)$ denote the number of partitions of the generic
integer $h$ into (at most) $\lambda+1-n$ parts, each $\leq (b(n,n-2)-
\hat{q})(\lambda+1)$.
In a given holonomy represention
$\Theta\colon\pi_1(M;\sigma^n_0)\to G$, and for a given value of the
parameter $b=b(n,n-2)$,
the generating function for the number
of distinct rooted dynamical
triangulations with $N_{n-2}(T_a^{(i)})=\lambda+1$
bones and given number
\begin{eqnarray}
q(0)=\frac{h+(b-\hat{q})(\lambda+1)}{n(\lambda+2-n)}+\hat{q}
\end{eqnarray}
of $n$-dimensional simplices
incident on the $n$ marked bones is given by
\vskip 0.5 cm
\begin{eqnarray}
{\cal G}[W(\Theta,\lambda,b;t)]=c_n\cdot\sum_{h\geq
0}p_{\lambda}(h)t^h=c_n\cdot
\left[ \begin{array}{c}
(b(n,n-2)-\hat{q})(\lambda+1)+(\lambda+1)-n\\ (\lambda+1)-n
\end{array}  \right]
\end{eqnarray}
\vskip 0.5 cm
\noindent where $0<c_n\leq 1$ is an exponential rescaling factor, (see
lemma
\ref{scalemma}
), and
\begin{eqnarray}
\left[ \begin{array}{c}
n\\ m
\end{array}  \right]=\frac{(1-t^n)(1-t^{n-1})\ldots(1-t^{m+1})}{(1-
t^{n-m})(1-t^{n-m-1})\ldots(1-t)}\doteq\frac{(t)_n}{(t)_{n-m}(t)_m}
\end{eqnarray}
is the Gauss polynomial in the variable $t$.
\label{bgenerating}
\end{Gaiat}
Note that the Gauss polynomial $\frac{(t)_n}{(t)_{n-m}(t)_m}
$ is a polynomial in $t$ of degree $(n-m)m$, and that
\begin{eqnarray}
\lim_{t\to 1}
\left[ \begin{array}{c}
n\\ m
\end{array}  \right]=\frac{n!}{m!(n-m)!}=
\left( \begin{array}{c}
n\\ m
\end{array}  \right)
\label{Gaussbinomial}
\end{eqnarray}
(see\cite{Andrews} for an extensive review of the properties of
Gaussian polynomials).
\vskip 0.5 cm
{\bf Proof}. The proof is a straightforward consequence of the
fact\cite{Andrews} that the Gauss polynomials are the generating
function of the partitions $p_{\lambda}(h)$. $\Box$.
\vskip 0.5 cm

From these observations it follows that the generating function for
dynamical triangulations with (spherical) boundaries is provided, up
to an exponential rescaling, (see lemma \ref{scalemma}), by a Gauss
polynomial of degree $(b-\hat{q})(\lambda+1)(\lambda+1-n)$, and we
expect that
their properties bear relevance to simplicial quantum gravity.
It is interesting to remark that these polynomials already play a
basic role in the celebrated solution of the {\it Hard Hexagon Model}
by R.J. Baxter\cite{Andrews2}. A first important property,
related to the theory of Gaussian polynomials, that we can exploit for
understanding the geometry of dynamical triangulations is related to
the {\it unimodality} of the Gaussian polynomials. Recall that a
polynomial $P(t)=a_0+a_1t+\ldots+a_nt^n$ is said to be
unimodal\cite{Andrews} if there exists $m$ such that
$a_0\leq a_1\leq a_2\leq\ldots\leq a_m\geq a_{m+1}\geq
a_{m+2}\geq\ldots\geq a_n$. A simple applications of the unimodal
properties to our case yields the following
\begin{Gaiat}
For all $(b(n,n-2)-\hat{q})(\lambda+1)$, $\lambda$, $h\geq 0$
\begin{eqnarray}
W(\Theta,\lambda,b,h)=W(\Theta,(b-\hat{q})(\lambda+1),b,h)
\end{eqnarray}
where $W(\Theta,(b-\hat{q})(\lambda+1),b,h)=c_n\cdot
p_{(b-\hat{q})(\lambda+1)}(h)$, with $p_{(b-\hat{q})(\lambda+1)}(h)$
denoting the number of
partitions
of the integer $h$ into at most $(b-\hat{q})(\lambda+1)$ parts, each
$\leq\lambda+1-n$. Moreover
\begin{eqnarray}
W(\Theta,\lambda,b,h)=
W(\Theta,\lambda, b,(b-\hat{q})(\lambda+1)(\lambda+1-n)-h)
\end{eqnarray}
and, for $0< h\leq \frac{(b-\hat{q})(\lambda+1)(\lambda+1-n)}{2}$,
\begin{eqnarray}
W(\Theta,\lambda,b,h)-W(\Theta,\lambda,b,h-1)
\geq 0
\end{eqnarray}
\label{unimodal}
\end{Gaiat}

{\bf Proof}. The proof is obtained by the known proof\cite{Andrews} of
unimodality for $p_{\lambda}(h)$, by replacing $p_{\lambda}(h)$ with
$W(\Theta,\lambda,b,h)=c\cdot p_{\lambda}(h)$. $\Box$
\vskip 0.5 cm
Since $W(\Theta,\lambda,b,h)$ enumerates the distinct rooted dynamical
triangulations with a marked curvature assignement related to the
value of the parameter $h$ by
\begin{eqnarray}
h=[nQ(0)-(b-\hat{q})](\lambda+1)-n(n-1)Q(0)
\end{eqnarray}
the above unimodality properties tells us that {\it grosso modo} the
largest entropic contribution to the dynamical triangulation
generating function comes from triangulations such that
\begin{eqnarray}
q(0)\simeq \frac{(b-\hat{q})(\lambda+1)}{2n}
\end{eqnarray}
(this bound being attained on the $n$ marked bones).

\vskip 0.5 cm
\subsection{Asymptotics and Entropy estimates}

If in the generating function ${\cal G}[W(\Theta,\lambda,b;t)]$ we let
$t\to 1$ we get the enumerator of the distinct dynamical
triangulations, for a given representation of the fundamental group.
Roughly speaking, for $t\to 1$, ${\cal
G}[W(\Theta,\lambda,b;t)]$ reduces to the sum over all possible
curvature assignements $q(0)$ in the set of distinct  dynamical
triangulations in the given holonomy representation. We have the
following
\begin{Gaiat}
In a given  represention
$\Theta\colon\pi_1(M;\sigma^n_0)\to G$, and for a given value of the
parameter $b=b(n,n-2)$,
the number $W(\Theta,\lambda,b)$  of distinct dynamical triangulations
with $N_{n-2}(T_a^{(i)})=\lambda+1$
bones, $n$ of which are marked, is given by
\begin{eqnarray}
W(\Theta,\lambda,b)= c_n\cdot
\left( \begin{array}{c}
(b(n,n-2)-\hat{q})(\lambda+1)+(\lambda+1)-n\\ (\lambda+1)-n
\end{array}  \right)
\label{bound}
\end{eqnarray}
\label{prop}
\end{Gaiat}
\vskip 0.5 cm
{\bf Proof}. The sum $\lim_{t\to 1}{\cal G}[W(\Theta,\lambda,b;t)]$
involving the partions $p_{\lambda}(h)$
can be evaluated by exploiting the property (\ref{Gaussbinomial}
) of the Gaussian polynomials, {\it viz.},
\begin{eqnarray}
& & \lim_{t\to 1}{\cal G}[W(\Theta,\lambda,b;t)]
=c_n\cdot\sum_{h\geq
0}^{(\lambda+1-n)(\lambda+1)(b-\hat{q})}p_{\lambda}(h)\nonumber\\
&= & c_n\cdot
\left( \begin{array}{c}
(b(n,n-2)-\hat{q})(\lambda+1)+(\lambda+1)-n\\ (\lambda+1)-n
\end{array}  \right)
\end{eqnarray}
establishing the desired result. $\Box$.
\vskip 0.5 cm

On applying Stirling's formula, we easily get that the above
result
yields, for large $\lambda$, the asymptotics
\vskip 0.5 cm
\begin{eqnarray}
W(\Theta;\lambda;b)\simeq
\frac{c_n}{\sqrt{2\pi}}
\sqrt{\frac{(b-\hat{q}+1)^{1-2n}}{b-\hat{q}}}
{\left [ \frac{(b-\hat{q}+1)^{b-\hat{q}+1}}{(b-\hat{q})^{b-\hat{q}}}
\right ] }^{\lambda+1}
{\lambda}^{-\frac{1}{2}}
\left( 1+O({\lambda}^{-\frac{3}{2}}) \right)
\label{superiore}
\end{eqnarray}
\vskip 0.5 cm

In order to obtain from (\ref{superiore})the asymptotics of the
number of distinct
dynamical triangulations $T_a$, with $\lambda+1$ bones, we have to
factor out the particular rooting we exploited for our enumerative
purposes.
We have first of all to factor out the
marking of the $n$ bones $\sigma^{n-2}(0),\sigma^{n-
2}(1),\ldots,\sigma^{n-2}(n-1)$, and the marking of the base
simplex $\sigma^n_0$.
To factor out the marking of the $n$ bones, we have to divide
(\ref{superiore}) by ${\lambda}^n$. There can be
at most $[\frac{(b(n,n-2)-\hat{q})}{n}](\lambda+1)$ simplices
$\sigma^n$ sharing one of
the marked bones $\{\sigma^{n-
2}(i)\}_{i=0,\ldots,n-1}$, (actually the actual count is
$[\frac{(b(n,n-2)-\tilde{q})}{n}](\lambda+1)$ where $\tilde{q}$
can be larger than $\hat{q}$; we use $\hat{q}$ for
simplicity). Since the marked simplex $\sigma^n_0$ can be incident on
any of the $n$ marked bones, we still have to divide by $n$. For
$\lambda>>1$,
we get in this way a normalization factor
$[b(n,n-2)-\hat{q}]\lambda^{n+1}$.
\vskip 0.5 cm
We have also
marked  an orthogonal representation $\Theta_0$ of
$\pi_1(M;p_0)$ corresponding to the marked base
simplex $\sigma^n_0$, and as discussed in section IV, besides the
above purely combinatorial
factors,
we have to divide (\ref{superiore}) also by the volume of the moduli
space, $Vol[{\cal M}(T_a)]$,  associated with the possible non-trivial
deformations of the class of dynamical triangulations considered.
According to theorem \ref{modfactor} this volume is
provided by

\begin{eqnarray}
Vol[{\cal M}(T_a)]\simeq A({\cal V})
 \left( 4\pi{\cal V}vol(\sigma^n)\frac{b}{n(n+1)}
\lambda \right)^{-D/2}
\end{eqnarray}
where ${\cal V}{\doteq} Vol [\frac{Hom({\pi}_1(M),G)}{G}]$ and
$D{\doteq} dim [Hom({\pi}_1(M),G)]$.
Hence, in
order to get an asymptotic estimate
for the number of distinct dynamical
triangulations, we have to
divide (\ref{superiore}) by
\begin{eqnarray}
A({\cal V})[b(n,n-2)-\hat{q}]
\lambda^{n+1}
 \left( 4\pi{\cal V}vol(\sigma^n)\frac{b}{n(n+1)}
\lambda \right)^{-D/2}
\end{eqnarray}
\vskip 0.5 cm
\noindent {\bf The Entropy Function}. These observations establish the
following
\begin{Gaiat}
 The number of
distinct dynamical triangulations, with $\lambda+1$ bones,
and with an average number, $b\equiv b(n,n-2)$, of
$n$-simplices incident on a
bone, on an $n$-dimensional, ($n\leq 4$), PL-
manifold $M$ of given fundamental group $\pi_1(M)$, can be
asymptotically estimated
according to

\vskip 0.5 cm
\begin{eqnarray}
W(\lambda,b)\simeq\nonumber
\end{eqnarray}
\vskip 0.5 cm
\begin{eqnarray}
W_{\pi}\cdot
\frac{c_n}{\sqrt{2\pi}}
\sqrt{\frac{(b-\hat{q}+1)^{1-2n}}{(b-\hat{q})^3}}
{\left [ \frac{(b-\hat{q}+1)^{b-\hat{q}+1}}{(b-\hat{q})^{b-\hat{q}}}
\right ] }^{\lambda+1}
(\frac{b(n,n-2)}{n(n+1)} \lambda)^{D/2}
{\lambda}^{-\frac{2n+3}{2}}
\label{packings}
\end{eqnarray}
\vskip 0.5 cm
\noindent where $0<c_n\leq 1$ is the exponential  rescaling factor of
lemma
\ref{scalemma},
and where we have introduced the topology dependent parameter
\begin{eqnarray}
W_{\pi}{\doteq} A^{-1}({\cal V})
[4\pi{\cal V}vol(\sigma^n)]^{D/2}
\end{eqnarray}
$\Box$.
\end{Gaiat}

\vskip 0.5 cm
\noindent As we shall see momentarily, the estimate (\ref{packings})
fits extremely well with the data
coming from
numerical simulations in dimension $n\geq 3$, and in dimension $n=2$,
(where $b(n,n-2)=6$),
it is in remarkable
agreement with the known analytical estimates\cite{Pari}.
Notice in particular that for a two-dimensional sphere  the
integration over the representation variety drops, and
(\ref{packings}) yields
\vskip 0.5 cm
\begin{eqnarray}
W(\lambda;S^2)\simeq
\frac{c_2}{24\sqrt{6\pi}}
{\left [\frac{4^4}{3^3}\right ]
}^{v}
{v}^{-\frac{7}{2}}
\label{sphere2}
\end{eqnarray}
\vskip 0.5 cm
\noindent where we have denoted by $v=\lambda+1$ the number of
vertices of the triangulation.
This estimate  should be compared with the known result
provided
long ago by Tutte\cite{Pari}, {\it viz.},
\vskip 0.5 cm
\begin{eqnarray}
\frac{1}{64\sqrt{6\pi}}
{\left [\frac{4^4}{3^3}\right ]
}^{v}
{v}^{-\frac{7}{2}}
\label{Tutte}
\end{eqnarray}
\vskip 0.5 cm
Thus, if the exponential rescaling $0<c_2\leq 1$, characterized by
lemma \ref{scalemma}, just reduces to  a constant rescaling
($c_2\simeq 24/64$), the
agreement
between the two asymptotic analysis would be excellent.

\vskip 0.5 cm
\subsection{The $2$-dimensional case}

If we exploit the compact formula (\ref{compact}) of lemma
\ref{twocompact}, providing the asymptotics of the volume of
R\"{o}mer-Z\"{a}hringer deformations for surfaces, then the agreement
between our analysis and the known results in dimension $n=2$ is
complete.
For a surface $\Sigma_h$ of genus $h$, and for
representations of $\pi_1(\Sigma_h)$ in a general semi-simple Lie
group G, the properties
of the representation variety \cite{Goldman} exploited in lemma
\ref{twocompact}
imply that (\ref{packings}) takes the form
\begin{eqnarray}
W(\lambda)|_{n=2}\simeq
W_{\pi}\cdot\frac{c_2}{24\sqrt{6\pi}}
{\left [\frac{4^4}{3^3}\right ]
}^{\lambda+1}
{\lambda}^{-1-\frac{5}{4}\chi(\Sigma_h)}(1+\ldots)
\label{localize}
\end{eqnarray}
\vskip 0.5 cm
\noindent where  $\chi(\Sigma_h)$ is the Euler characteristic of the
surface
$\Sigma_h$, and  $\ldots$ stands for terms of higher order.
\vskip 0.5 cm
If we rewrite (\ref{localize}) as
\begin{eqnarray}
W_{\pi}\cdot(\frac{4^4}{3^3})^{\lambda}
{\lambda}^{\gamma_s-3}
\left( 1+\ldots \right)
\end{eqnarray}
\vskip 0.5 cm
\noindent where $\gamma_s$ is the {\it critical exponent}, then we get
\vskip 0.5 cm
\begin{eqnarray}
\gamma_s(M)=2-\frac{5}{4}\chi(M)
\label{critgravity}
\end{eqnarray}
The exponent $\gamma_s(M)$ defines the universal critical
exponent of pure $2D$-gravity. The expression (\ref{critgravity})
coincides with the well-known expression obtained  by the use of
Liouville theory.

\subsection{The $n\geq 3$-dimensional case}

In order to check our analytical results against the MonteCarlo
simulations which up to now are the only available results in
dimension
$n\geq 3$, let us rewrite the partition function for dynamical
triangulations (\ref{five}) explicitly
in terms of the entropy function $W(\lambda,b)$, as

\begin{eqnarray}
Z[k_{n-2},k_n]=\sum_{N}\left ( \sum_{\lambda_{min}}^{\lambda_{max}}
W[\lambda,N=\frac{2b(\lambda+1]}{n(n+1)})e^{k_{n-2}\lambda}\right )
e^{-k_nN_n}
\label{ventidue}
\end{eqnarray}
\vskip 0.5 cm
\noindent where a sum  over all possible $\lambda$ appears. This
sum is needed since triangulations with distinct values of
$\lambda$ contribute to the number
of distinct triangulations with a given volume $N$.
Notice that in dimension $n=2$ this further summation is absent
since, in that case, the average value of $2$-simplices
incident on a bone (vertex) is always $b=6$.
\vskip 0.5 cm

The presence of this summation over $\lambda$ shows that in
dimension
greater than $2$, what entropically characterizes the infinite volume
limit of a dynamically triangulated model as a
statistical system is not just the entropy function $W(\lambda,b)$ ,
(which can be identified with the entropy in the micro-canonical
ensamble) but rather the {\bf effective
entropy} provided by the canonical partition function
\begin{eqnarray}
W(N, k_{n-2})_{eff}\equiv\sum_{\lambda_{min}}^{\lambda_{max}}
W(\lambda,N)e^{k_{n-2}\lambda}
\label{ventitre}
\end{eqnarray}

In order to evaluate the canonical partition function and compare the
results
with the know numerical simulations we can limit ourselves to
considering dynamical triangulations of simply connected manifolds,
since numerical data are available just for the $3$-sphere ${\Bbb
S}^3$ and the $4$-sphere ${\Bbb S}^4$.
Already in this case computations are far from being trivial, and much
care is needed in handling the asymptotic estimation of the sum
(\ref{ventitre}).
\vskip 0.5 cm
According to proposition \ref{prop}, the entropy for
simply-connected PL-manifolds is
\begin{eqnarray}
W(\lambda,b)= \frac{c_n}{[(b-\hat{q})(\lambda+1)^{n+1}]}
\left( \begin{array}{c}
(b(n,n-2)-\hat{q})(\lambda+1)+(\lambda+1)-n\\ (\lambda+1)-n
\end{array}  \right)
\label{micro}
\end{eqnarray}
where the factor $[(b-\hat{q})(\lambda+1)^{n+1}]$ comes from
removing the labellings associated with the rootings, (see the
discussion following theorem \ref{prop}). For discussing the
asymptotics of (\ref{micro}), we need Stirling's formula in the form
$k!\simeq \sqrt{2\pi}(k+1)^{k+\frac{1}{2}}e^{-k-1}$, which is accurate
also for small $k$.
\vskip 0.5 cm
With these caveats along the way, and setting explicitly $\hat{q}=3$,
a long but elementary computation
provides the
asymptotics
\vskip 0.5 cm
\begin{eqnarray}
W(\lambda, N)\simeq \frac{c_n}{\sqrt{2\pi}}
\frac{\lambda^{-1/2}(\hat{N}-
2\lambda+1)^{-n}}{\hat{N}\sqrt{(\hat{N}-3\lambda+1)(\hat{N}-
2\lambda+1)}}e^{h(\hat{N},\lambda)}
 \end{eqnarray}
\vskip 0.5 cm
\noindent where $\hat{N}\doteq\frac{1}{2}n(n+1)N$, and where we have
set
\begin{eqnarray}
h(\hat{N},\lambda)\doteq
-\lambda\log\lambda+
\log\frac{[\hat{N}-
2\lambda+1]^{(\hat{N}-2\lambda+1)}}{[\hat{N}-3\lambda+1]^{(\hat{N}-
3\lambda+1)}}
\end{eqnarray}
Thus, (\ref{ventitre}) reduces to
\begin{eqnarray}
W(N, k_{n-2})_{eff}=\frac{c_n}{\sqrt{2\pi}}
\sum_{\lambda_{min}}^{\lambda_{max}}
\frac{\lambda^{-1/2}(\hat{N}-
2\lambda+1)^{-n}}{\hat{N}\sqrt{(\hat{N}-3\lambda+1)(\hat{N}-
2\lambda+1)}}e^{h(\hat{N},\lambda)+k_{n-2}\lambda}
\label{laplace}
\end{eqnarray}
where for $n=3$
\begin{eqnarray}
\frac{\hat{N}}{6}=\lambda_{min}\leq\lambda\leq\lambda_{max}=\frac{2}{9
}\hat{N}
\end{eqnarray}
and for $n=4$
\begin{eqnarray}
\frac{\hat{N}}{5}=\lambda_{min}\leq\lambda\leq\lambda_{max}=\frac{\hat
{N}}{4}
\end{eqnarray}
\vskip 0.5 cm
We can estimate the sum (\ref{laplace}) by noticing that the function
appearing in the exponent, {\it viz.},
\begin{eqnarray}
& & f(\hat{N},\lambda){\doteq} h(\hat{N},\lambda)+k_{n-
2}\lambda=\nonumber\\
& & -\lambda\log\lambda+(\hat{N}-2\lambda+1)\log(\hat{N}-2\lambda+1)-
(\hat{N}-3\lambda+1)\log(\hat{N}-3\lambda+1)+ k_{n-2}\lambda
\end{eqnarray}
has a sharp maximum in correspondence of the solution,
$\lambda=\lambda^*(k_{n-2})$,  of the equation
\begin{eqnarray}
\log\frac{[\hat{N}-3\lambda+1]^3}{\lambda[\hat{N}-2\lambda+1]^2}=-
k_{n-2}
\end{eqnarray}
\vskip 0.5 cm
A straightforward computation provides
\begin{eqnarray}
\lambda^*=\frac{\hat{N}+1}{3}(1-\frac{1}{A(k_{n-2})})
\label{solution}
\end{eqnarray}
where for notational convenience we have set
\begin{eqnarray}
& & A(k_{n-2})\doteq\nonumber\\
& & { \left [\frac{27}{2}e^{k_{n-2}}+1+
\sqrt{(\frac{27}{2}e^{k_{n-2}}+1)^2-1} \right ]}^{1/3}+
{ \left [\frac{27}{2}e^{k_{n-2}}+1-
\sqrt{(\frac{27}{2}e^{k_{n-2}}+1)^2-1} \right ]}^{1/3}-1
\end{eqnarray}
\vskip 0.5 cm
The structure of (\ref{solution}) suggests the change of variable
\begin{eqnarray}
\lambda=(\hat{N}+1)\eta
\label{change}
\end{eqnarray}
(we wish to thank G. Gionti for this remark), and
by replacing the sum (\ref{laplace}) with an integration,
we get
\begin{eqnarray}
W(N, k_{n-
2})_{eff}=\frac{c_n}{\sqrt{2\pi}}{\hat{N}}^{-n-1/2}
\int_{\eta_{min}}^{\eta_{max}}d\eta
\frac{\eta^{-1/2}(1-2\eta)^{-n}}{\sqrt{(1-3\eta)(1-
2\eta)}}e^{(\hat{N}+1)f(\eta)}
\label{integrale}
\end{eqnarray}
\vskip 0.5 cm
where
\begin{eqnarray}
f(\eta)\doteq
-\eta\log\eta+(1-2\eta)\log(1-2\eta)-
(1-3\eta)\log(1-3\eta)+ k_{n-2}\eta
\end{eqnarray}
and $\eta_{min}\doteq\eta(\lambda_{min})$,
$\eta_{max}\doteq\eta(\lambda_{max})$. For $n=4$, we get
 $\eta_{min}\simeq 1/5$, $\eta_{max}\simeq 1/4$, whereas for $n=3$,
$\eta_{min}\simeq 1/6$ and $\eta_{max}\simeq 2/9$.
\vskip 0.5 cm
The obvious strategy is to estimate (\ref{integrale}) with  Laplace
method, however attention must be paid to the possibility that, as
$k_{n-2}$ varies, the maximum
\begin{eqnarray}
\eta^*(k_{n-2})=\frac{1}{3}(1-\frac{1}{A(k_{n-2})})
\end{eqnarray}

\noindent crosses the integration limits $\eta_{min}$ and
$\eta_{max}$. It will become clear in a moment that
the quantities telling us when we are {\it
nearby} these particular regions are
\begin{eqnarray}
\psi_{min}(k_{n-2})\doteq -sgn(\eta^*-\eta_{min})
\sqrt{2[f(\eta^*)-f(\eta_{min})]}
\label{min}
\end{eqnarray}
and
\begin{eqnarray}
\psi_{max}(k_{n-2}){\doteq}  sgn(\eta^*-\eta_{max})
\sqrt{2[f(\eta^*)-f(\eta_{max})]}
\label{max}
\end{eqnarray}
This suggests that in order to control the large $N$ behavior of the
effective entropy as $k_{n-2}$ varies,  we have to use uniform Laplace
estimation in terms of $\psi$. The following general theorem provides
such uniform asymptotics. We state it first for the case involving the
crossing of the lower integration limit $\eta_{min}$. A completely
analogous result holds, {\it mutatis mutandis}, for the upper crossing
$\eta_{max}$, and we state it as an obvious corollary .
\vskip 0.5 cm
\begin{Gaiat}
Let us consider the set of all simply-connected
$n$-dimensional, ($n=3,4$), dynamically triangulated manifolds.
Let $k_{n-2}^{inf}$ denote the unique solution
of the equation
\begin{eqnarray}
\frac{1}{3}(1-\frac{1}{A(k_{n-2})})=\eta_{min}
\end{eqnarray}
Let $0<\epsilon<1$ small enough, then
for all values of  the inverse gravitational coupling $k_{n-2}$ such
that
\begin{eqnarray}
k^{inf}_{n-2}-\epsilon< k_{n-2}<k^{inf}_{n-2}+\epsilon,
\end{eqnarray}
the large $N$-behavior of the canonical partition function $W(N, k_{n-
2})_{eff}$
is given by
the uniform asymptotics
\vskip 0.5 cm
\begin{eqnarray}
\frac{c_n}{\sqrt{2\pi}}{\hat{N}}^{-n-1/2}
e^{(\hat{N}+1)f(\eta_{min})}
\left[\frac{\alpha_0}{\sqrt{\hat{N}}}
w_0(\psi_{min}(k_{n-2})
\sqrt{\hat{N}})+\frac{\alpha_1}{\hat{N}}
w_{-1}(\psi_{min}(k_{n-2})\sqrt{\hat{N}})
\right]
\label{polymer}
\end{eqnarray}
\vskip 0.5 cm
\noindent where $w_r(z)\doteq\Gamma(1-r)e^{z^2/4}D_{r-1}(z)$,
($r<1$), $D_{r-
1}(z)$ and $\Gamma(1-r)$ respectively denote the parabolic cylinder
functions and the Gamma function, and where the constants $\alpha_0$
and
$\alpha_1$ are given by
\vskip 0.5 cm
\begin{eqnarray}
\alpha_0\doteq -
\frac{\eta_{min}^{-1/2}(1-2\eta_{min})^{-n}}{\sqrt{(1-3\eta_{min})(1-
2\eta_{min})}}\frac{\psi_{min}(k_{n-2})}{f_{\eta}(\eta_{min})}
\end{eqnarray}
\vskip 0.5 cm
\begin{eqnarray}
\alpha_1\doteq \frac{\alpha_0}{\psi_{min}(k_{n-2})}
-\frac{\eta*^{-1/2}(1-2\eta^*)^{-n}}{\sqrt{(1-3\eta^*)(1-
2\eta^*)}} \left[
\frac{1}{\psi_{min}(k_{n-2})\sqrt{-
f_{\eta\eta}(\eta^*)}} \right]
\end{eqnarray}
\vskip 0.5 cm
\label{fasi}
\end{Gaiat}
\vskip 0.5 cm
Notice that  the above expression,
(in which $f_{\eta}{\doteq} df/d\eta$, and $f_{\eta\eta}{\doteq}
d^2f/d\eta^2$), provides the leading asymptotics. The full
asymptotics
is discussed during the proof of the theorem. Notice also that in some
circumstances, (notably in the $3$-dimensional case when employing as
variables $N_0$, and $N_3$), there can be an equally important
constribution to $W(N,k_{n-2})_{eff}$ due to the upper crossing
$\eta_{max}$. In that case $W(N,k_{n-2})_{eff}$ is, at leading order,
characterized by the sum of the uniform asymptotics around both
$f(\eta_{min})$ and $f(\eta_{max})$.
\vskip 0.5 cm
\noindent {\bf Proof}. We have to provide a uniform asymptotic
estimation for
large $\hat{N}$ of the integral appearing in (\ref{integrale}), {\it
viz.},
\vskip 0.5 cm
\begin{eqnarray}
I{\doteq}\int_{\eta_{min}}^{\eta_{max}}d\eta
\frac{\eta^{-1/2}(1-2\eta)^{-n}}{\sqrt{(1-3\eta)(1-
2\eta)}}e^{(\hat{N}+1)f(\eta)}
\label{minimo}
\end{eqnarray}
\vskip 0.5 cm
The uniformity requirement stated here refers to the existence of a
$\delta\in {\Bbb R}^+$ such that for all  $|\psi(k_{n-2})|\leq\delta$
the error  after a finite number of terms in the asymptotic expansion
should be smaller in $N$ than the last term which is
kept\cite{Bleistein}. As discussed above, the integral $I$ has various
distinct asymptotic regimes,  according to the
relative location of the maximum $\eta^*$ with respect to the lower
integration limit $\eta_{min}$ or to the upper limit $\eta_{max}$. We
explicitly discuss for simplicity the case where $\eta^*$ may approach
$\eta_{min}$. The case in which
$\eta^*\to\eta_{max}$, can be dealt
with similarly, and we shall indicate the necessary modifications in
the proof and state the final result.
\vskip 0.5 cm
In order to carry out the required asymptotic estimation, we first
transform the exponential in the integral $I$ by introducing the
variable $p=p(\eta)$ according to
\begin{eqnarray}
f(\eta)-f(\eta_{min})=-\left( \frac{p^2}{2}+\psi_{min}(k_{n-2})p
\right)
\end{eqnarray}
where for $\eta=\eta_{min}$ we assume $p=0$ and $(d\eta/dp)|_{p=0}>0$,
so
that the orientation of the path of integration remains unchanged.
Differentiating this expression we get
\begin{eqnarray}
\frac{df}{d\eta}=-(p+\psi_{min})\frac{dp}{d\eta}
\end{eqnarray}
In order to have $d\eta(p)/dp$ finite and non-zero everywhere we
require that $p\to -\psi_{min}$ as $\eta\to\eta^*$. Thus, by
L'H\^opital rule
\begin{eqnarray}
\lim_{p\to-\psi}\frac{d\eta}{dp}=\sqrt{-
\frac{1}{f_{\eta\eta}(\eta^*)}}
\end{eqnarray}
To determine the expression of the parameter $\psi_{min}(k_{n-2})$,
(see (\ref{min})), controlling the uniform expansion, we evaluate
$f(\eta)-f(\eta_{min})$ at $\eta=\eta^*$, ({\it viz.}, for $p=-
\psi_{min}$):
\begin{eqnarray}
f(\eta^*)-f(\eta_{min})=\frac{1}{2}\psi^2_{min}
\end{eqnarray}
whence
\begin{eqnarray}
\psi_{min}=\pm\sqrt{2[f(\eta^*)-f(\eta_{min})]}
\end{eqnarray}
The branch of the square root is selected by examining
$\frac{d}{dp}[f(\eta)-f(\eta_{min})]$ at $\eta=\eta_{min}$, ({\it
i.e.}, for $p=0$). We get
\begin{eqnarray}
\frac{df}{dp}|_{p=0}=\frac{df}{d\eta}|_{p=0}=-
\frac{d}{dp}(\frac{p^2}{2}+\psi_{min}p)|_{p=0}=-\psi_{min}
\end{eqnarray}
Since $(d\eta/dp)|_{p=0}>0$ by hypothesis, and
$(df/d\eta)|_{\eta_{min}}>0$, for $\eta^*>\eta_{min}$,
($(df/d\eta)|_{\eta_{min}}<0$, for $\eta^*<\eta_{min}$), we get that
\begin{eqnarray}
sgn \psi_{min}=- sgn (\eta^*-\eta_{min}).
\end{eqnarray}
Thus
\begin{eqnarray}
\psi_{min}(k_{n-2})=- sgn (\eta^*-\eta_{min})
\sqrt{2[f(\eta^*)-f(\eta_{min})]}
\end{eqnarray}
\vskip 0.5 cm
\noindent (note that $\psi_{min}(k_{n-2})$ is a monotonically
decreasing function of $k_{n-2}$).
\vskip 0.5 cm
With these remarks out the way, we  can write
\vskip 0.5 cm
\begin{eqnarray}
I=e^{(\hat{N}+1)f(\eta_{min})}\int_0^{\infty}G(p)e^{-
(\hat{N}+1)(\frac{1}{2}p^2+\psi_{min}p)}dp
\end{eqnarray}
\vskip 0.5 cm
\noindent where we have extended the upper limit of integration to
$\infty$, (on
introducing a Heaviside function), and where we have set
\begin{eqnarray}
G(p)\doteq
\left[ \frac{\eta^{-1/2}(1-2\eta)^{-n}}{\sqrt{(1-3\eta)(1-
2\eta)}} \right] \frac{d\eta}{dp}
\end{eqnarray}
\vskip 0.5 cm
If $p=-\psi_{min}\in [0,+\infty)$, then we can expand the function
$G(p)$ in a neighborhood of $p=-\psi_{min}$, (corresponding to
$\eta=\eta^*$). Otherwise, if $p=-
\psi_{min}\notin [0,+\infty)$ we need to expand around $p=0$,
(corresponding to $\eta=\eta_{min}$). According to a standard
procedure\cite{Bleistein}, we can take care of both cases by setting
\begin{eqnarray}
G(p)\doteq\alpha_0+\alpha_1p+p(p+\psi)G_1(p)
\end{eqnarray}
The integral $I$ now becomes
\vskip 0.5 cm
\begin{eqnarray}
I=I_0+I_1 &= &
e^{(\hat{N}+1)f(\eta_{min})}\int_0^{\infty}(\alpha_0+\alpha_1p)
e^{-(\hat{N}+1)(\frac{1}{2}p^2+\psi_{min}p)}dp \nonumber\\
&+ & e^{(\hat{N}+1)
f(\eta_{min})}\int_0^{\infty}p(p+\psi_{min})G_1(p)
e^{-(\hat{N}+1)(\frac{1}{2}p^2+\psi_{min}p)}dp
\end{eqnarray}
\vskip 0.5 cm
Recall that for $r<1$
\begin{eqnarray}
w_r(z)\doteq
\int_0^{\infty}p^{-r}
e^{-(\frac{1}{2}p^2+zp)}dp=\Gamma(1-r)e^{z^2/4}D_{r-1}(z)
\end{eqnarray}
where $\Gamma(1-r)$ is the Euler Gamma function and $D_{r-1}(z)$ is
the parabolic cylinder function\cite{Abramowitz}. Thus we get
\vskip 0.5 cm
\begin{eqnarray}
I_0=
e^{(\hat{N}+1)f(\eta_{min})}
\left[\frac{\alpha_0}{\sqrt{\hat{N}}}
w_0(\psi_{min}(k_{n-2})
\sqrt{\hat{N}})+\frac{\alpha_1}{\hat{N}}
w_{-1}(\psi_{min}(k_{n-2})\sqrt{\hat{N}})
\right]
\end{eqnarray}
\vskip 0.5 cm
\noindent where the constants $\alpha_0$ and
$\alpha_1$ are determined by
\begin{eqnarray}
\alpha_0= G(0)
\end{eqnarray}
and
\begin{eqnarray}
\alpha_1=\frac{G(0)-G(-\psi_{min})}{\psi_{min}}
\end{eqnarray}
Namely:
\vskip 0.5 cm
\begin{eqnarray}
\alpha_0=-
\frac{\eta_{min}^{-1/2}(1-2\eta_{min})^{-n}}{\sqrt{(1-3\eta_{min})(1-
2\eta_{min})}}\frac{\psi_{min}(k_{n-2})}{f_{\eta}(\eta_{min})}
\end{eqnarray}
\vskip 0.5 cm
\begin{eqnarray}
\alpha_1\doteq
\frac{\alpha_0}{\psi_{min}(k_{n-2})}
-\frac{\eta*^{-1/2}(1-2\eta)^{-n}}{\sqrt{(1-3\eta^*)(1-
2\eta^*)}} \left[
\frac{1}{\psi_{min}(k_{n-2})\sqrt{-
f_{\eta\eta}(\eta^*)}} \right]
\end{eqnarray}
\vskip 0.5 cm
The remaining integral $I_1$ can be reduced to the same structure
of $I$ and an inductive argument\cite{Bleistein} shows that the
asymptotics of $I$ is provided by
\vskip 0.5 cm
\begin{eqnarray}
I=
e^{(\hat{N}+1)f(\eta_{min})}\left\{ \frac{w_0(\psi_{min}(k_{n-
2})\sqrt{\hat{N}+1})}{\sqrt{\hat{N}+1}}\left[\sum_{i=0}^m
\frac{\alpha_{2i}}{(\hat{N}+1)^i}+O\left(\frac{1}{{\hat{N}}^{m+1}}
\right)    \right]\right\}\nonumber
\end{eqnarray}
\begin{eqnarray}
+e^{(\hat{N}+1)f(\eta_{min})}\left\{ \frac{w_{-1}(\psi_{min}(k_{n-
2})\sqrt{\hat{N}+1})}{\hat{N}+1}\left[\sum_{i=0}^m\frac{\alpha_{2i+1}}
{(\hat{N}+1)^i}+O\left(\frac{1}{{\hat{N}}^{m+1}} \right)
\right]\right\}
\end{eqnarray}
\vskip 0.5 cm
\noindent where the constants $\alpha_{2i}$ and $\alpha_{2i+1}$ are
determined recursively from $\alpha_0$ and $\alpha_1$ according to
\begin{eqnarray}
\alpha_{2i}=G_i(0)
\end{eqnarray}
\begin{eqnarray}
\alpha_{2i+1}=\frac{G_i(0)-G_i(-
\psi_{min})}{\psi_{min}}+\frac{dG_i}{dp}|_{p=-\psi}
\end{eqnarray}
and where the functions $G_i(p)$ satisfy the equation
\begin{eqnarray}
G_i(p)+p\frac{dG_i(p)}{dp}=\alpha_{2i}+p\alpha_{2i+1}+
p(p+\psi_{min})G_{i+1}(p).
\end{eqnarray}
\vskip 0.5 cm
\vskip 0.5 cm Thus, the leading uniform asymptotics is given by
\vskip 0.5 cm
\begin{eqnarray}
I=
e^{(\hat{N}+1)f(\eta_{min})}\left\{ \frac{w_0(\psi_{min}(k_{n-
2})\sqrt{\hat{N}+1})}{\sqrt{\hat{N}+1}}\left[
\alpha_0 +O\left(\frac{1}{\hat{N}} \right)    \right]\right\}\nonumber
\end{eqnarray}
\begin{eqnarray}
+e^{(\hat{N}+1)f(\eta_{min})}\left\{ \frac{w_{-1}(\psi_{min}(k_{n-
2})\sqrt{\hat{N}+1})}{\hat{N}+1}\left[ \alpha_1
+O\left(\frac{1}{\hat{N}} \right)\right]\right\}
\label{dominante}
\end{eqnarray}
\vskip 0.5 cm
\noindent which provides the stated result.
\vskip 0.5 cm

A similar analysis can be carried out when $\eta^*$ approaches
$\eta_{max}=\frac{1}{4}$.
In this case, it is convenient to make the preliminary change of
variable $y=1-3\eta$ thus reducing the integral $I$ to
\begin{eqnarray}
I_{\infty}=3^n\int_{1-3\eta_{max}}^{1-3\eta_{min}}dy\frac{(1-y)^{-
1/2}(1+2y)^{-n}}{\sqrt{y(1+2y)}}
e^{(\hat{N}+1)f[\eta(y)]}
\end{eqnarray}
If we set
\begin{eqnarray}
\psi_{max}(k_{n-2})=- sgn (y(\eta^*))
\sqrt{2[f(\eta^*)-f(\eta_{max})]}
\end{eqnarray}
and
introduce the variable $p=p(y)$ according to
\begin{eqnarray}
f[\eta(y)]-f[\eta(0)]=-\left( \frac{p^2}{2}+\psi_{max}(k_{n-2})p
\right)
\end{eqnarray}
we get again an integral whose asymptotic estimation  can be
reduced to the integral representation of the parabolic cylinder
function and to a set of parameters $\beta_{2i}$ and $\beta_{2i+1}$
recursively determined from the function $H(p)$ defined by
\begin{eqnarray}
H(p)p^{-1/2}\doteq \left[ \frac{(1-y(p))^{-
3/2}}{\sqrt{y(p)(1+2y(p))}} \right]
\frac{dy(p)}{dp}
\end{eqnarray}
\vskip 0.5 cm
\noindent Explicitly we obtain
\begin{eqnarray}
I_{\infty}=
e^{(\hat{N}+1)f(\eta_{max})}
\left\{ \frac{w_{0}(\psi_{max}(k_{n-
2})\sqrt{\hat{N}+1})}{(\hat{N}+1)^{1/2}}\left[\sum_{i=0}^m
\frac{\xi_{2i}}{(\hat{N}+1)^i}+O\left(\frac{1}{{\hat{N}}^{m+1}}
\right)    \right]\right\}\nonumber
\end{eqnarray}
\begin{eqnarray}
+e^{(\hat{N}+1)f(\eta_{max})}\left\{ \frac{w_{-1}(\psi_{max}(k_{n-
2})\sqrt{\hat{N}+1})}{(\hat{N}+1)}\left[\sum_{i=0}^m
\frac{\xi_{2i+1}}{(\hat{N}+1)^i}+
O\left(\frac{1}{{\hat{N}}^{m+1}} \right) \right]\right\}
\end{eqnarray}
\vskip 0.5 cm
\noindent where
\begin{eqnarray}
\xi_0= H(0)
\end{eqnarray}
and
\begin{eqnarray}
\xi_1=\frac{H(0)-H(-\psi_{max})}{\psi_{max}}
\end{eqnarray}
The constants $\xi_{2i}$ and $\xi_{2i+1}$ are determined
recursively from $\xi_0$ and $\xi_1$ according to
\begin{eqnarray}
\xi_{2i}=H_i(0)
\end{eqnarray}
\begin{eqnarray}
\xi_{2i+1}=\frac{H_i(0)-H_i(-
\psi_{max})}{\psi_{max}}+\frac{dH_i}{dp}|_{p=-\psi}
\end{eqnarray}
and the functions $H_i(p)$ satisfy the equation
\begin{eqnarray}
H_i(p)+p\frac{dH_i(p)}{dp}=\xi_{2i}+p\xi_{2i+1}+
p(p+\psi_{max})H_{i+1}(p)
\end{eqnarray}
From these expressions it is straightforward to
obtain the leading $\hat{N}$-asymptotics for $I$ as $k_{n-2}$ is such
that $\eta^*$ crosses $\eta_{max}$. $\Box$.
\vskip 0.5 cm
Summing up we have the following
\begin{Gaial}
Let us consider the set of all simply-connected
$n$-dimensional, ($n=3,4$), dynamically triangulated manifolds.
Let $k_{n-2}^{crit}$ denote the unique solution
of the equation
\begin{eqnarray}
\frac{1}{3}(1-\frac{1}{A(k_{n-2})})=\eta_{max}
\end{eqnarray}
Let $0<\epsilon<1$ small enough, then
for all values of  the inverse gravitational coupling $k_{n-2}$ such
that
\begin{eqnarray}
k^{crit}_{n-2}-\epsilon< k_{n-2}<+\infty,
\end{eqnarray}
the large $N$-behavior of the canonical partition function $W(N, k_{n-
2})_{eff}$
is given by
the uniform asymptotics
\vskip 0.5 cm
\begin{eqnarray}
\frac{c_n}{\sqrt{2\pi}}{\hat{N}}^{-n-1/2}
e^{(\hat{N}+1)f(\eta_{max})}
\left[\frac{\xi_0}{\sqrt{\hat{N}}}
w_0(\psi_{max}(k_{n-2})
\sqrt{\hat{N}})+\frac{\xi_1}{\hat{N}}
w_{-1}(\psi_{max}(k_{n-2})\sqrt{\hat{N}})
\right]
\label{critpolymer}
\end{eqnarray}
\label{regime}
\end{Gaial}
\vskip 0.5 cm
\noindent The notation $k_{n-2}^{crit}$ characterizing the value of
$k_{n-2}$ for which we have the upper limit crossing, (as compared
with the notation $k_{n-2}^{inf}$ appearing in theorem \ref{fasi}) is
justified by the observation, discussed in the next paragraph, that
corresponding to this values of $k_{n-2}$ we have the phase
transitions observed in numerical simulations of simplicial quantum
gravity.

\subsection{Distinct asymptotic regimes}

As $k_{n-2}$ varies around  $k^{inf}_{n-2}$ and $k^{crit}_{n-2}$,
the two parameters
$\psi_{min}(k_{n-2})$  and $\psi_{max}(k_{n-2})$
vary monotonically from positive values
to negative values. This monotonicity allows us to describe in details
the
asymptotic behavior of $I$ in terms of $k_{n-2}$.
Strictly speaking we should consider separately both a neighborhood of
$k_{n-2}^{inf}$ as well as a neighborhood of $k_{n-2}^{crit}$.
However, corresponding to the former, available numerical simulations
are so polluted by finite size effects that the final confrontation
between the analytical results and the numerical results would be
quite unreliable. Thus we confine ourself to a detailed analysis of
what happens nearby $k_{n-2}^{crit}$, this corresponds to the value of
the inverse gravitational couplings
probed, reliably, by the MonteCarlo simulations. In any case, it is
trivial to extend the conclusions we reach to  $k_{n-
2}^{inf}$.
Thus, limiting ourself to a neighborhood of $k_{n-2}^{crit}$
we can distinguish the following cases:
\vskip 0.5 cm
\noindent {\it (i)} {\bf Strong coupling}: $k_{n-2}<k^{crit}_{n-2}-
\epsilon$, where $\epsilon$ is implicitly characterized by the
condition $2\psi^2_{max}(k^{crit}_{n-2})\hat{N}\simeq 1$.\par
\noindent In dimension $n=4$, this implies
\begin{eqnarray}
k_2< 1.387-
\epsilon
\end{eqnarray}
(notice, that $1.387$, is a
theoretical bound-see the discussion in the next paragraph). Since
$k_{n-2}$ is proportional to the inverse gravitational coupling, this
asymptotic behavior describes a strong coupling regime for simplicial
quantum gravity. For such a range of values of $k_{n-2}$,
$\eta^*<\eta_{max}$ and  we get
\begin{eqnarray}
\psi_{max}(k_{n-2})= sgn (\eta^*-\eta_{max})
\sqrt{2[f(\eta^*)-f(\eta_{max})]}=\sqrt{2[f(\eta^*)-
f(\eta_{max})]}<0
\end{eqnarray}
In such a case the appropriate asymptotic
expansion for the parabolic cylinder function is
\vskip 0.5 cm
\begin{eqnarray}
D_r(z)&\simeq & e^{-z^2/4}z^r\left( 1-\frac{r(r-1)}{2z^2}+
\frac{r(r-1)(r-2)(r-3)}{2\cdot4z^4}-\ldots \right)-\nonumber\\
&- &\frac{\sqrt{2\pi}}{\Gamma(-r)}e^{r\pi{i}}e^{z^2/4}z^{-r-1}
\left(
1+\frac{(r+1)(r+2)}{2z^2}+\frac{(r+1)(r+2)(r+3)(r+4)}{2\cdot4z^4}
+\ldots  \right)
\end{eqnarray}
\vskip 0.5 cm
\noindent (which holds for all complex $z$ such that $|z|>>1$,
$|z|>>r$,
and $\frac{\pi}{4}<arg(z)<\frac{5}{4}\pi$). Setting
$z=\psi_{max}\sqrt{\hat{N}+1}$ and $r=-1,-2$ in (\ref{dominante})
we get that the leading asymptotics for $I$ is
\vskip 0.5 cm
\begin{eqnarray}
I\simeq
\frac{\eta*^{-1/2}(1-2\eta^*)^{-n}}{\sqrt{(1-3\eta^*)(1-
2\eta^*)}}\sqrt{-\frac{2\pi}{f_{\eta\eta}(\eta^*)}}
(\hat{N}+1)^{-1/2}e^{(\hat{N}+1)f(\eta^*)}
\end{eqnarray}

\noindent (this, as expected, is twice the expression obtained in the
limiting case $k_{n-2}=k_{n-2}^{crit}$-see below, {\it (ii)}, for the
discussion of this limiting case).
The corresponding expression for the effective entropy in this range
of $k_{n-2}$ is given by
\vskip 0.5 cm
\begin{eqnarray}
W(N, k_{n-2})_{eff}=
{c_n}\left( \frac{(A(k_{n-2})+2)}{3A(k_{n-2})} \right)^{-n} N^{-n-
1}\cdot\nonumber
\end{eqnarray}
\begin{eqnarray}
\cdot\exp\left[[\frac{1}{2}n(n+1)\ln\frac{A(k_{n-2})+2}{3}]N
\right](1+O(N^{-3/2}))
\label{regular}
\end{eqnarray}
\vskip 0.5 cm
We have  a ${\hat{N}}^{-n-1}$ subleading behavior which changes
as $k_{n-2}\to k_{n-2}^{crit}$.
\vskip 1 cm
\noindent {\it (ii)} {\bf Critical coupling}: $k^{crit}_{n-2}-
\epsilon<k_{n-2}<k^{crit}_{n-2}+\epsilon$.\par

The above asymptotic is no longer appropriate when, for a given
(large)
value of $N$, $\psi$ is so small that
\begin{eqnarray}
2\psi^2(k_{n-2}^{crit}\pm\epsilon)\hat{N}<<1
\label{rounding}
\end{eqnarray}
In such a regime, the parabolic cylinder function is more correctly
described by
\vskip 0.5 cm
\begin{eqnarray}
D_r(z)\simeq\frac{\sqrt{\pi}}{2^{-r/2}\Gamma((1-r)/2)}
\exp[-\sqrt{-r-1/2}z+v_1]
\label{paracritico}
\end{eqnarray}
\vskip 0.5 cm
\noindent where
\begin{eqnarray}
v_1\simeq-sgn{z}\frac{\frac{2}{3}(\frac{1}{2}z)^3}{2\sqrt{-r-1/2}}
-\frac{(\frac{1}{2}z)^2}{(2\sqrt{-r-1/2})^2}-\ldots
\label{vuno}
\end{eqnarray}
(this holds for $-r-\frac{1}{2}>>z^2$, see\cite{Abramowitz}
pp.689). Setting $r=-1$ and $z=\psi_{max}\sqrt{\hat{N}+1}$, we get for
the canonical partition function the expression
\vskip 0.5 cm
\begin{eqnarray}
W(N, k_{n-
2})_{eff}=\frac{c_n}{\sqrt{2\pi}}
(\psi_{max}\sqrt{\hat{N}+1})
\left[\ln
\frac{\eta_{max}(1-2\eta_{max})^2}{(1-3\eta_{max})^3}-k_{n-2}
 \right]^{-1}\cdot\nonumber
\end{eqnarray}
\begin{eqnarray}
\cdot\frac{\eta_{max}^{-1/2}(1-2\eta_{max})^{-n}}{\sqrt{(1-
3\eta_{max})(1-
2\eta_{max})}}
(\hat{N}+1)^{-n-3/2}e^{(\hat{N}+1)f(\eta_{max})-
\psi_{max}\sqrt{\frac{\hat{N}+1}{2}}-\ldots}
\label{crumple}
\end{eqnarray}
\vskip 0.5 cm
\noindent where $\ldots$ stands for higher order correction terms
(which can be
obtained by the appropriate substitutions from (\ref{vuno})).
\vskip 0.5 cm
Notice that in this latter regime, the subleading asymptotics of
(\ref{crumple}) is no longer polynomial, but rather exponential in
$\psi_{max}\sqrt{N}$. This transient in the subleading asymptotics is
necessary
for taking care of the fact that as $k_{n-2}\to k^{crit}_{n-2}$ the
term $(-1/f_{\eta})$ tends to diverge. This divergence is compensated
for by the vanishing of $\psi_{max}\sqrt{\hat{N}+1}$, and indeed, the
singularity does not manifest itself in the exponential leading
behavior of the effective entropy $W(N, k_{n-2})_{eff}$  which goes
sufficiently smootly from $e^{(\hat{N}+1)f(\eta_{max})}$ into
$e^{(\hat{N}+1)f(\eta^*)}$, but rather in a change of
the polynomial part ${\hat{N}}^{-n-3/2}$ of the subleading
asymptotics,
which changes by
a factor ${\hat{N}}^{1/2}$.
 This transient exponential
subleading asymptotics is of relevance to the MonteCarlo simulations
since it
appears consistent with the numerical data for $k_{n-2}$ nearby
$k^{crit}_{n-2}$.
\vskip 0.5 cm
Notice that the range of $k_{n-2}$,  for which the exponential
subleading
asymptotics is active depends on the actual value of $N$, and
as $N\to\infty$, the range
$k^{crit}_{n-2}-\epsilon<k_{n-2}<k^{crit}_{n-2}+\epsilon$
becomes smaller and smaller since $\epsilon=\epsilon(N)$ is
characterized by the
condition
$2\psi^2(k_{n-2}^{max}\pm\epsilon)\hat{N}\simeq 1$ which implies that
$\lim_{N\to\infty}\epsilon(N)=0$.
In this case we have
\begin{eqnarray}
\psi_{max}(k_{n-2})= sgn (\eta^*-\eta_{max})
\sqrt{2[f(\eta^*)-f(\eta_{min})]}= 0
\end{eqnarray}
and by setting $\psi_{max}=0$ in (\ref{dominante})
we get that the leading asymptotics for $I$ is
\begin{eqnarray}
I\simeq
\frac{\eta*^{-1/2}(1-2\eta^*)^{-n}}{\sqrt{(1-3\eta^*)(1-
2\eta^*)}}\sqrt{-\frac{\pi}{-2f_{\eta\eta}(\eta^*)}}
(\hat{N}+1)^{-1/2}e^{(\hat{N}+1)f(\eta^*)}
\end{eqnarray}
The corresponding expression for the canonical partition function is
easily seen
to be
\vskip 0.5 cm
\begin{eqnarray}
W(N, k_{n-2})_{eff}=
\frac{c_n}{2}\left( \frac{(A(k_{n-2})+2)}{3A(k_{n-2})} \right)^{-n}
N^{-n-1}\cdot\nonumber
\end{eqnarray}
\begin{eqnarray}
\cdot\exp\left[[\frac{1}{2}n(n+1)\ln\frac{A(k_{n-2})+2}{3}]N
\right](1+O(N^{-3/2}))
\label{transition}
\end{eqnarray}
\vskip 0.5 cm

\noindent {\it (iii)} {\bf Weak coupling}: $k^{crit}_{n-
2}+\epsilon<k_{n-
2}<+\infty$.\par
This is a weak coupling regime in a rather loose sense, since the
asymptotics we obtain holds both for $k_{n-2}$'s in a small right-
neighborhood of $k^{crit}_{n-2}$ and in a large left-neighborhood of
$k_{n-2}=+\infty$.
\vskip 0.5 cm
For such range of values of $k_{n-2}$,
$\eta^*>\eta_{max}$ and
\begin{eqnarray}
\psi_{max}(k_{n-2})= sgn (\eta^*-\eta_{max})
\sqrt{2[f(\eta^*)-f(\eta_{max})]}= \sqrt{2[f(\eta^*)-f(\eta_{max})]}>0
\end{eqnarray}
In such a case, and as long as $\psi^2\hat{N}>>1$,
the appropriate asymptotic
expansion for the parabolic cylinder function is
\begin{eqnarray}
D_r(z)\simeq e^{-z^2/4}z^r\left( 1-\frac{r(r-1)}{2z^2}+
\frac{r(r-1)(r-2)(r-3)}{2\cdot4z^4}-\ldots \right)
\end{eqnarray}
(which holds for all complex $z$ such that $|z|>>1$, $|z|>>r$,
and $|arg(z)|<\frac{3}{4}\pi$). Setting $z=\psi_{max}\sqrt{\hat{N}+1}$
and $r=-1,-2$ in (\ref{dominante})
we get that the leading asymptotics for $I$ is
\begin{eqnarray}
I\simeq
\frac{\eta_{max}^{-1/2}(1-2\eta_{max})^{-n}}{\sqrt{(1-3\eta_{max})(1-
2\eta_{max})}}\left(-\frac{1}{f_{\eta}(\eta_{max})} \right)
(\hat{N}+1)^{-1}e^{(\hat{N}+1)f(\eta_{max})}
\end{eqnarray}
\vskip 0.5 cm
\noindent which provides the following expression for the canonical
partition function in
this regime
\vskip 0.5 cm
\begin{eqnarray}
W(N, k_{n-
2})_{eff}=\frac{c_n}{\sqrt{2\pi}}\left[
\ln\frac{\eta_{max}(1-2\eta_{max})^2}{(1-3\eta_{max})^3}-k_{n-2}
 \right]^{-1}\cdot\nonumber
\end{eqnarray}
\begin{eqnarray}
\cdot\frac{\eta_{max}^{-1/2}(1-2\eta_{max})^{-n}}{\sqrt{(1-
3\eta_{max})(1-
2\eta_{max})}}
(\hat{N}+1)^{-n-3/2}e^{(\hat{N}+1)f(\eta_{max})}
\label{lower}
\end{eqnarray}
\vskip 0.5 cm

\noindent which clearly exhibits a different subleading asymptotic
behavior, ${\hat{N}}^{-n-3/2}$, with respect to the previous one
discussed in the strong coupling case {\it (i)}.

Obviously, these results could have been obtained by  standard Laplace
approximation working in the appropriate range of values of the
inverse gravitational coupling, but the expression
(\ref{critpolymer}), (and its related counterpart (\ref{polymer})),
has
the advantage of
describing in a uniform way all possible regimes
for $k_{n-2}$ in an open neighborhood of $k_{n-2}^{crit}$.
 Roughly speaking, the asymptotics (\ref{regular})
corresponds to the
location of maximum of $f(\eta)$ {\it well} inside the integration
interval $\eta_{min}<\eta<\eta_{max}$, whereas the
asymptotics (\ref{lower}) occurs when $\eta^*$ is larger than
$\eta_{max}$, (but sufficiently near to it).  Notice also that
$A(k_{n-2})>1$, $\forall
k_{n-2}\in{\Bbb R}$.
\vskip 0.5 cm
We wish to stress the fact that the transition between the
two regimes (\ref{regular}) towards (\ref{lower}) manifests itself
in a change of the subleading asymptotics: the growth
$N^{-n-1}$ which
characterizes the expression (\ref{regular}) for
$W(N, k_{n-2})_{eff}$ changes into the $N^{-n-3/2}$
subleading behavior
characterizing (\ref{lower}), and this change is {\it mediated} by
an exponential subleading asymptotics in the range
$k^{crit}_{n-2}-\epsilon\leq k_{n-2}\leq k^{crit}_{n-2}+\epsilon $.
\vskip 0.5 cm
\noindent {\bf The infinite volume limit}.
From the above asymptotics we can also read the critical
value of the coupling $k^{crit}_n$ corresponding to which the {\it
infinite-volume} limit may be taken. In general, {\it i.e.}, for
all values of the inverse gravitational coupling $k_{n-2}$, this
critical value is
provided by
\begin{eqnarray}
k_n^{crit}(k_{n-2})=\lim_{N\to\infty}\frac{\log{W(N, k_{n-
2})_{eff}}}{N}
\end{eqnarray}
where $W(N, k_{n-2})_{eff}$ is given by (\ref{critpolymer}). By
considering separately the various  ranges of $k_{n-2}$ we explicitly
get
\vskip 0.5 cm
\begin{eqnarray}
k_n^{crit}(k_{n-2})=\frac{1}{2}n(n+1)\left[ \ln
\frac{A(k_2)+2}{3}+\lim_{N\to\infty}
\frac{\ln{c_n}}{N} \right]
\label{cosmo}
\end{eqnarray}
\vskip 0.5 cm
\noindent which, for $0<\epsilon(N)<1$ characterized by the condition
$2\psi^2(k_{n-2}^{crit}\pm\epsilon)\hat{N}\simeq 1$,
holds for all $k_{n-2}<k_{n-
2}^{crit}-\epsilon$.
Whereas, for $k^{crit}_{n-2}+\epsilon\leq k_{n-2}<+\infty$ we get
\vskip 0.5 cm
\begin{eqnarray}
k_n^{crit}(k_{n-2})=\frac{1}{2}n(n+1)
\left\{
\ln\left[\frac{(1-
2\eta_{max})^{(1-2\eta_{max})}}{\eta_{max}^{\eta_{max}}(1-
3\eta_{max})^{(1-3\eta_{max})}} \right] +k_{n-2}\eta_{max} +
\lim_{N\to\infty}\frac{\ln c_n}{N}
\right\}
\label{cosmo1}
\end{eqnarray}
\vskip 0.5 cm
Notice that if one works at {\it finite} $N$, (as is the case in
MonteCarlo simulations), care must be taken in the transition region
$k^{crit}_{n-2}-\epsilon\leq k_{n-2}
\leq k^{crit}_{n-2}+\epsilon$ since in that case a non-trivial
exponential subleading asymptotics is present, (see {\it (ii)}).

\vskip 0.5 cm
With these elements in our hands, we can compare our analytical
estimates with some of the known numerical data coming from MonteCarlo
simulations. We start with the $4$-dimensional case.
\vfill\eject
\section{Analytical vs. numerical data}
The interplay between the asymptotic estimates arising from
(\ref{regular}) and the results coming from Montecarlo simulations
requires some care, since there are manifolds
which are {\it algorithmically unrecognizable}. Denote such a manifold
by $M_0$, assume that $n=4$, and let $M_0$ be finitely described by a
dynamical triangulation $T(M_0)$. Monte Carlo simulations make use  of
two kinds of moves altering the initial triangulation $T(M_0)$: a
finite set of {\it local moves}\cite{Migdal}, and global {\it baby
universe surgery} moves. The local moves are ergodic, since given any
two distinct triangulations of the manifold it is possible in a finite
number of moves, carried out successively, to change on triangulation
into the other. The algorithmic unrecognizability of $M_0$ means that
there exists no algorithm which allows us to decide wether another
manifold $M$, again finitely described by a
triangulation $T(M)$, is combinatorially equivalent (in the PL-sense),
to $M_0$. It has been proved\cite{Ben} that the number of ergodic
moves\cite{Varsted} needed to connect two triangulations of $M_0$, $T$
and $T^*$, with $N_4(T)=N_4(T^*)$, cannot be bounded by any recursive
function of $N_4$. This result implies that there can be very large
barriers between some classes of triangulations of $M_0$ and that
there would be triangulations which can never be reached in any
reasonable number of steps,
to the effect that the local moves, although ergodic, can be
computationally non-ergodic. The manifold used in actual Monte Carlo
simulations of four-dimensional simplicial gravity is the four-sphere
${\Bbb S}^4$, which is not known to be algorithmically  recognizable
or not. Up to now, there is no numerical evidence settling this issue,
even using the non-local moves associated with baby-universes
surgery\cite{ergo},\cite{Bakker}. This implies that either the class
of configurations separated by high barriers contains only very
special configurations which are of measure zero and not important for
numerical simulations, or the barriers expected are so high that they
have not be encountered yet in the actual simulations. It is important
to have these remarks clearly in mind when discussing the comparision
with the analytical data.
\vskip 0.5 cm

\subsection{The four-dimensional case}

In order to discuss how the asymptotics associated with
(\ref{regular}) compares with MonteCarlo simulations, let us rewrite
$W(N,k_2)_{eff}$ as
\vskip 0.5 cm
\begin{eqnarray}
W(N,k_2)_{eff}=10c_n\cdot\frac{e}{\sqrt{2\pi}}
e^{k_4^{crit}(k_2)N}N^{\gamma_s(k_2)-3}(1+\ldots)
\end{eqnarray}
\vskip 0.5 cm
\noindent where according to theorem \ref{fasi} and lemma
\ref{regime}, the
critical $k^{crit}_4$ is
provided by:
\vskip 0.5 cm
\begin{eqnarray}
k_4^{crit}(k_{n-2})=10\ln \frac{A(k_2)+2}{3}+
\lim_{N\to\infty}10\frac{\ln{c_n}}{N}
\end{eqnarray}
\vskip 0.5 cm
\noindent for $k_2\leq k^{crit}_2-\epsilon$, and
\begin{eqnarray}
k_4^{crit}(k_2)=10
\left\{
\log\left[\frac{(1-
2\eta_{max})^{(1-2\eta_{max})}}{\eta_{max}^{\eta_{max}}(1-
3\eta_{max})^{(1-3\eta_{max})}} \right] +k_2\eta_{max}+
\lim_{N\to\infty}\frac{\ln{c_n}}{N}
\right\}
\end{eqnarray}
\vskip 0.5 cm
\noindent for  $k_2> k_2^{crit}+\epsilon$ and
$\eta_{max}=\frac{1}{4}$, with
\begin{eqnarray}
A(k_2)\equiv{ \left [\frac{27}{2}e^{k_2}+1+
\sqrt{(\frac{27}{2}e^{k_2}+1)^2-1} \right ]}^{1/3}+
{ \left [\frac{27}{2}e^{k_2}+1-
\sqrt{(\frac{27}{2}e^{k_2}+1)^2-1} \right ]}^{1/3}
-1
\end{eqnarray}
\vskip 0.5 cm
\noindent and where the factor $c_n$ is the normalizing factor of
lemma \ref{scalemma}.
\vskip 0.5 cm
The critical exponent $\gamma_s(k_2)$ is also provided by lemma
\ref{critpolymer}, (see the discussion on the distinct asymptotic
regimes),
according to
\vskip 0.5 cm
\begin{eqnarray}
\gamma_s(k_2)=2-n
\label{strongexpo}
\end{eqnarray}
\vskip 0.5 cm
for  $k_2< k_2^{inf}-\epsilon$,
\vskip 0.5 cm
\begin{eqnarray}
\gamma_s(k_2)=-\infty
\end{eqnarray}
\vskip 0.5 cm
\noindent for  $k_2^{inf}-\epsilon<k_2<k_2^{inf}+\epsilon$,
( recall that in this range, for large but finite $N$,
(\ref{crumple}) provides a non-polynomial subleading asymptotics which
formally corresponds to such a critical exponent),
and finally
\begin{eqnarray}
\gamma_s(k_2)=\frac{3}{2}-n
\label{weakexpo}
\end{eqnarray}
\vskip 0.5 cm
for  $k_2^{inf}+\epsilon<k_2$.
\vskip 0.5 cm

Notice that these critical exponents can get an additive normalization
by the term $\tau(n)$, (depending only on the dimension $n$),
associated
with the rescaling factor $c_n$, (see (\ref{corrections})). Thus, the
actual
critical exponents are provided by
\begin{eqnarray}
{\tilde \gamma}_s(k_2)=\gamma_s(k_2)+\tau(n).
\end{eqnarray}
\vskip 0.5 cm
This remark reminds us that we still have to fix
the scale factor $c_n$, affecting
the canonical partition function $W(M;N,k_2)_{eff}$. According to
(\ref{rinorm}) and lemma \ref{scalemma} this factor
normalizes the counting of the partitions $p_{\lambda}$ to the
partitions $p_{\lambda}^{curv}$ coming from actual curvature
assignements. In dimension $n=4$, a simple exponential (down)shift
times a constant polynomial subleading correction
is all is needed, (later on we see that this latter  correction to the
subleading asymptotics is related to a polymerization mechanism), and
if we  set
\begin{eqnarray}
c_n=(\frac{e}{2.066})^{-10N}N^{5/2}
\label{ratio}
\end{eqnarray}
then we get an extremely good agreement with MonteCarlo simulations.
It is interesting to note that the exponential scaling factor
$(\frac{e}{2.066})^{-10N}$ is a very close
approximation to
\begin{eqnarray}
[\cos^{-1}(\frac{1}{n})|_{n=4}]^{-10N}
\label{coseno}
\end{eqnarray}
suggesting that the ratio $p^{curv}_{\lambda}/p_{\lambda}$ is, as
expected, of  geometrical origin. We wish to stress that we do not
have yet a convincing geometrical proof that
the scale factor $c_n$ is actually given by (\ref{coseno}), and leave
this as a conjecture to be proven by the interested reader.
\vskip 0.5 cm
The above choice (\ref{ratio}) for the rescaling factor $c_n$ provides
\vskip 0.5 cm
\begin{eqnarray}
{\tilde K}^{crit}_4=10 \ln\frac{A(k_2)+2}{3e/2.066}
\label{curva}
\end{eqnarray}
\vskip 0.5 cm
\noindent with $k_2< k_2^{inf}-\epsilon$, and
\vskip 0.5 cm
\begin{eqnarray}
{\tilde k}_4^{crit}(k_2)=10
\left\{
\log\left[\frac{(1-
2\eta_{max})^{(1-2\eta_{max})}}{\eta_{max}^{\eta_{max}}(1-
3\eta_{max})^{(1-3\eta_{max})}} \right] +k_2\eta_{max}
-\log\left( \frac{e}{2.066} \right)
\right\}
\label{retta}
\end{eqnarray}
\vskip 0.5 cm
\noindent (with $k_2> k_2^{inf}+\epsilon$).
\vskip 0.5 cm
\noindent (Note that the subleading correction $N^{5/2}$ is irrelevant
for the characterization of (\ref{curva}) and (\ref{retta})).
\vskip 0.5 cm

The rescaled critical value of  $k_4$ so obtained  gives the critical
$k_4$ for any simply-connected closed $4$-manifold, in particular for
the $4$-sphere ${\Bbb S}^4$. Indeed,
it provides an extremely good fit with the numerical data relative to
the characterization of the $\infty$-volume line in the $(k_4,k_2)$-
couplings space. Explicitly,
if
we denote by $K^{crit}_4(Mont.Car)$ the critical value of $k_4$
obtained from Montecarlo simulations\cite{Kristj} for dynamically
triangulated four-spheres, then the following table demonstrates how
well this agreement works out
\vskip 0.5 cm

\begin{eqnarray}
& & k_2=0.0\ldots K_4^{crit}(Mont.Car)=0.98\ldots
{\tilde K}_4^{crit}=1.07\nonumber\\
& & k_2=0.2\ldots K_4^{crit}(Mont.Car)=1.45\ldots
{\tilde K}_4^{crit}=1.47\nonumber\\
& & k_2=0.3\ldots K_4^{crit}(Mont.Car)=1.65\ldots
{\tilde K}_4^{crit}=1.67\nonumber\\
& & k_2=0.4\ldots K_4^{crit}(Mont.Car)=1.89\ldots
{\tilde K}_4^{crit}=1.87\nonumber\\
& & k_2=0.5\ldots K_4^{crit}(Mont.Car)=2.07\ldots
{\tilde K}_4^{crit}=2.07\nonumber\\
& & k_2=0.6\ldots K_4^{crit}(Mont.Car)=2.30\ldots
{\tilde K}_4^{crit}=2.33\nonumber\\
& & k_2=0.7\ldots K_4^{crit}(Mont.Car)=2.54\ldots
{\tilde K}_4^{crit}=2.55\nonumber\\
& & k_2=0.8\ldots K_4^{crit}(Mont.Car)=2.78\ldots
{\tilde K}_4^{crit}=2.78\nonumber\\
& & k_2=0.9\ldots K_4^{crit}(Mont.Car)=3.02\ldots
{\tilde K}_4^{crit}=3.01\nonumber\\
& & k_2=1.0\ldots K_4^{crit}(Mont.Car)=3.25\ldots
{\tilde K}_4^{crit}=3.25\nonumber\\
& & k_2=1.1\ldots K_4^{crit}(Mont.Car)=3.49\ldots
{\tilde K}_4^{crit}=3.49\nonumber\\
& & k_2=1.3\ldots K_4^{crit}(Mont.Car)=3.97\ldots
{\tilde K}_4^{crit}=3.97\nonumber\\
& & k_2=1.5\ldots K_4^{crit}(Mont.Car)=4.47\ldots
{\tilde K}_4^{crit}=4.46\nonumber\\
& & k_2=1.8\ldots K_4^{crit}(Mont.Car)=5.20\ldots
{\tilde K}_4^{crit}=5.19\nonumber\\
\label{table}
\end{eqnarray}
\vskip 0.5 cm
\noindent (note that the
numerical data are affected by an error of the order $\pm 0.02$, and
the analitical data are rounded to the largest second significant
figure).
According to the distinct asymptotic regimes provided by theorem
\ref{fasi} and lemma \ref{critpolymer}, we have used
(\ref{retta}), (evaluated for $\eta_{min}=1/5$), in the range
for $0\leq k_2\leq 0.5$ , (however we stress that in
such region, and in particular near $k_2\simeq 0$, the MonteCarlo data
are
rather unreliable being strongly polluted by finite size effects). For
$0.6\leq k_2\leq k_2^{crit}\simeq 1.38$ we have used
(\ref{curva}). Incidentally, it is not difficult to check that the use
of
(\ref{curva}) works quite well also for $0\leq k_2\leq 0.5$,
with just a sligth discrepancy between
the MonteCarlo simulations and the analytical data. Finally, for $k_2>
k_2^{crit}$ we have used (\ref{retta}).
The numerical data in  (\ref{table}) come from
simulations \cite{Kristj} which are already a few years old, however
we wish to stress that the agreement
with the analytical estimates  holds beautifully also  when using
more recent and accurate MonteCarlo data, a fact that confirms that
our
simple counting strategy based on the use of the generating function
of partitions of integers $p_{\lambda}$ has a sound foundation in the
geometry of dynamical triangulations.
\vskip 0.5 cm

\subsection {Polymerization}

A well known consequence of the numerical simulations in $4$-
dimensional simplicial quantum gravity is the onset of a branched
polymer phase\cite{Jurk} in the weak coupling regime, {\it i.e.}, for
$k_2> k_2^{crit}$. This
polymerization  is very effectively described by the
asymptotics of theorem \ref{fasi} and lemma
\ref{critpolymer}, and it is generated again by a mechanism of
entropic origin.
\vskip 0.5 cm
Let us start by noticing that if in the expression yielding for the
canonical partition function (\ref{integrale}) we replace the upper
integration limit $\eta_{max}=1/4$ with the limit
${\tilde{\eta}}=1/3$,
associated with the algebraic singularity of the integrand  at
$\eta=1/3$ (and corresponding to the unphysical
average incidence $b(n,n-2)=3$), then we would get a canonical
partition function $W(N,k_{n-2})_{poly}$ whose large $N$ limit is
provided by the asymptotics (\ref{regular}), {\it viz.},
\vskip 0.5 cm
\begin{eqnarray}
W(N, k_{n-2})_{poly}=
{c_n}\left( \frac{(A(k_{n-2})+2)}{3A(k_{n-2})} \right)^{-n}
N^{-n-1}\cdot\nonumber
\end{eqnarray}
\begin{eqnarray}
\cdot\exp\left[[\frac{1}{2}n(n+1)\ln\frac{A(k_{n-2})+2}{3}]N
\right](1+O(N^{-3/2}))
\label{polimero}
\end{eqnarray}
\vskip 0.5 cm
\noindent for all $k_2^{inf}+\delta\leq k_2<+\infty$, where
$k_2^{inf}$
corresponds to the lower crossing $\eta^*\to\eta_{min}=1/5$, and
$\delta\in{\Bbb R}^+$ is a suitably small number.
\vskip 0.5 cm
\noindent In the range
$k_2^{inf}+\delta<k_2<k_2^{crit}-\epsilon$ this is exactly the
expression for the asymptotics of the canonical partition function
$W(N, k_{n-2})_{eff}$, but this identification breaks down as soon as
$k_2$ approaches $k_2^{crit}$. The geometrical
reason for such a behavior is very simple: as $k_2\to k_2^{crit}$, the
possible maximum value of the (positive)
average curvature for a  given volume, ($vol(M)\propto{N}$), is
saturated: no more positive average curvature is allowed for that
volume, (think of the average curvature on a sphere of given volume:
it is maximized for the round metric corresponding to that volume). On
the other hand,
(\ref{polimero}) as it stands, would allow, as $k_2$ grows, for more
and more (positive) average
curvature at fixed volume, and in order to comply with the constraint
relating volume to positive curvature we have to turn to the correct
weak coupling description of the partition function, {\it viz.},
(\ref{lower}). However, (\ref{polimero}) can still be interpreted as
the partition function of a large positive curvature PL manifold of
given volume, but now the manifold in question is {\it disconnected}.
Such a disconnected manifold is generated by means of a {\it
polymerization mechanism}, namely it results as the (Gromov-Hausdorff)
limit of a network of many smaller manifolds carrying large
curvature. Such curvature {\it blobs} are connected to each other by
thin tubes of negligible volume so that the sum of the volumes of the
costituents manifolds adds up to the given total volume, (the standard
example is afforded by a collection of highly curved small spheres
connected by tubules of negligible area, {\it vs.} the standard round
sphere of the some total area). The basic observation is that for
$k_2^{crit}+\epsilon<k_2<+\infty$, the effective entropy
(\ref{polimero}) is {\it larger} than (\ref{lower}): thus in the weak
coupling region, for a dynamical triangulation is energetically
favourable to polymerize rather than to comply to the curvature-volume
constraint with a {\it large} connected manifold.
\vskip 0.5 cm
The above remarks can be easily formalized. According to
(\ref{curv}), the average curvature (around the collection of bones
$\{B\}$) on a dynamically triangulated
manifold is proportional to
\begin{eqnarray}
<K(B)>\doteq\frac{1}{\hat{N}}\sum_BK(B)q(B)
\propto\frac{1}{\hat{N}}\sum_B^{N_{n-2}}\left [ 2\pi-q(B)\cos^{-
1}\frac{1}{n} \right ]
\end{eqnarray}
Namely
\begin{eqnarray}
<K(B)>\propto \frac{2\pi-b(n,n-2)\cos^{-1}(\frac{1}{n})}{b}.
\end{eqnarray}
\vskip 0.5 cm
\noindent The canonical average, $<<K(B)>>$, of $<K(B)>$ over the
ensemble of dynamical triangulation considered is, up to an
inessential positive constant, provided by (see (\ref{integrale})),
\begin{eqnarray}
<<K(B)>>=\int_{\eta_{min}}^{\eta_{max}}d\eta
\frac{\eta^{-1/2}(1-2\eta)^{-n}}{\sqrt{(1-3\eta)(1-
2\eta)}}[2\pi\eta^{-1}-\cos^{-1}(\frac{1}{n})|_{n=4}]
e^{(\hat{N}+1)f(\eta)}\times\nonumber
\end{eqnarray}
\begin{eqnarray}
\times\left[
\int_{\eta_{min}}^{\eta_{max}}d\eta
\frac{\eta^{-1/2}(1-2\eta)^{-n}}{\sqrt{(1-3\eta)(1-
2\eta)}}e^{(\hat{N}+1)f(\eta)} \right]^{-1}
\label{avcurv}
\end{eqnarray}
\vskip 0.5 cm
For $k_2=k_2^{crit}-\epsilon$, with $\epsilon\in{\Bbb R}^+$
sufficiently small, the maximum for $f(\eta)$, (see
(\ref{integrale})), is attained for an $\eta^*$ which uniformly
approaches $\eta=\frac{1}{4}$ as $\epsilon\to 0$. Correspondingly, the
integrands in (\ref{avcurv}) are peaked around $\eta\simeq 1/4$ and
$<<K(B)>>$ is positive, (explicit expression can be easily obtained
from the uniform asymptotics of the previous paragraphs). Let us now
extend (\ref{avcurv}) to
$\eta_{max}=\frac{1}{3}$, namely consider the ensemble average,
$<<K(B)>>_{poly}$ of $<K(B)>$ over $W(N, k_{n-2})_{poly}$.
It is trivially checked that $<<K(B)>>_{poly}$ formally remains
positive for all $k_2^{crit}<k_2<+\infty$, (corresponding to the
maximum $\eta^*$ ranging from
$\frac{1}{4}\le\eta^*\leq\frac{1}{3}$). But according to lemma
\ref{walkup}, there cannot be connected four-manifolds (regardless of
topology) with $b<4$, ({\it i.e.}, $\eta>\frac{1}{4}$), because for
any dynamically triangulated $4$-manifold
\begin{eqnarray}
5N_4\geq 2N_2-5\chi
\label{vincolo}
\end{eqnarray}
An inequality which implies that as $\eta\to\frac{1}{4}$, the average
curvature $<K(B)>$ attains its maximum $\propto [2\pi-4\cos^{-
1}(\frac{1}{4})]/4$ compatibly with the given volume, ($\propto N$).
\vskip 0.5 cm
\noindent There is a natural way of circumventing the  implications
of the
constraint (\ref{vincolo}): we have to remove the connectivity
requirement
for $M$.
\vskip 0.5 cm
 Let us assume that we have a collection $\{M_{\alpha}\}$ of
dynamically triangulated manifolds of fixed volume $\propto
N_4(M_{\alpha})=N$, and where each $M_{\alpha}$ is generated by a
large number $m$ of dynamically triangulated $4$-manifolds
$\{\Xi(i)\}_{i=1,\ldots,m}(\alpha)\subset M_{\alpha}$, connected to
each other by tubes of
negligible volume. Such a sequence of manifolds naturally converges,
in the Gromov-Hausdorff topology, to the disjoint union of a set of
$m$ $4$-manifolds, {\it viz.},
\vskip 0.5 cm
\begin{eqnarray}
\{M_{\alpha}\}\to M\doteq\coprod_i^m\Xi(i)
\end{eqnarray}
\vskip 0.5 cm
\noindent such that $N_4(M)=N_4(M_{\alpha})=N$.
The constraint (\ref{vincolo}), which holds for each connected
manifold $M_{\alpha}$ of the sequence, carries over to the
disconnected limit space $M$ and implies that
\vskip 0.5 cm
\begin{eqnarray}
5\sum_i^mN_4(\Xi(i))\geq
2\sum_i^mN_2(\Xi(i))-5\sum_i^m\chi(\Xi(i))
\label{tubules}
\end{eqnarray}
\vskip 0.5 cm
\noindent namely,
\vskip 0.5 cm
\begin{eqnarray}
b_M\doteq 10\frac{\sum_i^mN_4(\Xi(i))}{\sum_i^mN_2(\Xi(i))}
\geq 4 -10\frac{\sum_i^m\chi(\Xi(i))}{\sum_i^mN_2(\Xi(i))}
\end{eqnarray}
\vskip 0.5 cm
\noindent Thus, if each $\Xi(i)$ is topologically a $4$-sphere, ${\Bbb
S}^4$, and $N_2(\Xi(i))$ is $O(1)$, we can make $b_M$ take any value
in the range $3\leq b \leq 4$. In particular, if
\begin{eqnarray}
{\tilde N}_2(\Xi(i))\doteq\frac{1}{m}\sum_i^mN_2(\Xi(i))=20
\end{eqnarray}
and
\begin{eqnarray}
{\tilde N}_4(\Xi(i))\doteq\frac{1}{m}\sum_i^mN_4(\Xi(i))=6
\end{eqnarray}
then for such $M=\coprod\Xi(i)$, we get
\begin{eqnarray}
b_M\to 3
\end{eqnarray}
Such an $M$ is the disjoint union of $m\simeq\frac{N_4(M)}{6}$ {\it
small}
$4$-dimensional spheres, (recall that the standard triangulation
of a $4$-sphere obtained as the boundary of the standard $5$-simplex
$\sigma^5$ is
characterized by the $f$-vector
$(N_0,N_1,N_2,N_3,N_4)=(6,15,20,15,6)$).
\vskip 0.5 cm
It is trivially checked that for
$k_2^{crit}+\epsilon<k_2<+\infty$, the effective entropy
(\ref{polimero}) is {\it larger} than (\ref{lower}). Thus in the
weak
coupling region, it
is energetically
favourable to generate dynamical triangulations associated with
a disconnected manifold $M=\coprod\Xi(i)$,
rather than to comply with the curvature-volume
constraint on a {\it large} connected manifold. From the
asymptotics (\ref{lower}) and (\ref{regular}) it follows that the
critical exponent associated with such a polymer phase,
$\gamma_{poly}$, is
related to $\gamma_s(k_2)=2-n$, (see
(\ref{strongexpo})), by the relation
\vskip 0.5 cm
\begin{eqnarray}
\gamma_{poly}=\gamma_s(k_2)+\frac{1}{2}
\label{polijump}
\end{eqnarray}
\vskip 0.5 cm
\noindent This jump of $\frac{1}{2}$ has been observed in the
numerical simulation \cite{Jurk} exactly around $k_2^{crit}\simeq
1.3$. In such simulations there is also some indication that
$\gamma_s(k_2\simeq
k_2^{crit})\simeq 0$, so that $\gamma_{poly}$ is  exactly the
critical exponent of branched polymers, but this evidence is  not yet
conclusive due to critical slowing down near the transition point.
\vskip 0.5 cm
According to our counting strategy, we still
have the freedom to add to the critical exponents the term $\tau(n)$
associated with the count of
distinct dynamical triangulations with the same curvature
assignements. The correct critical exponents to be confronted with the
MonteCarlo data are then provided by
\begin{eqnarray}
{\tilde \gamma}_s(k_2)=\gamma_s(k_2) +\tau(n)
\end{eqnarray}
and
\begin{eqnarray}
{\tilde \gamma}_{poly}=\gamma_{poly}+\tau(n)
\label{correxpo}
\end{eqnarray}
\vskip 0.5 cm
Since the critical exponent of a branched polymer is exactly
$\gamma_{poly}=\frac{1}{2}$, and as we have shown (\ref{polimero})
seems to geometrically
describe a polymer phase, we can use this result to fix the value of
$\tau(n)$. Indeed, if at least provisionally, we accept this heuristic
argument, from (\ref{strongexpo}) we find that
the critical exponent scaling factor $\tau(n)$ must
satisfy the relation
\begin{eqnarray}
2-n+\tau(n)|_{n=4}= \frac{1}{2}
\end{eqnarray}
namely, $\tau(n)|_{n=4}=\frac{5}{2}$, a result already  anticipated in
the expression
(\ref{ratio}) of the normalizing factor $c_n$. Not surprisingly, this
position provides
also the correct critical exponent near criticality, {\it viz.},
\begin{eqnarray}
\gamma_s(k_2)=0
\end{eqnarray}
\vskip 0.5 cm
\noindent These latter results, although not fully rigorous, seem to
be conforted by the numerical simulations.
\vskip 0.5 cm
We wish to stress that the {\it polimerization phase} is not
properly speaking described by the dominance of a disconnected phase
({\it i.e.}, by a {\it gas} of weakly interacting small $4$-spheres),
but rather by the
dominance, in the statistical sum, of the
dynamically triangulated $4$-manifolds $M_{\alpha}$ generated by a
large number of
curvature blobs $\{\Xi(i)\}_{i=1,\ldots,m}(\alpha)$, connected to each
other by tubes of negligible volume.
\vskip 0.5 cm
Few remarks are also in order  for what concerns the transition
between the various phases. The
transition threshold towards a polymer phase
occurs, in our modelling, somewhere around the value of
$k_2$ such that $\eta^*=\eta_{max}$. Explicitly, for a
$k_2\simeq k_2^{crit}$ such that
\begin{eqnarray}
\frac{A(k_2)-1}{A(k_2)}=\frac{3}{4}
\end{eqnarray}
This condition provides $k_2^{crit}\simeq 1.387$ which appears quite
in a good agreement with the value indicated by the numerical
simulations. For instance in \cite{Jurk}, the critical point is
located around $k_2^{crit}(Mont.Carl)\simeq 1.336 (\pm 0.006)$. More
recent simulations \cite{Bakker} tend to move this point towards
$k_2^{crit}\simeq 1.22$.
The explanation of this discrepancy is related to the
observation that $\eta^*=\eta_{max}$ is a theoretical lower
bound, and that
$k_2^{crit}$ changes very rapidly with small
variations of the ratio $\frac{A(k_2)-1}{A(k_2)}$. Thus for instance,
if we set $\eta_{max}=1/4.1$ we would get
$k_2^{crit}\simeq 1.2$. As is clearly indicated by our analysis, this
is simply a question related to the fact that it is very delicate to
exactly
locate numerically the true critical point.

\subsection{Summing over simply-connected $4$-dimensional manifolds}

It is important to stress that the asymptotic
estimates of theorem \ref{fasi} refers to {\bf all simply-connected
$4$-manifolds}, not just to the $4$-sphere case, the only  for which
numerical data are currently available.\par
Besides ${\Bbb S}^4$, well-known example of closed simply-connected
topological $4$-manifolds are provided by ${\Bbb CP}^2$, ${\Bbb
S}^2\times{\Bbb S}^2$, the Kummer surface
$K_3\doteq\{[Z_0,Z_1,Z_2,Z_3]\in CP^3|Z_0^4+Z_1^4+Z_2^4+Z_3^4=0 \}$,
and more in general by all nonsingular algebraic surfaces of degree
$n$ in ${\Bbb CP}^3$.
The basic topological invariant for closed simply-connected $4$-
manifolds is the
intersection form\cite{Freed}. This is a natural symmetric bilinear
form, $\omega_M(,)$, on the second homology group $H_2(M,{\Bbb Z})$ of
the manifold:
for $\alpha$, $\beta\in H_2(M,{\Bbb Z})$, let $\alpha^*$ be the
Poincar\'e dual of $\alpha\in H_2(M,{\Bbb Z})$, then
$\omega_M(\alpha,\beta)\doteq <\alpha^*,\beta>_K$, where $<,>_K$ is
the Kronecker pairing between the cohomology group and the homology
group.  The case of $2$-dimensional (connected) surfaces $\Sigma$
provides a direct illustration of what is meant by this construction.
Here the relevant homology group is $H_1(\Sigma,{\Bbb Z}_2)$, every
homology class in
$H_1(\Sigma,{\Bbb Z}_2)$ can be represented by a simple closed curve
$C\subset\Sigma$, and if $C_1$, $C_2$ are any two such curves, their
intersection number $C_1\cdot C_2=C_2\cdot C_1\in{\Bbb Z}_2$ is
defined. Note that the self-intersection number $C\cdot C$ is zero as
soon as the curve $C_1$ has a neighborhood which is orientable, (by
deforming the curve $C\to\hat{C}$, while remaining in the same
homology class, we can
make $\hat{C}$ disjoint from $C$), while if such a neighborhood is
non-orientable, the self-intersection number will be odd, (think of a
deformation of the curve on a M\"{o}bius band). Actually, by
Poincar\'e duality we can make  $H_1(\Sigma,{\Bbb Z}_2)$  an inner
product space over ${\Bbb Z}_2$ if we use the intersection number
$C_1\cdot C_2$ as inner product. It can be proven that any two closed
connected surfaces $\Sigma_1$ and $\Sigma_2$ are homeomorphic if and
only if the associated inner product spaces $H_1(\Sigma_1,{\Bbb Z}_2)$
and
$H_1(\Sigma_2,{\Bbb Z}_2)$ are isomorphic (see Milnor-Husemoller's
book\cite{Freed}), and that every inner product space over ${\Bbb
Z}_2$ is isomorphic to $H_1(\Sigma,{\Bbb Z}_2)$ for some closed
connected surface $\Sigma$.\par
In a similar fashion, the intersection form $\omega_M(\alpha,\beta)$
on $H_2(M,{\Bbb Z})$ may be thought of as representing the transversal
intersection of two oriented surfaces $\alpha$ and $\beta$ in the
compact, simply-connected $4$-manifold $M$, $\alpha$ and $\beta$ being
cycles representing elements in $H_2(M,{\Bbb Z})$. Again, we can think
of $H_2(M,{\Bbb Z})$ as  an inner product space over ${\Bbb Z}$ if we
use the intersection number $\omega_M(\alpha,\beta)$ as inner product.
Moreover, it can be proven that any two closed oriented simply-
connected $4$-dimensional manifolds $M_1$ and $M_2$ have the same
homotopy type, ({\it i.e.}, there exists an orientation preserving
homotopy equivalence $M_1\to M_2$), if and only if the associated
symmetric inner product spaces $H_2(M_1,{\Bbb Z})$ and
$H_2(M_2,{\Bbb Z})$ are isomorphic, (Whitehead's theorem\cite{Freed}).
Actually, as in the case of surfaces more can be stated in terms of
this isomorphism, since
 the set of simply-connected oriented closed $4$-manifolds is in
correspondence\cite{Freed} with the set of all unimodular, (any
matrix representing $\omega(,)$ has determinant $\pm1$),
bilinear symmetric form, (the manifold is unique when the form is
even, and there are exactly two distinct manifolds when the form is
odd), (Freedman's theorem\cite{Freed}). However, if the $4$-manifolds
in question carry differentiable
structures, the correspondence is with positive-definite intersection
forms, (Donaldson's theorem\cite{Freed}).
\vskip 0.5 cm

Let us denote by $W(M;N,k_2)_{eff}$ the
canonical partition function  enumerating the distinct triangulations
on  a $4$-dimensional simply-connected manifold $M$ of given topology
and given volume ($N$). The set of such $M$ is formally
labelled by the set of all positive-definite, (we are dealing with PL
$4$-manifods), unimodular bilinear symmetric forms
$\omega_M$. According to theorem \ref{fasi}, the asymptotics
(\ref{polymer}) refers to all simply-connected $4$-dimensional PL-
manifolds of fixed
volume, thus we can write
\vskip 0.5 cm
\begin{eqnarray}
W(N,k_2)_{eff}&= &
\sum_{\omega_M}W(M;N,k_2)_{eff}\nonumber\\
&= &W({\Bbb S}^4;N,k_2)_{eff}
+W({\Bbb S}^2\times{\Bbb S}^2;N,k_2)_{eff}
+W(K^3;N,k_2)_{eff}+\ldots
\label{connected}
\end{eqnarray}
\vskip 0.5 cm
\noindent For $N$ large but fixed, the dynamical triangulations
entering
(\ref{connected}) are of bounded geometry, thus according to theorem
\ref{homotopy}, there are only a finite number of distinct homotopy
types of $4$-manifolds contributing to the above sum, and Whitehead's
theorem mentioned above implies that likewise finite is the number of
distinct types of (isomorphism classes of) intersection forms $\omega$
entering (\ref{connected}).\par
\vskip 0.5 cm
Thus from
(\ref{connected}) we get
\vskip 0.5 cm
\begin{eqnarray}
& & 10c_n\cdot\frac{e}{\sqrt{2\pi}}
e^{k_4^{crit}(k_2)N}N^{{\tilde \gamma}_s(k_2)-3}\nonumber\\
&= &
W({\Bbb S}^4;N,k_2)_{eff}
+W({\Bbb S}^2\times{\Bbb S}^2;N,k_2)_{eff}
+W(K^3;N,k_2)_{eff}+\ldots
\label{term}
\end{eqnarray}
\vskip 0.5 cm
\noindent where ${\tilde \gamma}_s(k_2)$ is the scaled critical
exponent (\ref{correxpo}). The identity (\ref{term}) makes clear that
the expression (\ref{critpolymer}) for the canonical partition
function   being common to all
dynamically triangulated
$4$-manifold (closed) with trivial fundamental group, tends to {\it
overestimate}  the canonical partition function of any
$4$-manifold $M$ in (\ref{term}).
However,
guided by the analogy with the $2$-dimensional case, and by the
agreement between our analytical results and the MonteCarlo
simulations for the $4$-sphere,
we may tentatively assume that all simply-connected DT manifolds
contribute in (\ref{term}) to the total entropy $W(N,k_2)_{eff}$ with
the some
exponential behavior and that the topological  differences come about
only in the subleading asymptotics. Since (\ref{connected}) is a
finite sum, (\ref{term}) implies that
the subleading asymptotics in each entropy function $W(M;N,k_2)_{eff}$
is of polynomial type, (at least in the range of $k_2$ considered).
Thus, we are led to the following {\it ansatz}:
\vskip 0.5 cm
\begin{eqnarray}
W(M;N,k_2)_{eff}=
10c_n\cdot\frac{e}{\sqrt{2\pi}} e^{{\tilde
k}_4^{crit}(k_2)N}N^{{\tilde\gamma}_s(\omega,k_2)-3}(1+\ldots)
\label{ansatz}
\end{eqnarray}
\vskip 0.5 cm
\noindent where ${\tilde\gamma}_s(\omega,k_2)$ is the critical
exponent for the topological $4$-manifold characterized by the
intersection form $\omega$. Since by hypothesis the exponential
leading term does not depend on topology, the above ansatz allows us
to rewrite (\ref{term}) as
\vskip 0.5 cm
\begin{eqnarray}
\sum_{\omega}N^{{\tilde\gamma}_s(\omega,k_2)-3}
=N^{{\tilde \gamma}_s(k_2)-3}
\label{intersezione}
\end{eqnarray}
\vskip 0.5 cm
The above hypothesis is fully consistent with the finiteness result
for the topology of dynamical triangulations of bounded geometry
discussed in Section V, (theorem \ref{petersen1}), according to which
the number of distinct homotopy types for such triangulations grows at
most polynomially with $N$.
\vskip 0.5 cm
In order to be  more specific, let us recall that the {\it rank}
$r(\omega)$ of a unimodular symmetric
bilinear form is the dimension of the space on which $\omega$ is
defined, namely $r(\omega)=dim[H_2(M,{\Bbb Z})]$.
The {\it signature} $s(\omega)$ of $\omega$ is defined by choosing an
orthogonal
basis $b_1,\ldots,b_n$ for the (rational) inner product space ${\Bbb
Q}\otimes{H_2(M,{\Bbb Z})}$, then\cite{Freed}
$s(\omega)=s({\Bbb Q}\otimes{H_2(M,{\Bbb Z})})\in{\Bbb Z}$ is the
difference between the number of basis elements $b_i$ with
$\omega(b_i,b_i)>0$ and the number of basis elements $b_j$ with
$\omega(b_j,b_j)<0$. Roughly speaking, $s(\omega)$ is the number of
positive eigenvalues minus the number of negative eigenvalues of
$\omega$ (thought of as a real form). Finally, the {\it type}
of $\omega$ is defined by stating that $\omega$ is {\it even}
(or of type $II$) if $\omega(\alpha,\alpha)\in 2{\Bbb Z}$ for all
$\alpha\in H_2(M,{\Bbb Z})$, ({\it i.e.}, the diagonal entries of
$\omega$
are even), otherwise $\omega$ is {\it odd}, (or of type
$I$\cite{Freed}). The relevance of such definitions lies in the fact
that any two (indefinite) inner product spaces
$(H_2(M_1,{\Bbb Z}),\omega_1)$ and $(H_2(M_2,{\Bbb Z}),\omega_2)$
are isomorphic if $\omega_1$ and $\omega_2$ have the same rank, type,
and signature\cite{Freed}. According to this observation, we can
replace the sum over the possible intersection forms in
(\ref{intersezione}) with a sum over the possible ranks, signatures
and types. To this end, let us introduce the functions $A_{\pm}(r,s)$
enumerating the distinct (isomorphism classes of) intersection forms
$\omega$ of rank $r(\omega)=r$, signature
$s(\omega)=s$ of even type, ($A_+(r,s)$), and of odd type,
($A_-(r,s)$) on a $4$-manifold of bounded geometry. Then, we can write
(\ref{intersezione}) as
\vskip 0.5 cm
\begin{eqnarray}
\sum_{r=0}^{\hat{r}}\sum_{s=0}^{\hat{s}}[A_+(r,s)+A_-(r,s)]
N^{{\tilde\gamma}_s(\omega,k_2)}
=N^{{\tilde \gamma}_s(k_2)}
\label{rank}
\end{eqnarray}
\vskip 0.5 cm
\noindent where $\hat{r}<+\infty$ and $\hat{s}<+\infty$ are the
largest value for the rank and signature allowed on a $4$-manifold of
bounded geometry. In analogy with the $2$-dimensional case, (where the
critical exponent depend linearly on topology), let us make the
further natural ansatz
\vskip 0.5 cm
\begin{eqnarray}
{\tilde\gamma}_s(\omega,k_2)=\alpha{r}+\beta{s}+\delta
\end{eqnarray}
\vskip 0.5 cm
\noindent where $\alpha$, $\beta$, $\delta$ are constants possibly
depending on the inverse gravitational coupling $k_2$.
Notice that since $H_2({\Bbb S}^4,{\Bbb Z})=0$, $\delta$ is the
critical exponent for the $4$-sphere.
\vskip 0.5 cm

The ansatz
concerning the structure of the effective entropy of each simply-
connected $4$-manifold $M$, implies that
\vskip 0.5 cm
\begin{eqnarray}
\sum_{r=0}^{\hat{r}}\sum_{s=0}^{\hat{s}}[A_+(r,s)+A_-(r,s)]
N^{\alpha{r}+\beta{s}+\delta}
=N^{{\tilde \gamma}_s(k_2)}
\end{eqnarray}
\vskip 0.5 cm
\noindent
which is a very strong bound on the set of possible intersection form
for $4$-manifolds of bounded geometry. Indeed, without the control on
geometry, the number of inequivalent intersection forms, say of given
rank $r$, grows with $r$ at an extremely fast rate. To make a rather
well-known example, (see Milnor-Husemoller book quoted in
\cite{Freed}), consider the subcase of positive-definite intersection
forms of odd type. The number of such forms is known to grow with the
rank $r$ at least as
\vskip 0.5 cm
\begin{eqnarray}
c\left( \frac{r}{2\pi e\sqrt{e}}\right)^{r^2/4}
\left( \frac{8\pi e}{r}\right)^{r/4}r^{-1/24}
\end{eqnarray}
 Thus, the interplay between our analytical approach and MonteCarlo
simulations seems to indicate results of great potential interests in
the borderline between $4D$-simplicial quantum gravity and the
topology of $4$-manifolds.
\vskip 0.5 cm
\subsection{Higher order phase transitions for simply connected $4$-
manifolds}

To conclude, we will directly prove that for a dynamically
triangulated $4$-
dimensional simply-connected manifold $M$, ({\it e.g.}, the sphere
${\Bbb S}^4$), the jump in the critical exponents
${\tilde\gamma}(M,k_2)$, as the inverse bare gravitational coupling
$k_2$ approaches the value $k_2=k_2^{crit}$, corresponds to a higher
order phase transition of our statistical system. This result settles
down in the affermative a long standing debate about the real nature
of the critical point $k_2^{crit}$ characterized by the MonteCarlo
simulations. As we shall see, some of the contraddictory results of
the numerical simulations find their rationale in our analysis. We
wish to stress that
most of the material
in this section is based on joint work with D. Gabrielli and G.
Gionti, and we wish to thank them for granting us permission to
present part of their results here.
\vskip 0.5 cm
The optimal way for discussing the onset of  a phase transition for
simplicial quantum gravity on a $4$-manifold $M$, would require the
characterization of the two-point function associated with the set of
all dynamical triangulations of $M$. The evaluation of such
two-point function requires the enumeration of all distinct dynamical
triangulations on $M$ with two marked simplex whose distance
is declared to be some running parameter $r$. As
already remarked, this counting problem is, at the moment of writing
these notes, not yet solved. However, a reliable indication that our
system exhibits a phase transition may came also from exhamining the
simplicial version of the {\it curvature susceptibility}. This is
defined by
\vskip 0.5 cm
\begin{eqnarray}
\chi_R\doteq\frac{1}{\hat{N}}\frac{{\partial}^2\ln W({\Bbb
S}^4;N,k_2)_{eff}}{\partial k_2^2}=
\frac{1}{\hat{N}}\left[<\lambda^2>-<\lambda>^2 \right]
\label{suscept}
\end{eqnarray}
\vskip 0.5 cm
\noindent where $\hat{N}\doteq\frac{1}{2}n(n+1)N|_{n=4}=10N$.
Since we work at fixed
$N\doteq N_4$, ({\it i.e.}, at fixed volume), such a  $\chi_R$ can be
equivalently written as
\vskip 0.5 cm
\begin{eqnarray}
\chi_R=\langle (\frac{1}{b})^2\rangle-\langle\frac{1}{b}\rangle^2
\end{eqnarray}
\vskip 0.5 cm
\noindent where as usual $b(n,n-
2)\doteq\frac{1}{2}n(n+1)\frac{N}{\lambda+1}$. Thus $\chi_R$
characterizes the fluctuations in
the average incidence $b$ on the set of all possible dynamical
triangulations (at fixed volume) on ${\Bbb S}^4$. For this reason, it
is not difficult to show on heuristic grounds that such a $\chi_R$ can
be actually considered as the dynamical triangulation counterpart of a
connected curvature-curvature correlation function:
\vskip 0.5 cm
\begin{eqnarray}
\langle\int_{S^4}d^4\xi_1d^4\xi_2\sqrt{g}R(\xi_1)\sqrt{g}
R(\xi_2)\rangle_{connected}
\end{eqnarray}
\vskip 0.5 cm
It is worth noticing that  such a $\chi_R$ has been considered
for describing the onset of phase transitions also in
computer-assisted simulations. However, its direct use in such
a setting is hampered by the fact that in such computer simulations
one is forced to allow for volume fluctuations.
Moreover, it has been
observed that the peak of such $\chi_R$ is rather asymmetric and
difficult to locate. Thus, in MonteCarlo simulations it is difficult
to provide a clear cut interpretation of the behavior of $\chi_R$.
\vskip 0.5 cm
Such difficulties in using the curvature susceptibility have a clear
origin in the analytical computation of $\chi_R$. Indeed, contrary to
what one would like to think, at the critical point $\chi_R$ {\it does
not} diverge with $N$, it rather exhibits a finite jump. According to
standard statistical mechanics, such a jump may or may not be
associated with the onset of long range order, and not surprisingly,
for characterizing the order of the transition at $k_2^{crit}$, we
are naturally led to a finite scaling argument {\it a' la Fisher} not
much dissimilar from the kind of argument adopted in the actual
MonteCarlo simulations.
\vskip 0.5 cm
Adopting the notation iof theorem \ref{fasi} and lemma
\ref{regime},
the possible existence of long range order at the critical point
$k_2=k_2^{crit}$ rests on the following
\vskip 0.5 cm
\begin{Gaiat}
As $|k_2-k_2^{crit}|\to0$ and $N$ varies there exist  pseudo critical
points $k_2=k_2^{crit}(N)$ characterized by the condition
\vskip 0.5 cm
\begin{eqnarray}
\psi_{max}(k_2^{crit}(N))\sqrt{\hat{N}}=\delta
\label{roundinterval}
\end{eqnarray}
\vskip 0.5 cm
\noindent where $0<\delta\leq 1$, is a fixed ( suitably small)
constant (indipendent from $N$-see (\ref{rounding})),  such that
\vskip 0.5 cm
\begin{eqnarray}
|k_2^{crit}(N)-k_2^{crit}|\simeq_{N>>1} \delta\frac{\sqrt{-
f_{\eta\eta}(\eta^*)}}{\sqrt{\hat{N}}}
\label{finitesize}
\end{eqnarray}
\vskip 0.5 cm
\noindent and
\vskip 0.5 cm
\begin{eqnarray}
\lim_{N\to\infty}\chi_{R}(k_2=k_2^{crit})=\frac{1}{2}\left[-
\frac{1}{f_{\eta\eta}(\eta^*)} \right].
\label{correcrit}
\end{eqnarray}
\vskip 0.5 cm
\noindent This indicates that at the critical point
$k_2=k_2^{crit}\simeq 1.387$ the dynamically triangulated $4$-manifold
$M$ undergoes a
higher order phase transition possibly associated with the onset of
long range
order.
\end{Gaiat}
\vskip 0.5 cm
\noindent {\bf Proof}. According to the uniform asymptotic estimation
of the canonical partition function $W(N,k_2)_{eff}$ provided by
theorem \ref{fasi} and its corollaries, the approach to the critical
point $k_2^{crit}\simeq 1.387$ is characterized by the parameter (see
(\ref{max})),
\vskip 0.5 cm
\begin{eqnarray}
\psi_{max}(k_{2}){\doteq}  sgn(\eta^*-\eta_{max})
\sqrt{2[f(\eta^*)-f(\eta_{max})]}
\end{eqnarray}
\vskip 0.5 cm
\noindent through the condition
\begin{eqnarray}
\psi_{max}(k_2)\sqrt{\hat{N}}\to 0.
\end{eqnarray}
\vskip 0.5 cm
Whereas, the separation between the weak (strong) coupling regions and
the critical region corresponds to values of the inverse gravitational
coupling $k_2$ such that
\begin{eqnarray}
\psi_{max}(k_2)\sqrt{\hat{N}}\simeq 1.
\end{eqnarray}
\vskip 0.5 cm
It follows that if we
choose a $\delta$, with $0<\delta\leq 1$, sufficiently small, then
through the condition (\ref{roundinterval}), we select the {\it
rounding interval} $k_2^{crit}-k_2^{crit}(N)\leq k_2\leq
k_2^{crit}+k_2^{crit}(N)$ corresponding to which the critical behavior
occurs. As the notation suggests, this condition characterizes a
value of the coupling $k_2\doteq{k_2^{crit}(N)}$ according to
\vskip 0.5 cm
\begin{eqnarray}
\psi_{max}(k_2^{sup}(N))\sqrt{\hat{N}}=\delta
\end{eqnarray}
\vskip 0.5 cm
\noindent which localizes the maximum $\eta^*$ of the
function $f(\eta,k_2)$ to an exponentialy small neighborhood of the
transition point $\eta=\eta^{max}$. For a given $N$, this
$k_2^{crit}(N)$ maximizes the corresponding value of the curvature
susceptibility $\chi_R(N)$, and as such it defines a pseudo critical
point. Explicitly, from the definition of $\chi_R$ and from the proof
of theorem \ref{fasi}, we compute
\vskip 0.5 cm
\begin{eqnarray}
\chi_R=\frac{1}{\hat{N}}\frac{\partial^2}{\partial{k_2}^2}
\ln\int_0^{\infty}G(p)e^{-
(\hat{N}+1)(\frac{1}{2}p^2+\psi_{max}p)}dp
\label{semplice1}
\end{eqnarray}
\vskip 0.5 cm
\noindent By exploiting the uniform asymptotics estimation of the
integral appearing in the right member of (\ref{semplice1}), we can
write, up to order $O(N^{-m-1})$,
\vskip 0.5 cm
\begin{eqnarray}
\chi_R=\frac{1}{\hat{N}}\frac{\partial^2}{\partial{k_2}^2}
\ln
\left\{ \frac{w_0(z)}{(\hat{N})^{1/2}}\sum_{i=0}^m
\frac{\xi_{2i}}{(\hat{N})^i}+
\frac{w_{-1}(z)}{(\hat{N})}\sum_{i=0}^m
\frac{\xi_{2i+1}}{(\hat{N})^i}\right\}
\end{eqnarray}
\vskip 0.5 cm
\noindent where $z\doteq \psi_{max}(k_{n-
2})\sqrt{\hat{N}}$, (henceforth we will omit the annoying $1$ whenever
$\hat{N}+1$ appears). If we introduce explicitly the expression of the
Whittaker functions $w_0(z)$ and $w_{-1}(z)$ in terms of the parabolic
cylinder function $D_r(z)$, {\it viz.},
$w_r(z)=\Gamma(r-1)e^{z^2/4}D_{r-1}(z)$, then a long  but
straightforward computation provides
\vskip 0.5 cm
\begin{eqnarray}
\chi_R=\frac{1}{2}\left(\frac{d}{dk_2}\psi_{max}(k_2) \right)^2+
\frac{1}{2}\psi_{max}\frac{d^2\psi_{max}(k_2)}{dk_2^2}+\nonumber
\end{eqnarray}
\begin{eqnarray}
\frac{1}{\hat{N}}\frac{\partial^2}{\partial{k_2}^2}
\ln
\left\{ \frac{D_{-1}(z)}{(\hat{N})^{1/2}}\sum_{i=0}^m
\frac{\xi_{2i}}{(\hat{N})^i}+
\frac{D_{-2}(z)}{(\hat{N})}\sum_{i=0}^m
\frac{\xi_{2i+1}}{(\hat{N})^i}\right\}
\end{eqnarray}
\vskip 0.5 cm
In the critical region characterized by
$\psi_{max}(k_2)\sqrt{\hat{N}}<<1$, the asymptotics
(\ref{paracritico}) of the parabolic cylinder functions shows that, to
leading order in $N$, we have
\vskip 0.5 cm
\begin{eqnarray}
\chi_R\simeq\frac{1}{2}\left(\frac{d}{dk_2}\psi_{max}(k_2) \right)^2+
\frac{1}{2}\psi_{max}\frac{d^2\psi_{max}(k_2)}{dk_2^2}
\end{eqnarray}
\vskip 0.5 cm
\noindent In order to compute the above derivatives, let us consider
the following limit
\begin{eqnarray}
\lim_{k_2\to{k_2^{crit}}}\frac{|\psi_{max}(k_2)|}{|k_2-k_2^{crit}|}=
\lim_{k_2\to{k_2^{crit}}}\sqrt{\frac{2[f(\eta^*)-f(\eta^{max})]}{(k_2-
k_2^{crit})^2}}
\end{eqnarray}
By l'H\^opital rule one easily obtains
\begin{eqnarray}
\lim_{k_2\to{k_2^{crit}}}\frac{|\psi_{max}(k_2)|}{|k_2-k_2^{crit}|}=
\left( \frac{d\eta^*}{dk_2}|_{k_2=k_2^{crit}}
\right)=\sqrt{\eta^*(1-3\eta^*)(1-2\eta^*)}.
\end{eqnarray}
But this latter is exactly the expression of $-f_{\eta\eta}^{-
1}(\eta^*)$, thus one computes
\begin{eqnarray}
\lim_{k_2\to{k_2^{crit}}}\frac{|\psi_{max}(k_2)|}{|k_2-
k_2^{crit}|}=\sqrt{-\frac{1}{f_{\eta\eta}(\eta^*)}}
\end{eqnarray}
\vskip 0.5 cm
\noindent This latter result implies that for $|k_2-k_2^{crit}|$ small
enough we may write
\begin{eqnarray}
|\psi_{max}(k_2)|\simeq\sqrt{-\frac{1}{f_{\eta\eta}(\eta^*)}}|k_2-
k_2^{crit}|
\label{finiteffect}
\end{eqnarray}
and thus, not unexpectedly, we obtain
\vskip 0.5 cm
\begin{eqnarray}
\chi_R=\frac{1}{2}\left(\sqrt{-\frac{1}{f_{\eta\eta}(\eta^*)}} \right)
\label{jump}
\end{eqnarray}
which proves that at criticality the susceptibility does not diverge
with $N$, but rather it has a finite jump.
\vskip 0.5 cm
\noindent We can rewrite (\ref{jump}) as
\begin{eqnarray}
\chi_R=\frac{\psi_{max}^2}{2(k_2-k_2^{crit})^2}
\end{eqnarray}
Thus, if $k_2^{max}(N)$ is defined as the value of $k_2$ for which
\begin{eqnarray}
\psi_{max}(k_2=k_2(N))\sqrt{\hat{N}}=\delta
\label{intervallo}
\end{eqnarray}
we obtain
\begin{eqnarray}
\chi_R(\delta, N)=\left(\frac{\delta}{\hat{N}}
\right)\frac{1}{(k_2^{crit}(N)-k_2^{crit})^2}
\end{eqnarray}
which shows that {\it at fixed $N$}, and for a given $\delta$, (fixing
the rounding interval), $\chi_R$ attains its maximum at
$k_2^{crit}(N)$. This indicates  that $k_2^{crit}(N)$ is a pseudo
critical point for the given rounding interval and
(\ref{finiteffect}) provides the way such pseudo critical points
approach, as $N$ increases, the true critical point $k_2^{crit}$.
Indeed, from (\ref{finiteffect}) and (\ref{intervallo}) we obtain
\vskip 0.5 cm
\begin{eqnarray}
|k_2^{crit}(N)-k_2^{crit}|\simeq_{N>>1} \delta\frac{\sqrt{-
f_{\eta\eta}(\eta^*)}}{\sqrt{\hat{N}}}
\end{eqnarray}
\vskip 0.5 cm
\noindent as stated.
\vskip 0.5 cm
\noindent As is well known, the scale fixing the rounding of a first-
order transition takes place on a scale proportional to the inverse of
the total volume of the system. In our case, this rounding takes place
on a scale proportional to the square root of the volume of the
manifold. Thus the phase transition at $k_2^{crit}$ is {\bf not} a
first order transition and it is possibly associated with the onset of
long
range order in our statistical system. $\Box$
\vskip 0.5 cm
This result strongly suggests that the ensemble of $4$-manifolds will
develop a genuine extension when $k_2\to{k_2}^{crit}$, and make this
transition point a potential candidate for a non-perturbative theory
of quantum gravity. As already emphasized, in order to carry out this
program (and eventually remove all the {\it possibly}'s in the above
statements concerning the existence of long range order), we have to
compute the two point function and characterize its
scaling properties at criticality.
The almost spectacular agreement between the analytical approach and
the MonteCarlo data indicates that
the picture of $4$-dimensional simplicial quantum gravity which would
emerge from this program should be very similar to that one already
elaborated on the basis of computer simulations. However, a delicate
point emerges here for what concerns the assessment of the nature of
the transition at $k_2^{crit}$. Some recent MonteCarlo simulations
have cast doubts to the standard wisdom concerning its second order
nature. It has been observed \cite{Bakker} a double peak in the
distribution of order of vertices, and this has been interpreted of
some indication of a first order nature of the transition. On the
other hand, finite size scaling results, as above, point to a higher
order transition. From the numerical simulations point of view the
situation is not yet settled. In fact the situation might be quite
similar to the one we encounter for compact QED in four dimensions: we
have a strong coupling phase were the string tension is different from
zero (to be compared with the crumpled phase in $4$-dimensional
gravity). The string tension scales to zero at the critical point (in
$4$D-gravity this would correspond to the mass parameter scaling to
zero at the critical point). At this critical point it is unclear if
there is a two-phase signal, and if the transition is first or higher
order. A two-phase signal has first been observed, then it has
disappeared in a sort of see-saw game, and in a way akin to the
situation we are experiencing in $4$D-gravity. This similarity
persists in the weak coupling phase: for small coupling constant we
have the coulomb phase in QED, {\it viz.}, for all coupling constants
smaller than the critical one we have a continuum massless phase, and
indeed in $4$D-gravity for all (bare) gravitational coupling constants
smaller than the critical one we have branched polymers, which indeed
admit a continuum limit. Since, as we have shown, there is more than a
superficial agreement between MonteCarlo simulations and our
analytical approach, one wonders if there is a way of understanding
the origin of a possible two-phase signal and how this may be
compatible with the results above concerning the higher order nature
of the transition.
\vskip 0.5 cm
The explanation is that the uniform asymptotics controlling the
leading behavior of the canonical  partition function may generate a
bistable behaviour: we have discussed the approach to criticality as
if the only leading contribution to $W(k_2,N)_{eff}$ comes from
developing in a neighborhood of $\psi_{max}(k_2)\sqrt{N}\to0$.
However, as it is easily checked,
$f(\eta_{min})$ is not much different from $f(\eta_{max})$, and
$W(k_2,N)_{eff}$ may receive an important contribution from
developing  around $\psi_{min}(k_2)$, too. It is this further
contribution that may generate a double peak and similar hysteresis
effects, in a way completely analogous to what happens in the three-
dimensional case and which we discuss in details in the next section.
These effects tend to obliterate the nature of the transition. They
are more manifest and more important in the three-dimensional case
essentially because  the difference $|f(\eta_{max})-f(\eta_{min})|$ is
smaller than in the four-dimensional case.

\subsection{ Hysteresis effects in the three-dimensional case}

The confrontation between MonteCarlo simulations and the analytical
estimates is quite satisfactory also in the  three-dimensional case.
\vskip 0.5 cm
Numerical data in dimension $n=3$ often employ\cite{Varsted} the pair
of variables $N_3$ and $N_0$ rather than $N_3$ and $N_1$, thus we have
to adapt our asymptotic estimates to this new pair of variables before
comparing with the simulations. This can be trivially done by noticing
that the Dehn-Sommerville relations
$N_0-N_1+N_2-N_3=0$ and $N_2=2N_3$ imply $N_1=N_3+N_0$, so that
the partition function  can be written as
\begin{eqnarray}
Z[k_1(k_0),k_3]&= &\sum_{N}\left ( \sum_{\lambda=N+N_0}
W(\lambda=N+N_0,N)e^{k_1(\lambda=N+N_0)}\right )
e^{-k_3N}\nonumber\\
&= & \sum_{N}\left ( \sum_{\lambda=N+N_0}
W(\lambda=N+N_0,N)e^{k_1N_0}\right )
e^{(k_1-k_3)N}
\end{eqnarray}
where $N{\doteq} N_3$, and $\lambda= N_1$.
\vskip 0.5 cm
The delicate point to stress is that in dimension $n=3$, we
cannot any longer use the results of theorem \ref{fasi} and lemma
\ref{regime} as they stand. The basic observation is that the function
$f(\eta)$, (now expressed in terms of the variable $N_0$), is, in the
interval $\eta_{min}\leq\eta\leq\eta_{max}$, a concave function
symmetric with respect to the point $\eta^*$ where it attains its
maximum, {\it viz.},
\vskip 0.5 cm
\begin{eqnarray}
f(\eta_{min})\simeq f(\eta_{max})
\label{coincidence}
\end{eqnarray}
\vskip 0.5 cm
\noindent Since the parameters driving the uniform asymptotics of the
saddle-point evaluation of the canonical partition function are, (see
(\ref{min}) and (\ref{max})),
\vskip 0.5 cm
\begin{eqnarray}
\psi_{min}(k_{n-2})\doteq -sgn(\eta^*-\eta_{min})
\sqrt{2[f(\eta^*)-f(\eta_{min})]}
\end{eqnarray}
and
\begin{eqnarray}
\psi_{max}(k_{n-2}){\doteq}  sgn(\eta^*-\eta_{max})
\sqrt{2[f(\eta^*)-f(\eta_{max})]}
\end{eqnarray}
\vskip 0.5 cm
\noindent then (\ref{coincidence}) implies that the correct
asymptotics of the canonical partition function, in the range
$\eta_{min}\leq\eta\leq\eta_{max}$ is provided by a linear combination
of the two uniform asymptotics obtained by developing the integral
(\ref{minimo}) in terms of both $\psi_{min}(k_1)$ and
$\psi_{max}(k_1)$, {\it viz.},
\vskip 0.5 cm
\begin{Gaiat}
Let us consider the set of all simply-connected
$3$-dimensional  dynamically triangulated manifolds.
Let $k_{0}^{inf}$, and $k_0^{crit}$ respectively denote the unique
solutions
of the equations
\vskip 0.5 cm
\begin{eqnarray}
\frac{1}{3}(1-\frac{1}{A(k_{0})})+\frac{1}{6}=\eta_{min}=\frac{1}{6}
\end{eqnarray}
\begin{eqnarray}
\frac{1}{3}(1-\frac{1}{A(k_{0})})+\frac{1}{6}=\eta_{max}=\frac{2}{9}
\end{eqnarray}
\vskip 0.5 cm

\noindent Let $0<\epsilon<1$ small enough, then
for all values of  the inverse gravitational coupling $k_{0}$ such
that
\begin{eqnarray}
k^{inf}_{0}+\epsilon< k_{0}<k^{crit}_{0}-\epsilon,
\end{eqnarray}
the large $N$-behavior of the canonical partition function $W(N,
k_{0})_{eff}$ for $3$-dimensional simplicial quantum gravity on a
simply connected manifold
is given by
the uniform asymptotics
\vskip 0.5 cm
\begin{eqnarray}
\frac{c_n}{\sqrt{2\pi}}{\hat{N}}^{-7/2}
e^{(\hat{N}+1)f(\eta_{min})}
\left[\frac{\alpha_0}{\sqrt{\hat{N}}}
w_0(\psi_{min}(k_{0})
\sqrt{\hat{N}})+\frac{\alpha_1}{\hat{N}}
w_{-1}(\psi_{min}(k_{0})\sqrt{\hat{N}})
\right]+\nonumber
\end{eqnarray}
\begin{eqnarray}
+\frac{c_n}{\sqrt{2\pi}}{\hat{N}}^{-7/2}
e^{(\hat{N}+1)f(\eta_{max})}
\left[\frac{\xi_0}{\sqrt{\hat{N}}}
w_0(\psi_{max}(k_{0})
\sqrt{\hat{N}})+\frac{\xi_1}{\hat{N}}
w_{-1}(\psi_{max}(k_{0})\sqrt{\hat{N}})
\right]
\end{eqnarray}
\vskip 0.5 cm
\noindent  where the constants $\alpha_0$,
$\alpha_1$, and $\xi_0$, $\xi_1$ are given by
\vskip 0.5 cm
\begin{eqnarray}
\alpha_0\doteq -
\frac{\eta_{min}^{-1/2}(1-2\eta_{min})^{-n}}{\sqrt{(1-3\eta_{min})(1-
2\eta_{min})}}\frac{\psi_{min}(k_{0})}{f_{\eta}(\eta_{min})}
\end{eqnarray}
\vskip 0.5 cm
\begin{eqnarray}
\alpha_1\doteq \frac{\alpha_0}{\psi_{min}(k_{0})}
-\frac{\eta*^{-1/2}(1-2\eta^*)^{-n}}{\sqrt{(1-3\eta^*)(1-
2\eta^*)}} \left[
\frac{1}{\psi_{min}(k_{0})\sqrt{-
f_{\eta\eta}(\eta^*)}} \right]
\end{eqnarray}
\vskip 0.5 cm
\begin{eqnarray}
\xi_0\doteq -
\frac{\eta_{max}^{-1/2}(1-2\eta_{max})^{-n}}{\sqrt{(1-3\eta_{max})(1-
2\eta_{max})}}\frac{\psi_{max}(k_{0})}{f_{\eta}(\eta_{max})}
\end{eqnarray}
\vskip 0.5 cm
\begin{eqnarray}
\xi_1\doteq \frac{\xi_0}{\psi_{max}(k_{0})}
-\frac{\eta*^{-1/2}(1-2\eta^*)^{-n}}{\sqrt{(1-3\eta^*)(1-
2\eta^*)}} \left[
\frac{1}{\psi_{max}(k_{0})\sqrt{-
f_{\eta\eta}(\eta^*)}} \right]
\end{eqnarray}
\label{3dimensioni}
\end{Gaiat}
\vskip 0.5 cm
\noindent

From this result
it follows that
$k^{crit}_3(k_0)$ can be formally obtained from an  expression similar
to
(\ref{cosmo}) by the substitution $k_1\to \alpha{k_0}$ and
$k^{crit}_3(k_1)\to k^{crit}_3(k_1\to\alpha{k_0})-{\alpha}k_0$,
where $0<\alpha\leq 1$ is a suitable constant, {\it viz.},
\begin{eqnarray}
k_3^{crit}(k_0)=6\ln \frac{A(\alpha{k_0})+2}{3}-
\alpha{k_0}+6\ln_{N\to\infty}\frac{\ln{c_n}}{N}
\label{cosmo3}
\end{eqnarray}
where
\vskip 0.5 cm
\begin{eqnarray}
A(\alpha{k_0})\equiv{ \left [\frac{27}{2}e^{\alpha{k_0}}+1+
\sqrt{(\frac{27}{2}e^{\alpha{k_0}}+1)^2-1} \right ]}^{1/3}+\nonumber
\end{eqnarray}
\begin{eqnarray}
+{ \left [\frac{27}{2}e^{\alpha{k_0}}+1-
\sqrt{(\frac{27}{2}e^{\alpha{k_0}}+1)^2-1} \right ]}^{1/3}
-1
\end{eqnarray}
\vskip 0.5 cm

If we fix the rescaling factor $c_n$ according to
\vskip 0.5 cm
\begin{eqnarray}
c_n=[\cos^{-1}(\frac{1}{3})]^{-(6/5)N}
\end{eqnarray}
\vskip 0.5 cm
\noindent and set the constant $\alpha\simeq 0.16$,
then it is not difficult to check that the rescaled  $k_3^{crit}(k_0)$
has a
distribution of values, as $k_0$ increases, which is in very good
agreement with the corresponding values determined by MonteCarlo
simulations.
Explicitly we get

\begin{eqnarray}
& & k_0=0\ldots K_3^{crit}(Mont.Car)=2.050\ldots
{K}_3^{crit}=2.04\nonumber\\
& & k_0=1\ldots K_3^{crit}(Mont.Car)=2.055\ldots
{K}_3^{crit}=2.07\nonumber\\
& & k_0=2\ldots K_3^{crit}(Mont.Car)=2.070\ldots
{K}_3^{crit}=2.11\nonumber\\
& & k_0=3\ldots K_3^{crit}(Mont.Car)=2.100\ldots
{K}_3^{crit}=2.15\nonumber\\
& & k_0=4\ldots K_3^{crit}(Mont.Car)=2.220\ldots
{K}_3^{crit}=2.20\nonumber\\
\end{eqnarray}
\vskip 0.5 cm
The critical point (corresponding to the crossing of the upper limit
$\eta_{max}=\frac{2}{9})$ occurs for $k^{crit}_0\simeq 3.845$ which
agrees well with the location of the critical point
$k_0^{crit}(Mont.Car)\simeq 3.9$  obtained by the MonteCarlo
simulations.
\vskip 0.5 cm
Some aspects of the nature of the transition at the above critical
point $k_0^{crit}$
are also rather easily characterized. It is standard wisdom among
groups performing MonteCarlo simulations that this transition should
be a first order transition, a credence fostered, among other things,
also by the fact that in three-dimensional gravity we expect no
dynamical graviton to be present, (but this is a very misleading
concept in a non-perturbative approach to quantum gravity!). More
seriously, the assumed first order nature of the transition is
associated to quite pronounced hysteresis effects, the same hysteresis
effects that (even if at a much less pronounced level) seem to be
manifest also in some recent simulation of $4$-dimensional simplicial
quantum gravity.
\vskip 0.5 cm
From theorem \ref{3dimensioni} is clear that as $k_0\to{k_0}^{crit}$
the system shows a cross-over behavior with the phase connected with
small values of $k_0$, since as $k_0$ approaches $k_0^{crit}$ both
$\psi_{max}$ and $\psi_{min}$ tend to go to $0$ and the system shows
a bistable behavior yielding for  an hysteresis associated with the
distinct phases described  by $\psi_{max}$ and $\psi_{min}$. This
bistability, if strong enough, can be
consistent with a first order nature of the transition at
$k_0^{crit}$.
\vskip 0.5 cm
\noindent It is very important to develop the above analysis in order
to study in detail the neighborhood of the transition point either in
$3$-dimensions as well as in $4$-dimensions. The analytical approach
developed here  already provides useful indications, but it  can be
certainly improved, hopefully to bring to full power dynamical
triangulations as a most effective approach to deal with non-
perturbative issues in euclidean quantum gravity.

\vfill\eject

{\bf Aknowledgements}

We would like to extend special gratitude to B. Br\"{u}gmann, U.
Bruzzo, L. Crane, G. Jug, R. Loll, R. Penrose, C. Rovelli, and L.
Smolin for many interesting discussions and critical remarks which
motivated several improvements. We would especially single out
D.Gabrielli,G.Gionti, C.F.Kristjansen, J.Jurkiewicz, and
J. Lewandowski for detailed commentary on many subtle issues and
inspiring conversations.
\par
\vskip 0.5 cm
This work has been supported by the National Institute for Nuclear
Physics (INFN).

\vfill\eject
\section{Appendix: The tangent space to the representation variety}

Given the representation
${\Theta}\colon{\pi}_1(M)\to{G}$ let us consider the associated
representation $\theta$ on the Lie algebra ${\frak g}$ generated by
composing $\Theta$ with
 the adjoint action of G on ${\frak g}$, {\it viz.},
${\theta}\colon{\pi}_1(M)\to_{\Theta}{G}\to_{Ad} End({\frak g})$,
(henceforth we
will always  refer to this representation ).
The representation $\theta$ defines a
flat bundle, ${\frak g}_{\theta}$, over the underlying space $M$ of
the dynamical triangulation considered. This bundle is costructed by
exploiting  the adjoint representation of $G$ on its
Lie algebra ${\frak g}$, {\it i.e.},
$Ad\colon{G}\to{End({\frak g})}$, and by considering the action of
${\pi}_1(M)$ on ${\frak g}$ generated by composing the adjoint action
and the representation $\Theta$:
\begin{eqnarray}
{\frak g}_{\theta}=\hat{M}\times{\frak
g}/{\pi}_1\otimes[Ad({\Theta}(\cdot))]^{-1}
\end{eqnarray}
where ${\pi}_1\otimes[Ad({\Theta}(\cdot))]^{-1}$ acts, through
${\pi}_1(M)$, by deck transformations on $\hat{M}$ and by
$[Ad({\Theta}(\cdot))]^{-1}$ on the Lie algebra ${\frak g}$. More
explicitly, if $\hat{\sigma}_1$, $\hat{\sigma}_2$ are simplices in
$\hat{M}$, and $g_1$, $g_2$ are elements of ${\frak g}$, then
$\hat{\sigma}_1,g_1)\sim (\hat{\sigma}_2,g_2)$
if and only if $\hat{\sigma}_2=\hat{\sigma}_1a$, and
$g_2=[Ad({\Theta}(a))]^{-1}g_1$, for some $a\in{\pi}_1(M)$.\par
\vskip 0.5 cm

Given a dynamical triangulation $T_a$ and a flat bundle ${\frak
g}_{\theta}$ over $M=|T_a|$, we can define homology and cohomology
groups with value in ${\frak g}_{\theta}$, (namely homology and
cohomology with values in the sheave of Killing vectors of
$M_{riem}$).
 On $T_a$ we define a $p$-
{\it chain} with  values in ${\frak g}_{\theta}$ to be a formal sum
$\sum_{i=1}^{m}g_i{\sigma}^p_i$ where $\sigma^p_i\in T_a$ is an
ordered $p$-simplex and $g_i$ is an element of the fiber of ${\frak
g}_{\theta}$ over the first vertex of $\sigma^p_i$. Namely,
first we consider chains with coefficients in
the
Lie algebra ${\frak g}$, {\it viz.}, $\sum_jg_j\hat{\sigma}^i_j$
with $g_j\in{\frak g}$, and then quotient the resulting chain
complex $C(\hat{M})\otimes{\frak g}$ by the action of
${\pi}_1\otimes[Ad({\theta}(\cdot))]^{-1}$. We denote the group of
such chains by $C_p(M,{\frak g}_{\theta})$.  There is
an action of ${\pi}_1(M)$ on the above chains expressed by
\begin{eqnarray}
a(\sum_jg_j\hat{\sigma}^i_j)\to
\sum_j([Ad({\theta}(a))]^{-1}g_j)a(\hat{\sigma}^i_j)
\end{eqnarray}
for any $a\in{\pi}_1(M)$, ({\it i.e.}, we are considering
${\frak g}$ as a ${\pi}_1(M)$-module).
This action commutes with the boundary operator, and we may define
homology groups $H_*(M,{\frak g}_{\theta})$
with local coefficients  in the flat bundle
${\frak g}_{\theta}$. By dualizing one defines the
cohomology $H^*(M,{\frak g}_{\theta})$, which enjoys the usual
properties of a cohomology theory.
\vskip 0.5 cm
These homology groups are naturally related to the geometry of the
infinitesimal deformations of representations of the fundamental group
of $M$. Given the representation
${\theta}\colon{\pi}_1(M)\to{G}$ let us consider the associated
representation on the Lie algebra ${\frak g}$ generated by
composing $\theta$ with
 the adjoint action of G on ${\frak g}$, {\it viz.},
${\theta}\colon{\pi}_1(M)\to{G}\to End({\frak g})$, (henceforth we
will always refer to this representation). Then a $1$-{\it cocycle} on
$\pi_1(M)$ with coefficients in $\theta$ is a map
$c\colon\pi_1(M)\to{\frak g}$ such that for $\gamma$,
$\delta\in\pi_1(M)$ we have
\begin{eqnarray}
c(\gamma\delta)=c(\gamma)+\theta(\gamma)\cdot c(\delta)
\end{eqnarray}
The space of such cocycles (often called crossed homomorphisms) is
denoted by $Z^1(\pi_1(M),{\frak g})$. We denote by
$B^1(\pi_1(M),{\frak g})\subset Z^1(\pi_1(M),{\frak g})$, the subspace
generated by the cocycles which are $1$-coboundaries,
namely the set of $c\in Z^1(\pi_1(M),{\frak g})$ for which there
exists a $g\in {\frak g}$ such that
\begin{eqnarray}
c(\gamma)=\theta(\gamma)g-g
\end{eqnarray}
for all $\gamma\in\pi_1(M)$. The {\it first cohomology group} of
$\pi_1(M)$ with values in ${\frak g}$ is then defined by
\begin{eqnarray}
H^1(\pi_1(M),{\frak g})=\frac{Z^1(\pi_1(M),{\frak
g})}{B^1(\pi_1(M),{\frak g})}
\end{eqnarray}
The definition of the higher cohomology groups is somewhat more
complicated, and the interested reader may consult\cite{Eilenberg}. In
order to relate
$H^1(\pi_1(M),{\frak g})$ to the infinitesimal deformations of the
representations of $\pi_1(M)$,
let $\frac{Hom({\pi}_1(M),G)}{G}$ denote the set of all conjugacy
classes of representations of the fundamental group ${\pi}_1(M)$
into the Lie group $G$. Notice that   if $\theta$ and
$F{\theta}F^{-1}$ are two conjugate representations of ${\pi}_1(M)$ in
$G$, then
through the map $Ad(F)\colon{\frak g}\to{\frak g}$ we get a natural
isomorphism between the groups $H_{i}(\pi_1(M),{\frak g})$ and
$H_{i}(\pi_1(M),F{\frak g}F^{-1})$.
\vskip 0.5 cm
 We are interested in understanding the structure of
$\frac{Hom({\pi}_1(M),G)}{G}$
when deforming a particular
representation ${\theta}\colon{\pi}_1(M)\to{G}$ through a
differentiable one-parameter family of representations
${\theta}_t$ with ${\theta}_0=\theta$ which are not tangent to the
$G$-orbit of $\theta\in{Hom({\pi}_1(M),G)}$.\par
To this end,
let us rewrite, for $t$ near $0$, the given one-parameter family of
representations ${\theta}_t$ as\cite{Millson},\cite{Goldman}
\begin{eqnarray}
{\theta}_t=\exp[tu(a)+O(t^2)]{\theta}(a)
\end{eqnarray}
where $a\in {\pi}_1(M)$, and where $u\colon{\pi}_1(M)\to{\frak g}$. In
particular,
 given $a$ and $b$ in ${\pi}_1(M)$, if we differentiate the
homomorphism condition ${\theta}_t(ab)={\theta}_t(a){\theta}_t(b)$, we
get that
$u$ actually is a one-cocycle of ${\pi}_1(M)$ with coefficients
in the ${\pi}_1(M)$-module ${\frak g}_{\theta}$, {\it viz.},
\begin{eqnarray}
u(ab)=u(a)+ [Ad({\theta}(a))]u(b)
\end{eqnarray}
Moreover, any $u$ verifying the above cocycle condition leads to a map
${\theta}_t\colon{\pi}_1(M)\to{G}$ which, to first order in $t$,
satisfies the homomorphism condition. This remark implies that the
(Zariski) tangent space to $Hom{({\pi}_1(M),G)}$ at
$\theta$, can be identified with $Z^1(\pi_1(M),{\frak g})$. For this
reason, the space $Z^1(\pi_1(M),{\frak g})$ is called the space of
infinitesimal deformation of the representation $\theta$.\par
In a similar way, it can be shown that the tangent space to the
$Ad$-orbit through ${\theta}$ is
$B^1(\pi_1(M),{\frak g})$. Thus, the (Zariski) tangent space to
$\frac{Hom({\pi}_1(M),G)}{G}$ corresponding to the conjugacy
class of representations
$[\theta]$ is $H^1(\pi_1(M),{\frak g})$ which, therefore, can be
identified with the formal tangent
space to the representation space.
\vskip 0.5 cm
It is not difficult
to establish an explicit isomorphism between the tangent space
$H^1(\pi_1(M),{\frak g})$ and the first cohomology group with
coefficient in the flat bundle ${\frak g}_{\theta}$, $H^1(M,{\frak
g}_{\theta})$, associated with  the underlying space $M$ of the
dynamical triangulation $T_a$. We choose a base $x_0\in M$, (say a
vertex of the base simplex $\sigma^n_0$), and a corresponding
basepoint ${\tilde x}_0\in\tilde{M}$, the universal cover of $M$, such
that $\pi({\tilde x}_0)=x_0$, $\pi\colon\tilde{M}\to M$ being the
covering map. If $g_0$ is the fiber of ${\frak g}_{\theta}$ over
$x_0$, then $g_0$ is also the fiber of $\pi^*{\frak g}_{\theta}$ over
${\tilde x}_0$. Since parallel transport (for the flat connection) is
independent from the path on $\tilde{M}$, we can parallel transport
the fibers of $\pi^*{\frak g}_{\theta}$ to $g_0$, and this generates a
map
$\epsilon\colon C^0(\tilde{M},\pi^*{\frak g}_{\theta})\to g_0$,
where $C^0(\tilde{M},\pi^*{\frak g}_{\theta})$ is the space of
functions  assigning to each vertex of $\tilde{M}$ an element of the
fiber of $\pi^*{\frak g}_{\theta}$ over the vertex considered. By
exploiting this map, we can define, for every $1$-cocycle $\alpha$ on
$M$ with values in ${\frak g}_{\theta}$, a function
$\phi_{\alpha}\colon\pi_1(M,x_0)\to g_0$ by setting
\begin{eqnarray}
\phi_{\alpha}(\gamma)=\epsilon(\pi^*\alpha(\tilde{\gamma}))
\end{eqnarray}
where $\gamma\in\pi_1(M,x_0)$, and $\tilde{\gamma}$ is a simplicial
lift of $\gamma$ to $\tilde{M}$ (starting at $x_0$).
Such $\phi_{\alpha}$ is a $1$-cocycle on $\pi_1(M,x_0)$ with values in
${\frak g}$, and the mapping $\alpha\to\phi_{\alpha}$ establishes the
seeked isomorphism from the cohomology group
$H^1(M,{\frak g}_{\theta})$ to the cohomology group
$H^1(\pi_1(M),{\frak g})$.

\vfill\eject

\section*{References}

\begin{description}

\bibitem[1] {Houches}
J.Ambj\o rn , {\it Quantization of Geometry}, Lect. given at
Les Houches Nato A.S.I., {\it Fluctuating Geometries in
Statistical Mechanics and field Theory}, Session LXII, 1994.\par
F.~David, {\it Simplicial Quantum Gravity and Random Lattices},
Lectures given at Les Houches Nato A.S.I. {\it Gravitation and
Quantizations}, Saclay Prep. T93/028 (1992);\par
F.David, {\it Geometry and Field Theory of Random Surfaces and
Membranes}, in {\it Statistical Mechanics of Membranes and Surfaces}
eds. D.Nelson, T.Piran, S.Weinberg (World Sci.1989);\par
P.Menotti and P.P.Peirano {\it Functional Integration on two
dimensional Regge geometries}, (hep-th/9602002) to appear on
Nucl.Phys.B; {\it Diffeomorphism invariant measure for finite
dimensional geometries}, (hep-th/9607071).
\bibitem[2] {Dav}
F.~David, {\it Planar diagrams, two-dimensional lattice gravity
and surface models}, Nucl.~Phys.~{\bf B257} (1985) 45--58.\par
B.Durhuus, J.Fr\"{o}hlich, T.J\'onsson, {\it Critical behaviour
in a model of planar random surfaces} Nucl.Phys.~{\bf B240}
(1984) 453--480.\par
R.Fernandez, J.Fr\"{o}hlich, A.Sokal, {\it Random walks, critical
phenomena, and triviality in quantum field theory},
(Springer-Verlag, Berlin  1992).\par
V.A.~Kazakov, {\it The appearance of matter fields from quantum
fluctuations
of 2D gravity}, Mod.~Phys.~Lett.~A~{\bf 4} (1989)  2125--2139.\par
C.Itzykson, J-M.Drouffe, {\it Statistical field theory: $2$},
(Cambridge University Press, Cambridge  1989).

\bibitem[3] {Williams}
R.~Williams, P.A.~Tuckey, {\it Regge calculus: a brief review and
bibliography},  Class.~Quantum Grav.~{\bf 9}  (1992) 1409--1422;\par
H.W.~Hamber, R.M.~Williams, {\it Simplicial quantum gravity in three
dimensions: analytical and numerical results},  Phys.~Rev. D~{\bf
47} (1993) 510--532.

\bibitem[4] {Regge}
T.Regge, {\it General relativity without coordinates}, Nuovo
Cim.~{\bf 19} (1961) 558--571.

\bibitem[5] {Weingarten}
D.~Weingarten, {\it Euclidean Quantum Gravity on a Lattice},
Nucl. Phys. B210 (1982), 229.

\bibitem[6] {Distler}
F. David, Mod.Phys.Lett.{\bf A 3} (1988) 1651;\par
J.Distler, H.Kaway, Nucl.Phys. {\bf B321} (1989) 509.

\bibitem[7] {Carfora}
C.Bartocci, U.Bruzzo, M.Carfora, A.Marzuoli, {\it Entropy of random
coverings and $4$-D quantum gravity}, J. of Geom. and Phys. {\bf 18}
(1996), 247-294.\par
M.~Carfora, A.~Marzuoli, {\it Entropy estimates for simplicial
quantum gravity}, J. of Geom. and Phys. {\bf 16} (1995), 99-119.\par
M.Carfora, A.Marzuoli, {\it Holonomy and entropy estimates for
dynamically triangulated manifolds}, J. Math.Phys. {\bf 36} (1995)
6353-6376.

\bibitem[8] {Pari}
D.~Bessis, C.~Itzykson, J.B.~Zuber, {\it Quantum field theory
techniques in graphical enumeration}, Adv.~Appl.~Math.~{\bf 1}
 (1980) 109--157;\par
E.~Br\'ezin, C.~Itzykson, G.Parisi, J.B.~Zuber, {\it Planar diagrams}
Commun.~Math.~Phys. {\bf 59}  (1978) 35--51;\par
D.~Bessis, {\it A new method in the combinatorics of the
topological expansion}, Commun.~Math.~Phys.~{\bf 69} (1979) 147--
163;\par
W.J.~Tutte, {\it A census of planar triangulations},
Canad.~J.~Math.~{\bf
14} (1962) 21--38.\par
J.Ambj\o rn, J.Jurkiewicz, {\it On the exponential bound in four
dimensional simplicial gravity}, Preprint NBI-HE-94-29 (1994).\par
D.V.Boulatov, {\it On entropy of $3$-dimensional simplicial
complexes}, Preprint NBI-HE-94-37 (1994).

\bibitem[9] {Gromov}
M.~Gromov, {\it Structures m\'etriques pour les vari\'et\'es
Riemanniennes} (Conception Edition Diffusion Information
Communication Nathan, Paris 1981);
see also
S.~Gallot, D.~Hulin, J.~Lafontaine, {\it Riemannian Geometry}
(Springer Verlag, New York,1987). A particularly clear account
of the results connected with Gromov-Hausdorff convergence of
riemannian manifolds is provided by the paper of
K.~Fukaya, {\it Hausdorff convergence of riemannian manifolds and its
applications}, in
{\it Recent topics in differential and analytic geometry},
Adv.~Studies  Pure Math.~{\bf 18-I} (1990) 143--238.

\bibitem[10] {Grov}
K.~Grove, P.V.~Petersen, {\it Bounding homotopy types by
geometry},  Ann.~Math.~{\bf 128} (1988) 195--206.\par
K.~Grove, P.V.~Petersen, J.Y.~Wu, {\it Controlled
topology in geometry},  Bull.~Am. Math.~Soc.~{\bf 20}, (1989) 181--
183; {\it
Geometric finiteness theorems via controlled topology},
Invent.~Math.~{\bf 99} (1990)   205--213; erratum, Invent.~Math.~{\bf
104}(1991)   221--222.\par
K.~Grove, P.V.~Petersen, {\it Manifolds near the boundary of
existence}, J.Diff.Geom. {\bf 33}, (1991) 379-394.\par
T. Yamaguchi, {\it Homotopy type finiteness theorems for certain
precompact families of riemannian manifolds}, Proc.Amer.Math.Soc.

\bibitem[11] {Jurk}
J.Ambj\o rn, J.Jurkiewicz, {\it Scaling in four-dimensional quantum
gravity}, Nucl.Phys. {\bf B 451}  (1995) 643-676.

\bibitem[12] {Moise}
E. Moise, {\it Affine structures in $3$-manifolds I,II,III,IV,V},
Ann.of Math. {\bf 54}(1951), 506-533; {\bf 55}(1952), 172-176;
{\bf 55}(1952), 203-214; {\bf 55}(1952), 215-222; {\bf 56}(1952), 96-
114.

\bibitem[13] {Ferry}
S.C.Ferry, {\it Finiteness theorems for manifolds in Gromov-Hausdorff
space}, Preprint SUNY at Binghamton (1993);\par
S.C.Ferry, {\it Counting simple homotopy types in Gromov-Hausdorff
space}, Preprint SUNY at Binghamton (1991).

\bibitem[14] {Freed}
D.S. Freed, K.K. Uhlenbeck, {\it Instantons and Four-Manifolds},
2nd.ed., (Springer-Verlag, New York, 1991).\par
M. H. Freedman, F.Luo, {\it Selected applications of geometry to low-
dimensional topology}, Univ.Lect.Series {\bf 1}, Amer.Math.Soc.
(1989).\par
S.Donaldson, {\it An application of gauge theory to the topology of
$4$-manifolds}, J.of Diff.Geom. {\bf 18} (1983), 269-316.
{\it Connections, cohomology, and the intersection forms of $4$-
manifolds} Ibidem, {\bf 24} (1986), 275.\par
J.Milnor, D.Husemoller, {\it Symmetric Bilinear Forms}, (Springer-
Verlag, Berlin, 1973).

\bibitem[15] {Frohlich}
J.Fr\"{o}hlich,
{\it Regge calculus and discretized gravitational functional
integrals}, Preprint IHES (1981), reprinted in {\it Non-perturbative
quantum field theory --- mathematical aspects and applications},
Selected Papers of J.Fr\"{o}hlich (World Sci. Singapore 1992).\par
H. R\"{o}mer, M. Z\"{a}hringer, {\it Functional integration and the
diffeomorphism group in Euclidean lattice quantum gravity}, Class.
Quantum Grav. {\bf 3} (1986) 897-910.\par
W. K\"{u}hnel, {\it Triangulations of manifolds with few vertices}, in
{\it Advances in differential geometry and topology}, Eds. I.S.I.- F.
Tricerri, (World Scientific, Singapore, 1990).

\bibitem[16] {Rourke}
C.P. Rourke, B.J. Sanderson, {\it Introduction to Piecewise-Linear
Topology}, (Springer-Verlag, New York 1982).\par
W.P.Thurston, {\it Three-dimensional Geometry and Topology},  Lcture
Notes (December 1991 Version), Math. Sci. Research Inst.
(Berkeley).\par
R.Kirby, L.Siebenmann, {\it Foundational Essays on Topological
manifolds, Smoothings, and Triangulations}, Annals of Math. Studies
{\bf 88}, (Princeton Univ.Press, Princeton, N.J., 1977).

\bibitem[17]{Stanley}
R. Stanley, {\it The
number of faces of a simplicial convex polytope}, Advances in Math.
{\bf 35} (1980), 236-238; {\it Subdivision and local $h$-vectors}, J.
Amer. Math. Soc. {\bf 5} (1992), 805-851.

\bibitem[18]{McMullen}
P.McMullen, {\it The number of faces of simplicial polytopes}, Israel
Journ. Math. {\bf 9}, (1971), 559-570

\bibitem[19] {Billera}
L.J. Billera, C.W. Lee, {\it Sufficiency of
McMullen's conditions for f-vectors of simplicial polytopes}, Bull.
Amer. Math. Soc. {\bf 2}, (1980), 181-185;
{\it A proof of the sufficiency of McMullen's conditions for f-vectors
of symplicial convex polytopes}, J. Combin. Theory {\bf A31} (1981),
237-255

\bibitem[20] {Smullen}
P. McMullen, {\it On simple polytopes}, Inventiones Math.
{\bf 113} (1993), 419-444

\bibitem[21] {Sphere}
R. Stanley, {\it The number of faces of symplicial
politopes and spheres}, Discrete Geometry and convexity, 212-223, Ann.
NY Acad. Sci., New York 1985.

\bibitem[22] {Ruth}
H. Hamber, R. Williams, Nucl.Phys. {\bf B248} (1984) 392.

\bibitem[23] {Petersen}
P. Petersen V, {Gromov-Hausdorff convergence of metric spaces},
Proc. Symp. in Pure Math., 1990 Summer Institute on Differential
Geometry, {\bf 54}, Part 3, 489-504 (1993).

\bibitem[24] {Cohen}
M.M. Cohen, {\it A course in simple homotopy theory}, GTM 10,
(Springer Verlag, New York, 1973).

\bibitem[25] {Scott}
G.P. Scott, {\it The geometries of $3$-manifolds}, Bull.
Lond.Math.Soc. {\bf 15} (1983), 401-487

\bibitem[26] {Goldman}
W.~Goldman, {\it Geometric structures on manifolds and varieties
of representations},  in
AMS-IMS-SIAM Joint Summer Research Conference {\it Geometry of group
representations}, W.~Goldman, A.R.~Magid eds.,
 Contemp.~Math.~{\bf 74} (1988) 169--198.\par
W.M.~Goldman, J.J.~Millson
{\it Deformations of flat bundles over K\"{a}hler manifolds},
in {\it Geometry and Topology},   C.~McCrory, T.~Shifrin eds.,
Lect.~Notes Pure  Appl.~Math.~{\bf 105} (M.~Dekker, New York 1987,
pp.~129--145).\par
N.J.~Hitchin, {\it Lie groups and Teichm\"muller space},  Topology
{\bf 31}
(1992) 449--473.\par
K.~Morrison, {\it Connected components of representation varieties}
in AMS-IMS-SIAM Joint Summer Research Conference {\it Geometry of
group
representations}, W.~Goldman, A.R.~Magid eds.,
Contemp.~Math.~{\bf 74} (1988) 255--269.\par
K.~Walker, {\it An extension of Casson's invariant} (Princeton
Univ.~Press, Princeton 1992).

\bibitem[27] {Munkres}
J.R. Munkres, {\it Elementary Differential Topology}, Princeton
Univ.Press (1966)

\bibitem[28] {Koba}
S. Kobayashi, {\it Differential geometry of complex vector bundles},
(Princeton Univ. Press, 1987).

\bibitem[29] {Witten}
E.~Witten, {\it Two-dimensional gravity and intersection theory on
moduli space},  Surveys in Diff.~Geom.~{\bf 1} (1991)  243--310
(Lehigh
University, Bethlehem PA).\par
E.~Witten, {\it Two-dimensional gauge theories revisited},
J.~Geom.~Phys.
{\bf 9} (1992)  303--368.\par
L.C.~Jeffrey, J.~Weitsman, {\it Half density quantization of
the moduli space of flat connections and Witten's semiclassical
manifold invariants}, Preprint IASSNS-HEP-91/94.\par
M.~Kontsevitch, {\it Intersection theory on the moduli
space of curves and the matrix Airy functions},
Commun.~Math.~Phys.~{\bf
147} (1992) 1--23.

\bibitem[30] {Forman}
R. Forman {\it Small Volume limit of $2$-d Yang-Mills},
Commun.Math.Phys. {\bf 151} (1993), 39-52

\bibitem[31] {Davies}
E.B. Davies,  {\it Heat Kernels and spectral theory}, Cambridge Tracts
in Math. {\bf 92}, Cambridge Univ. Press (1989).

\bibitem[32] {Millson}
D. Johnson, J.J. Millson, {\it Deformation spaces associated to
compact hyperbolic manifolds} in {\it Discrete Groups in Geometry and
Analysis} Ed. R. Howe, (Birkh\"{a}user, Boston, 1987).

\bibitem[33] {Bergery}
L. Bernard Bergery, J.P. Bourguignon, J. Lafontaine, {\it
Deformationes localement triviales des varietes riemanniennes},
Proc. Symposia in Pure Math. {\bf 27}, (1975) 3-32.

\bibitem[34] {King}
C. King, A. Sengupta, {\it An explicit description of the symplectic
structure of moduli spaces of flat connection},
J. Math.Phys. {\bf 35} (1994), 5338-5353; {\it The semiclassical limit
of the two-dimensional Yang-Mills model}, Ibidem, 5354-5361.

\bibitem[35] {Besse}
A. Besse, {\it Einstein Manifolds}, (Springer-Verlag, New York 1986).

\bibitem[36] {Anandan}
J. Anandan, {\it Holonomy groups in gravity and gauge fields},
in {\it Conference on Differential Geometric Methods in Theoretical
Physics}, eds. G.DeNardo and H.D.Doebner, (World Sci. Singapore 1983).

\bibitem[37] {Andrews}
G. E. Andrews, {\it The theory of partitions}, Encyclopedia of Math.
and its Applic. {\bf Vol.2}, (1976) Addison-Wesley Pub.Co.

\bibitem[38] {Migdal}
M.E. Agishtein, A.A. Migdal, {\it Simulations of four-dimensional
simplicial quantum gravity as dynamical trinagulation},
Mod.Phys.Lett.A {\bf 7} (1992), 1039-1061.\par
U. Pachner, {\it Konstruktionsmethoden und das kombinatorische
Hom\"{o}omorphieproblem f\"{u}r Triangulationen kompakter semilinearer
Mannigfaltigkeiten}, Abh. Math. Sem.Univ. Hamburg {\bf 57} (1986)
69.\par
M. Gross, D.Varsted, {\it Elemtary moves and ergodicity in $D$-
dimensional quantum gravity}, Nucl.Phys. {\bf B378} (1992) 367.

\bibitem[39] {Walters}
P. Walters, {\it Ergodic theory-Introductory Lectures}, Lect.Notes in
Math. {\bf 458}, Springer-Verlag (1975).

\bibitem[40] {Hall}
M.Hall, {\it Combinatorial theory}, 2nd.ed. J.Wiley, (1986).

\bibitem[41] {Andrews2}
G. E. Andrews, {\it $q$-Series: Their development and applications in
analysis, number theory, combinatorics, Physics, and computer
algebra}, Regional Conferences in Math. CBMS, Vol. {\bf 66}, The Am.
Math. Soc. (1986).

\bibitem[42] {Ben}
A. Nabutowsky, R.Ben-Av, Commun.Math.Phys. {\bf 157} (1993) 93.

\bibitem[43] {Varsted}
J.Ambj\o rn, S.Varsted, {\it Three-dimensional simplicial quantum
gravity}, Nucl.Phys. {\bf B 373} (1992) 557-577.\par

\bibitem[44] {ergo}
J.Ambj\o rn, J. Jurkiewicz, {\it Computational ergodicity of ${\Bbb
S}^4$}, Phys.Lett. {\bf B345} (1995) 435.

\bibitem[45] {Bakker}
B.V. de Bakker, {\it Simplicial Quantum Gravity }, Academisch
Proefschrift, (Amsterdam, 1995), hep-lat/9508006.

\bibitem[46] {Kristj}
J. Ambj\o rn, J. Jurkiewicz, C.F. Kristjansen, {\it Quantum gravity,
dynamical triangulations and higher-derivative regularization},
Nucl.Phys.{\bf B393} (1993) 601-629.

\bibitem[47] {Eilenberg}
S.P. Eilenberg, S. MacLane, {\it Cohomology theory in abstract groups
I}, Annals of Math., {\bf 48} (1947), 51-78.

\bibitem[48] {Bleistein}
N. Bleistein, {\it Uniform asymptotic expansions of integrals with
stationary point near algebraic singularity}, Commun. On Pure and
Applied Math. {\bf 29} (1966), 353-370. See also,
L. Sirovich {\it Techniques of asymptotic analysis}, Springer-Verlag
(1971).

\bibitem[49] {Abramowitz}
M. Abramowitz, I.A.Stegun, {\it Handbook of mathematical functions},
Dover (1972).

\bibitem[50] {Ballmann}
W.Ballmann, {\it Lectures on Spaces of Nonpositive Curvature},
DMV Seminar Band {\bf 25}, (Birkh\"{a}user Verlag, Basel 1995).\par
E.Ghys, P.de la Harpe Eds., {\it Sur les groupes hyperboliques d'apres
Mikhael Gromov}, Progress in Math. {\bf 83},
(Birkh\"{a}user Verlag, Boston 1990).\par
T.Januszkiewicz, {\it Hyperbolizations}, preprint CPT-90/P.2391
(Marseille, 1990).\par
I.Chavel, {\it Riemannian geometry-A modern introduction}, Cambridge
Tracts in Math. {\bf 108}, (Cambridge Univ. Press, 1993).

\bibitem[51] {Alexandrov}
A.D.Alexandrov, V.N.Berestovski,I.G.Nikolaev, {\it Generalized
Riemannian Spaces}, Russian Math. Surveys {\bf 41:3} (1986) 1-54

\bibitem[52] {Wolf}
J.A.Wolf, {\it Spaces of constant curvature}, 5th ed., (Publish or
Perish, Wilmington, 1984).

\bibitem[53] {Gromovpansu}
M.Gromov, P.Pansu, {\it Rigidity of lattices: An introduction}, in
{\it Geometric topology: recent developments}, eds. P. de Bartolomeis,
F. Tricerri,  Lect. Notes in Math. {\bf 1504} (Fondazione C.I.M.E.,
Springer Verlag, 1991).

\bibitem[54] {Rong}
X. Rong, {\it Bounding homotopy and homology groups by curvature and
diameter}, Duke Math. Journ. {\bf 78} (1995) 427-435.

\bibitem[55] {Walkup}
D. Walkup, {\it The lower bound conjecture for $3$- and $4$-
manifolds}, Acta Math. {\bf 125} (1970) 75-107.

\bibitem[*]{dar}E-mail: carfora@pv.infn.it, carfora@gandalf.sissa.it
\bibitem[\dag]{mac}E-mail: marzuoli@pv.infn.it
\bibitem[**]{nbi}E-mail: ambjorn@nbivax.nbi.dk

\end{description}

\end{document}